\newif\ifoup \ouptrue
\oupfalse

\ifoup
\documentclass[webpdf,modern,large]{oup-authoring-template}%
\else
\documentclass[pdflatex,a4paper,DIV=15,11pt]{scrartcl}
\fi

\usepackage{graphicx}
\usepackage{amsmath,amssymb}
\allowdisplaybreaks
\usepackage{mathrsfs}
\usepackage{mleftright}
\usepackage{textcomp}
\usepackage{url}

\unless\ifoup
\usepackage{caption}
\usepackage{subcaption}
\fi

\DeclareEmphSequence{\itshape, \normalfont\sffamily\bfseries, \itshape} 

\usepackage{tikz}
\usetikzlibrary{positioning}
\usetikzlibrary{tikzmark,calc,topaths}
\usetikzlibrary{graphs, trees}

\usepackage[whole]{bxcjkjatype} 

\usepackage[hidelinks]{hyperref}
\hypersetup{
colorlinks=true,
citecolor=red,
linkcolor=blue,
urlcolor=blue,
}

\usepackage{amsthm}
\usepackage[nameinlink, noabbrev]{cleveref}

\crefname{section}{Section}{Sections}
\crefname{subsection}{Section}{Sections}
\crefname{appendix}{Appendix}{Appendices}
\ifoup\crefname{figure}{Fig.}{Figs.}\else\crefname{figure}{Figure}{Figures}\fi
\crefname{table}{Table}{Tables}
\crefname{equation}{}{}
\creflabelformat{equation}{#2(#1)#3}

\unless\ifoup
\usepackage{enumitem}
\setlist[description]{%
  font={\normalfont},
}
\fi
\renewcommand*{\theenumi}{\textup{(}\arabic{enumi}\textup{)}}
\renewcommand*{\labelenumi}{\textup{(}\arabic{enumi}\textup{)}}

\crefname{enumi}{}{}
\crefname{enumii}{}{}

\ifoup \theoremstyle{thmstyleone}
\else \theoremstyle{plain}
\fi 
\newtheorem{theorem}{Theorem}[section]
\newtheorem{lemma}[theorem]{Lemma}
\newtheorem{proposition}[theorem]{Proposition}
\newtheorem{corollary}[theorem]{Corollary}

\ifoup \theoremstyle{thmstylethree}%
\else \theoremstyle{definition}
\fi
\newtheorem{definition}[theorem]{Definition}

\ifoup \theoremstyle{thmstyletwo}%
\else \theoremstyle{remark}
\fi

\ifoup
\else

\fi

\crefname{theorem}{Theorem}{Theorems}
\crefname{lemma}{Lemma}{Lemmata}
\crefname{proposition}{Proposition}{Propositions}
\crefname{corollary}{Corollary}{Corollaries}

\crefname{definition}{Definition}{Definitions}
\crefname{example}{Example}{Examples}
\crefname{question}{Question}{Questions}

\usepackage{bussproofs, bussproofs-extra}
\newenvironment*{inlineprooftree}{\medskip}{\DisplayProof}%
\renewcommand{\fCenter}{\Rightarrow} 

\usepackage[l2tabu, orthodox]{nag}

\newcommand{\relmiddle}[1]{\mathrel{}\middle#1\mathrel{}}
\newcommand{\mathsetintension}[2]{%
\mleft\{ #1 \relmiddle| #2 \mright\}%
}
\newcommand{\mathsetextension}[1]{%
\mleft\{ #1 \mright\}%
}

\newcommand{\mathof}[2]{\mathop{#1}\mleft(#2\mright)}

\newcommand{\mathtuple}[1]{\mleft( #1 \mright)}
\newcommand{\mathsequence}[2]{\mleft\{ #1 \mright\}_{#2}}

\newcommand{\mathcardinality}[1]{\mathof{\#}{#1}}

\newcommand{\mathnat}{\mathbb{N}}
\newcommand{\mathnatpos}{\mathbb{N}_{> 0}}

\newcommand{\mathtree}[1]{\mathcal{#1}}

\newcommand{\mathcoloneqq}{\mathrel{::=}}

\newcommand{\mathlengthsy}{\mathop{\mathrm{len}}}
\newcommand{\mathlength}[1]{\mathof{\mathlengthsy}{#1}}

\newcommand{\rulename}[1]{(\textsc{#1})}

\newcommand{\maththeunderlyingset}[1]{\mleft|#1\mright|}

\newcommand{\mathseqtoform}[1]{\mleft\langle\!\mleft\lvert #1 \mright\rvert\!\mright\rangle}

\newcommand{\biglor}{\bigvee}
\newcommand{\bigland}{\bigwedge}

\newcommand{\mathiffsy}{\mathbin{\leftrightarrow}}

\newcommand{\PDL}{\ensuremath{\mathtt{PDL}}}
\newcommand{\GPDL}{\ensuremath{\mathtt{GPDL}}}

\newcommand{\TPDL}{\ensuremath{\mathtt{TPDL}}}

\newcommand{\GTPDL}{\ensuremath{\mathtt{GTPDL}}}

\newcommand{\CGTPDL}{\ensuremath{\mathtt{CGTPDL}}}
\newcommand{\CGPDL}{\ensuremath{\mathtt{CGPDL}}}

\newcommand{\LPD}{\ensuremath{\mathtt{LPD}}}

\newcommand{\DochertyInfinitarySystem}{\ensuremath{\mathbf{G3PDL}^{\infty}}}
\newcommand{\DochertyCyclicSystem}{\ensuremath{\mathbf{G3PDL}^{\omega}}}

\newcommand{\DasCyclicSystem}{cyclic-\ensuremath{\mathbf{LPD}}}

\newcommand{\LeivantSystem}{\ensuremath{\mathbf{D}}}

\newcommand{\maththesetofpropval}{\ensuremath{\mathsf{Prop}}}
\newcommand{\maththesetofatomicprog}{\ensuremath{\mathsf{AtProg}}}

\newcommand{\mathdynbox}[1]{\mleft[ #1 \mright]}
\newcommand{\mathdyndia}[1]{\mleft\langle #1 \mright\rangle}

\newcommand{\mathrevbox}[1]{\mleft[ #1 \mright]^{-1}}
\newcommand{\mathrevdia}[1]{\mleft\langle #1 \mright\rangle^{-1}}

\newcommand{\mathFLcl}[1]{\mathop{\mathrm{FL}}\mleft( #1 \mright)}
\newcommand{\mathFLboxcl}[1]{\mathop{\mathrm{FL}^{\mathdynbox{\;}}}\mleft( #1 \mright)}

\newcommand{\mathsubFLS}[1]{\mathop{\mathrm{Sub}}\mleft( #1 \mright)}
\newcommand{\mathreducible}[1]{\mathop{\mathrm{Red}}\mleft( #1 \mright)}

\newcommand{\htriple}[3]{\mleft\{#1\mright\}\,#2\,\mleft\{#3\mright\}}
\newcommand{\ihtriple}[3]{\mleft[#1\mright]\,#2\,\mleft[#3\mright]}
\newcommand{\tihtriple}[3]{\mleft[\mleft[#1\mright]\,#2\,\mleft[#3\mright]\mright]}
\newcommand{\thtriple}[3]{\mleft\{ \mleft\{ #1 \mright\}\,#2\, \mleft\{ #3 \mright\} \mright\}}

\newcommand{\triple}[3]{\mleft\langle\!\mleft\langle#1\mright\rangle\!\mright\rangle\,#2\,\mleft\langle\!\mleft\langle#3\mright\rangle\!\mright\rangle}

\newcommand{\mathcompanion}[1]{\mathop{\mathcal{#1}}}

\newcommand{\mathproofgraph}[1]{\mathop{\mathcal{G}}{#1}}

\newcommand{\mathdertree}[1]{\mathcal{#1}}

\newcommand{\mathcountitsy}{\mathrm{IT}}
\newcommand{\mathcountit}[3]{\mathof{\mathcountitsy}{#1, #2, #3}}

\newcommand{\mathcountitexsy}{\mathrm{IT}_{\mathrm{Ex}}}
\newcommand{\mathcountitex}[1]{\mathof{\mathcountitexsy}{#1}}

\newcommand{\mathcounterexsy}{\mathrm{Ex}}
\newcommand{\mathcounterex}[3]{\mathof{\mathcounterexsy_{#1}}{#2, #3}}

\newcommand{\mathiterationsy}{\mathrm{IT}}
\newcommand{\mathiteration}[3]{\mathof{\mathiterationsy_{#1}}{#2, #3}}

\newcommand{\mathmincounterexsy}{\mathrm{minEx}}
\newcommand{\mathmincounterex}[3]{\mathof{\mathmincounterexsy_{#1}}{#2, #3}}

\newcommand{\GameName}[1]{\ensuremath{\mathof{\mathfrak{G}}{#1}}}
\newcommand{\GameSize}{\ensuremath{\mathbf{N}_{\text{size}}}}
\newcommand{\Prover}{\ensuremath{\mathbf{P}}}
\newcommand{\Refuter}{\ensuremath{\mathbf{R}}}
\newcommand{\player}[1]{\ensuremath{\mathcal{#1}}}
\newcommand{\opponent}[1]{\ensuremath{\overline{#1}}}
\newcommand{\mathmaxform}[1]{\ensuremath{\mathof{\max}{#1}}}
\newcommand{\mathwinningstrategy}[1]{\mathscr{#1}}

\newcommand{\mathworld}[1]{\mathscr{#1}}
\newcommand{\mathworldfirst}[1]{\mathof{\mathrm{Fir}}{#1}}
\newcommand{\mathworldcore}[1]{\mathof{\mathrm{Core}}{#1}}
\newcommand{\mathancestorofcore}{\lhd}

\newcommand{\mathSeqsy}{\mathrm{Seq}}
\newcommand{\mathSeq}[1]{\mathop{\mathSeqsy}\mleft( #1 \mright)}

\newcommand{\mathAntsy}{\mathrm{Ant}}
\newcommand{\mathAnt}[1]{\mathop{\mathAntsy}\mleft( #1 \mright)}
\newcommand{\mathConsy}{\mathrm{Con}}
\newcommand{\mathCon}[1]{\mathop{\mathConsy}\mleft( #1 \mright)}

\newcommand{\mathTrsy}{\mathrm{Tr}}
\newcommand{\mathTr}[1]{\mathop{\mathTrsy}\mleft( #1 \mright)}
\newcommand{\maththesetofTr}{\mathscr{T}}

\newcommand{\mathUSsy}{\mathrm{US}}
\newcommand{\mathUS}[1]{\mathop{\mathUSsy}\mleft( #1 \mright)}
\newcommand{\maththesetofUS}{\mathfrak{U}}

\newcommand{\mathfreshnumsy}{\mathrm{fr}}
\newcommand{\mathfreshnum}[1]{\mathop{\mathfreshnumsy}\mleft( #1 \mright)}

\newcommand{\mathCycle}[1]{\mathop{\mathrm{Cycle}}\mleft( #1 \mright)}

\newcommand{\mathstgraph}[1]{\mathof{\mathcal{G}}{#1}}

\newcommand{\mathprogramitsy}{\ast}
\newcommand{\mathprogramtestsy}{?}

\newcommand{\mathprogramseq}[2]{#1;#2}
\newcommand{\mathprogramndc}[2]{#1\cup #2}
\newcommand{\mathprogramit}[1]{\mleft(#1\mright)^{\mathprogramitsy}}
\newcommand{\mathprogramtest}[1]{\mleft(#1\mright)\mathprogramtestsy}

\newcommand{\mathreduction}[1]{\xrightarrow{#1}}

\newcommand{\mathannotation}[1]{\mathrm{#1}}
\newcommand{\mathannotated}[2]{{#1}^{(\mathannotation{#2})}}

\newcommand{\mathrelbar}{\mathrel{|}}

\unless\ifoup
\newcommand{\keywords}[1]{\par \noindent \textbf{Keywords:} \emph{#1} \par}
\fi

\ifoup
\onecolumn
\fi

\usepackage{docmute}
\begin{document}

\ifoup
\title[A study of cut-elimination in a non-labelled cyclic proof system for propositional dynamic logics]{A study of cut-elimination in a non-labelled cyclic proof system for propositional dynamic logics}

\author[1,$\ast$]{Yukihiro Oda}

\address[1]{\orgdiv{Graduate School of Information Sciences}, \orgname{Tohoku University}, \orgaddress{\street{Aoba, Aramaki-aza Aoba-ku}, Sendai, \postcode{980-8579}, \state{Miyagi Prefecture}, \country{Japan}}}

\corresp[$\ast$]{Corresponding author. yukihiro.oda.e5 [at] tohoku.ac.jp}

\journaltitle{Journal of Logic and Computation}
\DOI{ }
\copyrightyear{2026}
\pubyear{ }
\access{ }
\appnotes{ }

\firstpage{1}

\abstract{
\unless\ifoup
\begin{abstract}
\fi
Dynamic logic is a modal logic for reasoning about programs. 
A cyclic proof system is a proof system that allows proofs containing cycles
and is an alternative to a proof system containing (co-)induction. 
This paper introduces a sequent calculus and a non-labelled cyclic proof system for 
an extension of propositional dynamic logic obtained by adding backwards modal operators. 
We prove the soundness and completeness of these systems and 
show that cut-elimination fails in both.
Moreover, we show the cut-elimination property of the cyclic proof system for 
propositional dynamic logic obtained by restricting ours. 


\unless\ifoup
\end{abstract}
\fi

 }

\else
\title{A study of cut-elimination in a non-labelled cyclic proof system for propositional dynamic logics}
\author{Yukihiro Oda\thanks{Graduate School of Information Sciences, Tohoku University. Mail: yukihiro3socrates6hilbert [at] gmail.com}}
\date{\today}
\fi

\maketitle

\unless\ifoup
\unless\ifoup
\begin{abstract}
\fi
Dynamic logic is a modal logic for reasoning about programs. 
A cyclic proof system is a proof system that allows proofs containing cycles
and is an alternative to a proof system containing (co-)induction. 
This paper introduces a sequent calculus and a non-labelled cyclic proof system for 
an extension of propositional dynamic logic obtained by adding backwards modal operators. 
We prove the soundness and completeness of these systems and 
show that cut-elimination fails in both.
Moreover, we show the cut-elimination property of the cyclic proof system for 
propositional dynamic logic obtained by restricting ours. 


\unless\ifoup
\end{abstract}
\fi

\fi

\section{Introduction}
\label{sec:intro}

\emph{Dynamic logic} is a modal logic for reasoning about programs
\cite{Pratt1976, Fischer1977, Harel1979, Kozen1981, Harel2001, Renardel2008}.
In the logic, there is a modal operator $\mathdynbox{\pi}$ for each program $\pi$.
Informally, a formula $\mathdynbox{\pi}\varphi$ means as follows:
\begin{quote}
$\varphi$ holds for all final states that $\pi$ can reach and terminate in 
from the current state.
\end{quote}
Usually, we define its dual $\mathdyndia{\pi}\varphi$ as $\lnot\mathdynbox{\pi}\lnot\varphi$, where $\lnot$ is negation.
Then, $\mathdyndia{\pi}\varphi$ means the following:
\begin{quote}
 There exists an execution of $\pi$ from the current state to a state $w$ such that 
 $\varphi$ holds in $w$.
\end{quote}
\emph{Propositional dynamic logic} (\PDL) is a propositional variant of dynamic logic
\cite{Pratt1976}.

S. Chopra and D. Zhang \cite{Chopra2001} introduce \TPDL \footnote{
In \cite{Chopra2001}, it is not written what \TPDL\ stands for.
It might be an acronym for \emph{Temporal Propositional Dynamic Logic}
because there are many references to temporal logic in \cite{Chopra2001}. 
However, a different system from \TPDL\ is called
``Temporal Dynamic Logic''\cite{Platzer2007}. 
In this article, 
we intend that \TPDL\ stands for \emph{Two-directed Propositional Dynamic Logic}. 
}, 
an extension of \PDL, obtained by adding a backwards modal operator $\mathrevbox{\pi}$.
A formula $\mathrevbox{\pi}\varphi$ expresses the following:
\begin{quote}
For every state $w$,
if $\pi$ can reach and terminate in the current state from $w$, 
then $\varphi$ holds in $w$.
\end{quote}
In a similar way to \PDL,
we define its dual $\mathrevdia{\pi}\varphi$ as $\lnot\mathrevbox{\pi}\lnot\varphi$.
Then, $\mathrevdia{\pi}\varphi$ indicates the following:
\begin{quote}
 There exists an execution of $\pi$ from some state $w$ to the current state such that 
 $\varphi$ holds in $w$.
\end{quote}
As we will see later,
these modal operators are related to \emph{reverse Hoare logic} \cite{DeVries2011} 
(or \emph{incorrectness logic} \cite{OHearn2019}) and 
\emph{partial incorrectness logic} \cite{Zhang2022}.
Although \TPDL\ appears similar to \emph{converse \PDL} \cite{Parikh1978, Fischer1979, Gore2010}, 
an expansion of \PDL\ obtained by adding \emph{converse programs},
\TPDL\ is not the same because \TPDL\ has no converse programs.

A \emph{cyclic proof system}, also known as a \emph{circular proof system}\footnote{
A.\ Simpson \cite{Simpson2017} wrote as follows:
\begin{quote}
We prefer cyclic for two reasons: proofs are given as cyclic graphs; and the phrase 
``circular proof'' is uncomfortably close to ``circular argument'', 
which means an argument that goes round in a circle without establishing anything.
\end{quote}
Taking this into account, this article mainly uses ``cyclic'' proofs.
}, is a type of proof system that allows proofs containing cycles \cite{Brotherston2011}. 
This type of system is a variation of a \emph{non-well-founded} (\emph{ill-founded}) 
\emph{proof system} that permits proofs containing infinite paths.
It is typically used as an alternative to a proof system 
that includes induction or coinduction. In cyclic proofs, 
each cycle represents an iteration of the induction process. 
From this perspective, cyclic proofs are said to be a formalisation of (Fermat's) 
\emph{infinite descent}.

Since there is the induction axiom for iteration programs in \PDL, 
it is reasonable to seek a cyclic proof system for it. 
Indeed, there are some cyclic proof systems for \PDL\ and its variants
\cite{Lange2003,Docherty2019,Das2023,Zhang2024}.

For a proof system,
if any provable sequent is provable without the cut-rule,
then we say that the proof system has the \emph{cut-elimination property}.
It is an attractive property. For example, 
the cut-elimination theorem of first-order logic immediately implies 
some important properties:
consistency, the subformula property, and the interpolation theorem
\cite{Buss1998}.

\emph{In this paper, we present a non-labelled cyclic proof system for \TPDL\ and
show that cut-elimination fails in it}.
We also show \emph{the cut-elimination theorem for the cyclic proof system of \PDL\
obtained by restricting ours}.

\subsection{Our contribution}
Our contribution is below:
\begin{itemize}
 \item We define a non-labelled cyclic proof system \CGTPDL\ for \TPDL.
 \item We also define a (non-labelled) sequent calculus \GTPDL\ for \TPDL.
 \item We prove the soundness and completeness of these two systems.
 \item We provide counterexamples to cut-elimination in both.
 \item We show the cut-elimination theorem of a non-labelled cyclic proof system \CGPDL\ 
       for \PDL\ obtained by restricting \CGTPDL.
 \item To show the cut-elimination theorem of \CGPDL,
       we prove its cut-free completeness by considering a finite, \emph{not infinite}, game.
\end{itemize}

There are three reasons why we focus on \TPDL.

Firstly, as we will see in \cref{subsec:related-work}, 
modal operators in \TPDL\ are related to four important \emph{Hoare-style logics}: 
(\emph{partial}) \emph{Hoare logic} \cite{Hoare1969}, \emph{total Hoare logic} 
\cite{Manna1974,Zhang2022},
\emph{reverse Hoare logic} (\emph{incorrectness logic}) \cite{DeVries2011, OHearn2019}, 
and \emph{partial incorrectness logic} \cite{Zhang2022}. 
Since reverse Hoare logic has been thoroughly investigated these days, 
\TPDL\ has become more important than before.

Secondly, \TPDL\ is more intuitive than converse \PDL\ \cite{Parikh1978, Fischer1979}, 
which is another expansion of \PDL\ used for backwards reasoning. 
Converse \PDL\ is obtained by adding converse programs to \PDL. 
The \emph{converse program} of a program $\pi$, written as $\pi^{-}$, is a program 
that runs the original program $\pi$ in reverse \cite{Fischer1979}. 
However, the converse of a program is sometimes intuitively unclear. 
For example, if $\pi$ is a "Hello, World!" program,
it is unclear what $\pi^{-}$ would be.
Such unclear programs do not exist in \TPDL. 
Nevertheless, \TPDL\ can be used for backwards reasoning 
because there is a backwards modal operator $\mathrevbox{\pi}$, 
whose meaning is intuitively clear.

Thirdly, as in this paper, it is possible to achieve a similar result for converse \PDL\ easily. 
We can define a non-labelled cyclic proof system for converse \PDL\ 
that is analogous to our cyclic proof system, \CGTPDL. Furthermore, 
the soundness, completeness and failure of cut-elimination can be proven 
in a manner analogous to that of \CGTPDL.

Our (non-labelled) sequent calculus, \GTPDL, is based on the sequent calculus \GPDL\ for 
converse \PDL\ without \emph{test programs} $\mathprogramtest{\varphi}$ 
in \cite{Nishimura1979}. 
\GTPDL\ is obtained by removing the rules for converse programs from \GPDL\ and 
adding rules for test programs and the backwards modal operator $\mathrevbox{\pi}$.

The completeness of \GTPDL\ is proven by constructing a counter model for 
an unprovable sequent. 
As a by-product of our proof of the completeness of \GTPDL, 
the finite model property of \TPDL\ is obtained.

Our non-labelled cyclic proof system, \CGTPDL, is obtained by removing one of the rules 
for \emph{iteration programs} $\mathprogramit{\pi}$ from \GTPDL\ and 
adding the \emph{case-split} rule for formulae of the form 
$\mathdynbox{\mathprogramit{\pi}}\varphi$, 
which splits into the base case $\varphi$ and 
the step case $\mathdynbox{\pi}\mathdynbox{\mathprogramit{\pi}}\varphi$.

The soundness of \CGTPDL\ is proven by contradiction. We show that a \CGTPDL\ proof of 
an invalid sequent contains an infinite decreasing sequence of natural numbers 
along a cycle.
The completeness of \CGTPDL\ is proven by giving a way to transform each \GTPDL\ proof 
into a \CGTPDL\ proof. Therefore, we have the completeness of \CGTPDL\ 
from that of \GTPDL.

We give two counterexamples to the cut-elimination of \GTPDL. One is provable in 
\CGTPDL\ without the cut rule, while the other is not. In other words, the latter also 
serves as a counterexample to the cut-elimination of \CGTPDL. We have not found a sequent 
provable in \GTPDL\ but not in \CGTPDL\ without the cut rule, and we are unsure 
whether such a sequent exists.

Despite the failure of cut-elimination for \CGTPDL,
we have the cut-elimination theorem for \CGPDL, obtained by removing the rules 
for backwards modal operators from \CGTPDL.
\CGPDL\ is essentially equivalent to the non-labelled cut-free cyclic proof system 
\DasCyclicSystem\ for one-sided sequents of \PDL\ in \cite{Das2023}\footnote{They suggested some proof ideas for the cut-free completeness of \DasCyclicSystem, but
did not provide a detailed proof in \cite{Das2023}. We hear that they think that the proof techniques are known among experts.}.

To show the cut-elimination property,
we show the cut-free completeness of \CGPDL. 
While the completeness of cyclic proof system of $\mu$-calculus is proved 
by considering an infinite game \cite{Niwinski1996}, we consider a \emph{finite} game,
called the \emph{proof search game}. 
Because the game is a two-player finite zero-sum game of perfect information,
we easily have its \emph{determinacy} by Zermelo's theorem \cite{Zermelo1913}.
In other words, one of the players must have a winning strategy in the proof search game.
If one player has a winning strategy in the game, 
then a cut-free \CGPDL\ proof of the sequent exists, and
if the other player has a winning strategy,
then a counter model of the sequent exists.

\subsection{Related work}
\label{subsec:related-work}
We introduce related work.

There are many applications and expansions of (propositional) dynamic logic, 
including converse \PDL\ \cite{Parikh1978, Fischer1979, Gore2010}, \TPDL\ \cite{Chopra2001}, 
dynamic separation logic \cite{DeBoer2023} and temporal dynamic logic \cite{Platzer2007}. 
The same modal operators as in \TPDL\ are considered in 
\cite{Hartonas2013,Hartonas2014,Frittella2016}.

H. Nishimura \cite{Nishimura1979} introduces a sequent calculus \GPDL\ for 
converse \PDL\ without test programs, on which our sequent calculus, \GTPDL, is based. 
Cut-elimination does not hold in \GPDL. 

D. Leivant \cite{Leivant1981} defines a sequent calculus \LeivantSystem\
for \PDL\ with a rule whose assumption is infinite. The rule is like the $\omega$-rule 
for iteration programs $\mathprogramit{\pi}$.
Interestingly, there is a finite variant of \LeivantSystem\
due to \PDL's finite model property.
A similar situation in $\mu$-calculus is discussed in \cite{Jager2008}.
D. Leivant \cite{Leivant1981} also discusses the Lyndon interpolation theorem,
but the proof is disputed in \cite{Kracht1999,Gattinger2013}.

M. Lange \cite{Lange2003} introduces a satisfiability game for \PDL.
The game can be understood as a cyclic tableau system. 
The completeness of this game depends on the finite model property of \PDL.

B. Hill and F. Poggiolesi \cite{Hill2010} define a contraction-free and 
cut-free hyper-sequent calculus for \PDL.

C. Hartonas \cite{Hartonas2013, Hartonas2014} introduces a Hilbert-style system and 
a sequent calculus for a variant of \PDL\ including the same modal operators as in \TPDL\
and shows their soundness and completeness.
The systems seem essentially equivalent to \TPDL\ and \GTPDL, 
but are more complicated than they are.

S. Docherty and R. Rowe \cite{Docherty2019} introduce 
a labelled non-well-founded infinitary proof system, \DochertyInfinitarySystem, and 
a labelled cyclic proof system, \DochertyCyclicSystem, for \PDL. 
\DochertyCyclicSystem\ is obtained by restricting the shape of proof figures 
to regular trees in \DochertyInfinitarySystem. 
Cut-elimination holds in \DochertyInfinitarySystem, but it is an open question 
whether cut-elimination holds in \DochertyCyclicSystem.
Our proof of the soundness of our cyclic proof system, \CGTPDL, is simpler and 
more intuitive than 
the proof of the soundness of \DochertyCyclicSystem\ presented in \cite{Docherty2019},
as we do not use the \emph{Dershowitz-Manna ordering} \cite{Dershowitz1979}.

A. Das and M. Girlando \cite{Das2023} discuss the relation between 
a cut-free cyclic system of transitive closure logic and that of a variation of \PDL, 
\DasCyclicSystem, and give some proof ideas of the cut-free completeness of 
\DasCyclicSystem.
However, no details of the proof of the cut-free completeness are given in \cite{Das2023}. They suggested three ideas:
(1) the game theoretic approach of Niwinski and Walukiewicz \cite{Niwinski1996},
as carried out by Lange in \cite{Lange2003},
(2) the purely proof theoretic techniques of Studer \cite{Studer2008}, and
(3) embedding each cyclic proof into the $\mu$-calculus.
We do not think that either direction implies the cut-free completeness of 
\DasCyclicSystem\ \emph{immediately}, as they say.
Considering (1), since the satisfiability game in \cite{Lange2003} is very different 
from \DasCyclicSystem, the transformation from Lange's game to a cyclic proof 
in \DasCyclicSystem\ without using the cut rules is not obvious.
For (2), the transformation from each proof in the proof system with $\omega$-rule 
(\LeivantSystem\ in \cite{Leivant1981}?\footnote{\cite{Leivant1981} is not cited in \cite{Das2023}. }) 
to that in \DasCyclicSystem\ without using the cut-rules is not obvious.
For (3), we agree that we can embed each cyclic proof into $\mu$-calculus, but
what we need is the transformation from a cyclic proof in $\mu$-calculus 
to a proof in \DasCyclicSystem, and the existence of such a transformation is not obvious.
We cannot discuss further details, as the proof details are not publicly known.
We do not use their proof ideas for the cut-free completeness in this article.

Y. Zhang \cite{Zhang2024} introduces a labelled cyclic proof system 
for a parameterised dynamic logic, which is semantically incomplete. 
This system differs significantly from  \DochertyCyclicSystem\ \cite{Docherty2019},
\DasCyclicSystem\ \cite{Das2023} and our system, \CGTPDL, 
because the programs in \cite{Zhang2024} are not the same as those in \PDL.

M. Borzechowski, M. Gattinger, H. H. Hansen, R. Ramanayake, V. T. Dalmas, and Y. Venema 
\cite{Borzechowski2025} introduce a cyclic tableau system for \PDL\ and show 
the Craig interpolation theorem for \PDL. 
Whether the theorem holds has been an open problem for a long time.

\emph{Hoare-style logics} offer an alternative approach to reasoning about programs. 
In these logics, 
the units of reasoning are \emph{triples} such as $\triple{\varphi}{\pi}{\psi}$, 
where $\varphi$ and $\psi$ are logical formulae,
and $\pi$ is a program. 
A detailed survey of Hoare-style logics is given by K. R. Apt and E. Olderog 
\cite{Apt2O19}. 

The first Hoare-style logic is \emph{partial Hoare logic} \cite{Hoare1969}. 
This logic guarantees the partial correctness of $\htriple{\varphi}{\pi}{\psi}$, 
meaning that if $\pi$ terminates from a state in which $\varphi$ holds, 
then $\psi$ holds in any final state. 
Hence, $\htriple{\varphi}{\pi}{\psi}$ is the same meaning as $\varphi\to\mathdynbox{\pi}\psi$.
\emph{Total Hoare logic} \cite{Manna1974} is an extension of partial Hoare logic 
that also guarantees termination. 

\emph{Incorrectness logic} \cite{OHearn2019}, 
also known as \emph{reverse Hoare logic} \cite{DeVries2011}, 
guarantees for $\tihtriple{\varphi}{\pi}{\psi}$ that if $\psi$ holds at a state $w$, 
then there is an execution to $w$ such that $\varphi$ holds in the initial state. 
Then, 
$\tihtriple{\varphi}{\pi}{\psi}$ is the same meaning as $\psi\to\mathrevdia{\pi}\varphi$.
This logic is called ``incorrectness'' logic because it is typically used to prove 
the existence of bugs \cite{OHearn2019,Raad2020}. 

L. Zhang and B. L. Kaminski \cite{Zhang2022} define \emph{partial incorrectness logic}\footnote{
We believe that this logic should be called ``partial reverse Hoare logic''. This belief is based on its primary application and the nature of the guarantees provided by this logic.
} 
as a by-product of investigating \emph{predicate transformers} such as 
\emph{weakest pre-condition}, \emph{weakest liberal pre-condition}, 
\emph{strongest post-condition}, and \emph{strongest liberal post-condition}. 
If $\ihtriple{\varphi}{\pi}{\psi}$ is proven in this logic, then it can be seen that 
``for any execution of $\pi$, if $\psi$ holds at the final state, 
then $\varphi$ holds in the initial state''.
Therefore, $\ihtriple{\varphi}{\pi}{\psi}$ is the same meaning as 
$\psi\to\mathrevbox{\pi}\varphi$.

\begin{table*}[t]
 \centering
 \begin{tabular}[t]{|c|c|}
  \hline
  Logic & \TPDL\ formula \\
  \hline
  \hline
  (Weak) total Hoare logic $\thtriple{\varphi}{\pi}{\psi}$ & $\varphi\to\mathdyndia{\pi}\psi$ \\
  Partial Hoare logic $\htriple{\varphi}{\pi}{\psi}$ & $\varphi\to\mathdynbox{\pi}\psi$  \\
  Partial incorrectness logic $\ihtriple{\varphi}{\pi}{\psi}$ & $\psi\to\mathrevbox{\pi}\varphi$ \\
  Reverse Hoare logic  $\tihtriple{\varphi}{\pi}{\psi}$ & $\psi\to\mathrevdia{\pi}\varphi$ \\
  \hline
 \end{tabular}
 \caption{Hoare-style logic triples and the formulae in \TPDL}
 \label{fig:dynamic-logic-and-Hoare-style-logic} 
\end{table*}

A partial Hoare logic triple $\htriple{\varphi}{\pi}{\psi}$, 
a total Hoare logic\footnote{
This ``total Hoare logic'' is not the logic for guaranteeing total correctness. 
It refers to ``total Hoare logic'' in \cite{Zhang2022}. This logic guarantees that, 
for $\thtriple{\varphi}{\pi}{\psi}$, ``if $\varphi$ holds in a state $w$, 
then there exists an execution from $w$ such that $\psi$ holds in the final state''. 
For a deterministic program, this property is equivalent to total correctness. However, 
this is not the case for a non-deterministic program. We propose calling this logic 
``\emph{weak}'' \emph{total Hoare logic} 
to distinguish it from the logic for total correctness.
} triple $\thtriple{\varphi}{\pi}{\psi}$,
a partial incorrectness logic triple $\ihtriple{\varphi}{\pi}{\psi}$, and 
a reverse Hoare logic triple $\tihtriple{\varphi}{\pi}{\psi}$ can be expressed 
with modal operators in \TPDL\ by 
$\varphi\to\mathdynbox{\pi}\psi$, $\varphi\to\mathdyndia{\pi}\psi$, 
$\psi\to\mathrevbox{\pi}\varphi$, and $\psi\to\mathrevdia{\pi}\varphi$, respectively.
\cref{fig:dynamic-logic-and-Hoare-style-logic} summarises this relation.
We note that any partial incorrectness logic triple and reverse Hoare logic triple cannot 
be expressed in converse \PDL\ since there is no modal $\mathrevbox{\pi}$.
$\mathdynbox{\pi^{-}}$ is not a backwards modal for a program $\pi$ 
but just a forward modal for a program $\pi^{-}$.

These days, cyclic proofs are being investigated in great detail. 
R. Rowe \cite{Rowes-list} summarises the extensive list of academic work on cyclic and 
non-well-founded proof theory. Many studies on cyclic (or circular) proofs 
in various logics and theories have been examined, including 
first-order logic with inductive definitions 
\cite{Brotherston2011,Berardi2017,Berardi2019,Oda2023}, 
arithmetic \cite{Simpson2017,Das2020,Beklemishev2025}, 
bunched logic \cite{Brotherston2007,Saotome2021},
fixed point logic \cite{Kori2021}, 
G\"{o}del's T \cite{Das2021,Kuperberg2021},
Hoare-style logics \cite{Brotherston2025},
Kleene algebra \cite{Das2017,Das2018},
propositional logic \cite{Atserias2023}, and
separation logic 
\cite{Brotherston2008,Brotherston2011a,Tatsuta2019,Kimura2020,Saotome2020,Saotome2024}.
Cyclic proofs are also useful in software verification, including 
abduction \cite{Brotherston2014}, 
termination of pointer programs \cite{Brotherston2008,Rowe2017}, 
temporal property verification\cite{Tellez2020},
solving horn clauses \cite{Unno2017}, 
model checking\cite{Tsukada2022}, and
decision procedures for symbolic heaps 
\cite{Brotherston2012, Chu2015, Ta2016, Ta2018, Tatsuta2019}.

Cyclic proofs have also been given and investigated for various modal logics such as 
the Grzegorczyk modal logic \cite{Savateev2021},
modal $\mu$-calculus \cite{Sprenger2003, Studer2008,Afshari2017}, 
provability logic \cite{Shamkanov2014,Shamkanov2020,Das2024,Aguilera2024,Shamkanov2025}, 
logic of Common Knowledge \cite{Alonderis2025},
linear time $\mu$-calculus  \cite{Brunnler2008, Doumane2016,Kokkinis2016}.
Each of these logics includes modal operators for fixed points. 
Such modal operators are also in \TPDL\ (and \PDL). 

Cut-elimination in non-well-founded infinitary proof systems and cyclic proof systems is 
usually challenging due to the global soundness condition and 
the presence of cycles or infinite paths in each proof. 
Cut-elimination holds in non-well-founded infinitary proof systems for 
first-order logic with inductive definitions \cite{Brotherston2011} and
for linear logic with fixed points \cite{Doumane2017}. 
In contrast, it does not hold in a cyclic proof system for
first-order logic with inductive definitions \cite{Oda2023}, 
first-order arithmetic \cite{Das2020},
bunched logic \cite{Saotome2021}, and
separation logic \cite{Kimura2020}.
Notably, there are cut-eliminable cyclic proof systems for some modal logics, including 
G\"{o}del-L\"{o}b provability logic \cite{Shamkanov2014},
linear time $\mu$-calculus  \cite{Brunnler2008}, and
modal $\mu$-calculus \cite{Afshari2017}.

\subsection{The outline of this paper}
We present an outline of the paper.
\cref{sec:syntax-semantics} introduces the syntax and semantics of dynamic logic \TPDL.
\cref{sec:GTPDL} presents a sequent calculus for \GTPDL, 
a well-founded proof system for \TPDL.
\cref{sec:comp-GTPDL} shows the completeness of \GTPDL.
\cref{sec:CGTPDL} defines a cyclic proof system, \CGTPDL, for \TPDL\ and 
shows its completeness.
\cref{sec:sound-CGTPDL} proves the soundness of \CGTPDL.
\cref{sec:failure-cut-elimination} shows the failure of cut-elimination 
in \GTPDL\ and \CGTPDL.
\cref{sec:cut-elimination} introduces a cyclic proof system \CGPDL\ for \PDL\ 
as a fragment of \CGTPDL, and proves its cut-elimination property.
\cref{sec:conc} concludes.

\section{Syntax and semantics for \TPDL}
\label{sec:syntax-semantics}
This section introduces \TPDL, an extension of dynamic logic.
This logic is described in \cite{Chopra2001}.
\TPDL\ is obtained from \PDL\ by adding the backwards modal operator $\mathrevbox{\pi}$.

Let $\maththesetofpropval$ and $\maththesetofatomicprog$  
be the set of \emph{propositional variables} and
of \emph{atomic programs}, respectively.
The syntax of \TPDL\ is as follows:
\begin{align*}
 p&\in\maththesetofpropval  & \text{(Propositional variable)} \\
 \alpha&\in\maththesetofatomicprog  & \text{(Atomic program)} \\
 \varphi&\mathcoloneqq
 {\bot} \mathrelbar 
 {p} \mathrelbar   
 {(\varphi\to\varphi)} \mathrelbar 
 {\mathdynbox{\pi}\varphi} \mathrelbar 
 {\mathrevbox{\pi}\varphi}
 &\text{(Formula)} \\ 
 \pi&\mathcoloneqq
 {\alpha} \mathrelbar
 {\mathprogramseq{\pi}{\pi}} \mathrelbar 
 {\mathprogramndc{\pi}{\pi}} \mathrelbar 
 {\mathprogramit{\pi}}       \mathrelbar 
 {\mathprogramtest{\varphi}}
 &\text{(Program)}
\end{align*}
The outermost parentheses are omitted from each formula for the remainder of this paper.
We call a \emph{well-formed expression} a string which is a formula or a program.

We use the following abbreviation:
\begin{gather*}
 \varphi\to\bot \text{ as } \lnot\varphi, \quad
 (\varphi\to\bot)\to\psi  \text{ as } \varphi\lor\psi, \quad
 (\varphi\to(\psi\to\bot))\to\bot \text{ as } \varphi\land\psi, \\
 (\varphi\to\psi)\land(\psi\to\varphi)  \text{ as } \varphi\mathiffsy\psi, \quad
 \lnot\mathdynbox{\pi}\lnot\varphi \text{ as } \mathdyndia{\pi}\varphi, \quad
 \lnot\mathrevbox{\pi}\lnot\varphi \text{ as } \mathrevdia{\pi}\varphi.
\end{gather*}

For a set of formulae $\Gamma$ and 
$\heartsuit\in\mathsetintension{\mathdynbox{\pi}, \mathrevbox{\pi}, \mathdyndia{\pi}, \mathrevdia{\pi}}{\pi\text{ is a program}}\cup\mathsetextension{\lnot}$, 
we write  $\mathsetintension{\heartsuit\varphi}{\varphi\in\Gamma}$ as $\heartsuit\Gamma$.
For example, $\mathdynbox{\mathprogramtest{\psi}}\Gamma=\mathsetintension{\mathdynbox{\mathprogramtest{\psi}}\varphi}{\varphi\in\Gamma}$.

\begin{definition}[Length of well-formed expression]
 \label[definition]{def:length-of-expression}
 For a well-formed expression $\lambda$,
 we inductively define the \emph{length} of $\lambda$, written as $\mathlength{\lambda}$, 
 as follows:
 \begin{align*}
 \mathlength{\bot}&=1, \\
 \mathlength{p}&=1 &\text{ for } p\in \maththesetofpropval, \\
 \mathlength{\varphi\to\psi}&=\mathlength{\varphi} + \mathlength{\psi} + 1, & \\
 \mathlength{\mathdynbox{\pi}\varphi}&=\mathlength{\mathrevbox{\pi}\varphi}=\mathlength{\pi} + \mathlength{\varphi}, & \\
 \mathlength{\alpha}&=1 &\text{ for } \alpha\in \maththesetofatomicprog, \\
 \mathlength{\mathprogramseq{\pi_{0}}{\pi_{1}}}&=\mathlength{\mathprogramndc{\pi_{0}}{\pi_{1}}}=\mathlength{\pi_{0}}+\mathlength{\pi_{1}}+1, \\
 \mathlength{\mathprogramit{\pi}}&=\mathlength{\pi}+1, \\
 \mathlength{\mathprogramtest{\varphi}}&=\mathlength{\varphi}+1. 
 \end{align*}
\end{definition}

\begin{definition}[Model for \TPDL]
 We define a \emph{model} (for \TPDL ) as a tuple $\mathtuple{W, \mathsequence{\mathreduction{\alpha}}{\alpha\in\maththesetofatomicprog}, V}$, where
 $W$ is a non-empty set,
 $\mathreduction{\alpha}$ is a binary relation on $W$ for each $\alpha\in\maththesetofatomicprog$, and
 $V$ is a map from $W$ to the power set of $\maththesetofpropval$.
 We call each element of $W$ a \emph{state}.

 Let $M=\mathtuple{W, \mathsequence{\mathreduction{\alpha}}{\alpha\in\maththesetofatomicprog}, V}$ be a model for \TPDL.
 For states $w, w_{0}, w_{1} \in W$,  a program $\pi$, and a formula $\varphi$,
 we inductively define two relations $w_{0}\mathreduction{\pi}w_{1}$ and 
 $M, w\models \varphi$ as follows:
 \begin{enumerate}
  \item $w_{0}\mathreduction{\pi}w_{1}$ if
	$\pi\equiv\alpha$ and $w_{0}\mathreduction{\alpha}w_{1}$ for
	$\mathsequence{\mathreduction{\alpha}}{\alpha\in\maththesetofatomicprog}$.
  \item $w_{0}\mathreduction{\mathprogramseq{\pi_{0}}{\pi_{1}}}w_{1}$ if
	$w_{0}\mathreduction{\pi_{0}}w_{2}$ and
	$w_{2}\mathreduction{\pi_{1}}w_{1}$ for some $w_{2}\in W$.
  \item $w_{0}\mathreduction{\mathprogramndc{\pi_{0}}{\pi_{1}}}w_{1}$ if 
	either $w_{0}\mathreduction{\pi_{0}}w_{1}$ or
	$w_{0}\mathreduction{\pi_{1}}w_{1}$.
  \item $w_{0}\mathreduction{\mathprogramit{\pi}}w_{1}$
	if $w_{0}=w_{1}$ holds or there exists a sequence of states $\mathsequence{v_{i}}{i\leq n}$
	for $n\geq 0$ such that
	$v_{0}=w_{0}$, $v_{n}=w_{1}$, and 
	$v_{i}\mathreduction{\pi}v_{i+1}$ for $i={0, \dots, n-1}$.
  \item $w_{0}\mathreduction{\mathprogramtest{\psi}}w_{1}$
	if $w_{0}=w_{1}$ and $M, w_{0}\models \psi$.
  \item $M, w\not\models \bot$ for any $w\in W$.
  \item $M, w\models p$ for $p\in\maththesetofpropval$
	if $p\in\mathof{V}{w}$.
  \item $M, w\models \varphi_{0}\to\varphi_{1}$ 
	if $M, w\not\models \varphi_{0}$ or $M, w\models \varphi_{1}$.
  \item $M, w\models \mathdynbox{\pi}\varphi_{0}$
	if $M, v\models \varphi_{0}$ for any $v\in {W}$ with $w\mathreduction{\pi}v$.
  \item $M, w\models \mathrevbox{\pi}\varphi_{0}$
	if $M, v\models \varphi_{0}$ for any $v\in {W}$ with $v\mathreduction{\pi}w$.
 \end{enumerate}
\end{definition}

 We write $\maththeunderlyingset{M}$ for the set of states of $M$ for a model $M$.

 For any natural number $n$, we inductively $w\mathreduction{\pi}^{n}v$ as follows:
\begin{enumerate}
 \item $w\mathreduction{\pi}^{0}v$ if $w=v$.
 \item For $n>0$, $w\mathreduction{\pi}^{n}v$ if there exists $w'$ such that 
       $w\mathreduction{\pi}w'$ and $w'\mathreduction{\pi}^{n-1}v$.
\end{enumerate}

We note that $w\mathreduction{\pi}^{n}v$ for some $n$ 
if $w\mathreduction{\mathprogramit{\pi}}v$.

 For a model $M$,
 if $M, w\models \varphi$ for all $w\in \maththeunderlyingset{M}$,
 then we write $M\models \varphi$.
 For a set of formulae $\Gamma$, we write ${M, w}\models\Gamma$ 
 if ${M, w}\models\varphi$ for all $\varphi\in\Gamma$.
 We also write  $M\models\Gamma$ 
 if $M\models\varphi$ for all $\varphi\in\Gamma$.

S. Chopra and D. Zhang \cite{Chopra2001} had given 
a sound and complete Hilbert-style system for \TPDL.
It consists of the following axioms and inference rules:
\unless\ifoup
\begin{description}
\fi
 \ifoup \\ Axiom Taut: \else  \item[Axiom Taut:] \fi  All tautologies of propositional dynamic logic. 
 \ifoup \\ Axiom $\text{C}_{\text{P}}$: \else \item[Axiom $\text{C}_{\text{P}}$:] \fi  $\varphi\to\mathrevbox{\pi}\mathdyndia{\pi}\varphi$
 \ifoup \\ Axiom $\text{C}_{\text{F}}$: \else \item[Axiom $\text{C}_{\text{F}}$:] \fi  $\varphi\to\mathdynbox{\pi}\mathrevdia{\pi}\varphi$
 \ifoup \\ Axiom FK: \else \item[Axiom FK:] \fi $\mathdynbox{\pi}(\varphi \to \psi) \to (\mathdynbox{\pi}\varphi \to \mathdynbox{\pi}\psi)$
 \ifoup \\ Axiom PK: \else \item[Axiom PK:] \fi $\mathrevbox{\pi}(\varphi \to \psi) \to (\mathrevbox{\pi}\varphi \to \mathrevbox{\pi}\psi)$ 
 \ifoup \\ Inference rule FN: \else \item[Inference rule FN:] \fi \begin{inlineprooftree}
			  \AxiomC{$\varphi$}
			  \UnaryInfC{$\mathdynbox{\pi}\varphi$}
			  \end{inlineprooftree} 
 \ifoup \\ Inference rule PN: \else \item[Inference rule PN:] \fi \begin{inlineprooftree}
			  \AxiomC{$\varphi$}
			  \UnaryInfC{$\mathrevbox{\pi}\varphi$}
			 \end{inlineprooftree} 
 \ifoup \\ Inference rule MP: \else \item[Inference rule MP:] \fi \begin{inlineprooftree}
			  \AxiomC{$\varphi$}
			  \AxiomC{$\varphi\to\psi$}
			  \BinaryInfC{$\psi$}
			 \end{inlineprooftree} 
\unless\ifoup
\end{description}
\fi

We do not deal with the details of this proof system.

\section{\GTPDL : sequent calculus}
\label{sec:GTPDL}

\begin{figure}[tb]%
 \raggedright
 \emph{\emph{Logical Rules:}}

 \centering
 \begin{tabular}{cc}
  \begin{minipage}{0.5\hsize}
   \begin{prooftree}
    \AxiomC{}
    \LeftLabel{($\Gamma \cap \Delta \neq \emptyset$)}
    \RightLabel{\rulename{Ax}}
    \UnaryInfC{$\Gamma \fCenter \Delta$}
   \end{prooftree}
    \smallskip
  \end{minipage}
  &
   \begin{minipage}{0.5\hsize}
    \begin{prooftree}
     \AxiomC{}
     \RightLabel{\rulename{$\bot$}}
     \UnaryInfC{$\Gamma, \bot \fCenter \Delta$}
    \end{prooftree}
    \smallskip
   \end{minipage}
	  \\
   \begin{minipage}{0.5\hsize}
    \begin{prooftree}
     \AxiomC{$\Gamma \fCenter \varphi, \Delta$}
     \AxiomC{$\Gamma, \psi \fCenter \Delta $}
     \RightLabel{\rulename{$\to$ L}}
     \BinaryInfC{$ \Gamma, \varphi\to\psi \fCenter \Delta$}
    \end{prooftree}
    \smallskip
   \end{minipage}
   &
      \begin{minipage}{0.5\hsize}
       \begin{prooftree}
	\AxiomC{$\Gamma, \varphi \fCenter \psi, \Delta $}
	\RightLabel{\rulename{$\to$ R}}
	\UnaryInfC{$ \Gamma  \fCenter {\varphi\to \psi}, \Delta $}
       \end{prooftree}
      \end{minipage}
 \end{tabular}

 \raggedright
 \emph{\emph{Structural Rules:}}

 \centering

 \begin{tabular}{cc}
  \begin{minipage}{0.5\hsize}
   \begin{prooftree}
    \AxiomC{$ \Gamma \fCenter \Delta$}
    \LeftLabel{($\Gamma\subseteq\Gamma', \Delta\subseteq\Delta'$)}
    \RightLabel{\rulename{Wk}}
    \UnaryInfC{$\Gamma' \fCenter \Delta'$}
   \end{prooftree}
   \smallskip
  \end{minipage}
  &
  \begin{minipage}{0.5\hsize}
   \begin{prooftree}
    \AxiomC{$\Gamma \fCenter \varphi, \Delta$}
    \AxiomC{$\Gamma, \varphi \fCenter \Delta$}
    \RightLabel{\rulename{Cut}}
    \BinaryInfC{$\Gamma \fCenter \Delta$}
   \end{prooftree}
   \smallskip
  \end{minipage}
 \end{tabular}

 \raggedright
 \emph{\emph{Modal Rules:}}

 \centering
 \begin{tabular}{cc}
  \begin{minipage}{0.5\hsize}
   \begin{prooftree}
    \AxiomC{$\Gamma \fCenter \varphi, \mathrevbox{\pi}\Delta$}
    \RightLabel{\rulename{$\mathdynbox{ \, }$}}
    \UnaryInfC{$\mathdynbox{\pi}\Gamma \fCenter \mathdynbox{\pi}\varphi, \Delta$}
   \end{prooftree}
    \smallskip
  \end{minipage}
  &
   \begin{minipage}{0.5\hsize}
   \begin{prooftree}
    \AxiomC{$\Gamma \fCenter \varphi, \mathdynbox{\pi}\Delta$}
    \RightLabel{\rulename{$\mathrevbox{ \, }$}}
    \UnaryInfC{$\mathrevbox{\pi}\Gamma \fCenter \mathrevbox{\pi}\varphi, \Delta$}
   \end{prooftree}
    \smallskip
   \end{minipage}
	  \\
  \begin{minipage}{0.5\hsize}
   \begin{prooftree}
    \AxiomC{$\Gamma, \mathdynbox{\pi_{0}}\mathdynbox{\pi_{1}}\varphi \fCenter \Delta $}
    \RightLabel{\rulename{$\mathdynbox{\mathprogramseq{}{}}$ L}}
    \UnaryInfC{{$\Gamma, \mathdynbox{\mathprogramseq{\pi_{0}}{\pi_{1}}}\varphi \fCenter \Delta $}}
   \end{prooftree}
  \end{minipage}
  &
      \begin{minipage}{0.5\hsize}
       \begin{prooftree}
	\AxiomC{$\Gamma \fCenter \mathdynbox{\pi_{0}}\mathdynbox{\pi_{1}}\varphi, \Delta $}
	\RightLabel{\rulename{$\mathdynbox{\mathprogramseq{}{}}$ R}}
	\UnaryInfC{$\Gamma \fCenter \mathdynbox{\mathprogramseq{\pi_{0}}{\pi_{1}}}\varphi, \Delta $}
       \end{prooftree}
       \smallskip
      \end{minipage}
      \\
   \begin{minipage}{0.5\hsize}
    \begin{prooftree}
     \AxiomC{$\Gamma, \mathdynbox{ \pi_{0} }\varphi, \mathdynbox{ \pi_{1} }\varphi  \fCenter \Delta$}
     \RightLabel{\rulename{$\mathdynbox{ \mathprogramndc{}{} }$ L}}
     \UnaryInfC{$ \Gamma, \mathdynbox{ \mathprogramndc{\pi_{0}}{\pi_{1}} }\varphi \fCenter \Delta$}
    \end{prooftree}
    \smallskip
   \end{minipage}
   &
      \begin{minipage}{0.5\hsize}
       \begin{prooftree}
	\AxiomC{$\Gamma \fCenter \Delta, \mathdynbox{ \pi_{0} }\varphi $}
	\AxiomC{$\Gamma \fCenter \Delta, \mathdynbox{ \pi_{1} }\varphi $}
	\RightLabel{\rulename{$\mathdynbox{ \mathprogramndc{}{} }$ R}}
	\BinaryInfC{$\Gamma  \fCenter \mathdynbox{ \mathprogramndc{\pi_{0}}{\pi_{1}} }\varphi, \Delta$}
       \end{prooftree}
      \end{minipage}
 \end{tabular}

 \begin{tabular}{cc}
   \begin{minipage}{0.5\hsize}
    \begin{prooftree}
     \AxiomC{$\Gamma, \varphi, \mathdynbox{\pi}\mathdynbox{\mathprogramit{\pi}}\varphi \fCenter \Delta$}
     \RightLabel{\rulename{$\mathdynbox{\mathprogramitsy}$ L}}
     \UnaryInfC{$ \Gamma, \mathdynbox{\mathprogramit{\pi}}\varphi \fCenter \Delta$}
    \end{prooftree}
    \smallskip
   \end{minipage}
      & 
	  \begin{minipage}{0.5\hsize}
	   \begin{prooftree}
	    \AxiomC{$\Gamma, \varphi \fCenter \mathdynbox{\pi}\varphi$}
	    \RightLabel{\rulename{$\mathdynbox{\mathprogramitsy}$ R}}
	    \UnaryInfC{$\mathdynbox{\mathprogramit{\pi}}\Gamma, \varphi \fCenter \mathdynbox{\mathprogramit{\pi}}\varphi$}
	   \end{prooftree}
	   \smallskip
	  \end{minipage}
	  \\
 \end{tabular}

 \begin{tabular}{cc}
  \begin{minipage}{0.5\hsize}
   \begin{prooftree}
    \AxiomC{$\Gamma \fCenter \varphi, \Delta$}
    \AxiomC{$\Gamma, \psi \fCenter \Delta$}
    \RightLabel{\rulename{$\mathdynbox{ \mathprogramtestsy }$ L}}
    \BinaryInfC{$\Gamma, \mathdynbox{ \mathprogramtest{\varphi} }\psi \fCenter \Delta$}
   \end{prooftree}
    \smallskip
  \end{minipage}
  &
   \begin{minipage}{0.5\hsize}
   \begin{prooftree}
    \AxiomC{$\Gamma, \varphi \fCenter \psi, \Delta$}
    \RightLabel{\rulename{$\mathdynbox{ \mathprogramtestsy }$ R}}
    \UnaryInfC{$\Gamma \fCenter \mathdynbox{ \mathprogramtest{\varphi} }\psi, \Delta$}
   \end{prooftree}
    \smallskip
   \end{minipage}
 \end{tabular}

 \caption{Rules of \GTPDL}
 \label{fig:rules-GTPDL}
 \end{figure}

 In this section, we define a sequent calculus for \TPDL, named \GTPDL,
 which is a well-founded finitary proof system.

 A \emph{sequent} is a pair of finite sets of formulae, denoted by $\Gamma\fCenter\Delta$,
 where $\Gamma$ and $\Delta$ are finite sets of formulae. 
 For a sequent $\Gamma\fCenter\Delta$,
 we call $\Gamma$ the \emph{antecedent} of the sequent and 
 $\Delta$ the \emph{consequent} of the sequent.
 An \emph{inference rule} is written in the form:
 \begin{center}
  \begin{inlineprooftree}
   \AxiomC{$\Gamma_{1}\fCenter\Delta_{1}$ \, $\cdots$ \, $\Gamma_{n}\fCenter\Delta_{n}$} 
   \RightLabel{\rulename{R}}
   \UnaryInfC{$\Gamma\fCenter\Delta$}
  \end{inlineprooftree},
 \end{center}
 where $n$ is a natural number\footnote{If $n=0$, there is no assumption in the rule.};
 the sequents above the line are called the \emph{assumptions} of the rule, and
 the sequent below the line is called the \emph{conclusion} of the rule.
 A \emph{derivation tree} with a set of inference rules is defined as 
 a finite tree of sequents where each node is given by one of the inference rules.
 A leaf in a derivation tree that is not the conclusion of a rule is called a \emph{bud}.

 The inference rules of \GTPDL\ are in \cref{fig:rules-GTPDL}.
 \rulename{Ax} and \rulename{$\bot$} are the rules with no assumptions.
 We use commas in sequents to represent set union.
 We note that the contraction rule is implicit. 
 The \emph{principal formula} of a rule is the distinguished formula 
 introduced by the rule in its conclusion.
 The \emph{cut-formula} is the distinguished formula $\varphi$
 in the assumption of \rulename{Cut}.

 \begin{definition}[\GTPDL\ proof]
  We define a \emph{\GTPDL\ proof} to be a derivation tree 
  constructed from the rules in \cref{fig:rules-GTPDL}, without bud.
  If there is a \GTPDL\ proof of $\Gamma\fCenter\Delta$,
  then we say that $\Gamma\fCenter\Delta$ is \emph{provable} in \GTPDL.
 \end{definition}

 \cref{fig:example-stpdl-CF} and \cref{fig:example-stpdl-induction-ax} show 
 examples of \GTPDL\ proofs.

 \begin{figure*}[tb]
  \centering
  \begin{inlineprooftree}
   \AxiomC{}
   \RightLabel{\rulename{Ax}}
   \UnaryInfC{${p}\fCenter{p, \bot}$}

   \AxiomC{}
   \RightLabel{\rulename{$\bot$}}
   \UnaryInfC{${p, \bot}\fCenter{\bot}$}

   \RightLabel{\rulename{$\to$ L}}
   \BinaryInfC{${p, (p\to\bot)}\fCenter{\bot}$}
   \RightLabel{\rulename{$\to$ R}}
   \UnaryInfC{${p}\fCenter{(p\to\bot)\to\bot}$}
   \RightLabel{\rulename{Wk}}
   \UnaryInfC{${p}\fCenter{(p\to\bot)\to\bot, \mathrevbox{\alpha}\mathdyndia{\alpha}p}$}

   \AxiomC{}
   \RightLabel{\rulename{Ax}}
   \UnaryInfC{${\mathdynbox{\alpha}\lnot p}\fCenter{\bot, \mathdynbox{\alpha}\lnot p}$}
   \RightLabel{\rulename{$\to$ R}}
   \UnaryInfC{${}\fCenter{\mathdyndia{\alpha}p, \mathdynbox{\alpha}(p\to\bot)}$}
   \RightLabel{\rulename{$\mathrevbox{ \, }$}}
   \UnaryInfC{${}\fCenter{\mathrevbox{\alpha}\mathdyndia{\alpha}p, p\to\bot}$}

   \AxiomC{}
   \RightLabel{\rulename{$\bot$}}
   \UnaryInfC{${\bot}\fCenter{\mathrevbox{\alpha}\mathdyndia{\alpha}p}$}

   \RightLabel{\rulename{$\to$ L}}
   \BinaryInfC{${(p\to\bot)\to\bot}\fCenter{\mathrevbox{\alpha}\mathdyndia{\alpha}p}$}
   \RightLabel{\rulename{Wk}}
   \UnaryInfC{${p, (p\to\bot)\to\bot}\fCenter{\mathrevbox{\alpha}\mathdyndia{\alpha}p}$}
 
   \RightLabel{\rulename{Cut}}
   \BinaryInfC{${p}\fCenter{\mathrevbox{\alpha}\mathdyndia{\alpha} p}$}
   \RightLabel{\rulename{$\to$ R}}
   \UnaryInfC{${}\fCenter{p\to\mathrevbox{\alpha}\mathdyndia{\alpha}p}$}
  \end{inlineprooftree}

  \caption{A \GTPDL\ proof of ${}\fCenter{p\to\mathrevbox{\alpha}\mathdyndia{\alpha}p}$}
  \label{fig:example-stpdl-CF}

 \begin{inlineprooftree}
   \AxiomC{}
   \RightLabel{\rulename{Ax}}
   \UnaryInfC{$p \fCenter p, \mathdynbox{\alpha}p$}

   \AxiomC{}
   \RightLabel{\rulename{Ax}}
   \UnaryInfC{$p, \mathdynbox{\alpha} p \fCenter\mathdynbox{\alpha}p$}

   \RightLabel{\rulename{$\to$ L}}
   \BinaryInfC{$p, p\to\mathdynbox{\alpha} p \fCenter\mathdynbox{\alpha}p$}
   \RightLabel{\rulename{$\mathdynbox{\mathprogramitsy}$ R}}
   \UnaryInfC{$p, \mathdynbox{\mathprogramit{\alpha}}(p\to\mathdynbox{\alpha} p) \fCenter\mathdynbox{\mathprogramit{\alpha}}p$}
  \end{inlineprooftree}

  \caption{A \GTPDL\ proof of $p, \mathdynbox{\mathprogramit{\alpha}}(p\to\mathdynbox{\alpha} p) \fCenter\mathdynbox{\mathprogramit{\alpha}}p$}
  \label{fig:example-stpdl-induction-ax}
 \end{figure*}

 For a model $M$, a sequent $\Gamma\fCenter\Delta$ and $w\in \maththeunderlyingset{M}$,
 we write $M, w\models {\Gamma\fCenter\Delta}$ 
 if ${M, w}\models \Gamma$ implies ${M, w}\models \psi$ for some $\psi\in\Delta$.
 For a model $M$
 and a sequent $\Gamma\fCenter\Delta$,
 we write $M\models {\Gamma\fCenter\Delta}$ 
 if $M, w\models {\Gamma\fCenter\Delta}$ for all $w\in \maththeunderlyingset{M}$.
 We say that $\Gamma\fCenter\Delta$ is valid
 if $M\models {\Gamma\fCenter\Delta}$ holds for any model $M$.

 We have the soundness of \GTPDL\ easily.
 \begin{theorem}[Soundness of \GTPDL]
  \label{prop:soundness-GTPDL}
  If $\Gamma\fCenter\Delta$ is provable in \GTPDL, then $\Gamma\fCenter\Delta$ is valid.
 \end{theorem}
 
 \begin{proof}
  By induction on the construction of the \GTPDL\ proof.
 \end{proof}
 
 \GTPDL\ is sound and also complete. In other words, the following statement holds.
 \begin{theorem}[Completeness of \GTPDL]
  \label{prop:completeness-of-GTPDL}
  If $\Gamma\fCenter\Delta$ is valid, then $\Gamma\fCenter\Delta$ is provable in \GTPDL.
 \end{theorem}

 The proof of this theorem is in \cref{sec:comp-GTPDL}.
 As a by-product of the proof, we have the finite model property of \TPDL.

 \begin{theorem}[Finite model property]
  \label{prop:finite-model-property}
  If $M\not\models {\Gamma\fCenter\Delta}$ holds for some model $M$, 
  then there exists a model $M'$
  such that $M'\not\models {\Gamma\fCenter\Delta}$ holds and 
  $\maththeunderlyingset{M'}$ is finite. 
 \end{theorem}

 We also prove this theorem in \cref{sec:comp-GTPDL}.

\section{Completeness of \GTPDL}
\label{sec:comp-GTPDL}
This section shows the completeness of \GTPDL\ (\cref{prop:completeness-of-GTPDL}) and 
the finite model property of \TPDL\ (\cref{prop:finite-model-property}).

\ifoup
\noindent \emph{\emph{The road map of this section:}}
\else
\noindent \paragraph{The road map of this section:}
\fi
We prove the completeness of \GTPDL\ by constructing the canonical counter model of 
an unprovable sequent (\cref{def:canonical-counter-model}). To achieve this, we define 
the \emph{Fischer-Ladner closure} (\cref{def:Fischer-Ladner-closure}) and 
\emph{saturated sequents} (\cref{def:saturated-sequent}). 
Since the canonical counter model is finite, we can infer the finite model property of 
\TPDL. 
\cref{subsec:fundamental-properties-of-GTPDL} presents some fundamental properties of \GTPDL\ 
relevant to proving \cref{prop:completeness-of-GTPDL,prop:finite-model-property}. 
\cref{subsec:Fischer-Ladner-closure} introduces the Fischer-Ladner closure and 
demonstrates its properties, especially finiteness. 
\cref{subsec:canonical-counter-model} presents 
the canonical counter model and proves its property. 
Finally, \cref{subsec:proofs-completeness-fmp} proves 
\cref{prop:completeness-of-GTPDL,prop:finite-model-property}. 

\subsection{Fundamental properties of \GTPDL}
\label{subsec:fundamental-properties-of-GTPDL}
Before the main part of the proof of the completeness theorem,
we show some fundamental properties of \GTPDL.

 \begin{proposition}
  The following rules are derivable in \GTPDL:
  \begin{center}
   \begin{inlineprooftree}
    \AxiomC{$\Gamma\fCenter\Delta,\varphi$}
    \RightLabel{\rulename{$\lnot$ L}}
    \UnaryInfC{$\Gamma,\lnot\varphi\fCenter\Delta$}
   \end{inlineprooftree}
   \begin{inlineprooftree}
    \AxiomC{$\Gamma, \varphi\fCenter\Delta$}
    \RightLabel{\rulename{$\lnot$ R}}
    \UnaryInfC{$\Gamma\fCenter\Delta,\lnot\varphi$}
   \end{inlineprooftree}
   \begin{inlineprooftree}
    \AxiomC{$\Gamma,\varphi\fCenter\Delta$}
    \AxiomC{$\Gamma,\psi\fCenter\Delta$}
    \RightLabel{\rulename{$\lor$ L}}
    \BinaryInfC{$\Gamma, \varphi\lor\psi\fCenter\Delta$}
   \end{inlineprooftree} \ifoup \\ \fi
   \begin{inlineprooftree}
    \AxiomC{$\Gamma\fCenter\Delta, \varphi, \psi$}
    \RightLabel{\rulename{$\lor$ R}}
    \UnaryInfC{$\Gamma\fCenter\Delta, \varphi\lor\psi$}
   \end{inlineprooftree}
   \begin{inlineprooftree}
    \AxiomC{$\Gamma, \varphi, \psi\fCenter\Delta$}
    \RightLabel{\rulename{$\land$ L}}
    \UnaryInfC{$\Gamma, \varphi\land\psi\fCenter\Delta$}
   \end{inlineprooftree}
   \begin{inlineprooftree}
    \AxiomC{$\Gamma\fCenter\Delta,\varphi$}
    \AxiomC{$\Gamma\fCenter\Delta,\psi$}
    \RightLabel{\rulename{$\land$ R}}
    \BinaryInfC{$\Gamma\fCenter\Delta, \varphi\land\psi$}
   \end{inlineprooftree}
  \end{center}
 \end{proposition}

 \begin{proof}
  Easy.
 \end{proof}

 \begin{proposition}
  \label[proposition]{prop:old-rules}
  The following rules are derivable in \GTPDL\ without \rulename{$\mathdynbox{\mathprogramitsy}$ R}:
  \begin{center}
   \begin{inlineprooftree}
     \AxiomC{$\Gamma, \varphi \fCenter \Delta$}
     \RightLabel{\rulename{$\mathdynbox{\mathprogramitsy}$ L B}}
     \UnaryInfC{$ \Gamma, \mathdynbox{\mathprogramit{\pi}}\varphi \fCenter \Delta$}
   \end{inlineprooftree}
   \begin{inlineprooftree}
    \AxiomC{$\Gamma, \mathdynbox{\pi}\mathdynbox{\mathprogramit{\pi}}\varphi \fCenter \Delta$}
    \RightLabel{\rulename{$\mathdynbox{\mathprogramitsy}$ L S}}
    \UnaryInfC{$ \Gamma, \mathdynbox{\mathprogramit{\pi}}\varphi \fCenter \Delta$}
   \end{inlineprooftree}
   \begin{inlineprooftree}
    \AxiomC{$\Gamma, \mathdynbox{ \pi_{i} }\varphi  \fCenter \Delta$}
    \LeftLabel{($i=0, 1$)}
    \RightLabel{\rulename{$\mathdynbox{ \mathprogramndc{}{} }$ L + Wk}}
    \UnaryInfC{$ \Gamma, \mathdynbox{ \mathprogramndc{\pi_{0}}{\pi_{1}} }\varphi \fCenter \Delta$}
   \end{inlineprooftree}
  \end{center}
 \end{proposition}

 \begin{proof}
  Immediately.
 \end{proof}

 \begin{proposition}
  The following rule is derivable in \GTPDL:
  \begin{center}
   \begin{inlineprooftree}
    \AxiomC{$\Gamma\fCenter \varphi, \Delta$}
    \AxiomC{$\Gamma', \varphi\fCenter \Delta'$}
    \RightLabel{\rulename{Cut+Wk}}
    \BinaryInfC{$\Gamma, \Gamma'\fCenter\Delta, \Delta'$}
   \end{inlineprooftree}
  \end{center}
 \end{proposition}

 \begin{proof}
  Obvious.
 \end{proof}

 For a set of formulae $\Lambda=\mathsetextension{\varphi_{0}, \dots, \varphi_{n}}$,
 we define $\bigland\Lambda\equiv(\dots(\varphi_{0}\land\varphi_{1})\land\dots\land\varphi_{n})$
 and $\biglor\Lambda\equiv(\dots(\varphi_{0}\lor\varphi_{1})\lor\dots\lor\varphi_{n})$.
 We define $\bigland\emptyset\equiv \bot\to\bot$ and $\biglor\emptyset\equiv\bot$.

 \begin{proposition}
  The following rules are derivable in \GTPDL:
  \begin{center}
   \begin{inlineprooftree}
    \AxiomC{$\Gamma, \Gamma'\fCenter\Delta$}
    \RightLabel{\rulename{$\bigland$ L}}
    \UnaryInfC{$\Gamma, \bigland\Gamma'\fCenter\Delta$}
   \end{inlineprooftree}
   \begin{inlineprooftree}
    \AxiomC{$\mathsequence{\Gamma\fCenter\Delta, \varphi}{\varphi\in\Delta'}$}
    \RightLabel{\rulename{$\bigland$ R}}
    \UnaryInfC{$\Gamma\fCenter\Delta, \bigland\Delta'$}
   \end{inlineprooftree}
   \ifoup\nopagebreak \else \\ \fi
   \begin{inlineprooftree}
    \AxiomC{$\mathsequence{\Gamma, \psi\fCenter\Delta}{\psi\in\Gamma'}$}
    \RightLabel{\rulename{$\biglor$ L}}
    \UnaryInfC{$\Gamma, \biglor\Gamma'\fCenter\Delta$}
   \end{inlineprooftree}
   \begin{inlineprooftree}
    \AxiomC{$\Gamma\fCenter\Delta, \Delta'$}
    \RightLabel{\rulename{$\biglor$ R}}
    \UnaryInfC{$\Gamma\fCenter\Delta, \biglor\Delta'$}
   \end{inlineprooftree}
  \end{center}
 \end{proposition}

 \begin{proof}
  Straightforward.
 \end{proof}

 \begin{definition}
  For a sequent $\Gamma\fCenter\Delta$,
  we define a formula $\mathseqtoform{\Gamma\fCenter\Delta}$ as
  $\bigland\Gamma\to\biglor\Delta$.
 \end{definition}

 $\mathseqtoform{\Gamma\fCenter\Delta}$ is called the \emph{characteristic wff} of 
 a sequent $\Gamma\fCenter\Delta$ in \cite{Nishimura1979}.

 \begin{proposition}
  \label[proposition]{prop:hyper-sequent}
  $\Pi\fCenter\Sigma, \mathseqtoform{\Gamma\fCenter\Delta}$ is provable in \GTPDL\ 
  if and only if
  $\Gamma, \Pi\fCenter\Delta, \Sigma$ is provable in \GTPDL.
 \end{proposition}

 \begin{proof}
  Straightforward.
 \end{proof}

 \begin{proposition}
  \label[proposition]{prop:inner-cut-sequent}
  Let $\varphi$ be a formula.
  $\mathseqtoform{\Gamma\fCenter\Delta, \varphi}, \mathseqtoform{\Gamma, \varphi\fCenter\Delta}\fCenter\mathseqtoform{\Gamma\fCenter\Delta}$ is provable in \GTPDL.
 \end{proposition}

 \begin{proof}
  \cref{fig:prop-inner-cut-sequent}.
  \begin{figure}[tb]
   \centering
   \begin{prooftree}
    \footnotesize
    \AxiomC{}
    \RightLabel{\rulename{Ax}'s}
    \UnaryInfC{$\mathsequence{\Gamma, \mathseqtoform{\Gamma, \varphi\fCenter\Delta}\fCenter\Delta, \psi}{\psi\in\Gamma}$}
    \RightLabel{\rulename{$\bigland$ R}}
    \UnaryInfC{$\Gamma, \mathseqtoform{\Gamma, \varphi\fCenter\Delta}\fCenter\Delta, \bigland\Gamma$}

    \AxiomC{}
    \RightLabel{\rulename{Ax}'s}
    \UnaryInfC{$\mathsequence{\Gamma, \psi, \mathseqtoform{\Gamma, \varphi\fCenter\Delta}\fCenter\Delta}{\psi\in\Delta}$}

    \AxiomC{}
    \RightLabel{\rulename{Ax}}
    \UnaryInfC{$\mathseqtoform{\Gamma, \varphi\fCenter\Delta}\fCenter\mathseqtoform{\Gamma, \varphi\fCenter\Delta}$}
    \RightLabel{\rulename{*}}
    \doubleLine
    \UnaryInfC{$\Gamma, \varphi, \mathseqtoform{\Gamma, \varphi\fCenter\Delta}\fCenter\Delta$}
    \RightLabel{\rulename{$\biglor$ L}}
    \BinaryInfC{$\Gamma, \biglor\Delta\lor\varphi, \mathseqtoform{\Gamma, \varphi\fCenter\Delta}\fCenter\Delta$}

    \RightLabel{\rulename{$\to$ L}}
    \BinaryInfC{$\Gamma, \mathseqtoform{\Gamma\fCenter\Delta, \varphi}, \mathseqtoform{\Gamma, \varphi\fCenter\Delta}\fCenter\Delta$}
    \doubleLine
    \RightLabel{\rulename{*}}
    \UnaryInfC{$\mathseqtoform{\Gamma\fCenter\Delta, \varphi}, \mathseqtoform{\Gamma, \varphi\fCenter\Delta}\fCenter\mathseqtoform{\Gamma\fCenter\Delta}$}
   \end{prooftree}

   \rulename{*} by \cref{prop:hyper-sequent}
   \caption{Proof of \cref{prop:inner-cut-sequent}}
   \label{fig:prop-inner-cut-sequent}
  \end{figure}
 \end{proof}

\begin{definition}
 \label[definition]{def:the-set-of-divisions}
 For a finite set of formulae $\Xi$,
 we define 
 $\mathof{D}{\Xi}=\mathsetintension{\mathseqtoform{\Gamma\fCenter\Delta}}{\Xi=\Gamma\cup\Delta, \Gamma\cap\Delta=\emptyset}$.
\end{definition}

 For a set $A$, we write $\mathcardinality{A}$ for the number of elements in $A$.

 \begin{proposition}
  \label[proposition]{prop:divisions-in-antecedent}
  For any finite set of formulae $\Xi$, $\mathof{D}{\Xi}\fCenter$ is provable in \GTPDL.
 \end{proposition}

 \begin{proof}
  We show that $\mathof{D}{\Xi}\fCenter$ is provable in \GTPDL\ by induction
  on $\mathcardinality{\Xi}$.

  In case $\mathcardinality{\Xi}=0$, $\mathof{D}{\Xi}\fCenter$ is provable in \GTPDL,
  obviously.
  
  Consider the case $\mathcardinality{\Xi}=n$.
  Let $\varphi\in\Xi$ and $\Xi'=\Xi\setminus\mathsetextension{\varphi}$.
  By the induction hypothesis,
  $\mathof{D}{\Xi'}\fCenter$ is provable in \GTPDL.
  We note that 
  \[
  \mathof{D}{\Xi}=\mathsetintension{\mathseqtoform{\Gamma, \varphi\fCenter\Delta}}{\mathseqtoform{\Gamma\fCenter\Delta}\in\mathof{D}{\Xi'}}\cup\mathsetintension{\mathseqtoform{\Gamma\fCenter\Delta, \varphi}}{\mathseqtoform{\Gamma\fCenter\Delta}\in\mathof{D}{\Xi'}}.
  \]
  By \cref{prop:inner-cut-sequent}, 
  $\mathseqtoform{\Gamma\fCenter\Delta, \varphi}, \mathseqtoform{\Gamma, \varphi\fCenter\Delta}\fCenter\mathseqtoform{\Gamma\fCenter\Delta}$
  is provable in \GTPDL\ for all $\mathseqtoform{\Gamma\fCenter\Delta}\in\mathof{D}{\Xi'}$.
  We have $\mathof{D}{\Xi}\fCenter$
  from $\mathof{D}{\Xi'}\fCenter$ by repeating application of \rulename{Cut+Wk} with
  \ifoup \linebreak[3] \else \linebreak[4] \fi
  $\mathseqtoform{\Gamma\fCenter\Delta, \varphi}, \mathseqtoform{\Gamma, \varphi\fCenter\Delta}\fCenter\mathseqtoform{\Gamma\fCenter\Delta}$
  for all $\mathseqtoform{\Gamma\fCenter\Delta}\in\mathof{D}{\Xi'}$.
 \end{proof}

 \begin{lemma}
  \label[lemma]{lemma:switching-lemma}
  Let $\pi$ be a program.
  $\Gamma\fCenter\Delta, \mathrevbox{\pi}\mathseqtoform{\Pi\fCenter\Sigma}$
  is  provable in \GTPDL\
  if and only if
  $\Pi\fCenter\Sigma, \mathdynbox{\pi}\mathseqtoform{\Gamma\fCenter\Delta}$
  is provable in \GTPDL.
 \end{lemma}

 \begin{proof}
  Let $\pi$ be a program.

  \noindent `if' part: 
  Assume that 
  $\Pi\fCenter\Sigma, \mathdynbox{\pi}\mathseqtoform{\Gamma\fCenter\Delta}$
  is provable in \GTPDL.
  By \cref{prop:hyper-sequent},
  $\fCenter\mathseqtoform{\Pi\fCenter\Sigma}, \mathdynbox{\pi}\mathseqtoform{\Gamma\fCenter\Delta}$
  is provable in \GTPDL.
  By \rulename{$\mathrevbox{ \, }$}, we see that 
  $\fCenter\mathrevbox{\pi}\mathseqtoform{\Pi\fCenter\Sigma}, \mathseqtoform{\Gamma\fCenter\Delta}$
  is provable in \GTPDL.
  By \cref{prop:hyper-sequent},
  $\Gamma\fCenter\Delta, \mathrevbox{\pi}\mathseqtoform{\Pi\fCenter\Sigma}$
  is provable in \GTPDL.

  \noindent `only if' part:
  Assume that 
  $\Gamma\fCenter\Delta, \mathrevbox{\pi}\mathseqtoform{\Pi\fCenter\Sigma}$
  is provable in \GTPDL.
  By \cref{prop:hyper-sequent},
  $\fCenter\mathseqtoform{\Gamma\fCenter\Delta}, \mathrevbox{\pi}\mathseqtoform{\Pi\fCenter\Sigma}$
  is provable in \GTPDL.
  By \rulename{$\mathdynbox{ \, }$}, we see that
  $\fCenter\mathdynbox{\pi}\mathseqtoform{\Gamma\fCenter\Delta}, \mathseqtoform{\Pi\fCenter\Sigma}$
  is provable in \GTPDL.
  By \cref{prop:hyper-sequent},
  $\Pi\fCenter\Sigma, \mathdynbox{\pi}\mathseqtoform{\Gamma\fCenter\Delta}$
  is provable in \GTPDL.
 \end{proof}





 \begin{lemma}
  \label[lemma]{lemma:weakening-in-box}
  If $\Pi\fCenter\Sigma, \mathdynbox{\pi}\mathseqtoform{\Gamma\fCenter\Delta}$ is provable 
  in \GTPDL,
  then $\Pi\fCenter\Sigma, \mathdynbox{\pi}\mathseqtoform{\Gamma, \Phi\fCenter\Delta, \Psi}$ 
  is provable in \GTPDL\ with sets of formulae $\Phi$ and $\Psi$.
 \end{lemma}

 \begin{proof}
  Assume that 
  $\Pi\fCenter\Sigma, \mathdynbox{\pi}\mathseqtoform{\Gamma\fCenter\Delta}$ is provable 
  in \GTPDL.
  By \cref{lemma:switching-lemma},
  $\Gamma\fCenter\Delta, \mathrevbox{\pi}\mathseqtoform{\Pi\fCenter\Sigma}$
  is provable in \GTPDL.
  By \rulename{Wk},
  $\Gamma, \Phi\fCenter\Delta, \Psi, \mathrevbox{\pi}\mathseqtoform{\Pi\fCenter\Sigma}$
  is provable in \GTPDL.
  By \cref{lemma:switching-lemma},
  $\Pi\fCenter\Sigma, \mathdynbox{\pi}\mathseqtoform{\Gamma, \Phi\fCenter\Delta, \Psi}$ 
  is provable in \GTPDL.
 \end{proof}




 \begin{lemma}
  \label[lemma]{lemma:weakening-in-revbox}
  If $\Pi\fCenter\Sigma, \mathrevbox{\pi}\mathseqtoform{\Gamma\fCenter\Delta}$ is provable 
  in \GTPDL,
  then $\Pi\fCenter\Sigma, \mathrevbox{\pi}\mathseqtoform{\Gamma, \Phi\fCenter\Delta, \Psi}$ 
  is provable in \GTPDL\ with sets of formulae $\Phi$ and $\Psi$.
 \end{lemma}

 \begin{proof}
  In the similar way to the proof of \cref{lemma:weakening-in-box}.
 \end{proof}

 \begin{lemma}
  \label[lemma]{lemma:cut-in-box}
  If $\Pi\fCenter\Sigma, \mathdynbox{\pi}\mathseqtoform{\Gamma\fCenter\varphi, \Delta}$ and
  $\Pi\fCenter\Sigma, \mathdynbox{\pi}\mathseqtoform{\Gamma, \varphi\fCenter\Delta}$
  are provable in \GTPDL,
  then $\Pi\fCenter\Sigma, \mathdynbox{\pi}\mathseqtoform{\Gamma\fCenter\Delta}$ 
  is provable in \GTPDL.
 \end{lemma}

 \begin{proof}
  Assume that 
  $\Pi\fCenter\Sigma, \mathdynbox{\pi}\mathseqtoform{\Gamma\fCenter\varphi, \Delta}$ and
  $\Pi\fCenter\Sigma, \mathdynbox{\pi}\mathseqtoform{\Gamma, \varphi\fCenter\Delta}$
  are provable in \GTPDL.
  By \cref{lemma:switching-lemma},
  $\Gamma\fCenter\varphi, \Delta, \mathrevbox{\pi}\mathseqtoform{\Pi\fCenter\Sigma}$ and
  $\Gamma, \varphi\fCenter\Delta, \mathrevbox{\pi}\mathseqtoform{\Pi\fCenter\Sigma}$
  are provable in \GTPDL.
  By \rulename{Cut},
  $\Gamma\fCenter\Delta, \mathrevbox{\pi}\mathseqtoform{\Pi\fCenter\Sigma}$ 
  is provable in \GTPDL.
  By \cref{lemma:switching-lemma},
  $\Pi\fCenter\Sigma, \mathdynbox{\pi}\mathseqtoform{\Gamma\fCenter\Delta}$ 
  is provable in \GTPDL.
 \end{proof}

 \begin{lemma}
  \label[lemma]{lemma:cut-in-revbox}
  If $\Pi\fCenter\Sigma, \mathrevbox{\pi}\mathseqtoform{\Gamma\fCenter\varphi, \Delta}$ and
  $\Pi\fCenter\Sigma, \mathrevbox{\pi}\mathseqtoform{\Gamma, \varphi\fCenter\Delta}$
  are provable in \GTPDL,
  then $\Pi\fCenter\Sigma, \mathrevbox{\pi}\mathseqtoform{\Gamma\fCenter\Delta}$ 
  is provable in \GTPDL.
 \end{lemma}

 \begin{proof}
  In the similar way to the proof of \cref{lemma:cut-in-box}.
 \end{proof}

 \begin{lemma}
  \label[lemma]{lemma:not-provable-sequent-can-expand}
  For a formula $\varphi$ and a sequent $\Gamma\fCenter\Delta$,
  if $\Gamma\fCenter\Delta$ is not provable in \GTPDL,
  then either $\Gamma\fCenter\varphi, \Delta$ or $\Gamma, \varphi\fCenter\Delta$
  is not provable in \GTPDL.
 \end{lemma}

 \begin{proof}
  Assume $\varphi\in\Lambda$.
  We show the contrapositive of the statement.

  Suppose that both $\Gamma\fCenter\varphi, \Delta$ and $\Gamma, \varphi\fCenter\Delta$
  are provable in \GTPDL.
  Then, $\Gamma\fCenter\Delta$ is also proved in \GTPDL\ as follows:
  \begin{center}
   \begin{inlineprooftree}
    \AxiomC{$\Gamma\fCenter\varphi, \Delta$}
    \AxiomC{$\Gamma, \varphi\fCenter\Delta$}
    \RightLabel{\rulename{Cut}}
    \BinaryInfC{$\Gamma\fCenter\Delta$}
   \end{inlineprooftree}.
  \end{center}
 \end{proof}

 \begin{lemma}
  \label[lemma]{lemma:not-provable-sequent-can-expand-set}
  Let $\Lambda$ be a finite set of formulae.
  Assume that $\Gamma\fCenter\Delta$ is not provable in \GTPDL\ and
  ${\Gamma\cup\Delta}\subseteq\Lambda$.
  There is a sequent $\tilde{\Gamma}\fCenter\tilde{\Delta}$ not provable in \GTPDL,
  where $\Gamma\subseteq\tilde{\Gamma}$, $\Delta\subseteq\tilde{\Delta}$, and
  $\tilde{\Gamma}\cup\tilde{\Delta}=\Lambda$.
 \end{lemma}

 \begin{proof}
  Let $\Lambda$ be a finite set of formulae.
  Assume that $\Gamma\fCenter\Delta$ is not provable in \GTPDL\ and 
  ${\Gamma\cup\Delta}\subseteq\Lambda$ holds.
  If $\Lambda={\Gamma\cup\Delta}$ holds,
  then $\Gamma\fCenter\Delta$ is a sequent that we want.

  Assume ${\Gamma\cup\Delta}\subsetneq\Lambda$ and
  $\Lambda\setminus{(\Gamma\cup\Delta)}=\mathsetextension{\varphi_{1}, \dots, \varphi_{n}}$.
  We inductively define a sequence 
  $\mathsequence{\Gamma_{i}\fCenter\Delta_{i}}{0\leq i\leq n}$
  of sequents satisfying the following properties:
  \begin{enumerate}
   \item $\Gamma_{0}=\Gamma$ and $\Delta_{0}=\Delta$, 
	 \label{item:1-lemma-not-provable-sequent-can-expand-set}
   \item $\Gamma_{i}\fCenter\Delta_{i}$ is not provable in \GTPDL\ 
	 for each $0\leq i\leq n$,
	 \label{item:2-lemma-not-provable-sequent-can-expand-set}
   \item either $\varphi_{i}\in\Gamma_{i}$ or $\varphi_{i}\in\Delta_{i}$ holds 
	 for each $1\leq i\leq n$, and
	 \label{item:3-lemma-not-provable-sequent-can-expand-set}
   \item $\Gamma_{i-1}\subseteq\Gamma_{i}$ and $\Delta_{i-1}\subseteq\Delta_{i}$
	 for each $1\leq i\leq n$.
	 \label{item:4-lemma-not-provable-sequent-can-expand-set}
  \end{enumerate}

  Define $\Gamma_{0}=\Gamma$ and $\Delta_{0}=\Delta$.
  
  Consider the case $i>0$. By \cref{lemma:not-provable-sequent-can-expand},
  either $\Gamma_{i-1}\fCenter\varphi_{i}, \Delta_{i-1}$ or 
  $\Gamma_{i-1}, \varphi_{i}\fCenter\Delta_{i-1}$
  is not provable in \GTPDL.
  Define $\Gamma_{i}=\Gamma_{i-1}$ and $\Delta_{i}=\varphi_{i}, \Delta_{i-1}$
  if $\Gamma_{i-1}\fCenter\varphi_{i}, \Delta_{i-1}$ is not provable in \GTPDL; 
  otherwise $\Gamma_{i}=\Gamma_{i-1}, \varphi_{i}$ and $\Delta_{i}=\Delta_{i-1}$.

  By definition of $\mathsequence{\Gamma_{i}\fCenter\Delta_{i}}{0\leq i\leq n}$,
  we have
  \cref{item:1-lemma-not-provable-sequent-can-expand-set},
  \cref{item:2-lemma-not-provable-sequent-can-expand-set},
  \cref{item:3-lemma-not-provable-sequent-can-expand-set}, and
  \cref{item:4-lemma-not-provable-sequent-can-expand-set}, immediately.
  Let $\tilde{\Gamma}=\Gamma_{n}$ and $\tilde{\Delta}=\Delta_{n}$.
  By  \cref{item:1-lemma-not-provable-sequent-can-expand-set} and 
  \cref{item:4-lemma-not-provable-sequent-can-expand-set},
  $\Gamma\subseteq\tilde{\Gamma}$ and $\Delta\subseteq\tilde{\Delta}$.
  From \cref{item:3-lemma-not-provable-sequent-can-expand-set} and
  \cref{item:4-lemma-not-provable-sequent-can-expand-set},
  either $\varphi_{i}\in\tilde{\Gamma}$ or $\varphi_{i}\in\tilde{\Delta}$ holds 
  for all $i\in\mathsetextension{0, \dots, n}$.
  Hence, we have ${\tilde{\Gamma}\cup\tilde{\Delta}}=\Lambda$.
  By \cref{item:2-lemma-not-provable-sequent-can-expand-set},
  $\tilde{\Gamma}\fCenter\tilde{\Delta}$ is a sequent which we want.
 \end{proof}

 \begin{lemma}
  \label[lemma]{lemma:not-provable-sequent-can-expand-set-in-box}
  Let $\Lambda$ be sets of formulae.
  Assume that $\Pi\fCenter\Sigma, \mathdynbox{\pi}\mathseqtoform{\Gamma\fCenter\Delta}$ 
  is not provable in \GTPDL\ and
  ${\Gamma\cup\Delta}\subseteq\Lambda$.
  There is a sequent $\tilde{\Gamma}\fCenter\tilde{\Delta}$
  such that $\Gamma\subseteq\tilde{\Gamma}$, $\Delta\subseteq\tilde{\Delta}$ and
  $\tilde{\Gamma}\cup\tilde{\Delta}=\Lambda$ hold, and both
  $\Pi\fCenter\Sigma, \mathdynbox{\pi}\mathseqtoform{\tilde{\Gamma}\fCenter\tilde{\Delta}}$
  and $\tilde{\Gamma}\fCenter\tilde{\Delta}$ are not provable in \GTPDL.
 \end{lemma}

 \begin{proof}
  Let $\Lambda$ be sets of formulae.
  Assume that $\Pi\fCenter\Sigma, \mathdynbox{\pi}\mathseqtoform{\Gamma\fCenter\Delta}$ 
  is not provable in \GTPDL\ and ${\Gamma\cup\Delta}\subseteq\Lambda$.
  By \cref{lemma:switching-lemma},
  $\Gamma\fCenter\Delta, \mathrevbox{\pi}\mathseqtoform{\Pi\fCenter\Sigma}$ 
  is not provable in \GTPDL.
  By \cref{lemma:not-provable-sequent-can-expand-set},
  There is a sequent $\tilde{\Gamma}\fCenter\tilde{\Delta}, \mathrevbox{\pi}\mathseqtoform{\Pi\fCenter\Sigma}$ not provable in \GTPDL\
  such that the following holds:
  \begin{enumerate}
   \item $\Gamma\subseteq\tilde{\Gamma}$,
   \item $(\Delta, \mathrevbox{\pi}\mathseqtoform{\Pi\fCenter\Sigma})\subseteq(\tilde{\Delta}, \mathrevbox{\pi}\mathseqtoform{\Pi\fCenter\Sigma})$, 
	 and
   \item $\tilde{\Gamma}\cup\tilde{\Delta}\cup\mathsetextension{\mathrevbox{\pi}\mathseqtoform{\Pi\fCenter\Sigma}}=\Lambda\cup\mathsetextension{\mathrevbox{\pi}\mathseqtoform{\Pi\fCenter\Sigma}}$.
  \end{enumerate}
  By \cref{lemma:switching-lemma}, 
  $\Pi\fCenter\Sigma, \mathdynbox{\pi}\mathseqtoform{\tilde{\Gamma}\fCenter\tilde{\Delta}}$
  is not provable in \GTPDL.

  Assume, for contradiction, that 
  $\tilde{\Gamma}\fCenter\tilde{\Delta}$ is provable in \GTPDL,
  By \cref{prop:hyper-sequent},
  $\fCenter\mathseqtoform{\tilde{\Gamma}\fCenter\tilde{\Delta}}$ is provable in \GTPDL.
  By \rulename{$\mathdynbox{ \, }$} and \rulename{Wk},
  $\Pi\fCenter\Sigma, \mathdynbox{\pi}\mathseqtoform{\tilde{\Gamma}\fCenter\tilde{\Delta}}$
  is provable in \GTPDL, which is a contradiction.
  Hence, $\tilde{\Gamma}\fCenter\tilde{\Delta}$ is not provable in \GTPDL.
 \end{proof}

 \begin{lemma}
  \label[lemma]{lemma:not-provable-sequent-can-expand-set-in-revbox}
  Let $\Lambda$ be sets of formulae.
  Assume that $\Pi\fCenter\Sigma, \mathrevbox{\pi}\mathseqtoform{\Gamma\fCenter\Delta}$ 
  is not provable in \GTPDL\ and
  ${\Gamma\cup\Delta}\subseteq\Lambda$.
  There is a sequent $\tilde{\Gamma}\fCenter\tilde{\Delta}$
  such that $\Gamma\subseteq\tilde{\Gamma}$, $\Delta\subseteq\tilde{\Delta}$ and
  $\tilde{\Gamma}\cup\tilde{\Delta}=\Lambda$ hold, and both
  $\Pi\fCenter\Sigma, \mathrevbox{\pi}\mathseqtoform{\tilde{\Gamma}\fCenter\tilde{\Delta}}$
  and $\tilde{\Gamma}\fCenter\tilde{\Delta}$ are not provable in \GTPDL.
 \end{lemma}

 \begin{proof}
  In the similar way to \cref{lemma:not-provable-sequent-can-expand-set-in-box}.
 \end{proof}

 \subsection{Fischer-Ladner closure}
 \label{subsec:Fischer-Ladner-closure}
 This section introduces the Fischer-Ladner closure and proves its properties.

 \begin{definition}[Fischer-Ladner closure]
  \label[definition]{def:Fischer-Ladner-closure}
  For a formula $\varphi$ and a program $\pi$, 
  we inductively define $\mathFLcl{\varphi}$ and $\mathFLboxcl{\mathdynbox{\pi}\varphi}$
  as follows:
  \begin{enumerate}
   \item $\mathFLcl{\bot}=\mathsetextension{\bot}$.
   \item $\mathFLcl{p}=\mathsetextension{p}$ for $p\in\maththesetofpropval$.
   \item $\mathFLcl{\psi_{0}\to\psi_{1}}=\mathsetextension{\psi_{0}\to\psi_{1}}\cup\mathFLcl{\psi_{0}}\cup\mathFLcl{\psi_{1}}$.
   \item $\mathFLcl{\mathdynbox{\pi}\psi}=\mathFLboxcl{\mathdynbox{\pi}\psi}\cup\mathFLcl{\psi}$.
   \item $\mathFLcl{\mathrevbox{\pi}\psi}=\mathFLboxcl{\mathrevbox{\pi}\psi}\cup\mathFLcl{\psi}$.
   \item $\mathFLboxcl{\varphi}=\emptyset$
	 if $\varphi$ is of the form 
	 neither $\mathdynbox{\pi}\psi$ nor $\mathrevbox{\pi}\psi$.
   \item $\mathFLboxcl{\mathdynbox{\alpha}\psi}=\mathsetextension{\mathdynbox{\alpha}\psi}$
	 for $\alpha\in\maththesetofatomicprog$.
   \item $\mathFLboxcl{\mathdynbox{\mathprogramseq{\pi_{0}}{\pi_{1}}}\psi}=\mathsetextension{\mathdynbox{\mathprogramseq{\pi_{0}}{\pi_{1}}}\psi}\cup\mathFLboxcl{\mathdynbox{\pi_{0}}\mathdynbox{\pi_{1}}\psi}\cup\mathFLboxcl{\mathdynbox{\pi_{1}}\psi}$.
   \item $\mathFLboxcl{\mathdynbox{\mathprogramndc{\pi_{0}}{\pi_{1}}}\psi}=\mathsetextension{\mathdynbox{\mathprogramndc{\pi_{0}}{\pi_{1}}}\psi}\cup\mathFLboxcl{\mathdynbox{\pi_{0}}\psi}\cup\mathFLboxcl{\mathdynbox{\pi_{1}}\psi}$.
   \item $\mathFLboxcl{\mathdynbox{\mathprogramit{\pi}}\psi}=\mathsetextension{\mathdynbox{\mathprogramit{\pi}}\psi}\cup\mathFLboxcl{\mathdynbox{\pi}\mathdynbox{\mathprogramit{\pi}}\psi}$.
   \item $\mathFLboxcl{\mathdynbox{\mathprogramtest{\psi_{0}}}\psi_{1}}=\mathsetextension{\mathdynbox{\mathprogramtest{\psi_{0}}}\psi_{1}}\cup\mathFLcl{\psi_{0}}$.
   \item $\mathFLboxcl{\mathrevbox{\alpha}\psi}=\mathsetextension{\mathrevbox{\alpha}\psi}$
	 for $\alpha\in\maththesetofatomicprog$.
   \item $\mathFLboxcl{\mathrevbox{\mathprogramseq{\pi_{0}}{\pi_{1}}}\psi}=\mathsetextension{\mathrevbox{\mathprogramseq{\pi_{0}}{\pi_{1}}}\psi}\cup\mathFLboxcl{\mathrevbox{\pi_{1}}\mathrevbox{\pi_{0}}\psi}\cup\mathFLboxcl{\mathrevbox{\pi_{0}}\psi}$.
   \item $\mathFLboxcl{\mathrevbox{\mathprogramndc{\pi_{0}}{\pi_{1}}}\psi}=\mathsetextension{\mathrevbox{\mathprogramndc{\pi_{0}}{\pi_{1}}}\psi}\cup\mathFLboxcl{\mathrevbox{\pi_{0}}\psi}\cup\mathFLboxcl{\mathrevbox{\pi_{1}}\psi}$.
   \item $\mathFLboxcl{\mathrevbox{\mathprogramit{\pi}}\psi}=\mathsetextension{\mathrevbox{\mathprogramit{\pi}}\psi}\cup\mathFLboxcl{\mathrevbox{\pi}\mathrevbox{\mathprogramit{\pi}}\psi}$.
   \item $\mathFLboxcl{\mathrevbox{\mathprogramtest{\psi_{0}}}\psi_{1}}=\mathsetextension{\mathrevbox{\mathprogramtest{\psi_{0}}}\psi_{1}}\cup\mathFLcl{\psi_{0}}$.
  \end{enumerate}

  $\mathFLcl{\varphi}$ is called \emph{the Fischer–Ladner closure of $\varphi$}.
  For a set of formulae $\Lambda$, we define 
  $\mathFLcl{\Lambda}={\bigcup_{\varphi\in\Lambda}\mathFLcl{\varphi}}$.
 \end{definition}

 \begin{proposition}
  \label[proposition]{prop:FL-size}
  For any formula $\varphi$, 
  $\mathcardinality{\mathFLcl{\varphi}}\leq\mathlength{\varphi}$.
 \end{proposition}

 \begin{proof}
  Let $\varphi$ be a formula.
  We show the statements by induction on construction of $\varphi$.
  Consider cases according to the form of $\varphi$.
  
  Consider the case $\varphi\equiv\bot$.
  Then,
  \[
   \mathcardinality{\mathFLcl{\bot}}=\mathcardinality{\mathsetextension{\bot}}=1=\mathlength{\bot}.
  \]
  Specifically, $\mathcardinality{\mathFLcl{\bot}}\leq\mathlength{\bot}$.

  Consider the case $\varphi\equiv p$ for $p\in\maththesetofpropval$.
  Then,
  \[
   \mathcardinality{\mathFLcl{p}}=\mathcardinality{\mathsetextension{p}}=1=\mathlength{p}.
  \]
  Specifically, $\mathcardinality{\mathFLcl{p}}\leq\mathlength{p}$.

  Consider the case $\varphi\equiv \psi_{0}\to\psi_{1}$.
  Then, $\mathFLcl{\psi_{0}\to\psi_{1}}=\mathsetextension{\psi_{0}\to\psi_{1}}\cup\mathFLcl{\psi_{0}}\cup\mathFLcl{\psi_{1}}$.
  By the induction hypothesis, we have
  $\mathcardinality{\mathFLcl{\psi_{0}}}\leq\mathlength{\psi_{0}}$ and
  $\mathcardinality{\mathFLcl{\psi_{0}}}\leq\mathlength{\psi_{1}}$.
  Hence,
  \begin{align*}
   \mathcardinality{\mathFLcl{\psi_{0}\to\psi_{1}}}&\leq\mathcardinality{\mathsetextension{\psi_{0}\to\psi_{1}}}+\mathcardinality{\mathFLcl{\psi_{0}}}+\mathcardinality{\mathFLcl{\psi_{1}}} \\
   &\leq 1+\mathlength{\psi_{0}}+\mathlength{\psi_{1}} \\
   &\leq \mathlength{\varphi}.
  \end{align*}

  Consider the case $\varphi\equiv \mathdynbox{\pi}\psi$.
  We show  
  $\mathcardinality{\mathFLboxcl{\mathdynbox{\pi}\rho}}\leq\mathlength{\pi}$
  for any formula $\rho$ by construction of $\pi$.
  There are five cases.

  Consider the case $\pi\equiv\alpha$ for $\alpha\in\maththesetofatomicprog$.
  Then, $\mathFLboxcl{\mathdynbox{\alpha}\rho}=\mathsetextension{\mathdynbox{\alpha}\rho}$.
  Hence, we have
  \[
   \mathcardinality{\mathFLboxcl{\mathdynbox{\alpha}\rho}}=\mathcardinality{\mathsetextension{\mathdynbox{\alpha}\rho}}=1=\mathlength{\alpha}.
  \]
  Specifically, $\mathcardinality{\mathFLboxcl{\mathdynbox{\alpha}\rho}}\leq\mathlength{\alpha}$.

  Consider the case $\pi\equiv\mathprogramseq{\pi_{0}}{\pi_{1}}$.
  Then, 
  $\mathFLboxcl{\mathdynbox{\mathprogramseq{\pi_{0}}{\pi_{1}}}\rho}=\mathsetextension{\mathdynbox{\mathprogramseq{\pi_{0}}{\pi_{1}}}\rho}\cup\mathFLboxcl{\mathdynbox{\pi_{0}}\mathdynbox{\pi_{1}}\rho}\cup\mathFLboxcl{\mathdynbox{\pi_{1}}\rho}$.
  By the induction hypothesis, we have 
  $\mathcardinality{\mathFLboxcl{\mathdynbox{\pi_{0}}\mathdynbox{\pi_{1}}\rho}}\leq\mathlength{\pi_{0}}$ and
  $\mathcardinality{\mathFLboxcl{\mathdynbox{\pi_{1}}\rho}}\leq\mathlength{\pi_{1}}$.
  Hence, we have
  \begin{align*}
   \mathcardinality{\mathFLboxcl{\mathdynbox{\mathprogramseq{\pi_{0}}{\pi_{1}}}\rho}}
   &\leq\mathcardinality{\mathsetextension{\mathdynbox{\mathprogramseq{\pi_{0}}{\pi_{1}}}\rho}}+\mathcardinality{\mathFLboxcl{\mathdynbox{\pi_{0}}\mathdynbox{\pi_{1}}\rho}}+\mathcardinality{\mathFLboxcl{\mathdynbox{\pi_{1}}\rho}} \\
   &\leq 1+\mathlength{\pi_{0}}+\mathlength{\pi_{1}} \\
   &\leq \mathlength{\mathprogramseq{\pi_{0}}{\pi_{1}}}.
  \end{align*}

  Consider the case $\pi\equiv\mathprogramndc{\pi_{0}}{\pi_{1}}$.
  Then, 
  $\mathFLboxcl{\mathdynbox{\mathprogramndc{\pi_{0}}{\pi_{1}}}\rho}=\mathsetextension{\mathdynbox{\mathprogramndc{\pi_{0}}{\pi_{1}}}\rho}\cup\mathFLboxcl{\mathdynbox{\pi_{0}}\rho}\cup\mathFLboxcl{\mathdynbox{\pi_{1}}\rho}$.
  By the induction hypothesis, we have 
  $\mathcardinality{\mathFLboxcl{\mathdynbox{\pi_{0}}\rho}}\leq\mathlength{\pi_{0}}$ and
  $\mathcardinality{\mathFLboxcl{\mathdynbox{\pi_{1}}\rho}}\leq\mathlength{\pi_{1}}$.
  Hence, we have
  \begin{align*}
   \mathcardinality{\mathFLboxcl{\mathdynbox{\mathprogramndc{\pi_{0}}{\pi_{1}}}\rho}}
   &\leq\mathcardinality{\mathsetextension{\mathdynbox{\mathprogramndc{\pi_{0}}{\pi_{1}}}\rho}}+\mathcardinality{\mathFLboxcl{\mathdynbox{\pi_{0}}\rho}}+\mathcardinality{\mathFLboxcl{\mathdynbox{\pi_{1}}\rho}} \\
   &\leq 1+\mathlength{\pi_{0}}+\mathlength{\pi_{1}} \\
   &\leq \mathlength{\mathprogramndc{\pi_{0}}{\pi_{1}}}.
  \end{align*}

  Consider the case $\pi\equiv\mathprogramit{\pi_{0}}$.
  Then,
  $\mathFLboxcl{\mathdynbox{\mathprogramit{\pi_{0}}}\rho}=\mathsetextension{\mathdynbox{\mathprogramit{\pi_{0}}}\rho}\cup\mathFLboxcl{\mathdynbox{\pi_{0}}\mathdynbox{\mathprogramit{\pi_{0}}}\rho}$.
  By the induction hypothesis, we have 
  $\mathcardinality{\mathFLboxcl{\mathdynbox{\pi_{0}}\mathdynbox{\mathprogramit{\pi_{0}}}\rho}}\leq\mathlength{\pi_{0}}$.
  Hence, we have
  \begin{align*}
   \mathcardinality{\mathFLboxcl{\mathdynbox{\mathprogramit{\pi_{0}}}\rho}}
   &\leq\mathcardinality{\mathsetextension{\mathdynbox{\mathprogramit{\pi_{0}}}\rho}}+\mathcardinality{\mathFLboxcl{\mathdynbox{\pi_{0}}\mathdynbox{\mathprogramit{\pi_{0}}}\rho}} \\
   &\leq 1+\mathlength{\pi_{0}} \\
   &\leq \mathlength{\mathprogramit{\pi_{0}}}.
  \end{align*}

  Consider the case $\pi\equiv\mathprogramtest{\psi'}$.
  Then,
  $\mathFLboxcl{\mathdynbox{\mathprogramtest{\psi'}}\rho}=\mathsetextension{\mathdynbox{\mathprogramtest{\psi'}}\rho}\cup\mathFLcl{\psi'}$.
  By the induction hypothesis on $\varphi$, we have 
  $\mathcardinality{\mathFLcl{\psi'}}\leq\mathlength{\psi'}$.
  Hence, we have
  \begin{align*}
   \mathcardinality{\mathFLboxcl{\mathdynbox{\mathprogramtest{\psi'}}\rho}}
   &\leq\mathcardinality{\mathsetextension{\mathdynbox{\mathprogramtest{\psi'}}\rho}}+\mathcardinality{\mathFLcl{\psi'}} \\
   &\leq 1+\mathlength{\psi'} \\
   &\leq \mathlength{\mathprogramtest{\psi'}}.
  \end{align*}

  Then, we have $\mathcardinality{\mathFLboxcl{\mathdynbox{\pi}\psi}}\leq\mathlength{\pi}$.
  By the induction hypothesis, we have
  $\mathcardinality{\mathFLcl{\psi}}\leq\mathlength{\psi}$.
  Because of 
  $\mathFLcl{\mathdynbox{\pi}\psi}=\mathFLboxcl{\mathdynbox{\pi}\psi}\cup\mathFLcl{\psi}$, 
  we see
  \begin{align*}
   \mathcardinality{\mathFLcl{\mathdynbox{\pi}\psi}}&\leq\mathcardinality{\mathFLboxcl{\mathdynbox{\pi}\psi}}+\mathcardinality{\mathFLcl{\psi}} \\
   &\leq\mathlength{\pi}+\mathlength{\psi} \\
   &\leq\mathlength{\mathdynbox{\pi}\psi}.
  \end{align*}  
  
  In the case $\varphi\equiv \mathrevbox{\pi}\psi$,
  we can show the statement in the similar way to the case
  $\varphi\equiv \mathdynbox{\pi}\psi$.
 \end{proof}

 \begin{corollary}
  \label[corollary]{cor:Fischer-Ladner-closure-is-finite}
  For any formula $\varphi$, $\mathFLcl{\varphi}$ is finite.
 \end{corollary} 

 \begin{proof}
  By \cref{prop:FL-size}.
 \end{proof}

 \begin{proposition}
  \label[proposition]{prop:A-in-FL-A}
  $\varphi\in\mathFLcl{\varphi}$ for a formula $\varphi$.
 \end{proposition}

 \begin{proof}
  By induction on construction of $\varphi$.
 \end{proof}

 \begin{lemma}
  \label[lemma]{lemma:FL-and-box}
  $\mathFLboxcl{\varphi}\subseteq\mathFLcl{\varphi}$ for a formula $\varphi$.
 \end{lemma}

 \begin{proof}
  By induction on construction of $\varphi$.
 \end{proof}

 \begin{proposition}
  \label[proposition]{prop:FL-closure}
  For formulae $\varphi$ and $\psi$,
  if $\psi\in\mathFLcl{\varphi}$, then $\mathFLcl{\psi}\subseteq\mathFLcl{\varphi}$.
 \end{proposition}

 \begin{proof}
  \ifoup
  Straightforward.
  \else
  Assume $\psi\in\mathFLcl{\varphi}$.
  We show $\mathFLcl{\psi}\subseteq\mathFLcl{\varphi}$ 
  by induction on construction of $\varphi$.
  Consider cases according to the form of $\varphi$.
  
  Consider the case $\varphi\equiv\bot$.
  Then, $\mathFLcl{\varphi}=\mathsetextension{\bot}$.
  Because of $\psi\in\mathFLcl{\varphi}$, we have $\psi\equiv\bot$.
  Hence, $\mathFLcl{\psi}=\mathFLcl{\varphi}$.
  Specifically, $\mathFLcl{\psi}\subseteq\mathFLcl{\varphi}$.

  Consider the case $\varphi\equiv p$ for $p\in\maththesetofpropval$.
  Then, $\mathFLcl{\varphi}=\mathsetextension{p}$.
  Because of $\psi\in\mathFLcl{\varphi}$, we have $\psi\equiv p$.
  Hence, $\mathFLcl{\psi}=\mathFLcl{\varphi}$.
  Specifically, $\mathFLcl{\psi}\subseteq\mathFLcl{\varphi}$.

  Consider the case $\varphi\equiv \psi_{0}\to\psi_{1}$.
  Then, $\mathFLcl{\varphi}=\mathsetextension{\psi_{0}\to\psi_{1}}\cup\mathFLcl{\psi_{0}}\cup\mathFLcl{\psi_{1}}$.
  Hence, either $\psi\in\mathsetextension{\psi_{0}\to\psi_{1}}$ or
  $\psi\in\mathFLcl{\psi_{i}}$ for $i=0, 1$.
  If $\psi\in\mathsetextension{\psi_{0}\to\psi_{1}}$, 
  then $\mathFLcl{\psi}=\mathFLcl{\varphi}$ and therefore 
  $\mathFLcl{\psi}\subseteq\mathFLcl{\varphi}$.
  Assume $\psi\in\mathFLcl{\psi_{i}}$ for $i=0, 1$.
  By the induction hypothesis, $\mathFLcl{\psi}\subseteq\mathFLcl{\psi_{i}}$.
  Hence, we have $\mathFLcl{\psi}\subseteq\mathFLcl{\varphi}$.

  Consider the case $\varphi\equiv \mathdynbox{\pi}\psi'$.
  We show that $\rho'\in\mathFLboxcl{\mathdynbox{\pi}\rho}$ implies
  $\mathFLboxcl{\rho'}\subseteq\mathFLboxcl{\mathdynbox{\pi}\rho}$
  for any formula $\rho$ by construction of $\pi$.
  Assume $\rho'\in\mathFLboxcl{\mathdynbox{\pi}\rho}$. There are five cases.

  Consider the case $\pi\equiv\alpha$ for $\alpha\in\maththesetofatomicprog$.
  Then, $\mathFLboxcl{\mathdynbox{\alpha}\rho}=\mathsetextension{\mathdynbox{\alpha}\rho}$.
  Because of $\rho'\in\mathFLboxcl{\mathdynbox{\alpha}\rho}$, 
  we have $\rho'\equiv \mathdynbox{\alpha}\rho$.
  Hence, $\mathFLboxcl{\rho'}=\mathFLboxcl{\mathdynbox{\alpha}\rho}$.
  Specifically, $\mathFLboxcl{\rho'}\subseteq\mathFLboxcl{\mathdynbox{\alpha}\rho}$.

  Consider the case $\pi\equiv\mathprogramseq{\pi_{0}}{\pi_{1}}$.
  Then, 
  $\mathFLboxcl{\mathdynbox{\mathprogramseq{\pi_{0}}{\pi_{1}}}\rho}=\mathsetextension{\mathdynbox{\mathprogramseq{\pi_{0}}{\pi_{1}}}\rho}\cup\mathFLboxcl{\mathdynbox{\pi_{0}}\mathdynbox{\pi_{1}}\rho}\cup\mathFLboxcl{\mathdynbox{\pi_{1}}\rho}$.
  Because of $\rho'\in\mathFLboxcl{\mathdynbox{\mathprogramseq{\pi_{0}}{\pi_{1}}}\rho}$, 
  we have $\rho'\in\mathsetextension{\mathdynbox{\mathprogramseq{\pi_{0}}{\pi_{1}}}\rho}$
  or $\rho'\in\mathFLboxcl{\mathdynbox{\pi_{0}}\mathdynbox{\pi_{1}}\rho}$ or
  $\rho'\in\mathFLboxcl{\mathdynbox{\pi_{1}}\rho}$.
  If $\rho'\in\mathsetextension{\mathdynbox{\mathprogramseq{\pi_{0}}{\pi_{1}}}\rho}$,
  then $\mathFLboxcl{\rho'}=\mathFLboxcl{\mathdynbox{\mathprogramseq{\pi_{0}}{\pi_{1}}}\rho}$
  and therefore $\mathFLboxcl{\rho'}\subseteq\mathFLboxcl{\mathdynbox{\mathprogramseq{\pi_{0}}{\pi_{1}}}\rho}$.
  Assume $\rho'\in\mathFLboxcl{\mathdynbox{\pi_{0}}\mathdynbox{\pi_{1}}\rho}$.
  By the induction hypothesis,
  $\mathFLboxcl{\rho'}\subseteq\mathFLboxcl{\mathdynbox{\pi_{0}}\mathdynbox{\pi_{1}}\rho}$.
  Hence, we have  $\mathFLboxcl{\rho'}\subseteq\mathFLboxcl{\mathdynbox{\mathprogramseq{\pi_{0}}{\pi_{1}}}\rho}$.
  Assume $\rho'\in\mathFLboxcl{\mathdynbox{\pi_{1}}\rho}$.
  By the induction hypothesis,
  $\mathFLboxcl{\rho'}\subseteq\mathFLboxcl{\mathdynbox{\pi_{1}}\rho}$.
  Hence, we have  $\mathFLboxcl{\rho'}\subseteq\mathFLboxcl{\mathdynbox{\mathprogramseq{\pi_{0}}{\pi_{1}}}\rho}$.

  Consider the case $\pi\equiv\mathprogramndc{\pi_{0}}{\pi_{1}}$.
  Then, 
  $\mathFLboxcl{\mathdynbox{\mathprogramndc{\pi_{0}}{\pi_{1}}}\rho}=\mathsetextension{\mathdynbox{\mathprogramndc{\pi_{0}}{\pi_{1}}}\rho}\cup\mathFLboxcl{\mathdynbox{\pi_{0}}\rho}\cup\mathFLboxcl{\mathdynbox{\pi_{1}}\rho}$.
  Because of $\rho'\in\mathFLboxcl{\mathdynbox{\mathprogramndc{\pi_{0}}{\pi_{1}}}\rho}$, 
  we have $\rho'\in\mathsetextension{\mathdynbox{\mathprogramndc{\pi_{0}}{\pi_{1}}}\rho}$
  or $\rho'\in\mathFLboxcl{\mathdynbox{\pi_{i}}\rho}$ for $i=0, 1$.
  If $\rho'\in\mathsetextension{\mathdynbox{\mathprogramndc{\pi_{0}}{\pi_{1}}}\rho}$,
  then $\mathFLboxcl{\rho'}=\mathFLboxcl{\mathdynbox{\mathprogramndc{\pi_{0}}{\pi_{1}}}\rho}$
  and therefore $\mathFLboxcl{\rho'}\subseteq\mathFLboxcl{\mathdynbox{\mathprogramndc{\pi_{0}}{\pi_{1}}}\rho}$.
  Assume $\rho'\in\mathFLboxcl{\mathdynbox{\pi_{i}}\rho}$ for $i=0, 1$.
  By the induction hypothesis,
  $\mathFLboxcl{\rho'}\subseteq\mathFLboxcl{\mathdynbox{\pi_{i}}\rho}$.
  Hence, we have  $\mathFLboxcl{\rho'}\subseteq\mathFLboxcl{\mathdynbox{\mathprogramndc{\pi_{0}}{\pi_{1}}}\rho}$.

  Consider the case $\pi\equiv\mathprogramit{\pi_{0}}$.
  Then,
  $\mathFLboxcl{\mathdynbox{\mathprogramit{\pi_{0}}}\rho}=\mathsetextension{\mathdynbox{\mathprogramit{\pi_{0}}}\rho}\cup\mathFLboxcl{\mathdynbox{\pi_{0}}\mathdynbox{\mathprogramit{\pi_{0}}}\rho}$.
  Because of $\rho'\in\mathFLboxcl{\mathdynbox{\mathprogramndc{\pi_{0}}{\pi_{1}}}\rho}$, 
  we have $\rho'\in\mathsetextension{\mathdynbox{\mathprogramit{\pi_{0}}}\rho}$ or
  $\rho'\in\mathFLboxcl{\mathdynbox{\pi_{0}}\mathdynbox{\mathprogramit{\pi_{0}}}\rho}$.
  If $\rho'\in\mathsetextension{\mathdynbox{\mathprogramit{\pi_{0}}}\rho}$
  then $\mathFLboxcl{\rho'}=\mathFLboxcl{\mathdynbox{\mathprogramit{\pi_{0}}}\rho}$
  and therefore $\mathFLboxcl{\rho'}\subseteq\mathFLboxcl{\mathdynbox{\mathprogramit{\pi_{0}}}\rho}$.
  Assume 
  $\rho'\in\mathFLboxcl{\mathdynbox{\pi_{0}}\mathdynbox{\mathprogramit{\pi_{0}}}\rho}$.
  By the induction hypothesis,
  $\mathFLboxcl{\rho'}\subseteq\mathFLboxcl{\mathdynbox{\pi_{0}}\mathdynbox{\mathprogramit{\pi_{0}}}\rho}$.
  Hence, we have
  $\mathFLboxcl{\rho'}\subseteq\mathFLboxcl{\mathdynbox{\mathprogramit{\pi_{0}}}\rho}$.

  Consider the case $\pi\equiv\mathprogramtest{\psi''}$.
  Then,
  $\mathFLboxcl{\mathdynbox{\mathprogramtest{\psi''}}\rho}=\mathsetextension{\mathdynbox{\mathprogramtest{\psi''}}\rho}\cup\mathFLcl{\psi''}$.
  Because of $\rho'\in\mathFLboxcl{\mathdynbox{\mathprogramtest{\psi''}}\rho}$, 
  we have
  $\rho'\in\mathsetextension{\mathdynbox{\mathprogramtest{\psi''}}\rho}$ or
  $\rho'\in\mathFLcl{\psi''}$.
  If $\rho'\in\mathsetextension{\mathdynbox{\mathprogramtest{\psi''}}\rho}$,
  then $\mathFLboxcl{\rho'}=\mathFLboxcl{\mathdynbox{\mathprogramtest{\psi''}}\rho}$
  and therefore $\mathFLboxcl{\rho'}\subseteq\mathFLboxcl{\mathdynbox{\mathprogramtest{\psi''}}\rho}$.
  Assume $\rho'\in\mathFLcl{\psi''}$.
  By the induction hypothesis on $\varphi$,
  we have $\mathFLcl{\rho'}\subseteq\mathFLcl{\psi''}$.
  By \cref{lemma:FL-and-box},
  $\mathFLboxcl{\rho'}\subseteq\mathFLcl{\rho'}$.
  Therefore, we have 
  $\mathFLboxcl{\rho'}\subseteq\mathFLboxcl{\mathdynbox{\mathprogramtest{\psi''}}\rho}$.
  
  Now, we show $\mathFLcl{\psi}\subseteq\mathFLcl{\mathdynbox{\pi}\psi'}$
  by construction on $\psi$.

  If $\psi\equiv\bot$ or $\psi\equiv p$ for $p\in\maththesetofpropval$,
  then we have $\mathsetextension{\psi}=\mathFLcl{\psi}\subseteq\mathFLboxcl{\mathdynbox{\pi}\psi'}$.

  Assume $\psi\equiv \psi'_{0}\to\psi'_{1}$.
  Then, $\mathFLcl{\psi}=\mathsetextension{ \psi'_{0}\to\psi'_{1}}\cup\mathFLcl{\psi'_{0}}\cup\mathFLcl{\psi'_{1}}$.
  By the induction hypothesis, we have
  $\mathFLcl{\psi'_{i}}\subseteq\mathFLcl{\mathdynbox{\pi}\psi'}$ for $i=0, 1$.
  Hence,
  $\mathFLcl{\psi}\subseteq\mathFLcl{\mathdynbox{\pi}\psi'}$.

  Assume $\psi\equiv \mathdynbox{\pi'}\psi'''$.
  By the induction hypothesis,
  $\mathFLcl{\psi'''}\subseteq\mathFLcl{\mathdynbox{\pi}\psi'}$.

  Because of 
  $\mathFLcl{\mathdynbox{\pi}\psi'}=\mathFLboxcl{\mathdynbox{\pi}\psi'}\cup\mathFLcl{\psi'}$,
  we have $\psi\in\mathFLboxcl{\mathdynbox{\pi}\psi'}$ or
  $\psi\in\mathFLcl{\psi'}$.

  Assume $\psi\in\mathFLcl{\psi'}$.
  By the induction hypothesis on $\varphi$,
  $\mathFLcl{\psi}\subseteq\mathFLcl{\psi'}$.
  Hence, we have $\mathFLcl{\psi}\subseteq\mathFLcl{\varphi}$.

  Assume $\psi\in\mathFLboxcl{\mathdynbox{\pi}\psi'}$.
  Because of $\psi\in\mathFLboxcl{\mathdynbox{\pi}\psi'}$,
  we have $\mathFLboxcl{\mathdynbox{\pi'}\psi'''}\subseteq\mathFLboxcl{\mathdynbox{\pi}\psi'}$.
  Then, we have
  \begin{align*}
   \mathFLcl{\mathdynbox{\pi'}\psi'''}&=\mathFLboxcl{\mathdynbox{\pi'}\psi'''}\cup\mathFLcl{\psi'''} \\
   &\subseteq\mathFLboxcl{\mathdynbox{\pi}\psi'}\cup\mathFLcl{\mathdynbox{\pi}\psi'} \\
   &\subseteq\mathFLcl{\mathdynbox{\pi}\psi'}.
  \end{align*}

  In the case $\psi\equiv \mathrevbox{\pi'}\psi'''$,
  we can show
  $\mathFLcl{\psi}\subseteq\mathFLcl{\mathdynbox{\pi}\psi'}$
  in the similar way to the case $\psi\equiv \mathdynbox{\pi'}\psi'''$.
  Thus, we have $\mathFLcl{\psi}\subseteq\mathFLcl{\varphi}$ in the case
  $\varphi\equiv \mathdynbox{\pi}\psi'$.

  In the case $\varphi\equiv \mathrevbox{\pi}\psi'$,
  we can show $\mathFLcl{\psi}\subseteq\mathFLcl{\varphi}$ in the similar way to the case
  $\varphi\equiv \mathdynbox{\pi}\psi'$.
  \fi
 \end{proof}

 \begin{corollary}
  \label[corollary]{cor:Fischer-Ladner-property-formula}
  Let $\varphi$ be a formula.
  \begin{enumerate}
   \item If ${\psi_{0}\to\psi_{1}}\in\mathFLcl{\varphi}$, then ${\psi_{0}}\in\mathFLcl{\varphi}$ and ${\psi_{1}}\in\mathFLcl{\varphi}$.
   \item If ${\mathdynbox{\pi}\psi}\in \mathFLcl{\varphi}$, then ${\psi}\in\mathFLcl{\varphi}$.
   \item If ${\mathdynbox{\mathprogramseq{\pi_{0}}{\pi_{1}}}\psi}\in\mathFLcl{\varphi}$, 
	 then ${\mathdynbox{\pi_{0}}\mathdynbox{\pi_{1}}\psi}\in\mathFLcl{\varphi}$.
   \item If ${\mathdynbox{\mathprogramndc{\pi_{0}}{\pi_{1}}}\psi}\in\mathFLcl{\varphi}$, 
	 then ${\mathdynbox{\pi_{i}}\psi}\in\mathFLcl{\varphi}$ for all $i\in\mathsetextension{0, 1}$.
   \item If ${\mathdynbox{\mathprogramit{\pi}}\psi}\in\mathFLcl{\varphi}$, 
	 then ${\mathdynbox{\pi}\mathdynbox{\mathprogramit{\pi}}\psi}\in\mathFLcl{\varphi}$.
   \item If ${\mathdynbox{\mathprogramtest{\psi_{0}}}\psi_{1}}\in \mathFLcl{\varphi}$, 
	 then $\psi_{0}\in\mathFLcl{\varphi}$.
   \item If ${\mathrevbox{\pi}\psi}\in\mathFLcl{\varphi}$, then ${\psi}\in\mathFLcl{\varphi}$.
   \item If ${\mathrevbox{\mathprogramseq{\pi_{0}}{\pi_{1}}}\psi}\in\mathFLcl{\varphi}$, 
	 then ${\mathrevbox{\pi_{1}}\mathrevbox{\pi_{0}}\psi}\in \mathFLcl{\varphi}$.
   \item If ${\mathrevbox{\mathprogramndc{\pi_{0}}{\pi_{1}}}\psi}\in \mathFLcl{\varphi}$, 
	 then ${\mathrevbox{\pi_{i}}\psi}\in\mathFLcl{\varphi}$ for all $i\in\mathsetextension{0, 1}$.
   \item If ${\mathrevbox{\mathprogramit{\pi}}\psi}\in\mathFLcl{\varphi}$, 
	 then ${\mathrevbox{\pi}\mathrevbox{\mathprogramit{\pi}}\psi}\in\mathFLcl{\varphi}$.
   \item If ${\mathrevbox{\mathprogramtest{\psi_{0}}}\psi_{1}}\in\mathFLcl{\varphi}$, 
	 then $\psi_{0}\in\mathFLcl{\varphi}$.
  \end{enumerate}
 \end{corollary} 

 \begin{proof}
  By \cref{prop:A-in-FL-A} and \cref{prop:FL-closure}.
 \end{proof}

 \begin{corollary}
  \label[corollary]{cor:Fischer-Ladner-property-set}
  Let $\Lambda$ be a set of formulae.
  \begin{enumerate}
   \item If ${\psi_{0}\to\psi_{1}}\in\mathFLcl{\Lambda}$, then ${\psi_{0}}\in\mathFLcl{\Lambda}$ and ${\psi_{1}}\in\mathFLcl{\Lambda}$.
	 \label{item:implication-cor-Fischer-Ladner-property-set}
   \item If ${\mathdynbox{\pi}\psi}\in \mathFLcl{\Lambda}$, then ${\psi}\in\mathFLcl{\Lambda}$.
	 \label{item:dynbox-cor-Fischer-Ladner-property-set}
   \item If ${\mathdynbox{\mathprogramseq{\pi_{0}}{\pi_{1}}}\psi}\in\mathFLcl{\Lambda}$, 
	 then ${\mathdynbox{\pi_{0}}\mathdynbox{\pi_{1}}\psi}\in\mathFLcl{\Lambda}$.
	 \label{item:dynbox-seq-cor-Fischer-Ladner-property-set}
   \item If ${\mathdynbox{\mathprogramndc{\pi_{0}}{\pi_{1}}}\psi}\in\mathFLcl{\Lambda}$, 
	 then ${\mathdynbox{\pi_{i}}\psi}\in\mathFLcl{\Lambda}$ for all $i\in\mathsetextension{0, 1}$.
	 \label{item:dynbox-ndc-cor-Fischer-Ladner-property-set}
   \item If ${\mathdynbox{\mathprogramit{\pi}}\psi}\in\mathFLcl{\Lambda}$, 
	 then ${\mathdynbox{\pi}\mathdynbox{\mathprogramit{\pi}}\psi}\in\mathFLcl{\Lambda}$.
	 \label{item:dynbox-it-cor-Fischer-Ladner-property-set}
   \item If ${\mathdynbox{\mathprogramtest{\psi_{0}}}\psi_{1}}\in \mathFLcl{\Lambda}$, 
	 then $\psi_{0}\in\mathFLcl{\Lambda}$.
	 \label{item:dynbox-test-cor-Fischer-Ladner-property-set},
   \item If ${\mathrevbox{\pi}\psi}\in\mathFLcl{\Lambda}$, then ${\psi}\in\mathFLcl{\Lambda}$.
	 \label{item:revbox-cor-Fischer-Ladner-property-set}
   \item If ${\mathrevbox{\mathprogramseq{\pi_{0}}{\pi_{1}}}\psi}\in\mathFLcl{\Lambda}$, 
	 then ${\mathrevbox{\pi_{1}}\mathrevbox{\pi_{0}}\psi}\in \mathFLcl{\Lambda}$.
	 \label{item:revbox-seq-cor-Fischer-Ladner-property-set}
   \item If ${\mathrevbox{\mathprogramndc{\pi_{0}}{\pi_{1}}}\psi}\in \mathFLcl{\Lambda}$, 
	 then ${\mathrevbox{\pi_{i}}\psi}\in\mathFLcl{\Lambda}$ for all $i\in\mathsetextension{0, 1}$.
	 \label{item:revbox-ndc-cor-Fischer-Ladner-property-set}
   \item If ${\mathrevbox{\mathprogramit{\pi}}\psi}\in\mathFLcl{\Lambda}$, 
	 then ${\mathrevbox{\pi}\mathrevbox{\mathprogramit{\pi}}\psi}\in\mathFLcl{\Lambda}$.
	 \label{item:revbox-it-cor-Fischer-Ladner-property-set}
   \item If ${\mathrevbox{\mathprogramtest{\psi_{0}}}\psi_{1}}\in\mathFLcl{\Lambda}$, 
	 then $\psi_{0}\in\mathFLcl{\Lambda}$.
	 \label{item:revbox-test-cor-Fischer-Ladner-property-set},
  \end{enumerate}
 \end{corollary} 

 \begin{proof}
  By \cref{cor:Fischer-Ladner-property-formula}.
 \end{proof}

  \subsection{The canonical counter model}
  \label{subsec:canonical-counter-model}
  This section defines the canonical counter model of an unprovable sequent and 
  shows its property.

 \begin{definition}[Saturated sequent]
  \label[definition]{def:saturated-sequent}
  We say that $\Gamma\fCenter\Delta$ is a saturated sequent
  if the following conditions hold:
  \begin{enumerate}
   \item ${\Gamma\cup\Delta}=\mathFLcl{\Gamma\cup\Delta}$.
   \item $\Gamma\fCenter\Delta$ is not provable in \GTPDL.
  \end{enumerate}
 \end{definition}

 We note that $\Gamma\cap\Delta=\emptyset$ holds 
 for a saturated sequent $\Gamma\fCenter\Delta$
 since $\Gamma\fCenter\Delta$ is not provable in \GTPDL.
 
 The following lemma means that an unprovable sequent can be expanded to 
 a saturated sequent.
 \begin{lemma}
  \label[lemma]{lemma:not-provable-sequent-to-saturated-sequent}
  Assume that $\Gamma\fCenter\Delta$ is not provable in \GTPDL.
  There is a saturated sequent $\tilde{\Gamma}\fCenter\tilde{\Delta}$ 
  such that $\Gamma\subseteq\tilde{\Gamma}$ and $\Delta\subseteq\tilde{\Delta}$.
 \end{lemma}

 \begin{proof}
  By \cref{cor:Fischer-Ladner-closure-is-finite},
  $\mathFLcl{\Gamma\cup\Delta}$ is finite.
  Then, ${\Gamma\cup\Delta}\subseteq\mathFLcl{\Gamma\cup\Delta}$ and 
  \cref{lemma:not-provable-sequent-can-expand-set} imply the statement.
 \end{proof}

 Noting that we can expand the sequent to the saturated sequent,
 it suffices to define the canonical counter model of only the saturated sequent.

 \begin{definition}[Canonical counter model]
  \label[definition]{def:canonical-counter-model}
  For a saturated sequent $\Gamma\fCenter\Delta$, we define a model 
  $M_{\Gamma\fCenter\Delta}=\mathtuple{W_{\Gamma\fCenter\Delta}, \mathsequence{{\mathreduction{\alpha}}_{\Gamma\fCenter\Delta}}{\alpha\in\maththesetofatomicprog}, V_{\Gamma\fCenter\Delta}}$,
  called the \emph{canonical counter model} of $\Gamma\fCenter\Delta$,
  as follows:
  \begin{enumerate}
   \item $W_{\Gamma\fCenter\Delta}=\mathsetintension{(\Pi\fCenter\Sigma)}{\Pi\cup\Sigma=\Gamma\cup\Delta, \text{ and } \Pi\fCenter\Sigma \text{ is not provable in \GTPDL}}$.
   \item $(\Pi_{0}\fCenter\Sigma_{0}){\mathreduction{\alpha}}_{\Gamma\fCenter\Delta}(\Pi_{1}\fCenter\Sigma_{1})$ 
	 if and only if
	 $\Pi_{0}\fCenter\Sigma_{0}, \mathdynbox{\alpha}\mathseqtoform{\Pi_{1}\fCenter\Sigma_{1}}$ 
	 is not provable in \GTPDL.
	 \label{item:reduction-def-canonical-counter-model}
   \item $\mathof{V_{\Gamma\fCenter\Delta}}{(\Pi\fCenter\Sigma)}=\mathsetintension{p\in\maththesetofpropval}{p\in\Pi}$.
  \end{enumerate}
 \end{definition}

 It seems strange that there is no direct condition for backwards modal operators. 
 Noting that the provability of 
 $\Pi_{0}\fCenter\Sigma_{0}, \mathdynbox{\alpha}\mathseqtoform{\Pi_{1}\fCenter\Sigma_{1}}$ 
 is equivalent to that of 
 $\Pi_{1}\fCenter\Sigma_{1}, \mathrevbox{\alpha}\mathseqtoform{\Pi_{0}\fCenter\Sigma_{0}}$,
 we can consider \cref{item:reduction-def-canonical-counter-model} in 
 \cref{def:canonical-counter-model} to be a condition for backwards modal operators.

 We note that $\Pi\fCenter\Sigma$ is a saturated sequent for all 
 $(\Pi\fCenter\Sigma)\in W_{\Gamma\fCenter\Delta}$ because 
 ${\Pi\cup\Sigma}={\Gamma\cup\Delta}=\mathFLcl{\Gamma\cup\Delta}=\mathFLcl{\Pi\cup\Sigma}$ 
 holds. 

 \begin{lemma}
  \label[lemma]{lemma:saturated-sequents-in-antecedent}
  Let $\Gamma\fCenter\Delta$ be a saturated sequent and
  $M=\mathtuple{W, \mathsequence{{\mathreduction{\alpha}}}{\alpha\in\maththesetofatomicprog}, V}$
  be the canonical counter model of $\Gamma\fCenter\Delta$.
  Let $\Xi=\mathsetintension{\mathseqtoform{\Pi\fCenter\Sigma}}{(\Pi\fCenter\Sigma)\in W}$.
  $\Xi\fCenter$ is provable in \GTPDL.
 \end{lemma}

 \begin{proof}
  Let $\Gamma\fCenter\Delta$ be a saturated sequent and
  $M=\mathtuple{W, \mathsequence{{\mathreduction{\alpha}}}{\alpha\in\maththesetofatomicprog}, V}$
  be the canonical counter model of $\Gamma\fCenter\Delta$.
  Let $\Xi=\mathsetintension{\mathseqtoform{\Pi\fCenter\Sigma}}{(\Pi\fCenter\Sigma)\in W}$.

  From \cref{prop:divisions-in-antecedent},
  $\mathof{D}{\Gamma\cup\Delta}\fCenter$ is provable in \GTPDL.
  By \cref{def:canonical-counter-model},
  $\Pi\fCenter\Sigma$ is provable in \GTPDL\
  for each $\mathseqtoform{\Pi\fCenter\Sigma}\in\mathof{D}{\Gamma\cup\Delta}\setminus\Xi$.
  Hence, \cref{prop:hyper-sequent} implies that
  $\fCenter\mathseqtoform{\Pi\fCenter\Sigma}$ is provable in \GTPDL\
  for each $\mathseqtoform{\Pi\fCenter\Sigma}\in\mathof{D}{\Gamma\cup\Delta}\setminus\Xi$.
  We have $\Xi\fCenter$
  from $\mathof{D}{\Gamma\cup\Delta}\fCenter$ by repeating application of \rulename{Cut} 
  with $\fCenter\mathseqtoform{\Pi\fCenter\Sigma}$
  for all $\Pi\fCenter\Sigma\in\mathof{D}{\Gamma\cup\Delta}\setminus\Xi$.
 \end{proof}

 \begin{lemma}
  \label[lemma]{lemma:key-lemma-completeness-GTPDL}
  Let $\Gamma\fCenter\Delta$ be a saturated sequent and
  $M=\mathtuple{W, \mathsequence{{\mathreduction{\alpha}}}{\alpha\in\maththesetofatomicprog}, V}$
  be the canonical counter model of $\Gamma\fCenter\Delta$.
  Let $\lambda$ be a well-formed expression such that $\lambda\in{\Gamma\cup\Delta}$ or
  either $\mathdynbox{\lambda}\varphi\in{\Gamma\cup\Delta}$ or 
  $\mathrevbox{\lambda}\varphi\in{\Gamma\cup\Delta}$ with some formula $\varphi$. 
  The following statements hold:
  \begin{enumerate}
   \item For ${(\Pi\fCenter\Sigma)}\in W$,
	 if $\lambda\in\Pi$, then $M, {(\Pi\fCenter\Sigma)}\models\lambda$.
	 \label{item:antecedent-lemma-key-lemma-completeness-GTPDL}
   \item For ${(\Pi\fCenter\Sigma)}\in W$,
	 if $\lambda\in\Sigma$, then $M, {(\Pi\fCenter\Sigma)}\not\models\lambda$.
	 \label{item:consequent-lemma-key-lemma-completeness-GTPDL}
   \item For ${(\Pi_{0}\fCenter\Sigma_{0})}$, ${(\Pi_{1}\fCenter\Sigma_{1})}\in W$,
	 if $\mathdynbox{\lambda}\varphi\in{\Pi_{0}}$ and 
	 $(\Pi_{0}\fCenter\Sigma_{0}){\mathreduction{\lambda}}(\Pi_{1}\fCenter\Sigma_{1})$,
	 then $\varphi\in{\Pi_{1}}$.
	 \label{item:dynbox-lemma-key-lemma-completeness-GTPDL}
   \item For ${(\Pi_{0}\fCenter\Sigma_{0})}$, ${(\Pi_{1}\fCenter\Sigma_{1})}\in W$,
	 if $\mathrevbox{\lambda}\varphi\in{\Pi_{1}}$ and 
	 $(\Pi_{0}\fCenter\Sigma_{0}){\mathreduction{\lambda}}(\Pi_{1}\fCenter\Sigma_{1})$,
	 then $\varphi\in{\Pi_{0}}$.
	 \label{item:revbox-lemma-key-lemma-completeness-GTPDL}
   \item For ${(\Pi_{0}\fCenter\Sigma_{0})}$, ${(\Pi_{1}\fCenter\Sigma_{1})}\in W$, if
	 $\Pi_{0}\fCenter\Sigma_{0}, \mathdynbox{\lambda}\mathseqtoform{\Pi_{1}\fCenter\Sigma_{1}}$
	 is not provable in \GTPDL, then
	 $(\Pi_{0}\fCenter\Sigma_{0}){\mathreduction{\lambda}}(\Pi_{1}\fCenter\Sigma_{1})$.
	 \label{item:well-reduction-lemma-key-lemma-completeness-GTPDL}
  \end{enumerate}
 \end{lemma}

 \begin{proof}
  Let $\Gamma\fCenter\Delta$ be a saturated sequent and
  $M=\mathtuple{W, \mathsequence{{\mathreduction{\alpha}}}{\alpha\in\maththesetofatomicprog}, V}$
  be the canonical counter model of $\Gamma\fCenter\Delta$.
  Let $\lambda$ be a well-formed expression such that $\lambda\in{\Gamma\cup\Delta}$ or
  either $\mathdynbox{\lambda}\varphi\in{\Gamma\cup\Delta}$ or 
  $\mathrevbox{\lambda}\varphi\in{\Gamma\cup\Delta}$ with some formula $\varphi$. 
  We show the statements by induction on construction of $\lambda$.

  Consider the case $\lambda\equiv\bot$.
  In this case, there is nothing to prove for 
  \cref{item:dynbox-lemma-key-lemma-completeness-GTPDL},
  \cref{item:revbox-lemma-key-lemma-completeness-GTPDL}, and
  \cref{item:well-reduction-lemma-key-lemma-completeness-GTPDL}.

  \noindent \cref{item:antecedent-lemma-key-lemma-completeness-GTPDL}
  Since $\Pi\fCenter\Sigma$ is not provable in \GTPDL\ for all ${(\Pi\fCenter\Sigma)}\in W$,
  we have $\bot\notin\Pi$ for all ${(\Pi\fCenter\Sigma)}\in W$.
  Hence, there is nothing to prove.

  \noindent \cref{item:consequent-lemma-key-lemma-completeness-GTPDL}
  Let ${(\Pi\fCenter\Sigma)}\in W$. Assume $\lambda\in\Sigma$.
  By $\lambda\equiv\bot$, $M, {(\Pi\fCenter\Sigma)}\not\models\lambda$.

  Consider the case $\lambda\equiv p$ for $p\in\maththesetofpropval$.
  In this case, there is nothing to prove for 
  \cref{item:dynbox-lemma-key-lemma-completeness-GTPDL},
  \cref{item:revbox-lemma-key-lemma-completeness-GTPDL}, and
  \cref{item:well-reduction-lemma-key-lemma-completeness-GTPDL}.

  \noindent \cref{item:antecedent-lemma-key-lemma-completeness-GTPDL}
  Let ${(\Pi\fCenter\Sigma)}\in W$. Assume $\lambda\in\Pi$. 
  By \cref{def:canonical-counter-model},
  $\mathof{V}{(\Pi\fCenter\Sigma)}=\mathsetintension{p\in\maththesetofpropval}{p\in\Pi}$.
  Hence, $p\in\mathof{V}{(\Pi\fCenter\Sigma)}$.
  Thus, $M, {(\Pi\fCenter\Sigma)}\models\lambda$.

  \noindent \cref{item:consequent-lemma-key-lemma-completeness-GTPDL}
  Let ${(\Pi\fCenter\Sigma)}\in W$. Assume $\lambda\in\Sigma$.
  Since $\Pi\fCenter\Sigma$ is not provable in \GTPDL,
  we have $\lambda\notin\Pi$.
  Hence, $p\notin\mathof{V}{(\Pi\fCenter\Sigma)}$.
  Thus, $M, {(\Pi\fCenter\Sigma)}\not\models\lambda$.

  Consider the case $\lambda\equiv \psi_{0}\to\psi_{1}$. 
  In this case, there is nothing to prove for 
  \cref{item:dynbox-lemma-key-lemma-completeness-GTPDL},
  \cref{item:revbox-lemma-key-lemma-completeness-GTPDL}, and
  \cref{item:well-reduction-lemma-key-lemma-completeness-GTPDL}.

  \noindent \cref{item:antecedent-lemma-key-lemma-completeness-GTPDL}
  Let ${(\Pi\fCenter\Sigma)}\in W$. Assume $\lambda\in\Pi$.
  Then, $\lambda\in\mathFLcl{\Pi\cup\Sigma}$.
  By \cref{cor:Fischer-Ladner-property-set} \cref{item:implication-cor-Fischer-Ladner-property-set},
  $\psi_{0}, \psi_{1}\in\mathFLcl{\Pi\cup\Sigma}$. 
  Since $\Pi\fCenter\Sigma$ is a saturated sequent, 
  we have $\psi_{0}, \psi_{1}\in\mathFLcl{\Pi\cup\Sigma}$. 

  By the induction hypothesis \cref{item:consequent-lemma-key-lemma-completeness-GTPDL},
  $M, {(\Pi\fCenter\Sigma)}\not\models\psi_{0}$ if $\psi_{0}\in\Sigma$.
  Hence, $M, {(\Pi\fCenter\Sigma)}\models\lambda$ if $\psi_{0}\in\Sigma$.

  Assume $\psi_{0}\notin\Sigma$. Then, $\psi_{0}\in\Pi$.
  If $\psi_{1}\in\Sigma$, then $\Pi\fCenter\Sigma$ is provable in \GTPDL\ as follows:
  \begin{center}
   \begin{inlineprooftree}
    \AxiomC{}
    \RightLabel{\rulename{Ax}}
    \UnaryInfC{$\bar{\Pi}, \psi_{1},\psi_{0} \fCenter \psi_{0}, \bar{\Sigma}, \psi_{1}$}
    \AxiomC{}
    \RightLabel{\rulename{Ax}}
    \UnaryInfC{$\bar{\Pi},\psi_{0}, \psi_{1}\fCenter\bar{\Sigma}, \psi_{1}$}
    \RightLabel{\rulename{$\to$ L}}
    \BinaryInfC{$\bar{\Pi}, \psi_{0}, \psi_{0}\to\psi_{1}\fCenter\bar{\Sigma}, \psi_{1}$}
   \end{inlineprooftree},
  \end{center}
  where $\Pi=\bar{\Pi}, \psi_{0}, \psi_{0}\to\psi_{1}$ and $\Sigma=\bar{\Sigma}, \psi_{1}$.
  Hence, $\psi_{1}\notin\Sigma$. Therefore, $\psi_{1}\in\Pi$.
  By the induction hypothesis \cref{item:antecedent-lemma-key-lemma-completeness-GTPDL}, 
  $M, {(\Pi\fCenter\Sigma)}\models\psi_{1}$.
  Hence, $M, {(\Pi\fCenter\Sigma)}\models\lambda$.

  \noindent \cref{item:consequent-lemma-key-lemma-completeness-GTPDL}
  Let ${(\Pi\fCenter\Sigma)}\in W$. Assume $\lambda\in\Sigma$.
  Then, $\lambda\in\mathFLcl{\Pi\cup\Sigma}$.
  By \cref{cor:Fischer-Ladner-property-set},
  $\psi_{0}, \psi_{1}\in\mathFLcl{\Pi\cup\Sigma}$. 
  Since $\Pi\fCenter\Sigma$ is a saturated sequent, 
  we have $\psi_{0}, \psi_{1}\in\mathFLcl{\Pi\cup\Sigma}$. 

  If $\psi_{0}\in\Sigma$, then $\Pi\fCenter\Sigma$ is provable in \GTPDL\ as follows:
  \begin{center}
   \begin{inlineprooftree}
    \AxiomC{}
    \RightLabel{\rulename{Ax}}
    \UnaryInfC{$\Pi, \psi_{0}\fCenter\psi_{1}, \bar{\Sigma}, \psi_{0}$}
    \RightLabel{\rulename{$\to$ R}}
    \UnaryInfC{$\Pi\fCenter\psi_{0}\to\psi_{1}, \bar{\Sigma}, \psi_{0}$}
   \end{inlineprooftree},
  \end{center}
  where $\Sigma=\psi_{0}\to\psi_{1}, \bar{\Sigma}, \psi_{0}$.
  Because $\Pi\fCenter\Sigma$ is not provable in \GTPDL, we have $\psi_{0}\notin\Sigma$. 
  Then, $\psi_{0}\in\Pi$.
  If $\psi_{1}\in\Pi$, then $\Pi\fCenter\Sigma$ is provable in \GTPDL\ as follows:
  \begin{center}
   \begin{inlineprooftree}
    \AxiomC{}
    \RightLabel{\rulename{Ax}}
    \UnaryInfC{$\bar{\Pi}, \psi_{1}, \psi_{0} \fCenter \psi_{1}, \bar{\Sigma}$}
    \RightLabel{\rulename{$\to$ R}}
    \UnaryInfC{$\bar{\Pi}, \psi_{1}\fCenter \psi_{0}\to\psi_{1}, \bar{\Sigma}$}
   \end{inlineprooftree},
  \end{center}
  where $\Pi=\bar{\Pi}, \psi_{1}$ and $\Sigma=\psi_{0}\to\psi_{1}, \bar{\Sigma}$.
  Hence, $\psi_{1}\notin\Pi$. Therefore, $\psi_{1}\in\Sigma$.
  By the induction hypothesis \cref{item:antecedent-lemma-key-lemma-completeness-GTPDL}, 
  $M, {(\Pi\fCenter\Sigma)}\models\psi_{0}$.
  From the induction hypothesis \cref{item:consequent-lemma-key-lemma-completeness-GTPDL}, 
  $M, {(\Pi\fCenter\Sigma)}\not\models\psi_{1}$.
  Hence, $M, {(\Pi\fCenter\Sigma)}\not\models\lambda$.

  Consider the case $\lambda\equiv \mathdynbox{\pi}\psi$.
  In this case, there is nothing to prove for 
  \cref{item:dynbox-lemma-key-lemma-completeness-GTPDL},
  \cref{item:revbox-lemma-key-lemma-completeness-GTPDL}, and
  \cref{item:well-reduction-lemma-key-lemma-completeness-GTPDL}.

  \noindent \cref{item:antecedent-lemma-key-lemma-completeness-GTPDL}
  Let ${(\Pi\fCenter\Sigma)}\in W$. Assume $\lambda\in\Pi$.
  By the induction hypothesis \cref{item:dynbox-lemma-key-lemma-completeness-GTPDL}, 
  $\psi\in{\Pi'}$ for all $\Pi'\fCenter\Sigma'$ with
  $(\Pi\fCenter\Sigma){\mathreduction{\pi}}(\Pi'\fCenter\Sigma')$.
  By the induction hypothesis \cref{item:antecedent-lemma-key-lemma-completeness-GTPDL},
  $M, {(\Pi'\fCenter\Sigma')}\models\psi$ for all  $\Pi'\fCenter\Sigma'$ with
  $(\Pi\fCenter\Sigma){\mathreduction{\pi}}(\Pi'\fCenter\Sigma')$.
  Hence, we have $M, {(\Pi\fCenter\Sigma)}\models\lambda$.

  \noindent \cref{item:consequent-lemma-key-lemma-completeness-GTPDL}
  Let ${(\Pi\fCenter\Sigma)}\in W$. Assume $\lambda\in\Sigma$.
  Since $\mathdynbox{\pi}\mathseqtoform{\fCenter\psi}\fCenter\mathdynbox{\pi}\psi$ 
  is provable in \GTPDL,
  we see that $\Pi\fCenter\Sigma, \mathdynbox{\pi}\mathseqtoform{\fCenter\psi}$ 
  is not provable in \GTPDL.
  By \cref{lemma:not-provable-sequent-can-expand-set-in-box},
  there exists $(\tilde{\Pi}\fCenter\tilde{\Sigma})\in W$
  such that $\mathsetextension{\psi}\subseteq\tilde{\Sigma}$ holds and
  $\Pi\fCenter\Sigma, \mathdynbox{\pi}\mathseqtoform{\tilde{\Pi}\fCenter\tilde{\Sigma}}$
  is not provable in \GTPDL.
  By the induction hypothesis \cref{item:well-reduction-lemma-key-lemma-completeness-GTPDL},
  we have $(\Pi\fCenter\Sigma){\mathreduction{\pi}}(\tilde{\Pi}\fCenter\tilde{\Sigma})$.
  By the induction hypothesis \cref{item:consequent-lemma-key-lemma-completeness-GTPDL},
  we see $M, {(\tilde{\Pi}\fCenter\tilde{\Sigma})}\not\models\psi$.
  Thus, $M, {(\Pi\fCenter\Sigma)}\not\models\lambda$. 

  Consider the case $\lambda\equiv \mathrevbox{\pi}\psi$.
  In this case, there is nothing to prove for 
  \cref{item:dynbox-lemma-key-lemma-completeness-GTPDL},
  \cref{item:revbox-lemma-key-lemma-completeness-GTPDL}, and
  \cref{item:well-reduction-lemma-key-lemma-completeness-GTPDL}.

  \noindent \cref{item:antecedent-lemma-key-lemma-completeness-GTPDL}
  Let ${(\Pi\fCenter\Sigma)}\in W$. Assume $\lambda\in\Pi$.
  By the induction hypothesis \cref{item:revbox-lemma-key-lemma-completeness-GTPDL}, 
  $\psi\in{\Pi'}$ for all $\Pi'\fCenter\Sigma'$ with
  $(\Pi'\fCenter\Sigma'){\mathreduction{\pi}}(\Pi\fCenter\Sigma)$.
  By the induction hypothesis \cref{item:antecedent-lemma-key-lemma-completeness-GTPDL},
  $M, {(\Pi'\fCenter\Sigma')}\models\psi$ for all $\Pi'\fCenter\Sigma'$ with
  $(\Pi'\fCenter\Sigma'){\mathreduction{\pi}}(\Pi\fCenter\Sigma)$.
  Hence, we have $M, {(\Pi\fCenter\Sigma)}\models\lambda$.
  
  \noindent \cref{item:consequent-lemma-key-lemma-completeness-GTPDL}
  Let ${(\Pi\fCenter\Sigma)}\in W$. Assume $\lambda\in\Sigma$.
  Since $\mathrevbox{\pi}\mathseqtoform{\fCenter\psi}\fCenter\mathrevbox{\pi}\psi$ 
  is provable in \GTPDL,
  we see that $\Pi\fCenter\Sigma, \mathrevbox{\pi}\mathseqtoform{\fCenter\psi}$ 
  is not provable in \GTPDL.
  By \cref{lemma:not-provable-sequent-can-expand-set-in-revbox},
  there exists $(\tilde{\Pi}\fCenter\tilde{\Sigma})\in W$
  such that $\mathsetextension{\psi}\subseteq\tilde{\Sigma}$ holds and
  $\Pi\fCenter\Sigma, \mathrevbox{\pi}\mathseqtoform{\tilde{\Pi}\fCenter\tilde{\Sigma}}$
  is not provable in \GTPDL.
  From  \cref{lemma:switching-lemma},
  $\tilde{\Pi}\fCenter\tilde{\Sigma}, \mathdynbox{\pi}\mathseqtoform{\Pi\fCenter\Sigma}$
  is not provable in \GTPDL.
  By the induction hypothesis \cref{item:well-reduction-lemma-key-lemma-completeness-GTPDL},
  we have $(\tilde{\Pi}\fCenter\tilde{\Sigma}){\mathreduction{\pi}}(\Pi\fCenter\Sigma)$.
  By the induction hypothesis \cref{item:consequent-lemma-key-lemma-completeness-GTPDL},
  we see $M, {(\tilde{\Pi}\fCenter\tilde{\Sigma})}\not\models\psi$.
  Thus, $M, {(\Pi\fCenter\Sigma)}\not\models\lambda$.   

  Consider the case $\lambda\equiv\alpha$ for $\alpha\in\maththesetofatomicprog$.
  In this case, there is nothing to prove for
  \cref{item:antecedent-lemma-key-lemma-completeness-GTPDL} and
  \cref{item:consequent-lemma-key-lemma-completeness-GTPDL}.

  \noindent \cref{item:dynbox-lemma-key-lemma-completeness-GTPDL}
  Assume $\mathdynbox{\lambda}\varphi\in{\Pi_{0}}$ and 
  $(\Pi_{0}\fCenter\Sigma_{0}){\mathreduction{\lambda}}(\Pi_{1}\fCenter\Sigma_{1})$
  for ${(\Pi_{0}\fCenter\Sigma_{0})}$, ${(\Pi_{1}\fCenter\Sigma_{1})}\in W$.
  By \cref{def:canonical-counter-model},
  $\Pi_{0}\fCenter\Sigma_{0}, \mathdynbox{\alpha}\mathseqtoform{\Pi_{1}\fCenter\Sigma_{1}}$
  is not provable in \GTPDL.
  Assume, for contradiction, $\varphi\notin\Pi_{1}$. Then, $\varphi\in\Sigma_{1}$.
  Since $\Pi_{0}\fCenter\Sigma_{0}, \mathdynbox{\alpha}\varphi$ is provable in \GTPDL,
  $\Pi_{0}\fCenter\Sigma_{0}, \mathdynbox{\alpha}\mathseqtoform{\fCenter\varphi}$ 
  is provable in \GTPDL.
  By $\varphi\in\Sigma_{1}$ and \cref{lemma:weakening-in-box},
  $\Pi_{0}\fCenter\Sigma_{0}, \mathdynbox{\alpha}\mathseqtoform{\Pi_{1}\fCenter\Sigma_{1}}$
  is provable in \GTPDL, which is a contradiction.
  Thus, $\varphi\in\Pi_{1}$.

  \noindent \cref{item:revbox-lemma-key-lemma-completeness-GTPDL}
  Assume $\mathrevbox{\lambda}\varphi\in{\Pi_{1}}$ and 
  $(\Pi_{0}\fCenter\Sigma_{0}){\mathreduction{\lambda}}(\Pi_{1}\fCenter\Sigma_{1})$
  for ${(\Pi_{0}\fCenter\Sigma_{0})}$, ${(\Pi_{1}\fCenter\Sigma_{1})}\in W$.
  By \cref{def:canonical-counter-model},
  $\Pi_{0}\fCenter\Sigma_{0}, \mathdynbox{\alpha}\mathseqtoform{\Pi_{1}\fCenter\Sigma_{1}}$
  is not provable in \GTPDL.
  Assume, for contradiction, $\varphi\notin\Pi_{0}$. Then, $\varphi\in\Sigma_{0}$.
  Since $\Pi_{1}\fCenter\Sigma_{1}, \mathrevbox{\alpha}\varphi$ is provable in \GTPDL,
  $\Pi_{1}\fCenter\Sigma_{1}, \mathrevbox{\alpha}\mathseqtoform{\fCenter\varphi}$ 
  is provable in \GTPDL.
  By $\varphi\in\Sigma_{0}$ and \cref{lemma:weakening-in-revbox},
  $\Pi_{1}\fCenter\Sigma_{1}, \mathrevbox{\alpha}\mathseqtoform{\Pi_{0}\fCenter\Sigma_{0}}$
  is provable in \GTPDL.
  By \cref{lemma:switching-lemma},
  $\Pi_{0}\fCenter\Sigma_{0}, \mathdynbox{\alpha}\mathseqtoform{\Pi_{1}\fCenter\Sigma_{1}}$
  is provable in \GTPDL, which is a contradiction.
  Thus, $\varphi\in\Pi_{0}$.

  \noindent \cref{item:well-reduction-lemma-key-lemma-completeness-GTPDL}
  Assume that
  $\Pi_{0}\fCenter\Sigma_{0}, \mathdynbox{\lambda}\mathseqtoform{\Pi_{1}\fCenter\Sigma_{1}}$
  is not provable in \GTPDL\
  for ${(\Pi_{0}\fCenter\Sigma_{0})}$, ${(\Pi_{1}\fCenter\Sigma_{1})}\in W$.
  By \cref{def:canonical-counter-model}, we have
  $(\Pi_{0}\fCenter\Sigma_{0}){\mathreduction{\lambda}}(\Pi_{1}\fCenter\Sigma_{1})$.

  Consider the case $\lambda\equiv\mathprogramseq{\pi_{0}}{\pi_{1}}$.
  In this case, there is nothing to prove for
  \cref{item:antecedent-lemma-key-lemma-completeness-GTPDL} and
  \cref{item:consequent-lemma-key-lemma-completeness-GTPDL}.

  \noindent \cref{item:dynbox-lemma-key-lemma-completeness-GTPDL}
  Assume $\mathdynbox{\lambda}\varphi\in{\Pi_{0}}$ and 
  $(\Pi_{0}\fCenter\Sigma_{0}){\mathreduction{\lambda}}(\Pi_{1}\fCenter\Sigma_{1})$
  for ${(\Pi_{0}\fCenter\Sigma_{0})}$, ${(\Pi_{1}\fCenter\Sigma_{1})}\in W$.
  By \cref{cor:Fischer-Ladner-property-set} 
  \cref{item:dynbox-seq-cor-Fischer-Ladner-property-set},
  $\mathdynbox{\pi_{0}}\mathdynbox{\pi_{1}}\varphi\in\mathFLcl{\Gamma\cup\Delta}$.
  If $\mathdynbox{\pi_{0}}\mathdynbox{\pi_{1}}\varphi\in{\Sigma_{0}}$,
  then $\Pi_{0}\fCenter\Sigma_{0}$ is provable in \GTPDL\ as follows:
  \begin{center}
   \begin{inlineprooftree}
    \AxiomC{}
    \RightLabel{\rulename{Ax}}
    \UnaryInfC{$\bar{\Pi}_{0}, \mathdynbox{\pi_{0}}\mathdynbox{\pi_{1}}\varphi \fCenter\bar{\Sigma}_{0}, \mathdynbox{\pi_{0}}\mathdynbox{\pi_{1}}\varphi$}
    \RightLabel{\rulename{$\mathdynbox{\mathprogramseq{}{}}$ L}}
    \UnaryInfC{$\bar{\Pi}_{0}, \mathdynbox{\mathprogramseq{\pi_{0}}{\pi_{1}}}\varphi \fCenter\bar{\Sigma}_{0}, \mathdynbox{\pi_{0}}\mathdynbox{\pi_{1}}\varphi$}
   \end{inlineprooftree},
  \end{center}
  where $\Pi_{0}=\bar{\Pi}_{0}, \mathdynbox{\mathprogramseq{\pi_{0}}{\pi_{1}}}\varphi$ and $\Sigma_{0}=\bar{\Sigma}_{0}, \mathdynbox{\pi_{0}}\mathdynbox{\pi_{1}}\varphi$.
  Hence, $\mathdynbox{\pi_{0}}\mathdynbox{\pi_{1}}\varphi\notin{\Sigma_{0}}$,
  that is, $\mathdynbox{\pi_{0}}\mathdynbox{\pi_{1}}\varphi\in{\Pi_{0}}$.
  Because of 
  $(\Pi_{0}\fCenter\Sigma_{0}){\mathreduction{\mathprogramseq{\pi_{0}}{\pi_{1}}}}(\Pi_{1}\fCenter\Sigma_{1})$,
  there exists $(\Pi_{2}\fCenter\Sigma_{2})$ such that
  $(\Pi_{0}\fCenter\Sigma_{0}){\mathreduction{\pi_{0}}}(\Pi_{2}\fCenter\Sigma_{2})$ 
  and
  $(\Pi_{2}\fCenter\Sigma_{2}){\mathreduction{\pi_{1}}}(\Pi_{1}\fCenter\Sigma_{1})$.
  By the induction hypothesis \cref{item:dynbox-lemma-key-lemma-completeness-GTPDL},
  we have $\mathdynbox{\pi_{1}}\varphi\in{\Pi_{2}}$.
  Then, the induction hypothesis \cref{item:dynbox-lemma-key-lemma-completeness-GTPDL} implies
  $\varphi\in{\Pi_{1}}$.

  \noindent \cref{item:revbox-lemma-key-lemma-completeness-GTPDL}
  Assume $\mathrevbox{\lambda}\varphi\in{\Pi_{1}}$ and 
  $(\Pi_{0}\fCenter\Sigma_{0}){\mathreduction{\lambda}}(\Pi_{1}\fCenter\Sigma_{1})$
  for ${(\Pi_{0}\fCenter\Sigma_{0})}$, ${(\Pi_{1}\fCenter\Sigma_{1})}\in W$.
  By \cref{cor:Fischer-Ladner-property-set} 
  \cref{item:revbox-seq-cor-Fischer-Ladner-property-set},
  $\mathrevbox{\pi_{1}}\mathrevbox{\pi_{0}}\varphi\in\mathFLcl{\Gamma\cup\Delta}$.
  If $\mathrevbox{\pi_{1}}\mathrevbox{\pi_{0}}\varphi\in{\Sigma_{1}}$,
  then $\Pi_{1}\fCenter\Sigma_{1}$ is provable in \GTPDL\ as follows:
  \begin{center}
   \begin{inlineprooftree}
    \AxiomC{\cref{fig:revbox-seq-left}}
    \AxiomC{\cref{fig:revbox-seq-right}}
    \RightLabel{\rulename{Cut+Wk}}
    \BinaryInfC{$\mathrevbox{\mathprogramseq{\pi_{0}}{\pi_{1}}}\varphi \fCenter \mathrevbox{\pi_{1}}\mathrevbox{\pi_{0}}\varphi$}
    \RightLabel{\rulename{Wk}}
    \UnaryInfC{$\bar{\Pi}_{1}, \mathrevbox{\mathprogramseq{\pi_{0}}{\pi_{1}}}\varphi \fCenter\bar{\Sigma}_{1}, \mathrevbox{\pi_{1}}\mathrevbox{\pi_{0}}\varphi$}
   \end{inlineprooftree},
  \end{center}
  where $\Pi_{1}=\bar{\Pi}_{1}, \mathrevbox{\mathprogramseq{\pi_{0}}{\pi_{1}}}\varphi$ and $\Sigma_{1}=\bar{\Sigma}_{1}, \mathrevbox{\pi_{1}}\mathrevbox{\pi_{0}}\varphi$.
  Hence, $\mathrevbox{\pi_{1}}\mathrevbox{\pi_{0}}\varphi\notin{\Sigma_{1}}$,
  that is, $\mathrevbox{\pi_{1}}\mathrevbox{\pi_{0}}\varphi\in{\Pi_{1}}$.
  Because of 
  $(\Pi_{0}\fCenter\Sigma_{0}){\mathreduction{\mathprogramseq{\pi_{0}}{\pi_{1}}}}(\Pi_{1}\fCenter\Sigma_{1})$,
  there exists $(\Pi_{2}\fCenter\Sigma_{2})$ such that
  $(\Pi_{0}\fCenter\Sigma_{0}){\mathreduction{\pi_{0}}}(\Pi_{2}\fCenter\Sigma_{2})$ 
  and
  $(\Pi_{2}\fCenter\Sigma_{2}){\mathreduction{\pi_{1}}}(\Pi_{1}\fCenter\Sigma_{1})$.
  By the induction hypothesis \cref{item:revbox-lemma-key-lemma-completeness-GTPDL},
  we have $\mathrevbox{\pi_{0}}\varphi\in{\Pi_{2}}$.
  Then, the induction hypothesis \cref{item:revbox-lemma-key-lemma-completeness-GTPDL} implies
  $\varphi\in{\Pi_{0}}$.
 
  \begin{figure}[tb]
   \begin{prooftree}
    \AxiomC{}
    \RightLabel{\rulename{Ax}}
    \UnaryInfC{$\mathrevbox{\mathprogramseq{\pi_{0}}{\pi_{1}}}\varphi \fCenter \mathrevbox{\mathprogramseq{\pi_{0}}{\pi_{1}}}\varphi$}
    \RightLabel{\rulename{$\lnot$ L}}
    \UnaryInfC{$\mathrevbox{\mathprogramseq{\pi_{0}}{\pi_{1}}}\varphi, \lnot\mathrevbox{\mathprogramseq{\pi_{0}}{\pi_{1}}}\varphi \fCenter$}
    \RightLabel{\rulename{$\lnot$ R}}
    \UnaryInfC{$\mathrevbox{\mathprogramseq{\pi_{0}}{\pi_{1}}}\varphi \fCenter \lnot\lnot\mathrevbox{\mathprogramseq{\pi_{0}}{\pi_{1}}}\varphi$}

    \AxiomC{}
    \RightLabel{\rulename{Ax}}
    \UnaryInfC{$\mathdynbox{\pi_{0}}\mathdynbox{\pi_{1}}\lnot\mathrevbox{\mathprogramseq{\pi_{0}}{\pi_{1}}}\varphi \fCenter \mathdynbox{\pi_{0}}\mathdynbox{\pi_{1}}\lnot\mathrevbox{\mathprogramseq{\pi_{0}}{\pi_{1}}}\varphi$}
    \RightLabel{\rulename{$\mathdynbox{\mathprogramseq{}{}}$ L}}
    \UnaryInfC{$\mathdynbox{\mathprogramseq{\pi_{0}}{\pi_{1}}}\lnot\mathrevbox{\mathprogramseq{\pi_{0}}{\pi_{1}}}\varphi \fCenter \mathdynbox{\pi_{0}}\mathdynbox{\pi_{1}}\lnot\mathrevbox{\mathprogramseq{\pi_{0}}{\pi_{1}}}\varphi$}
    \RightLabel{\rulename{$\lnot$ R}}
    \UnaryInfC{$\fCenter \mathdyndia{\mathprogramseq{\pi_{0}}{\pi_{1}}}\mathrevbox{\mathprogramseq{\pi_{0}}{\pi_{1}}}\varphi, \mathdynbox{\pi_{0}}\mathdynbox{\pi_{1}}\lnot\mathrevbox{\mathprogramseq{\pi_{0}}{\pi_{1}}}\varphi$}
    \RightLabel{\rulename{$\mathrevbox{ \, }$}}
    \UnaryInfC{$\fCenter \mathrevbox{\pi_{0}}\mathdyndia{\mathprogramseq{\pi_{0}}{\pi_{1}}}\mathrevbox{\mathprogramseq{\pi_{0}}{\pi_{1}}}\varphi, \mathdynbox{\pi_{1}}\lnot\mathrevbox{\mathprogramseq{\pi_{0}}{\pi_{1}}}\varphi$}
    \RightLabel{\rulename{$\mathrevbox{ \, }$}}
    \UnaryInfC{$\fCenter \mathrevbox{\pi_{1}}\mathrevbox{\pi_{0}}\mathdyndia{\mathprogramseq{\pi_{0}}{\pi_{1}}}\mathrevbox{\mathprogramseq{\pi_{0}}{\pi_{1}}}\varphi, \lnot\mathrevbox{\mathprogramseq{\pi_{0}}{\pi_{1}}}\varphi$}
    \RightLabel{\rulename{$\lnot$ L}}
    \UnaryInfC{$\lnot\lnot\mathrevbox{\mathprogramseq{\pi_{0}}{\pi_{1}}}\varphi \fCenter \mathrevbox{\pi_{1}}\mathrevbox{\pi_{0}}\mathdyndia{\mathprogramseq{\pi_{0}}{\pi_{1}}}\mathrevbox{\mathprogramseq{\pi_{0}}{\pi_{1}}}\varphi$}

    \RightLabel{\rulename{Cut+Wk}}
    \BinaryInfC{$\mathrevbox{\mathprogramseq{\pi_{0}}{\pi_{1}}}\varphi \fCenter \mathrevbox{\pi_{1}}\mathrevbox{\pi_{0}}\mathdyndia{\mathprogramseq{\pi_{0}}{\pi_{1}}}\mathrevbox{\mathprogramseq{\pi_{0}}{\pi_{1}}}\varphi$}
   \end{prooftree}

   \caption{A proof of $\mathrevbox{\mathprogramseq{\pi_{0}}{\pi_{1}}}\varphi \fCenter \mathrevbox{\pi_{1}}\mathrevbox{\pi_{0}}\mathdyndia{\mathprogramseq{\pi_{0}}{\pi_{1}}}\mathrevbox{\mathprogramseq{\pi_{0}}{\pi_{1}}}\varphi$}
   \label{fig:revbox-seq-left}

   \begin{prooftree}
    \AxiomC{}
    \UnaryInfC{$\mathrevbox{\mathprogramseq{\pi_{0}}{\pi_{1}}}\varphi \fCenter \mathrevbox{\mathprogramseq{\pi_{0}}{\pi_{1}}}\varphi$}
    \RightLabel{\rulename{$\lnot$ R}}
    \UnaryInfC{$\fCenter \mathrevbox{\mathprogramseq{\pi_{0}}{\pi_{1}}}\varphi, \lnot\mathrevbox{\mathprogramseq{\pi_{0}}{\pi_{1}}}\varphi$}
    \RightLabel{\rulename{$\mathdynbox{ \, }$}}
    \UnaryInfC{$\fCenter \varphi, \mathdynbox{\mathprogramseq{\pi_{0}}{\pi_{1}}}\lnot\mathrevbox{\mathprogramseq{\pi_{0}}{\pi_{1}}}\varphi$}
    \RightLabel{\rulename{$\lnot$ L}}
    \UnaryInfC{$\mathdyndia{\mathprogramseq{\pi_{0}}{\pi_{1}}}\mathrevbox{\mathprogramseq{\pi_{0}}{\pi_{1}}}\varphi \fCenter \varphi$}
    \RightLabel{\rulename{$\mathrevbox{ \, }$}}
    \UnaryInfC{$\mathrevbox{\pi_{0}}\mathdyndia{\mathprogramseq{\pi_{0}}{\pi_{1}}}\mathrevbox{\mathprogramseq{\pi_{0}}{\pi_{1}}}\varphi \fCenter \mathrevbox{\pi_{0}}\varphi$}
    \RightLabel{\rulename{$\mathrevbox{ \, }$}}
    \UnaryInfC{$\mathrevbox{\pi_{1}}\mathrevbox{\pi_{0}}\mathdyndia{\mathprogramseq{\pi_{0}}{\pi_{1}}}\mathrevbox{\mathprogramseq{\pi_{0}}{\pi_{1}}}\varphi \fCenter \mathrevbox{\pi_{1}}\mathrevbox{\pi_{0}}\varphi$}
   \end{prooftree}
   \caption{A proof of $\mathrevbox{\pi_{1}}\mathrevbox{\pi_{0}}\mathdyndia{\mathprogramseq{\pi_{0}}{\pi_{1}}}\mathrevbox{\mathprogramseq{\pi_{0}}{\pi_{1}}}\varphi \fCenter \mathrevbox{\pi_{1}}\mathrevbox{\pi_{0}}\varphi$}
   \label{fig:revbox-seq-right}
  \end{figure}

  \noindent \cref{item:well-reduction-lemma-key-lemma-completeness-GTPDL}
  Assume that
  $\Pi_{0}\fCenter\Sigma_{0}, \mathdynbox{\lambda}\mathseqtoform{\Pi_{1}\fCenter\Sigma_{1}}$
  is not provable in \GTPDL\
  for ${(\Pi_{0}\fCenter\Sigma_{0})}$, ${(\Pi_{1}\fCenter\Sigma_{1})}\in W$.
  Since $\Pi_{0}\fCenter\Sigma_{0}, \mathdynbox{\mathprogramseq{\pi_{0}}{\pi_{1}}}\mathseqtoform{\Pi_{1}\fCenter\Sigma_{1}}$
  is not provable in \GTPDL,
  $\Pi_{0}\fCenter\Sigma_{0}, \mathdynbox{\pi_{0}}\mathdynbox{\pi_{1}}\mathseqtoform{\Pi_{1}\fCenter\Sigma_{1}}$
  is not provable in \GTPDL.
  Hence,
  $\Pi_{0}\fCenter\Sigma_{0}, \mathdynbox{\pi_{0}}\mathseqtoform{\fCenter\mathdynbox{\pi_{1}}\mathseqtoform{\Pi_{1}\fCenter\Sigma_{1}}}$
  is not provable in \GTPDL.
  By \cref{lemma:not-provable-sequent-can-expand-set-in-box},
  there is a sequent $\tilde{\Pi}\fCenter\tilde{\Sigma}, \mathdynbox{\pi_{1}}\mathseqtoform{\Pi_{1}\fCenter\Sigma_{1}}$
  such that 
  $\tilde{\Pi}\cup\tilde{\Sigma}=\Gamma\cup\Delta$ holds, and
  $\Pi_{0}\fCenter\Sigma_{0}, \mathdynbox{\pi_{0}}\mathseqtoform{\tilde{\Pi}\fCenter\tilde{\Sigma}, \mathdynbox{\pi_{1}}\mathseqtoform{\Pi_{1}\fCenter\Sigma_{1}}}$
  and $\tilde{\Pi}\fCenter\tilde{\Sigma}, \mathdynbox{\pi_{1}}\mathseqtoform{\Pi_{1}\fCenter\Sigma_{1}}$ are not provable in \GTPDL.
  If $\Pi_{0}\fCenter\Sigma_{0}, \mathdynbox{\pi_{0}}\mathseqtoform{\tilde{\Pi}\fCenter\tilde{\Sigma}}$
  is provable in \GTPDL,
  \cref{lemma:weakening-in-box} implies that
  $\Pi_{0}\fCenter\Sigma_{0}, \mathdynbox{\pi_{0}}\mathseqtoform{\tilde{\Pi}\fCenter\tilde{\Sigma}, \mathdynbox{\pi_{1}}\mathseqtoform{\Pi_{1}\fCenter\Sigma_{1}}}$
  is provable in \GTPDL.
  Hence, $\Pi_{0}\fCenter\Sigma_{0}, \mathdynbox{\pi_{0}}\mathseqtoform{\tilde{\Pi}\fCenter\tilde{\Sigma}}$
  is not provable in \GTPDL.
  By the induction hypothesis \cref{item:well-reduction-lemma-key-lemma-completeness-GTPDL},
  we have
  $(\Pi_{0}\fCenter\Sigma_{0}){\mathreduction{\pi_{0}}}(\tilde{\Pi}\fCenter\tilde{\Sigma})$
  and
  $(\tilde{\Pi}\fCenter\tilde{\Sigma}){\mathreduction{\pi_{1}}}(\Pi_{1}\fCenter\Sigma_{1})$.
  Thus, $(\Pi_{0}\fCenter\Sigma_{0}){\mathreduction{\lambda}}(\Pi_{1}\fCenter\Sigma_{1})$.

  Consider the case $\lambda\equiv\mathprogramndc{\pi_{0}}{\pi_{1}}$.
  In this case, there is nothing to prove for
  \cref{item:antecedent-lemma-key-lemma-completeness-GTPDL} and
  \cref{item:consequent-lemma-key-lemma-completeness-GTPDL}.
 
  \noindent \cref{item:dynbox-lemma-key-lemma-completeness-GTPDL}
  Assume $\mathdynbox{\lambda}\varphi\in{\Pi_{0}}$ and 
  $(\Pi_{0}\fCenter\Sigma_{0}){\mathreduction{\lambda}}(\Pi_{1}\fCenter\Sigma_{1})$
  for ${(\Pi_{0}\fCenter\Sigma_{0})}$, ${(\Pi_{1}\fCenter\Sigma_{1})}\in W$.
  By \cref{cor:Fischer-Ladner-property-set} 
  \cref{item:dynbox-ndc-cor-Fischer-Ladner-property-set},
  $\mathdynbox{\pi_{i}}\varphi\in\mathFLcl{\Gamma\cup\Delta}$
  for all $i\in\mathsetextension{0, 1}$.
  If $\mathdynbox{\pi_{i}}\varphi\in{\Sigma_{0}}$ for some $i\in\mathsetextension{0, 1}$,
  then $\Pi_{0}\fCenter\Sigma_{0}$ is provable in \GTPDL\ as follows: 
  \begin{center} 
   \begin{inlineprooftree}
    \AxiomC{}
    \RightLabel{\rulename{Ax}}
    \UnaryInfC{$\bar{\Pi}_{0}, \mathdynbox{\pi_{i}}\varphi \fCenter\bar{\Sigma}_{0}, \mathdynbox{\pi_{i}}\varphi$}
    \RightLabel{\rulename{$\mathdynbox{\mathprogramseq{}{}}$ L}}
    \UnaryInfC{$\bar{\Pi}_{0}, \mathdynbox{\mathprogramndc{\pi_{0}}{\pi_{1}}}\varphi \fCenter\bar{\Sigma}_{0}, \mathdynbox{\pi_{i}}\varphi$}
   \end{inlineprooftree},
  \end{center}
  where $\Pi_{0}=\bar{\Pi}_{0}, \mathdynbox{\mathprogramndc{\pi_{0}}{\pi_{1}}}\varphi$ and $\Sigma_{0}=\bar{\Sigma}_{0}, \mathdynbox{\pi_{i}}\varphi$.
  Hence, $\mathdynbox{\pi_{i}}\varphi\notin{\Sigma_{0}}$,
  that is, $\mathdynbox{\pi_{i}}\varphi\in{\Pi_{0}}$ for all $i\in\mathsetextension{0, 1}$.
  Because of 
  $(\Pi_{0}\fCenter\Sigma_{0}){\mathreduction{\mathprogramndc{\pi_{0}}{\pi_{1}}}}(\Pi_{1}\fCenter\Sigma_{1})$,
  we have
  $(\Pi_{0}\fCenter\Sigma_{0}){\mathreduction{\pi_{0}}}(\Pi_{1}\fCenter\Sigma_{1})$ or
  $(\Pi_{0}\fCenter\Sigma_{0}){\mathreduction{\pi_{1}}}(\Pi_{1}\fCenter\Sigma_{1})$.
  By the induction hypothesis \cref{item:dynbox-lemma-key-lemma-completeness-GTPDL},
  we have $\varphi\in{\Pi_{1}}$.

  \noindent \cref{item:revbox-lemma-key-lemma-completeness-GTPDL}
  Assume $\mathrevbox{\lambda}\varphi\in{\Pi_{1}}$ and 
  $(\Pi_{0}\fCenter\Sigma_{0}){\mathreduction{\lambda}}(\Pi_{1}\fCenter\Sigma_{1})$
  for ${(\Pi_{0}\fCenter\Sigma_{0})}$, ${(\Pi_{1}\fCenter\Sigma_{1})}\in W$.
  By \cref{cor:Fischer-Ladner-property-set} 
  \cref{item:revbox-ndc-cor-Fischer-Ladner-property-set},
  $\mathrevbox{\pi_{i}}\varphi\in\mathFLcl{\Gamma\cup\Delta}$ 
  for all $i\in\mathsetextension{0, 1}$.
  If $\mathrevbox{\pi_{i}}\varphi\in{\Sigma_{1}}$ 
  for some $i\in\mathsetextension{0, 1}$,
  then $\Pi_{1}\fCenter\Sigma_{1}$ is provable in \GTPDL\ as follows:
  \begin{center}
   \begin{inlineprooftree}
    \AxiomC{\cref{fig:revbox-ndc-left}}
    \AxiomC{\cref{fig:revbox-ndc-right}}
    \RightLabel{\rulename{Cut+Wk}}
    \BinaryInfC{$\mathrevbox{\mathprogramndc{\pi_{0}}{\pi_{1}}}\varphi \fCenter \mathrevbox{\pi_{i}}\varphi$}
    \RightLabel{\rulename{Wk}}
    \UnaryInfC{$\bar{\Pi}_{1}, \mathrevbox{\mathprogramndc{\pi_{0}}{\pi_{1}}}\varphi \fCenter\bar{\Sigma}_{1}, \mathrevbox{\pi_{i}}\varphi$}
   \end{inlineprooftree},
  \end{center}
  where $\Pi_{1}=\bar{\Pi}_{1}, \mathrevbox{\mathprogramndc{\pi_{0}}{\pi_{1}}}\varphi$ and $\Sigma_{1}=\bar{\Sigma}_{1}, \mathrevbox{\pi_{i}}\varphi$.
  Hence, $\mathrevbox{\pi_{i}}\varphi\notin{\Sigma_{1}}$,
  that is, $\mathrevbox{\pi_{i}}\varphi\in{\Pi_{1}}$
  for all $i\in\mathsetextension{0, 1}$.
  Because of 
  $(\Pi_{0}\fCenter\Sigma_{0}){\mathreduction{\mathprogramndc{\pi_{0}}{\pi_{1}}}}(\Pi_{1}\fCenter\Sigma_{1})$,
  we have
  $(\Pi_{0}\fCenter\Sigma_{0}){\mathreduction{\pi_{0}}}(\Pi_{1}\fCenter\Sigma_{1})$ or
  $(\Pi_{0}\fCenter\Sigma_{0}){\mathreduction{\pi_{1}}}(\Pi_{1}\fCenter\Sigma_{1})$.
  By the induction hypothesis \cref{item:revbox-lemma-key-lemma-completeness-GTPDL},
  we have $\varphi\in{\Pi_{0}}$.
 
  \begin{figure}[tb]
   \begin{prooftree}
    \AxiomC{}
    \RightLabel{\rulename{Ax}}
    \UnaryInfC{$\mathrevbox{\mathprogramndc{\pi_{0}}{\pi_{1}}}\varphi \fCenter \mathrevbox{\mathprogramndc{\pi_{0}}{\pi_{1}}}\varphi$}
    \RightLabel{\rulename{$\lnot$ L}}
    \UnaryInfC{$\mathrevbox{\mathprogramndc{\pi_{0}}{\pi_{1}}}\varphi, \lnot\mathrevbox{\mathprogramndc{\pi_{0}}{\pi_{1}}}\varphi \fCenter$}
    \RightLabel{\rulename{$\lnot$ R}}
    \UnaryInfC{$\mathrevbox{\mathprogramndc{\pi_{0}}{\pi_{1}}}\varphi \fCenter \lnot\lnot\mathrevbox{\mathprogramndc{\pi_{0}}{\pi_{1}}}\varphi$}

    \AxiomC{}
    \RightLabel{\rulename{Ax}}
    \UnaryInfC{$\mathdynbox{\pi_{i}}\lnot\mathrevbox{\mathprogramndc{\pi_{0}}{\pi_{1}}}\varphi \fCenter \mathdynbox{\pi_{i}}\lnot\mathrevbox{\mathprogramndc{\pi_{0}}{\pi_{1}}}\varphi$}
    \RightLabel{\rulename{$\mathdynbox{\mathprogramndc{}{}}$ L + Wk}}
    \UnaryInfC{$\mathdynbox{\mathprogramndc{\pi_{0}}{\pi_{1}}}\lnot\mathrevbox{\mathprogramndc{\pi_{0}}{\pi_{1}}}\varphi \fCenter \mathdynbox{\pi_{i}}\lnot\mathrevbox{\mathprogramndc{\pi_{0}}{\pi_{1}}}\varphi$}
    \RightLabel{\rulename{$\lnot$ R}}
    \UnaryInfC{$\fCenter \mathdyndia{\mathprogramndc{\pi_{0}}{\pi_{1}}}\mathrevbox{\mathprogramndc{\pi_{0}}{\pi_{1}}}\varphi, \mathdynbox{\pi_{i}}\lnot\mathrevbox{\mathprogramseq{\pi_{0}}{\pi_{1}}}\varphi$}
    \RightLabel{\rulename{$\mathrevbox{ \, }$}}
    \UnaryInfC{$\fCenter \mathrevbox{\pi_{i}}\mathdyndia{\mathprogramndc{\pi_{0}}{\pi_{1}}}\mathrevbox{\mathprogramndc{\pi_{0}}{\pi_{1}}}\varphi, \lnot\mathrevbox{\mathprogramndc{\pi_{0}}{\pi_{1}}}\varphi$}
    \RightLabel{\rulename{$\lnot$ L}}
    \UnaryInfC{$\lnot\lnot\mathrevbox{\mathprogramndc{\pi_{0}}{\pi_{1}}}\varphi \fCenter \mathrevbox{\pi_{i}}\mathdyndia{\mathprogramndc{\pi_{0}}{\pi_{1}}}\mathrevbox{\mathprogramndc{\pi_{0}}{\pi_{1}}}\varphi$}

    \RightLabel{\rulename{Cut+Wk}}
    \BinaryInfC{$\mathrevbox{\mathprogramndc{\pi_{0}}{\pi_{1}}}\varphi \fCenter \mathrevbox{\pi_{i}}\mathdyndia{\mathprogramndc{\pi_{0}}{\pi_{1}}}\mathrevbox{\mathprogramndc{\pi_{0}}{\pi_{1}}}\varphi$}
   \end{prooftree}

   \caption{A proof of $\mathrevbox{\mathprogramndc{\pi_{0}}{\pi_{1}}}\varphi \fCenter \mathrevbox{\pi_{i}}\mathdyndia{\mathprogramndc{\pi_{0}}{\pi_{1}}}\mathrevbox{\mathprogramndc{\pi_{0}}{\pi_{1}}}\varphi$ with $i=0, 1$}
   \label{fig:revbox-ndc-left}

   \begin{prooftree}
    \AxiomC{}
    \UnaryInfC{$\mathrevbox{\mathprogramndc{\pi_{0}}{\pi_{1}}}\varphi \fCenter \mathrevbox{\mathprogramndc{\pi_{0}}{\pi_{1}}}\varphi$}
    \RightLabel{\rulename{$\lnot$ R}}
    \UnaryInfC{$\fCenter \mathrevbox{\mathprogramndc{\pi_{0}}{\pi_{1}}}\varphi, \lnot\mathrevbox{\mathprogramndc{\pi_{0}}{\pi_{1}}}\varphi$}
    \RightLabel{\rulename{$\mathdynbox{ \, }$}}
    \UnaryInfC{$\fCenter \varphi, \mathdynbox{\mathprogramndc{\pi_{0}}{\pi_{1}}}\lnot\mathrevbox{\mathprogramndc{\pi_{0}}{\pi_{1}}}\varphi$}
    \RightLabel{\rulename{$\lnot$ L}}
    \UnaryInfC{$\mathdyndia{\mathprogramndc{\pi_{0}}{\pi_{1}}}\mathrevbox{\mathprogramndc{\pi_{0}}{\pi_{1}}}\varphi \fCenter \varphi$}
    \RightLabel{\rulename{$\mathrevbox{ \, }$}}
    \UnaryInfC{$\mathrevbox{\pi_{i}}\mathdyndia{\mathprogramndc{\pi_{0}}{\pi_{1}}}\mathrevbox{\mathprogramndc{\pi_{0}}{\pi_{1}}}\varphi \fCenter \mathrevbox{\pi_{i}}\varphi$}
   \end{prooftree}
   \caption{A proof of $\mathrevbox{\pi_{i}}\mathdyndia{\mathprogramndc{\pi_{0}}{\pi_{1}}}\mathrevbox{\mathprogramndc{\pi_{0}}{\pi_{1}}}\varphi \fCenter \mathrevbox{\pi_{i}}\varphi$ with $i=0, 1$}
   \label{fig:revbox-ndc-right}
  \end{figure}

  \noindent \cref{item:well-reduction-lemma-key-lemma-completeness-GTPDL}
  Assume that
  $\Pi_{0}\fCenter\Sigma_{0}, \mathdynbox{\lambda}\mathseqtoform{\Pi_{1}\fCenter\Sigma_{1}}$
  is not provable in \GTPDL\
  for ${(\Pi_{0}\fCenter\Sigma_{0})}$, ${(\Pi_{1}\fCenter\Sigma_{1})}\in W$.
  Since $\Pi_{0}\fCenter\Sigma_{0}, \mathdynbox{\mathprogramndc{\pi_{0}}{\pi_{1}}}\mathseqtoform{\Pi_{1}\fCenter\Sigma_{1}}$
  is not provable in \GTPDL,
  $\Pi_{0}\fCenter\Sigma_{0}, \mathdynbox{\pi_{i}}\mathseqtoform{\Pi_{1}\fCenter\Sigma_{1}}$
  is not provable in \GTPDL\ for all $i\in\mathsetextension{0, 1}$.
  By the induction hypothesis \cref{item:well-reduction-lemma-key-lemma-completeness-GTPDL},
  we have
  $(\Pi_{0}\fCenter\Sigma_{0}){\mathreduction{\pi_{i}}}(\Pi_{1}\fCenter\Sigma_{1})$
  for $i={0, 1}$.
  Thus, $(\Pi_{0}\fCenter\Sigma_{0}){\mathreduction{\lambda}}(\Pi_{1}\fCenter\Sigma_{1})$.

  Consider the case $\lambda\equiv\mathprogramit{\pi}$.
  In this case, there is nothing to prove for
  \cref{item:antecedent-lemma-key-lemma-completeness-GTPDL} and
  \cref{item:consequent-lemma-key-lemma-completeness-GTPDL}.

  \noindent \cref{item:dynbox-lemma-key-lemma-completeness-GTPDL}
  Assume $\mathdynbox{\lambda}\varphi\in{\Pi_{0}}$ and 
  $(\Pi_{0}\fCenter\Sigma_{0}){\mathreduction{\lambda}}(\Pi_{1}\fCenter\Sigma_{1})$
  for ${(\Pi_{0}\fCenter\Sigma_{0})}$, ${(\Pi_{1}\fCenter\Sigma_{1})}\in W$.
  By \cref{cor:Fischer-Ladner-property-set}
  \cref{item:dynbox-cor-Fischer-Ladner-property-set}
  \cref{item:dynbox-it-cor-Fischer-Ladner-property-set},
  $\varphi\in\mathFLcl{\Gamma\cup\Delta}$ and
  $\mathdynbox{\pi}\mathdynbox{\mathprogramit{\pi}}\varphi\in\mathFLcl{\Gamma\cup\Delta}$.

  If $\varphi\in\Sigma_{0}$, then
  $\Pi_{0}\fCenter\Sigma_{0}$ is provable in \GTPDL\ as follows:
  \begin{center}
   \begin{inlineprooftree}
    \AxiomC{}
    \RightLabel{\rulename{Ax}}
    \UnaryInfC{$\bar{\Pi}_{0}, \varphi \fCenter\bar{\Sigma}_{0},\varphi$}
    \RightLabel{\rulename{$\mathdynbox{\mathprogramit{}}$ L B}}
    \UnaryInfC{$\bar{\Pi}_{0}, \mathdynbox{\mathprogramit{\pi}}\varphi \fCenter\bar{\Sigma}_{0}, \varphi$}
   \end{inlineprooftree},
  \end{center}
  where $\Pi_{0}=\bar{\Pi}_{0}, \mathdynbox{\mathprogramit{\pi}}\varphi$ and 
  $\Sigma_{0}=\bar{\Sigma}_{0}, \varphi$.
  Hence, $\varphi\notin{\Sigma_{0}}$,
  that is, $\varphi\in{\Pi_{0}}$.

  If $\mathdynbox{\pi}\mathdynbox{\mathprogramit{\pi}}\varphi\in\Sigma_{0}$, then
  $\Pi_{0}\fCenter\Sigma_{0}$ is provable in \GTPDL\ as follows:
  \begin{center}
   \begin{inlineprooftree}
    \AxiomC{}
    \RightLabel{\rulename{Ax}}
    \UnaryInfC{$\bar{\Pi}_{0}, \mathdynbox{\pi}\mathdynbox{\mathprogramit{\pi}}\varphi \fCenter\bar{\Sigma}_{0}, \mathdynbox{\pi}\mathdynbox{\mathprogramit{\pi}}\varphi$}
    \RightLabel{\rulename{$\mathdynbox{\mathprogramit{}}$ L S}}
    \UnaryInfC{$\bar{\Pi}_{0}, \mathdynbox{\mathprogramit{\pi}}\varphi \fCenter\bar{\Sigma}_{0}, \mathdynbox{\pi}\mathdynbox{\mathprogramit{\pi}}\varphi$}
   \end{inlineprooftree},
  \end{center}
  where $\Pi_{0}=\bar{\Pi}_{0}, \mathdynbox{\mathprogramit{\pi}}\varphi$ and 
  $\Sigma_{0}=\bar{\Sigma}_{0}, \mathdynbox{\pi}\mathdynbox{\mathprogramit{\pi}}\varphi$.
  Hence, $\mathdynbox{\pi}\mathdynbox{\mathprogramit{\pi}}\varphi\notin{\Sigma_{0}}$,
  that is, $\mathdynbox{\pi}\mathdynbox{\mathprogramit{\pi}}\varphi\in{\Pi_{0}}$.

  Because of 
  $(\Pi_{0}\fCenter\Sigma_{0}){\mathreduction{\lambda}}(\Pi_{1}\fCenter\Sigma_{1})$,
  there exists a natural number $n$ such that
  $(\Pi_{0}\fCenter\Sigma_{0}){\mathreduction{\pi}^{n}}(\Pi_{1}\fCenter\Sigma_{1})$.
  We show $\varphi\in\Pi_{1}$ by induction on $n$.

  Consider the case $n=0$.
  Then, $(\Pi_{1}\fCenter\Sigma_{1})=(\Pi_{0}\fCenter\Sigma_{0})$.
  Hence, $\varphi\in\Pi_{0}=\Pi_{1}$.

  Consider the case $n>0$.
  Then, there exists $(\Pi_{2}\fCenter\Sigma_{2})\in W$ such that
  $(\Pi_{0}\fCenter\Sigma_{0}){\mathreduction{\pi}}(\Pi_{2}\fCenter\Sigma_{2})$ and
  $(\Pi_{2}\fCenter\Sigma_{2}){\mathreduction{\pi}^{n-1}}(\Pi_{1}\fCenter\Sigma_{1})$.
  We note that $\mathdynbox{\pi}\mathdynbox{\mathprogramit{\pi}}\varphi\in{\Pi_{0}}$.
  By the induction hypothesis \cref{item:dynbox-lemma-key-lemma-completeness-GTPDL}, 
  $\mathdynbox{\mathprogramit{\pi}}\varphi\in{\Pi_{2}}$.
  By the induction hypothesis on $n$,we have $\varphi\in{\Pi_{1}}$.

  \noindent \cref{item:revbox-lemma-key-lemma-completeness-GTPDL}
  Assume $\mathrevbox{\lambda}\varphi\in{\Pi_{1}}$ and 
  $(\Pi_{0}\fCenter\Sigma_{0}){\mathreduction{\lambda}}(\Pi_{1}\fCenter\Sigma_{1})$
  for ${(\Pi_{0}\fCenter\Sigma_{0})}$, ${(\Pi_{1}\fCenter\Sigma_{1})}\in W$.
  By \cref{cor:Fischer-Ladner-property-set} 
  \cref{item:revbox-cor-Fischer-Ladner-property-set}
  \cref{item:revbox-it-cor-Fischer-Ladner-property-set},
  $\varphi\in\mathFLcl{\Gamma\cup\Delta}$ and
  $\mathrevbox{\pi}\mathrevbox{\mathprogramit{\pi}}\varphi\in\mathFLcl{\Gamma\cup\Delta}$.

  If $\varphi\in{\Sigma_{1}}$,
  then $\Pi_{1}\fCenter\Sigma_{1}$ is provable in \GTPDL\ as follows:
  \begin{center}
   \begin{inlineprooftree}
    \AxiomC{\cref{fig:revbox-it-base-left}}

    \AxiomC{\cref{fig:revbox-it-base-right}}

    \RightLabel{\rulename{Cut+Wk}}
    \BinaryInfC{$\mathrevbox{\mathprogramit{\pi}}\varphi \fCenter \varphi$}
    \RightLabel{\rulename{Wk}}
    \UnaryInfC{$\bar{\Pi}_{1}, \mathrevbox{\mathprogramit{\pi}}\varphi \fCenter\bar{\Sigma}_{1}, \varphi$}
   \end{inlineprooftree},
  \end{center}
  where $\Pi_{1}=\bar{\Pi}_{1}, \mathrevbox{\mathprogramit{\pi}}\varphi$ and $\Sigma_{1}=\bar{\Sigma}_{1}, \varphi$.
  Hence, $\varphi\notin{\Sigma_{1}}$,
  that is, $\varphi\in{\Pi_{1}}$.

  \begin{figure}[tb]
   \begin{prooftree}
    \AxiomC{}
    \RightLabel{\rulename{Ax}}
    \UnaryInfC{$\mathrevbox{\mathprogramit{\pi}}\varphi \fCenter \mathrevbox{\mathprogramit{\pi}}\varphi$}
    \RightLabel{\rulename{$\lnot$ L}}
    \UnaryInfC{$\mathrevbox{\mathprogramit{\pi}}\varphi, \lnot\mathrevbox{\mathprogramit{\pi}}\varphi \fCenter$}
    \RightLabel{\rulename{$\lnot$ R}}
    \UnaryInfC{$\mathrevbox{\mathprogramit{\pi}}\varphi \fCenter \lnot\lnot\mathrevbox{\mathprogramit{\pi}}\varphi$}

    \AxiomC{}
    \RightLabel{\rulename{Ax}}
    \UnaryInfC{$\lnot\mathrevbox{\mathprogramit{\pi}}\varphi\fCenter\lnot\mathrevbox{\mathprogramit{\pi}}\varphi$}
    \RightLabel{\rulename{$\mathdynbox{\mathprogramit{}}$ L B}}
    \UnaryInfC{$\mathdynbox{\mathprogramit{\pi}}\lnot\mathrevbox{\mathprogramit{\pi}}\varphi\fCenter \lnot\mathrevbox{\mathprogramit{\pi}}\varphi$}
    \RightLabel{\rulename{$\lnot$ R}}
    \UnaryInfC{$\fCenter\mathdyndia{\mathprogramit{\pi}}\mathrevbox{\mathprogramit{\pi}}\varphi, \lnot\mathrevbox{\mathprogramit{\pi}}\varphi$}
    \RightLabel{\rulename{$\lnot$ L}}
    \UnaryInfC{$\lnot\lnot\mathrevbox{\mathprogramit{\pi}}\varphi\fCenter\mathdyndia{\mathprogramit{\pi}}\mathrevbox{\mathprogramit{\pi}}\varphi$}

    \RightLabel{\rulename{Cut+Wk}}
    \BinaryInfC{$\mathrevbox{\mathprogramit{\pi}}\varphi\fCenter\mathdyndia{\mathprogramit{\pi}}\mathrevbox{\mathprogramit{\pi}}\varphi$}
   \end{prooftree}

   \caption{A proof of $\mathrevbox{\mathprogramit{\pi}}\varphi\fCenter\mathdyndia{\mathprogramit{\pi}}\mathrevbox{\mathprogramit{\pi}}\varphi$}
   \label{fig:revbox-it-base-left}

   \begin{prooftree}
    \AxiomC{}
    \UnaryInfC{$\mathrevbox{\mathprogramit{\pi}}\varphi \fCenter \mathrevbox{\mathprogramit{\pi}}\varphi$}
    \RightLabel{\rulename{$\lnot$ R}}
    \UnaryInfC{$\fCenter \mathrevbox{\mathprogramit{\pi}}\varphi, \lnot\mathrevbox{\mathprogramit{\pi}}\varphi$}
    \RightLabel{\rulename{$\mathdynbox{ \, }$}}
    \UnaryInfC{$\fCenter \varphi, \mathdynbox{\mathprogramit{\pi}}\lnot\mathrevbox{\mathprogramit{\pi}}\varphi$}
    \RightLabel{\rulename{$\lnot$ L}}
    \UnaryInfC{$\mathdyndia{\mathprogramit{\pi}}\mathrevbox{\mathprogramit{\pi}}\varphi \fCenter \varphi$}
   \end{prooftree}
   \caption{A proof of $\mathdyndia{\mathprogramit{\pi}}\mathrevbox{\mathprogramit{\pi}}\varphi \fCenter \varphi$}
   \label{fig:revbox-it-base-right}
  \end{figure}

  If $\mathrevbox{\pi}\mathrevbox{\mathprogramit{\pi}}\varphi\in{\Sigma_{1}}$,
  then $\Pi_{1}\fCenter\Sigma_{1}$ is provable in \GTPDL\ as follows:
  \begin{center}
   \begin{inlineprooftree}
    \AxiomC{\cref{fig:revbox-it-step-left}}

    \AxiomC{\cref{fig:revbox-it-step-right}}

    \RightLabel{\rulename{Cut+Wk}}
    \BinaryInfC{$\mathrevbox{\mathprogramit{\pi}}\varphi \fCenter \mathrevbox{\pi}\mathrevbox{\mathprogramit{\pi}}\varphi$}
    \RightLabel{\rulename{Wk}}
    \UnaryInfC{$\bar{\Pi}_{1}, \mathrevbox{\mathprogramit{\pi}}\varphi \fCenter\bar{\Sigma}_{1}, \mathrevbox{\pi}\mathrevbox{\mathprogramit{\pi}}\varphi$}
   \end{inlineprooftree},
  \end{center}
  where $\Pi_{1}=\bar{\Pi}_{1}, \mathrevbox{\mathprogramit{\pi}}\varphi$ and $\Sigma_{1}=\bar{\Sigma}_{1}, \mathrevbox{\pi}\mathrevbox{\mathprogramit{\pi}}\varphi$.
  Hence, $\mathrevbox{\pi}\mathrevbox{\mathprogramit{\pi}}\varphi\notin{\Sigma_{1}}$,
  that is, $\mathrevbox{\pi}\mathrevbox{\mathprogramit{\pi}}\varphi\in{\Pi_{1}}$.

  \begin{figure}[tb]
   \centering
   \begin{prooftree}
    \AxiomC{}
    \RightLabel{\rulename{Ax}}
    \UnaryInfC{$\mathrevbox{\mathprogramit{\pi}}\varphi \fCenter \mathrevbox{\mathprogramit{\pi}}\varphi$}
    \RightLabel{\rulename{$\lnot$ L}}
    \UnaryInfC{$\mathrevbox{\mathprogramit{\pi}}\varphi, \lnot\mathrevbox{\mathprogramit{\pi}}\varphi \fCenter$}
    \RightLabel{\rulename{$\lnot$ R}}
    \UnaryInfC{$\mathrevbox{\mathprogramit{\pi}}\varphi \fCenter \lnot\lnot\mathrevbox{\mathprogramit{\pi}}\varphi$}

    \AxiomC{\cref{fig:revbox-it-step-left-sub}}
    \RightLabel{\rulename{$\mathdynbox{\mathprogramit{}}$ L S}}
    \UnaryInfC{$\mathdynbox{\mathprogramit{\pi}}\lnot\mathrevbox{\mathprogramit{\pi}}\varphi\fCenter\mathdynbox{\mathprogramit{\pi}}\mathdynbox{\pi}\lnot\mathrevbox{\mathprogramit{\pi}}\varphi$}
    \RightLabel{\rulename{$\lnot$ R}}
    \UnaryInfC{$\fCenter\mathdyndia{\mathprogramit{\pi}}\mathrevbox{\mathprogramit{\pi}}\varphi, \mathdynbox{\mathprogramit{\pi}}\mathdynbox{\pi}\lnot\mathrevbox{\mathprogramit{\pi}}\varphi$}
    \RightLabel{\rulename{$\mathrevbox{ \, }$}}
    \UnaryInfC{$\fCenter\mathrevbox{\mathprogramit{\pi}}\mathdyndia{\mathprogramit{\pi}}\mathrevbox{\mathprogramit{\pi}}\varphi, \mathdynbox{\pi}\lnot\mathrevbox{\mathprogramit{\pi}}\varphi$}
    \RightLabel{\rulename{$\mathrevbox{ \, }$}}
    \UnaryInfC{$\fCenter\mathrevbox{\pi}\mathrevbox{\mathprogramit{\pi}}\mathdyndia{\mathprogramit{\pi}}\mathrevbox{\mathprogramit{\pi}}\varphi, \lnot\mathrevbox{\mathprogramit{\pi}}\varphi$}
    \RightLabel{\rulename{$\lnot$ L}}
    \UnaryInfC{$\lnot\lnot\mathrevbox{\mathprogramit{\pi}}\varphi\fCenter\mathrevbox{\pi}\mathrevbox{\mathprogramit{\pi}}\mathdyndia{\mathprogramit{\pi}}\mathrevbox{\mathprogramit{\pi}}\varphi$}

    \RightLabel{\rulename{Cut+Wk}}
    \BinaryInfC{$\mathrevbox{\mathprogramit{\pi}}\varphi\fCenter\mathrevbox{\pi}\mathrevbox{\mathprogramit{\pi}}\mathdyndia{\mathprogramit{\pi}}\mathrevbox{\mathprogramit{\pi}}\varphi$}
   \end{prooftree}

   \caption{A proof of $\mathrevbox{\mathprogramit{\pi}}\varphi\fCenter\mathdyndia{\mathprogramit{\pi}}\mathrevbox{\mathprogramit{\pi}}\varphi$}
   \label{fig:revbox-it-step-left}

   \begin{prooftree}
    \AxiomC{}
    \RightLabel{\rulename{Ax}}
    \UnaryInfC{$\mathrevbox{\mathprogramit{\pi}}\varphi \fCenter \mathrevbox{\mathprogramit{\pi}}\varphi$}
    \RightLabel{\rulename{$\lnot$ R}}
    \UnaryInfC{$\fCenter \mathrevbox{\mathprogramit{\pi}}\varphi, \lnot\mathrevbox{\mathprogramit{\pi}}\varphi$}
    \RightLabel{\rulename{$\mathdynbox{ \, }$}}
    \UnaryInfC{$\fCenter \varphi, \mathdynbox{\mathprogramit{\pi}}\lnot\mathrevbox{\mathprogramit{\pi}}\varphi$}
    \RightLabel{\rulename{$\lnot$ L}}
    \UnaryInfC{$\mathdyndia{\mathprogramit{\pi}}\mathrevbox{\mathprogramit{\pi}}\varphi \fCenter \varphi$}
    \RightLabel{\rulename{$\mathrevbox{ \, }$}}
    \UnaryInfC{$\mathrevbox{\mathprogramit{\pi}}\mathdyndia{\mathprogramit{\pi}}\mathrevbox{\mathprogramit{\pi}}\varphi \fCenter \mathrevbox{\mathprogramit{\pi}}\varphi$}
    \RightLabel{\rulename{$\mathrevbox{ \, }$}}
    \UnaryInfC{$\mathrevbox{\pi}\mathrevbox{\mathprogramit{\pi}}\mathdyndia{\mathprogramit{\pi}}\mathrevbox{\mathprogramit{\pi}}\varphi \fCenter \mathrevbox{\pi}\mathrevbox{\mathprogramit{\pi}}\varphi$}
   \end{prooftree}
   \caption{A proof of $\mathrevbox{\pi}\mathrevbox{\mathprogramit{\pi}}\mathdyndia{\mathprogramit{\pi}}\mathrevbox{\mathprogramit{\pi}}\varphi \fCenter \mathrevbox{\pi}\mathrevbox{\mathprogramit{\pi}}\varphi$}
   \label{fig:revbox-it-step-right}

   \begin{inlineprooftree}
    \AxiomC{}
    \RightLabel{\rulename{Ax}}
    \UnaryInfC{$\mathdynbox{\pi}\mathdynbox{\mathprogramit{\pi}}\psi\fCenter\mathdynbox{\pi}\mathdynbox{\mathprogramit{\pi}}\psi$}
    \RightLabel{\rulename{$\mathdynbox{\mathprogramit{}}$ L S}}
   \UnaryInfC{$\mathdynbox{\mathprogramit{\pi}}\psi\fCenter\mathdynbox{\pi}\mathdynbox{\mathprogramit{\pi}}\psi$}
    \RightLabel{\rulename{$\mathdynbox{ \, }$}}
    \UnaryInfC{$\mathdynbox{\pi}\mathdynbox{\mathprogramit{\pi}}\psi\fCenter\mathdynbox{\pi}\mathdynbox{\pi}\mathdynbox{\mathprogramit{\pi}}\psi$}
    \RightLabel{\rulename{$\mathdynbox{\mathprogramit{}}$ R}}
    \UnaryInfC{$\mathdynbox{\pi}\mathdynbox{\mathprogramit{\pi}}\psi\fCenter\mathdynbox{\mathprogramit{\pi}}\mathdynbox{\pi}\mathdynbox{\mathprogramit{\pi}}\psi$}

    \AxiomC{}
    \UnaryInfC{$\psi\fCenter\psi$}
    \RightLabel{\rulename{$\mathdynbox{\mathprogramit{}}$ L B}}
    \UnaryInfC{$\mathdynbox{\mathprogramit{\pi}}\psi\fCenter\psi$}
    \RightLabel{\rulename{$\mathdynbox{ \, }$}}
    \UnaryInfC{$\mathdynbox{\pi}\mathdynbox{\mathprogramit{\pi}}\psi\fCenter\mathdynbox{\pi}\psi$}
    \RightLabel{\rulename{$\mathdynbox{ \, }$}}
    \UnaryInfC{$\mathdynbox{\mathprogramit{\pi}}\mathdynbox{\pi}\mathdynbox{\mathprogramit{\pi}}\psi\fCenter\mathdynbox{\mathprogramit{\pi}}\mathdynbox{\pi}\psi$}
    
    \RightLabel{\rulename{Cut+Wk}}
    \BinaryInfC{$\mathdynbox{\pi}\mathdynbox{\mathprogramit{\pi}}\psi\fCenter\mathdynbox{\mathprogramit{\pi}}\mathdynbox{\pi}\psi$}
   \end{inlineprooftree},
   
   \medskip
   where $\psi\equiv\lnot\mathrevbox{\mathprogramit{\pi}}\varphi$ 

   \caption{A proof of $\mathdynbox{\pi}\mathdynbox{\mathprogramit{\pi}}\lnot\mathrevbox{\mathprogramit{\pi}}\varphi\fCenter\mathdynbox{\mathprogramit{\pi}}\mathdynbox{\pi}\lnot\mathrevbox{\mathprogramit{\pi}}\varphi$}
   \label{fig:revbox-it-step-left-sub}   
  \end{figure}

  Because of 
  $(\Pi_{0}\fCenter\Sigma_{0}){\mathreduction{\lambda}}(\Pi_{1}\fCenter\Sigma_{1})$,
  there exists a natural number $n$ such that
  $(\Pi_{0}\fCenter\Sigma_{0}){\mathreduction{\pi}^{n}}(\Pi_{1}\fCenter\Sigma_{1})$.
  We show $\varphi\in\Pi_{0}$ by induction on $n$.

  Consider the case $n=0$.
  Then, $(\Pi_{0}\fCenter\Sigma_{0})=(\Pi_{1}\fCenter\Sigma_{1})$.
  Hence, $\varphi\in\Pi_{1}=\Pi_{0}$.

  Consider the case $n>0$.
  Then, there exists $(\Pi_{2}\fCenter\Sigma_{2})\in W$ such that
  $(\Pi_{0}\fCenter\Sigma_{0}){\mathreduction{\pi}^{n-1}}(\Pi_{2}\fCenter\Sigma_{2})$ and
  $(\Pi_{2}\fCenter\Sigma_{2}){\mathreduction{\pi}}(\Pi_{1}\fCenter\Sigma_{1})$.
  We note that $\mathrevbox{\pi}\mathrevbox{\mathprogramit{\pi}}\varphi\in{\Pi_{1}}$.
  By the induction hypothesis \cref{item:revbox-lemma-key-lemma-completeness-GTPDL}, 
  $\mathrevbox{\mathprogramit{\pi}}\varphi\in{\Pi_{2}}$.
  By the induction hypothesis on $n$,we have $\varphi\in{\Pi_{0}}$.
  
  \noindent \cref{item:well-reduction-lemma-key-lemma-completeness-GTPDL}
  Fix $(\Pi_{0}\fCenter\Sigma_{0})\in W$. Let 
  \begin{align*}
   R &=\mathsetintension{\mathseqtoform{\Phi\fCenter\Psi}}{(\Pi_{0}\fCenter\Sigma_{0}){\mathreduction{\mathprogramit{\pi}}}(\Phi\fCenter\Psi)} \text{ and } \\
   U &=\mathsetintension{\mathseqtoform{\Xi\fCenter\Theta}}{\Xi\fCenter\Theta\notin R \text{ and } (\Xi\fCenter\Theta)\in W }. 
  \end{align*}
  It suffices to show that $\mathseqtoform{\Pi_{1}\fCenter\Sigma_{1}}\in R$ holds
  if $\Pi_{0}\fCenter\Sigma_{0}, \mathdynbox{\mathprogramit{\pi}}\mathseqtoform{\Pi_{1}\fCenter\Sigma_{1}}$
  is not provable in \GTPDL.

  If $(\Phi\fCenter\Psi){\mathreduction{\pi}}(\Xi\fCenter\Theta)$
  holds for $\mathseqtoform{\Phi\fCenter\Psi}\in R$,
  then we have
  $(\Pi_{0}\fCenter\Sigma_{0}){\mathreduction{\mathprogramit{\pi}}}(\Xi\fCenter\Theta)$.
  Hence, 
  $(\Phi\fCenter\Psi){\mathreduction{\pi}}(\Xi\fCenter\Theta)$
  does not hold for 
  $\mathseqtoform{\Phi\fCenter\Psi}\in R$ and
  $\mathseqtoform{\Xi\fCenter\Theta}\in U$.
  By the contrapositive of the induction hypothesis 
  \cref{item:well-reduction-lemma-key-lemma-completeness-GTPDL}, we see that 
  $\Phi\fCenter\Psi, \mathdynbox{\pi}\mathseqtoform{\Xi\fCenter\Theta}$
  is provable in \GTPDL\ for each
  $\mathseqtoform{\Phi\fCenter\Psi}\in R$ and each
  $\mathseqtoform{\Xi\fCenter\Theta}\in U$.
  By \cref{prop:hyper-sequent} and \rulename{$\lnot$ L},
  $\lnot\mathseqtoform{\Phi\fCenter\Psi}\fCenter\mathdynbox{\pi}\mathseqtoform{\Xi\fCenter\Theta}$
  is provable in \GTPDL\ for each
  $\mathseqtoform{\Phi\fCenter\Psi}\in R$ and each
  $\mathseqtoform{\Xi\fCenter\Theta}\in U$.
  By \rulename{$\bigland$ R},
  $\lnot\mathseqtoform{\Phi\fCenter\Psi}\fCenter \bigland\mathdynbox{\pi}U$
  is provable in \GTPDL\ for each
  $\mathseqtoform{\Phi\fCenter\Psi}\in R$.
  By \rulename{$\biglor$ L},
  $\biglor\lnot R\fCenter \bigland\mathdynbox{\pi}U$ is provable in \GTPDL.
  \cref{lemma:saturated-sequents-in-antecedent} implies that
  $R, U\fCenter$ is provable in \GTPDL.
  Hence, $\bigland U\fCenter\biglor\lnot R$ is provable in \GTPDL.
  Then, we have the following proof:
  \begin{center}
   \begin{inlineprooftree}
    \AxiomC{$\bigland U\fCenter\biglor\lnot R$}
    \AxiomC{$\biglor\lnot R\fCenter\bigland\mathdynbox{\pi}U$}
    \RightLabel{\rulename{Cut+Wk}}
    \BinaryInfC{$\bigland U\fCenter\bigland\mathdynbox{\pi}U$}
    
    \AxiomC{}
    \RightLabel{\rulename{Ax}'s}
    \UnaryInfC{$\mathsequence{U\fCenter \mathseqtoform{\Xi\fCenter\Theta}}{\mathseqtoform{\Xi\fCenter\Theta}\in U}$}
    \RightLabel{\rulename{$\bigland$ R}}
    \UnaryInfC{$U\fCenter\bigland U$}
    \RightLabel{\rulename{$\mathdynbox{ \, }$}}
    \UnaryInfC{$\mathdynbox{\pi}U\fCenter\mathdynbox{\pi}\bigland U$}
    \RightLabel{\rulename{$\bigland$ L}}
    \UnaryInfC{$\bigland\mathdynbox{\pi}U\fCenter\mathdynbox{\pi}\bigland U$}

    \RightLabel{\rulename{Cut+Wk}}
    \BinaryInfC{$\bigland U\fCenter\mathdynbox{\pi}\bigland U$}
    \RightLabel{\rulename{$\mathdynbox{\mathprogramitsy}$ R}}
    \UnaryInfC{$\bigland U\fCenter\mathdynbox{\mathprogramit{\pi}}\bigland U$}
   \end{inlineprooftree}.
  \end{center}
  
  Because of $\mathseqtoform{\Pi_{0}\fCenter\Sigma_{0}}\in R$,
  we have $\Pi_{0}\neq\Xi$ and $\Sigma_{0}\neq\Theta$ for each
  $\mathseqtoform{\Xi\fCenter\Theta}\in U$.
  Hence, we see
  $\Pi_{0}\cap\Theta\neq\emptyset$ and $\Sigma_{0}\cap\Xi\neq\emptyset$ for each
  $\mathseqtoform{\Xi\fCenter\Theta}\in U$.
  Therefore, $\Pi_{0}, \Xi\fCenter\Sigma_{0}, \Theta$
  is provable in \GTPDL\ for each $\mathseqtoform{\Xi\fCenter\Theta}\in U$.
  \cref{prop:hyper-sequent} implies that
  $\Pi_{0}\fCenter\Sigma_{0}, \mathseqtoform{\Xi\fCenter\Theta}$
  is provable in \GTPDL\ for each $\mathseqtoform{\Xi\fCenter\Theta}\in U$.
  By \rulename{$\bigland$ R},
  $\Pi_{0}\fCenter\Sigma_{0}, \bigland U$ is provable in \GTPDL.
  Then, we have the following proof for each $\mathseqtoform{\Xi\fCenter\Theta}\in U$:
  \begin{center}
   \begin{inlineprooftree}
    \AxiomC{$\Pi_{0}\fCenter\Sigma_{0}, \bigland U$}
    \AxiomC{$\bigland U\fCenter\mathdynbox{\mathprogramit{\pi}}\bigland U$}
    \RightLabel{\rulename{Cut+Wk}}
    \BinaryInfC{$\Pi_{0}\fCenter\Sigma_{0}, \mathdynbox{\mathprogramit{\pi}}\bigland U$}

    \AxiomC{}
    \RightLabel{\rulename{Ax}}
    \UnaryInfC{$ U\fCenter\mathseqtoform{\Xi\fCenter\Theta}$}
    \RightLabel{\rulename{$\bigland$ L}}
    \UnaryInfC{$\bigland U\fCenter\mathseqtoform{\Xi\fCenter\Theta}$}
    \RightLabel{\rulename{$\mathdynbox{ \, }$}}
    \UnaryInfC{$\mathdynbox{\mathprogramit{\pi}}\bigland U\fCenter \mathdynbox{\mathprogramit{\pi}}\mathseqtoform{\Xi\fCenter\Theta}$}
    \RightLabel{\rulename{Cut+Wk}}
    \BinaryInfC{$\Pi_{0}\fCenter\Sigma_{0}, \mathdynbox{\mathprogramit{\pi}}\mathseqtoform{\Xi\fCenter\Theta}$}
   \end{inlineprooftree}.
  \end{center}
 
  Thus, if $\Pi_{0}\fCenter\Sigma_{0}, \mathdynbox{\mathprogramit{\pi}}\mathseqtoform{\Pi_{1}\fCenter\Sigma_{1}}$
  is not provable in \GTPDL, then
  $\mathseqtoform{\Pi_{1}\fCenter\Sigma_{1}}\notin U$, that is,
  $\mathseqtoform{\Pi_{1}\fCenter\Sigma_{1}}\in R$.

  Consider the case $\lambda\equiv\mathprogramtest{\psi}$.
  In this case, there is nothing to prove for
  \cref{item:antecedent-lemma-key-lemma-completeness-GTPDL} and
  \cref{item:consequent-lemma-key-lemma-completeness-GTPDL}.

  \noindent \cref{item:dynbox-lemma-key-lemma-completeness-GTPDL}
  Assume $\mathdynbox{\lambda}\varphi\in{\Pi_{0}}$ and 
  $(\Pi_{0}\fCenter\Sigma_{0}){\mathreduction{\lambda}}(\Pi_{1}\fCenter\Sigma_{1})$
  for ${(\Pi_{0}\fCenter\Sigma_{0})}$, ${(\Pi_{1}\fCenter\Sigma_{1})}\in W$.
  By \cref{cor:Fischer-Ladner-property-set} 
  \cref{item:dynbox-cor-Fischer-Ladner-property-set} and
  \cref{item:dynbox-test-cor-Fischer-Ladner-property-set},
  we have $\varphi\in\mathFLcl{\Lambda}$ and $\psi\in\mathFLcl{\Lambda}$.
  If $\varphi\in\Sigma_{0}$ and $\psi\in\Pi_{0}$, then 
  $\Pi_{0}\fCenter\Sigma_{0}$ is provable in \GTPDL\ as follows:
  \begin{center}
   \begin{inlineprooftree}
    \AxiomC{}
    \RightLabel{\rulename{Ax}}
    \UnaryInfC{$\bar{\Pi}_{0}, \psi \fCenter\psi, \bar{\Sigma}_{0}, \varphi$}
    \AxiomC{}
    \RightLabel{\rulename{Ax}}
    \UnaryInfC{$\bar{\Pi}_{0}, \psi, \varphi \fCenter\bar{\Sigma}_{0}, \varphi$}

    \RightLabel{\rulename{$\mathdynbox{ \mathprogramtestsy }$ L}}
    \BinaryInfC{$\bar{\Pi}_{0}, \psi, \mathdynbox{\mathprogramtest{\psi}}\varphi \fCenter\bar{\Sigma}_{0}, \varphi$}
   \end{inlineprooftree},
  \end{center}
  where $\Pi_{0}=\bar{\Pi}_{0}, \mathdynbox{\mathprogramtest{\psi}}\varphi$ and $\Sigma_{0}=\bar{\Sigma}_{0}, \varphi$.
  Hence, either $\varphi\notin\Sigma_{0}$ or $\psi\notin\Pi_{0}$, that is,
  either $\varphi\in\Pi_{0}$ or $\psi\in\Sigma_{0}$.
  Since $(\Pi_{0}\fCenter\Sigma_{0}){\mathreduction{\mathprogramtest{\psi}}}(\Pi_{1}\fCenter\Sigma_{1})$ 
  holds,
  we have $\Pi_{1}=\Pi_{0}$, $\Sigma_{1}=\Sigma_{0}$, and 
  $M, {(\Pi_{1}\fCenter\Sigma_{1})}\models\psi$.
  If $\psi\in\Sigma_{1}$,
  then the induction hypothesis \cref{item:consequent-lemma-key-lemma-completeness-GTPDL}
  implies $M, {(\Pi_{1}\fCenter\Sigma_{1})}\not\models\psi$.
  Hence, we have $\psi\notin\Sigma_{1}=\Sigma_{0}$.
  Since either $\varphi\in\Pi_{0}$ or $\psi\in\Sigma_{0}$ holds,
  we have $\varphi\in\Pi_{0}$.
  Therefore, $\varphi\in\Pi_{1}$.

  \noindent \cref{item:revbox-lemma-key-lemma-completeness-GTPDL}
  Assume $\mathrevbox{\lambda}\varphi\in{\Pi_{1}}$ and 
  $(\Pi_{0}\fCenter\Sigma_{0}){\mathreduction{\lambda}}(\Pi_{1}\fCenter\Sigma_{1})$
  for ${(\Pi_{0}\fCenter\Sigma_{0})}$, ${(\Pi_{1}\fCenter\Sigma_{1})}\in W$.
  By \cref{cor:Fischer-Ladner-property-set} 
  \cref{item:revbox-cor-Fischer-Ladner-property-set} and
  \cref{item:revbox-test-cor-Fischer-Ladner-property-set},
  we have $\varphi\in\mathFLcl{\Lambda}$ and $\psi\in\mathFLcl{\Lambda}$.
  If $\varphi\in\Sigma_{1}$ and $\psi\in\Pi_{1}$, then 
  $\Pi_{1}\fCenter\Sigma_{1}$ is provable in \GTPDL\ as follows:
  \begin{center}
   \footnotesize
   \begin{inlineprooftree}
    \AxiomC{}
    \RightLabel{\rulename{Ax}}
    \UnaryInfC{$\psi, \mathrevbox{\mathprogramtest{\psi}}\varphi \fCenter \psi $}

    \AxiomC{}
    \RightLabel{\rulename{Ax}}
    \UnaryInfC{$\psi, \mathrevbox{\mathprogramtest{\psi}}\varphi \fCenter \mathrevbox{\mathprogramtest{\psi}}\varphi $}
    \RightLabel{\rulename{$\lnot$ L}}
    \UnaryInfC{$\psi, \mathrevbox{\mathprogramtest{\psi}}\varphi, \lnot\mathrevbox{\mathprogramtest{\psi}}\varphi \fCenter $}

    \RightLabel{\rulename{$\mathdynbox{ \mathprogramtestsy }$ L}}
    \BinaryInfC{$\psi, \mathrevbox{\mathprogramtest{\psi}}\varphi, \mathdynbox{\mathprogramtest{\psi}}\lnot\mathrevbox{\mathprogramtest{\psi}}\varphi \fCenter $}
    \RightLabel{\rulename{$\lnot$ R}}
    \UnaryInfC{$\psi, \mathrevbox{\mathprogramtest{\psi}}\varphi \fCenter \mathdyndia{\mathprogramtest{\psi}}\mathrevbox{\mathprogramtest{\psi}}\varphi$}

    \AxiomC{}
    \RightLabel{\rulename{Ax}}
    \UnaryInfC{$\mathrevbox{\mathprogramtest{\psi}}\varphi\fCenter \mathrevbox{\mathprogramtest{\psi}}\varphi$}
    \RightLabel{\rulename{$\lnot$ R}}
    \UnaryInfC{$\fCenter \mathrevbox{\mathprogramtest{\psi}}\varphi, \lnot\mathrevbox{\mathprogramtest{\psi}}\varphi $}
    \RightLabel{\rulename{$\mathdynbox{ \, }$}}
    \UnaryInfC{$\fCenter \varphi, \mathdynbox{\mathprogramtest{\psi}}\lnot\mathrevbox{\mathprogramtest{\psi}}\varphi $}
    \RightLabel{\rulename{$\lnot$ L}}
    \UnaryInfC{$\mathdyndia{\mathprogramtest{\psi}}\mathrevbox{\mathprogramtest{\psi}}\varphi \fCenter \varphi $}

    \RightLabel{\rulename{Cut+Wk}}
    \BinaryInfC{$\psi, \mathrevbox{\mathprogramtest{\psi}}\varphi \fCenter \varphi$}
    \RightLabel{\rulename{Wk}}
    \UnaryInfC{$\bar{\Pi}_{1}, \psi, \mathrevbox{\mathprogramtest{\psi}}\varphi \fCenter\bar{\Sigma}_{1}, \varphi$}
   \end{inlineprooftree},
  \end{center}
  where $\Pi_{1}=\bar{\Pi}_{1}, \mathrevbox{\mathprogramtest{\psi}}\varphi$ and $\Sigma_{1}=\bar{\Sigma}_{1}, \varphi$.
  Hence, either $\varphi\notin\Sigma_{1}$ or $\psi\notin\Pi_{1}$, that is,
  either $\varphi\in\Pi_{1}$ or $\psi\in\Sigma_{1}$.
  Since $(\Pi_{0}\fCenter\Sigma_{0}){\mathreduction{\mathprogramtest{\psi}}}(\Pi_{1}\fCenter\Sigma_{1})$ 
  holds,
  we have $\Pi_{0}=\Pi_{1}$, $\Sigma_{0}=\Sigma_{1}$, and 
  $M, {(\Pi_{0}\fCenter\Sigma_{0})}\models\psi$.
  If $\psi\in\Sigma_{1}=\Sigma_{0}$,
  then the induction hypothesis \cref{item:consequent-lemma-key-lemma-completeness-GTPDL}
  implies $M, {(\Pi_{0}\fCenter\Sigma_{0})}\not\models\psi$.
  Hence, we have $\psi\in\Pi_{0}=\Pi_{1}$.
  Since either $\varphi\in\Pi_{1}$ or $\psi\in\Sigma_{1}$ holds,
  we have $\varphi\in\Pi_{1}$.
  Therefore, $\varphi\in\Pi_{0}$.

  \noindent \cref{item:well-reduction-lemma-key-lemma-completeness-GTPDL}
  We show the contrapositive.

  Assume that
  $(\Pi_{0}\fCenter\Sigma_{0}){\mathreduction{\mathprogramtest{\psi}}}(\Pi_{1}\fCenter\Sigma_{1})$
  does not hold.
  Then, either $\Pi_{0}\neq\Pi_{1}$ and $\Sigma_{0}\neq\Sigma_{1}$, or
  $M, {(\Pi_{0}\fCenter\Sigma_{0})}\not\models\psi$.
  
  Consider the case $\Pi_{0}\neq\Pi_{1}$ and $\Sigma_{0}\neq\Sigma_{1}$.
  Then, $\Pi_{0}\cap\Sigma_{1}\neq\emptyset$ or $\Pi_{1}\cap\Sigma_{0}\neq\emptyset$.
  We can prove
  $\Pi_{0}\fCenter\Sigma_{0}, \mathdynbox{\mathprogramtest{\psi}}\mathseqtoform{\Pi_{1}\fCenter\Sigma_{1}}$
  in \GTPDL\ as follows:
  \begin{center}
   \begin{inlineprooftree}
    \AxiomC{}
    \RightLabel{\rulename{Ax}}
    \UnaryInfC{$\Pi_{0}, \Pi_{1}, \psi\fCenter\Sigma_{0}, \Sigma_{1}$}
    \doubleLine
    \RightLabel{\cref{prop:hyper-sequent}}
    \UnaryInfC{$\Pi_{0}, \psi\fCenter\Sigma_{0}, \mathseqtoform{\Pi_{1}\fCenter\Sigma_{1}}$}
    \RightLabel{\rulename{$\mathdynbox{ \mathprogramtestsy }$ R}}
    \UnaryInfC{$\Pi_{0}\fCenter\Sigma_{0}, \mathdynbox{\mathprogramtest{\psi}}\mathseqtoform{\Pi_{1}\fCenter\Sigma_{1}}$}
   \end{inlineprooftree}.
  \end{center}

  Consider the case $M, {(\Pi_{0}\fCenter\Sigma_{0})}\not\models\psi$.
  By the contrapositive of the induction hypothesis \cref{item:antecedent-lemma-key-lemma-completeness-GTPDL},
  $\psi\notin\Pi_{0}$, that is, $\psi\in\Sigma_{0}$.
  We can prove
  $\Pi_{0}\fCenter\Sigma_{0}, \mathdynbox{\mathprogramtest{\psi}}\mathseqtoform{\Pi_{0}\fCenter\Sigma_{0}}$
  in \GTPDL\ as follows:
  \begin{center}
   \begin{inlineprooftree}
    \AxiomC{}
    \RightLabel{\rulename{Ax}}
    \UnaryInfC{$\Pi_{0}, \psi\fCenter\Sigma_{0}, \mathseqtoform{\Pi_{1}\fCenter\Sigma_{1}}$}
    \RightLabel{\rulename{$\mathdynbox{ \mathprogramtestsy }$ R}}
    \UnaryInfC{$\Pi_{0}\fCenter\Sigma_{0}, \mathdynbox{\mathprogramtest{\psi}}\mathseqtoform{\Pi_{1}\fCenter\Sigma_{1}}$}
   \end{inlineprooftree}.
  \end{center}

  Thus, if $(\Pi_{0}\fCenter\Sigma_{0}){\mathreduction{\mathprogramtest{\psi}}}(\Pi_{1}\fCenter\Sigma_{1})$
  does not hold,
  $\Pi_{0}\fCenter\Sigma_{0}, \mathdynbox{\mathprogramtest{\psi}}\mathseqtoform{\Pi_{1}\fCenter\Sigma_{1}}$
  is provable in \GTPDL.
 \end{proof}

  \subsection{Proofs of the completeness of \GTPDL\ and the finite model property of \TPDL}
  \label{subsec:proofs-completeness-fmp}
 Now, we show \cref{prop:completeness-of-GTPDL} (the completeness of \GTPDL) and \cref{prop:finite-model-property} (the finite model property of \TPDL).

\begin{proof}[Proof of \cref{prop:completeness-of-GTPDL}]
 We show the contrapositive of \cref{prop:completeness-of-GTPDL}.

 Assume that $\Gamma\fCenter\Delta$ is not provable in \GTPDL.
 By \cref{lemma:not-provable-sequent-to-saturated-sequent},
 there is a saturated sequent $\tilde{\Gamma}\fCenter\tilde{\Delta}$ 
 such that $\Gamma\subseteq\tilde{\Gamma}$ and $\Delta\subseteq\tilde{\Delta}$.
 Let $M$ be the canonical counter model of $\tilde{\Gamma}\fCenter\tilde{\Delta}$.
 By \cref{lemma:key-lemma-completeness-GTPDL} \cref{item:antecedent-lemma-key-lemma-completeness-GTPDL} 
 and \cref{item:consequent-lemma-key-lemma-completeness-GTPDL},
 we have $M, (\tilde{\Gamma}\fCenter\tilde{\Delta})\not\models \Gamma\fCenter\Delta$.
\end{proof}

\begin{proof}[Proof of \cref{prop:finite-model-property}]
 Assume that $M\not\models {\Gamma\fCenter\Delta}$ for some model $M$.
 By \cref{prop:completeness-of-GTPDL},
 $\Gamma\fCenter\Delta$ is not provable in \GTPDL.
 By \cref{lemma:not-provable-sequent-to-saturated-sequent},
 there is a saturated sequent $\tilde{\Gamma}\fCenter\tilde{\Delta}$ 
 such that $\Gamma\subseteq\tilde{\Gamma}$ and $\Delta\subseteq\tilde{\Delta}$.
 Let $M'$ be the canonical counter model of $\tilde{\Gamma}\fCenter\tilde{\Delta}$.
 By \cref{lemma:key-lemma-completeness-GTPDL} \cref{item:antecedent-lemma-key-lemma-completeness-GTPDL} 
 and \cref{item:consequent-lemma-key-lemma-completeness-GTPDL},
 we have $M', (\tilde{\Gamma}\fCenter\tilde{\Delta})\not\models \Gamma\fCenter\Delta$.
 Then, $\mathcardinality{\maththeunderlyingset{M'}}\leq\mathcardinality{\mathof{D}{\tilde{\Gamma}\cup\tilde{\Delta}}}\leq 2^{\mathcardinality{\tilde{\Gamma}\cup\tilde{\Delta}}}$.
 Since $\tilde{\Gamma}\cup\tilde{\Delta}$ is finite,
 $\maththeunderlyingset{M'}$ is finite. Thus, $M'$ is the model which we want.
\end{proof}

In the proof of \cref{prop:completeness-of-GTPDL},
we have the size of a counter model of an invalid sequent $\Gamma\fCenter\Delta$.
Hence, the following corollary holds.

\begin{corollary}
 \label{prop:finite-model-property-size}
 If $M\not\models {\Gamma\fCenter\Delta}$ holds for some model $M$, 
 then there exists a model 
 $M'$
 such that $M'\not\models {\Gamma\fCenter\Delta}$ and 
 $\maththeunderlyingset{M'}\leq 2^{\mathcardinality{\mathFLcl{\Gamma\cup\Delta}}}$. 
\end{corollary}

\section{\CGTPDL : a non-labelled cyclic proof system}
\label{sec:CGTPDL}
This section introduces a non-labelled cyclic proof system \CGTPDL\ and 
shows its completeness.

 The inference rules of \CGTPDL\ are the same as those of \GTPDL\ except for
 \rulename{$\mathdynbox{\mathprogramitsy}$ R}.
 The rule is replaced by the case-split rule, 
 the following rule:
 \begin{center}
  \begin{inlineprooftree}
   \AxiomC{$\Gamma \fCenter \varphi, \Delta$}
   \AxiomC{$\Gamma \fCenter \mathdynbox{\pi}\mathdynbox{\mathprogramit{\pi}}\varphi, \Delta$}
   \RightLabel{\rulename{C-s}}
   \BinaryInfC{$\Gamma \fCenter \mathdynbox{\mathprogramit{\pi}}\varphi, \Delta$}
  \end{inlineprooftree}.
 \end{center}
 
 \begin{definition}[Companion]
   For a derivation tree,
   we define a \emph{companion} for a bud 
   as an inner node labelled by the sequent same as the bud.
 \end{definition}

  \begin{definition}[\CGTPDL\ pre-proof]
   We define a \CGTPDL\ \emph{pre-proof}
   to be a pair $\mleft(\mathdertree{D},  \mathcompanion{C}\mright)$
   such that $\mathdertree{D}$ is a derivation tree and 
  $\mathcompanion{C}$ is a function mapping each bud to its companion.
  \end{definition}  
  The \emph{derivation graph} of a pre-proof 
  $\mleft(\mathdertree{D}, \mathcompanion{C}\mright)$, 
  written as $\mathproofgraph{\mleft(\mathdertree{D},  \mathcompanion{C}\mright)}$,
  is the graph obtained from $\mathdertree{D}$ by identifying each bud $\mu$ in $\mathdertree{D}$
  with its companion $\mathof{\mathcompanion{C}}{\mu}$.

 \begin{figure}[tb]
  \centering
  \begin{inlineprooftree}
   \AxiomC{\tikzmark{node-bud-CGTPDL-empty}($\surd$)\qquad $ \fCenter $ \quad}
   \RightLabel{\rulename{Wk}}
   \LeftLabel{\tikzmark{node-M-CGTPDL-empty}}
   \UnaryInfC{\tikzmark{node-companion-CGTPDL-empty}($\surd$)\qquad $ \fCenter $ \quad}
  \end{inlineprooftree}
  \begin{tikzpicture}[remember picture, overlay, thick, relative, auto, rounded corners, line width=1.0pt]
   \coordinate (b) at ({pic cs:node-bud-CGTPDL-empty});
   \coordinate (c) at ({pic cs:node-companion-CGTPDL-empty});
   \coordinate (M) at ({pic cs:node-M-CGTPDL-empty});
   
   \draw[->] let \p1=($(M)-(b)$), \p2=($(c)-(b)$) in
   (b) -- ++(\x1 - 1.0em,0) -- ++(0,\y2) -- (c);
  \end{tikzpicture}
  \caption{\CGTPDL\ pre-proof of the empty sequent}
  \label{fig:circular-der-tree-of-empty}
 \end{figure}
 
 \ifoup
    \begin{figure*}[tb]
   \centering
   \begin{inlineprooftree}
    \scriptsize
    \AxiomC{}
    \RightLabel{\rulename{Ax}}
    \UnaryInfC{$p, \mathdynbox{\mathprogramit{\alpha}}(p\to\mathdynbox{\alpha} p) \fCenter p$}

    \AxiomC{}
    \RightLabel{\rulename{Ax}}
    \UnaryInfC{$p, \mathdynbox{\mathprogramit{\alpha}}(p\to\mathdynbox{\alpha} p), \mathdynbox{\alpha} p \fCenter \mathdynbox{\alpha}\mathdynbox{\mathprogramit{\alpha}}p, p$}

    \AxiomC{($\spadesuit$) $p, \mathdynbox{\mathprogramit{\alpha}}(p\to\mathdynbox{\alpha} p) \fCenter \underline{\mathdynbox{\mathprogramit{\alpha}}p}$ \tikzmark{node-bud-not-cut-free-proof-1}}
    \RightLabel{\rulename{$\mathdynbox{ \, }$}}
    \UnaryInfC{$\mathdynbox{\alpha}p, \mathdynbox{\alpha}\mathdynbox{\mathprogramit{\alpha}}(p\to\mathdynbox{\alpha} p) \fCenter \underline{\mathdynbox{\alpha}\mathdynbox{\mathprogramit{\alpha}}p}$}
    \RightLabel{\rulename{$\mathdynbox{\mathprogramitsy}$ L S} \tikzmark{node-M-not-cut-free-proof-1}}
   \UnaryInfC{$\mathdynbox{\alpha}p, \mathdynbox{\mathprogramit{\alpha}}(p\to\mathdynbox{\alpha} p) \fCenter \underline{\mathdynbox{\alpha}\mathdynbox{\mathprogramit{\alpha}}p}$}
    \RightLabel{\rulename{Wk}}
   \UnaryInfC{$p, \mathdynbox{\mathprogramit{\alpha}}(p\to\mathdynbox{\alpha} p), \mathdynbox{\alpha}p \fCenter \underline{\mathdynbox{\alpha}\mathdynbox{\mathprogramit{\alpha}}p}$}

    \RightLabel{\rulename{$\to$ L}}
   \BinaryInfC{$p, \mathdynbox{\mathprogramit{\alpha}}(p\to\mathdynbox{\alpha} p), p\to\mathdynbox{\alpha} p \fCenter \underline{\mathdynbox{\alpha}\mathdynbox{\mathprogramit{\alpha}}p}$}
    \RightLabel{\rulename{$\mathdynbox{\mathprogramitsy}$ L B}}
   \UnaryInfC{$p, \mathdynbox{\mathprogramit{\alpha}}(p\to\mathdynbox{\alpha} p) \fCenter \underline{\mathdynbox{\alpha}\mathdynbox{\mathprogramit{\alpha}}p}$}

   \RightLabel{\rulename{C-s}}
   \BinaryInfC{($\spadesuit$) $p, \mathdynbox{\mathprogramit{\alpha}}(p\to\mathdynbox{\alpha} p) \fCenter \underline{\mathdynbox{\mathprogramit{\alpha}}p}$ \tikzmark{node-companion-not-cut-free-proof-1}}
   \end{inlineprooftree}
   \begin{tikzpicture}[remember picture, overlay, thick, relative, auto, rounded corners, line width=1.0pt]
    \coordinate (b) at ({pic cs:node-bud-not-cut-free-proof-1});
    \coordinate (c) at ({pic cs:node-companion-not-cut-free-proof-1});
    \coordinate (M) at ({pic cs:node-M-not-cut-free-proof-1});
   
    \draw[->] let \p1=($(M)-(b)$), \p2=($(c)-(b)$) in
    (b) -- ++(\x1 + 0.125em,0) -- ++(0,\y2) -- (c);
   \end{tikzpicture}

   \caption{A \CGTPDL\ proof of $p, \mathdynbox{\mathprogramit{\alpha}}(p\to\mathdynbox{\alpha} p) \fCenter \mathdynbox{\mathprogramit{\alpha}}p$}
   \label{fig:example-CGTPDL-proof}
  \end{figure*}

  \fi

 Because there is a \CGTPDL\ pre-proof of an invalid sequent (\cref{fig:circular-der-tree-of-empty}),
 each proof in this system must satisfy the condition for soundness, 
 \emph{the global trace condition}. 
 To describe the condition, we define some concepts.

  \begin{definition}[Path] 
   We define a \emph{path} in a derivation graph of a pre-proof 
   $\mathproofgraph{\mleft(\mathdertree{D},  \mathcompanion{C}\mright)}$ to be 
   a (possibly infinite) sequence 
   $\mathsequence{\Gamma_{i} \fCenter \Delta_{i}}{0\leq i < \alpha}$ of nodes in 
   $\mathproofgraph{\mleft(\mathdertree{D},  \mathcompanion{C}\mright)}$
   such that $\Gamma_{i+1} \fCenter \Delta_{i+1}$ is an assumption of  
   $\Gamma_{i} \fCenter \Delta_{i}$
   and $\alpha\in \mathnat \cup \mathsetextension{\omega}$, where 
   $\mathnat$ is the set of natural numbers and $\omega$ is the least infinite ordinal.
  \end{definition}

 \begin{definition}[Trace] 
  \label[definition]{def:Trace}
  For a path $\mathsequence{\Gamma_{i} \fCenter \Delta_{i}}{0\leq i < \alpha}$
  in a derivation graph of a pre-proof 
  $\mathproofgraph{\mleft(\mathdertree{D},  \mathcompanion{C}\mright)}$,
   we define a \emph{trace following}  
  $\mathsequence{\Gamma_{i} \fCenter \Delta_{i}}{0\leq i < \alpha}$ 
  to be a sequence of formulae
  $\mathsequence{\tau_{i}}{0\leq i}$
  such that the following conditions hold:
   \begin{enumerate}
    \item $\tau_{i}\in\Delta_{i}$. 
    \item If $\Gamma_{i}\fCenter \Delta_{i}$ is the conclusion of \rulename{$\to$ R}, 
	  then either
	  \begin{itemize}
	   \item[(r)]  $\tau_{i}$ is the principal formula 
		  $\phi\to\psi$ of the rule and 
		  $\tau_{i+1}$ is $\psi$, or
	   \item  $\tau_{i+1}$ is the same as $\tau_{i}$. 
	  \end{itemize}
    \item If $\Gamma_{i}\fCenter \Delta_{i}$ is the conclusion of \rulename{$\mathdynbox{ \, }$}, 
	  then 
	  \begin{itemize}
	   \item[(r)]  $\tau_{i}$ is the principal formula 
		  $\mathdynbox{\pi}\varphi$ of the rule and 
		  $\tau_{i+1}$ is $\varphi$.
	  \end{itemize}
    \item If $\Gamma_{i}\fCenter \Delta_{i}$ is the conclusion of \rulename{$\mathrevbox{ \, }$}, 
	  then 
	  \begin{itemize}
	   \item[(r)]  $\tau_{i}$ is the principal formula 
		  $\mathrevbox{\pi}\varphi$ of the rule and 
		  $\tau_{i+1}$ is $\varphi$.
	  \end{itemize}
    \item If $\Gamma_{i}\fCenter \Delta_{i}$ is the conclusion of \rulename{$\mathdynbox{\mathprogramseq{}{}}$ R}, 
	  then either
	  \begin{itemize}
	   \item[(r)]  $\tau_{i}$ is the principal formula 
		  $\mathdynbox{\mathprogramseq{\pi_{0}}{\pi_{1}}}\varphi$ of the rule and 
		  $\tau_{i+1}$ is $\mathdynbox{\pi_{0}}\mathdynbox{\pi_{1}}\varphi$, or
	   \item  $\tau_{i+1}$ is the same as $\tau_{i}$. 
	  \end{itemize}
    \item If $\Gamma_{i}\fCenter \Delta_{i}$ is the conclusion of \rulename{$\mathdynbox{ \mathprogramndc{}{} }$ R}
	  with
	  $\Delta_{i}=\Delta, \mathdynbox{ \mathprogramndc{\pi_{0}}{\pi_{1}} }\varphi$ 
	  and $\Delta_{i+1}=\Delta, \mathdynbox{ \pi_{j} }\varphi$ 
	  for some $j=0, 1$ and some $\Delta$, 
	  then either
	  \begin{itemize}
	   \item[(r)] $\tau_{i}$ is the principal formula 
		      $\mathdynbox{ \mathprogramndc{\pi_{0}}{\pi_{1}} }\varphi$ of 
		      the rule, and
		      $\tau_{i+1}$ is $\mathdynbox{ \pi_{j} }\varphi$, or
	   \item  $\tau_{i+1}$ is the same as $\tau_{i}$. 
	  \end{itemize}
    \item If $\Gamma_{i}\fCenter \Delta_{i}$ is the conclusion of \rulename{$\mathdynbox{ \mathprogramtestsy }$ R}, 
	  then either
	  \begin{itemize}
	   \item[(r)]  $\tau_{i}$ is the principal formula 
		  $\mathdynbox{ \mathprogramtest{\varphi} }\psi$ of the rule and
		  $\tau_{i+1}$ is $\psi$, or
	   \item  $\tau_{i+1}$ is the same as $\tau_{i}$. 
	  \end{itemize}
    \item If $\Gamma_{i}\fCenter \Delta_{i}$ is  the conclusion of \rulename{C-s}
	  with
	  $\Delta_{i}=\Delta, \mathdynbox{\mathprogramit{\pi}}\varphi$ and 
	  $\Delta_{i+1}=\Delta, \varphi$ for some $\Delta$, 
	  then either
	  \begin{itemize}
	   \item[(r)]  $\tau_{i}$ is the principal formula 
		  $\mathdynbox{\mathprogramit{\pi}}\varphi$ of the rule and 
		  $\tau_{i+1}$ is $\varphi$, or
	   \item  $\tau_{i+1}$ is the same as $\tau_{i}$. 
	  \end{itemize}
    \item If $\Gamma_{i}\fCenter \Delta_{i}$ is  the conclusion of \rulename{C-s}
	  with
	  $\Delta_{i}=\Delta, \mathdynbox{\mathprogramit{\pi}}\varphi$ and 
	  $\Delta_{i+1}=\Delta, \mathdynbox{\pi}\mathdynbox{\mathprogramit{\pi}}\varphi$ for some $\Delta$, 
	  then either
	  \begin{itemize}
	   \item[(r)]  $\tau_{i}$ is the principal formula 
		  $\mathdynbox{\mathprogramit{\pi}}\varphi$ of the rule and 
		  $\tau_{i+1}$ is 
		  $\mathdynbox{\pi}\mathdynbox{\mathprogramit{\pi}}\varphi$, or
	   \item  $\tau_{i+1}$ is the same as $\tau_{i}$. 

		  In the former case, $i$ is said to be a \emph{progress point} of the trace.
	  \end{itemize}
    \item If $\Gamma_{i} \fCenter \Delta_{i}$ is the conclusion of any other rule, 
	  then $\tau_{i+1}\equiv\tau_{i}$.
   \end{enumerate}

  In the cases annotated by (r), 
  $i$ is said to be a \emph{rewriting point} of the trace.
  If a trace has infinitely many progress points,
   we call the trace an \emph{infinitely progressing trace}.
 \end{definition}
  
  We note that an element of a trace which is a conclusion of 
  \rulename{$\mathdynbox{ \, }$} or \rulename{$\mathrevbox{ \, }$}
  must be the principal formula.

  \begin{definition}[Global trace condition]
   For a derivation graph,
   if, for every infinite path $\mathsequence{\Gamma_{i} \fCenter \Delta_{i}}{i\geq 0}$
   in the derivation graph, 
   there exists an infinitely progressing trace following a tail of the path
   $\mathsequence{\Gamma_{i} \fCenter \Delta_{i}}{i\geq k}$ with some $k \geq 0$,
   we say that the derivation graph satisfies \emph{the global trace condition}.
  \end{definition}

  \begin{definition}[\CGTPDL\ proof]
   We define a \emph{\CGTPDL\ proof} of a sequent $\Gamma \fCenter \Delta$ 
   to be a \CGTPDL\ pre-proof of $\Gamma \fCenter \Delta$ 
   whose graph satisfies the global trace condition.
   If a \CGTPDL\ proof of $\Gamma \fCenter \Delta$ exists, 
   we say $\Gamma \fCenter \Delta$ is \emph{provable} in \CGTPDL.
  \end{definition}

  \unless\ifoup
  
  \fi

  An example of \CGTPDL\ proof is \cref{fig:example-CGTPDL-proof}, where
  ($\spadesuit$) indicates the pairing of the companion with the bud and 
  the underlined formulae denote
  the infinitely progressing trace for some tails of the infinite path.

  The soundness theorem of \CGTPDL\ is the following.
  \begin{theorem}[Soundness of \CGTPDL]
   \label{theorem:soundness-CGTPDL}
  If $\Gamma\fCenter\Delta$ is provable in \CGTPDL,
  then $\Gamma\fCenter\Delta$ is valid.
  \end{theorem}

  This theorem is proven in \cref{sec:sound-CGTPDL} since the proof is long.

  We show the completeness of \CGTPDL\ under \cref{theorem:soundness-CGTPDL}.

  \begin{theorem}[Completeness of \CGTPDL]
   \label{thm:completeness-of-CGTPDL}
   For any sequent $\Gamma\fCenter\Delta$, the following three statements are equivalent: 
   \begin{enumerate}
    \item $\Gamma\fCenter\Delta$ is provable in \GTPDL.    
	  \label{item:GTPDL-thm-completeness-of-CGTPDL}
    \item $\Gamma\fCenter\Delta$ is provable in \CGTPDL.
	  \label{item:CGTPDL-thm-completeness-of-CGTPDL}
    \item $\Gamma\fCenter\Delta$ is valid.
	  \label{item:validity-thm-completeness-of-CGTPDL}
   \end{enumerate}
  \end{theorem}

  \begin{proof}
   We show \cref{item:GTPDL-thm-completeness-of-CGTPDL} $\Rightarrow$
   \cref{item:CGTPDL-thm-completeness-of-CGTPDL} $\Rightarrow$
   \cref{item:validity-thm-completeness-of-CGTPDL} $\Rightarrow$
   \cref{item:GTPDL-thm-completeness-of-CGTPDL}.

   \begin{figure*}[tb]
    \centering
    \begin{inlineprooftree}
    \small
    \AxiomC{\phantom{$\mathdynbox{\mathprogramit{\pi}}\Gamma, \varphi \fCenter \varphi$}}
    \RightLabel{\rulename{Ax}}
    \UnaryInfC{$\mathdynbox{\mathprogramit{\pi}}\Gamma, \varphi \fCenter \varphi$}

    \AxiomC{$\Gamma, \varphi \fCenter \mathdynbox{\pi}\varphi$}
    \RightLabel{\rulename{$\mathdynbox{\mathprogramitsy}$ L B}'s}
    
    \DeduceC{$\mathdynbox{\mathprogramit{\pi}}\Gamma, \varphi \fCenter \mathdynbox{\pi}\varphi$}
    \RightLabel{\rulename{Wk}}
    \UnaryInfC{$\mathdynbox{\mathprogramit{\pi}}\Gamma, \varphi \fCenter \mathdynbox{\pi}\varphi, \mathdynbox{\pi}\mathdynbox{\mathprogramit{\pi}}\varphi$}

    \AxiomC{($\dagger$) $\mathdynbox{\mathprogramit{\pi}}\Gamma, \varphi \fCenter \underline{\mathdynbox{\mathprogramit{\pi}}\varphi}$ \tikzmark{node-bud-CGTPDL-sim}}
    \RightLabel{\rulename{$\mathdynbox{ \, }$}}
    \UnaryInfC{$\mathdynbox{\pi}\mathdynbox{\mathprogramit{\pi}}\Gamma, \mathdynbox{\pi}\varphi \fCenter \underline{\mathdynbox{\pi}\mathdynbox{\mathprogramit{\pi}}\varphi}$}

    \RightLabel{\rulename{$\mathdynbox{\mathprogramitsy}$ L S}'s}
    \DeduceC{$\mathdynbox{\mathprogramit{\pi}}\Gamma, \mathdynbox{\pi}\varphi \fCenter \underline{\mathdynbox{\pi}\mathdynbox{\mathprogramit{\pi}}\varphi}$}
    \RightLabel{\rulename{Wk}}
    \UnaryInfC{$\mathdynbox{\mathprogramit{\pi}}\Gamma, \varphi, \mathdynbox{\pi}\varphi \fCenter \underline{\mathdynbox{\pi}\mathdynbox{\mathprogramit{\pi}}\varphi}$}

    \RightLabel{\rulename{Cut} \tikzmark{node-M-CGTPDL-sim}}
    \BinaryInfC{$\mathdynbox{\mathprogramit{\pi}}\Gamma, \varphi \fCenter \underline{\mathdynbox{\pi}\mathdynbox{\mathprogramit{\pi}}\varphi}$}
    \RightLabel{\rulename{C-s}}
    \BinaryInfC{($\dagger$) $\mathdynbox{\mathprogramit{\pi}}\Gamma, \varphi \fCenter \underline{\mathdynbox{\mathprogramit{\pi}}\varphi}$ \tikzmark{node-companion-CGTPDL-sim}}
    \end{inlineprooftree}
   \begin{tikzpicture}[remember picture, overlay, thick, relative, auto, rounded corners, line width=1.0pt]
    \coordinate (b) at ({pic cs:node-bud-CGTPDL-sim});
    \coordinate (c) at ({pic cs:node-companion-CGTPDL-sim});
    \coordinate (M) at ({pic cs:node-M-CGTPDL-sim});
   
    \draw[->] let \p1=($(M)-(b)$), \p2=($(c)-(b)$) in
    (b) -- ++(\x1 + 0.125em,0) -- ++(0,\y2) -- (c);
   \end{tikzpicture}

    \caption{A simulation of \rulename{$\mathdynbox{\mathprogramitsy}$ R} in \CGTPDL}
    \label{fig:CGTPDL-sim}
   \end{figure*}

   \noindent
   \cref{item:GTPDL-thm-completeness-of-CGTPDL} $\Rightarrow$
   \cref{item:CGTPDL-thm-completeness-of-CGTPDL}: 
   It suffices to show that \rulename{$\mathdynbox{\mathprogramitsy}$ R} 
   is derivable in \CGTPDL.
   Note that \rulename{$\mathdynbox{\mathprogramitsy}$ L B} and 
   \rulename{$\mathdynbox{\mathprogramitsy}$ L S} in \cref{prop:old-rules} are derivable 
   in \CGPDL\ since both rules are derivable in \GTPDL\ without 
   \rulename{$\mathdynbox{\mathprogramitsy}$ R}.
   \rulename{$\mathdynbox{\mathprogramitsy}$ R} can be derivable in \CGTPDL\ as 
   \cref{fig:CGTPDL-sim},
   where ($\dagger$) indicates the pairing of the companion with the bud and 
   the underlined formulae denote
   the infinitely progressing trace for some tails of the infinite path.

   \noindent
   \cref{item:CGTPDL-thm-completeness-of-CGTPDL} $\Rightarrow$
   \cref{item:validity-thm-completeness-of-CGTPDL}: By \cref{theorem:soundness-CGTPDL}.

   \noindent
   \cref{item:validity-thm-completeness-of-CGTPDL} $\Rightarrow$
   \cref{item:GTPDL-thm-completeness-of-CGTPDL}: By \cref{prop:completeness-of-GTPDL}.
  \end{proof}

  \section{Soundness of \CGTPDL}
  \label{sec:sound-CGTPDL}
  This section proves the soundness of \CGTPDL, \cref{theorem:soundness-CGTPDL}.
  To show the soundness of \CGTPDL, we define some concepts and show some lemmata.
  
  We define an \emph{annotated program} as a program annotated with $(f)$ or $(b)$.

  \begin{definition}[Execution]
   Let $M$ be a model.

   An \emph{execution} (of length $k$) on $M$ is defined as an alternating sequence
   $w_{0}\xi_{0}w_{1}\pi_{1}\dots \xi_{k-1}w_{k}$ of states and annotated programs 
   such that,
   for all $i=0, 1, \dots, {k-1}$,
   $\xi_{i}=\mathannotated{\pi}{f}$ implies $w_{i}\mathreduction{\pi}w_{i+1}$ and
   $\xi_{i}=\mathannotated{\pi}{b}$ implies $w_{i+1}\mathreduction{\pi}w_{i}$.
  \end{definition}

  We write $\omega$ for the least infinite ordinal.
  We note that $\omega$ is greater than every natural number.

  \begin{definition}
   \label[definition]{def:counting-func}
   Let $M$ be a model.
   For $w_{0}, w_{1}\in \maththeunderlyingset{M}$ and an annotated program $\xi$,
   we inductively define $\mathcountit{w_{0}}{\xi}{w_{1}}$ on construction of a program 
   $\xi$ as follows:
   \begin{enumerate}
    \item $\mathcountit{w_{0}}{\mathannotated{\pi}{f}}{w_{1}}=\omega$ 
	  if $w_{0}\mathreduction{\pi}w_{1}$ does not hold.
    \item $\mathcountit{w_{0}}{\mathannotated{\pi}{b}}{w_{1}}=\omega$ 
	  if $w_{1}\mathreduction{\pi}w_{0}$ does not hold.
    \item $\mathcountit{w_{0}}{\mathannotated{\alpha}{f}}{w_{1}}=0$ 
	  for $\alpha\in\maththesetofatomicprog$ if $w_{0}\mathreduction{\alpha}w_{1}$.
    \item $\mathcountit{w_{0}}{\mathannotated{\alpha}{b}}{w_{1}}=0$ 
	  for $\alpha\in\maththesetofatomicprog$ if $w_{1}\mathreduction{\alpha}w_{0}$.
    \item $\mathcountit{w_{0}}{\mathannotated{(\mathprogramseq{\pi_{0}}{\pi_{1}})}{f}}{w_{1}}=\mathof{\min}{\mathsetintension{\mathcountit{w_{0}}{\mathannotated{\pi_{0}}{f}}{w'}+\mathcountit{w'}{\mathannotated{\pi_{1}}{f}}{w_{1}}}{w_{0}\mathreduction{\pi_{0}}w' \text{ and } w'\mathreduction{\pi_{1}}w_{1}}}$
	  if $w_{0}\mathreduction{\mathprogramseq{\pi_{0}}{\pi_{1}}}w_{1}$.
    \item $\mathcountit{w_{0}}{\mathannotated{(\mathprogramseq{\pi_{0}}{\pi_{1}})}{b}}{w_{1}}=\mathof{\min}{\mathsetintension{\mathcountit{w_{0}}{\mathannotated{\pi_{0}}{b}}{w'}+\mathcountit{w'}{\mathannotated{\pi_{1}}{b}}{w_{1}}}{w_{1}\mathreduction{\pi_{0}}w' \text{ and } w'\mathreduction{\pi_{1}}w_{0}}}$
	  if $w_{1}\mathreduction{\mathprogramseq{\pi_{0}}{\pi_{1}}}w_{0}$.
    \item $\mathcountit{w_{0}}{\mathannotated{(\mathprogramndc{\pi_{0}}{\pi_{1}})}{f}}{w_{1}}=\mathof{\min}{\mathcountit{w_{0}}{\mathannotated{\pi_{0}}{f}}{w_{1}}, \mathcountit{w_{0}}{\mathannotated{\pi_{1}}{f}}{w_{1}}}$
	  if $w_{0}\mathreduction{\mathprogramndc{\pi_{0}}{\pi_{1}}}w_{1}$.
    \item $\mathcountit{w_{0}}{\mathannotated{(\mathprogramndc{\pi_{0}}{\pi_{1}})}{b}}{w_{1}}=\mathof{\min}{\mathcountit{w_{0}}{\mathannotated{\pi_{0}}{b}}{w_{1}}, \mathcountit{w_{0}}{\mathannotated{\pi_{1}}{b}}{w_{1}}}$
	  if $w_{1}\mathreduction{\mathprogramndc{\pi_{0}}{\pi_{1}}}w_{0}$.
    \item $\mathcountit{w_{0}}{\mathannotated{(\mathprogramit{\pi})}{a}}{w_{1}}=0$
   	  if $w_{0}=w_{1}$ and $\mathannotation{a}=\mathannotation{f}, \mathannotation{b}$.
    \item $\mathcountit{w_{0}}{\mathannotated{(\mathprogramit{\pi})}{f}}{w_{1}}=n+\mathof{\min}{\mathsetintension{\sum_{i=0}^{n-1}\mathcountit{v_{i}}{\mathannotated{\pi}{f}}{v_{i+1}}}{ 
\begin{gathered}
 v_{0}=w_{0}, v_{n}=w_{1} \text{ and } \\ 
 v_{i}\mathreduction{\pi}v_{i+1} \text{ for all } i < n
\end{gathered} 
}}$, 
   	  where
	  $n=\mathof{\min}{\mathsetintension{m}{ w_{0}\mathreduction{\pi}^{m}w_{1} }}$
	  if $w_{0}\mathreduction{\mathprogramit{\pi}}w_{1}$ and $w_{0}\neq w_{1}$.
    \item $\mathcountit{w_{0}}{\mathannotated{(\mathprogramit{\pi})}{b}}{w_{1}}=n+\mathof{\min}{\mathsetintension{\sum_{i=0}^{n-1}\mathcountit{v_{i}}{\mathannotated{\pi}{b}}{v_{i+1}}}{ 
	  \begin{gathered}
	  v_{0}=w_{1}, v_{n}=w_{0} \text{ and } \\
  v_{i+1}\mathreduction{\pi}v_{i} \text{ for all } i < n
	  \end{gathered} 
}}$, 
   	  where 
	  $n=\mathof{\min}{\mathsetintension{m}{ w_{1}\mathreduction{\pi}^{m}w_{0} }}$
	  if $w_{1}\mathreduction{\mathprogramit{\pi}}w_{0}$ and $w_{0}\neq w_{1}$.
    \item $\mathcountit{w_{0}}{\mathannotated{(\mathprogramtest{\varphi})}{a}}{w_{1}}=0$
   	  if $w_{0}\mathreduction{\mathprogramtest{\varphi}}w_{1}$
	  and $\mathannotation{a}=\mathannotation{f}, \mathannotation{b}$.
   \end{enumerate}
  \end{definition}

  For an execution $w_{0}\dots \xi_{k-1}w_{k}$,
  we define $\mathcountitex{w_{0}\dots \xi_{k-1}w_{k}}$ by
  \[
  \mathcountitex{w_{0}\dots \xi_{k-1}w_{k}}=
   \begin{cases}
    0, & k=0, \\
    \sum_{i=0}^{k-1}\mathcountit{w_{i}}{\xi_{i}}{w_{i+1}}, & k>0.
   \end{cases}
   \]
   
  For sets of executions $A$ and $B$,
  we define $A++B$ by
  \[
   A++B=\mathsetintension{v_{0}\dots \xi_{k-1}w_{0}\dots \xi'_{k'-1}w_{k'}}{v_{0}\dots \xi_{k-1}v_{k}\in A \text{ and } w_{0}\dots \xi'_{k'-1}w_{k'} \in B \text{ with } v_{k}=w_{0}}.
  \]

  \begin{definition}
   Let $M$ be a model. For $w\in \maththeunderlyingset{M}$,
   we inductively define a set of executions $\mathcounterex{M}{w}{\varphi}$ 
   on construction of a formula $\varphi$ as follows:
   \begin{enumerate}
    \item $\mathcounterex{M}{w}{\varphi}=\emptyset$ if ${M, w}\models{\varphi}$.
    \item $\mathcounterex{M}{w}{\bot}=\mathsetextension{w}$.
    \item $\mathcounterex{M}{w}{p}=\mathsetextension{w}$ for $p\in\maththesetofpropval$
	  if ${M, w}\not\models{p}$.
    \item $\mathcounterex{M}{w}{\psi_{0}\to\psi_{1}}=\mathcounterex{M}{w}{\psi_{1}}$
	  if ${M, w}\not\models{\psi_{0}\to\psi_{1}}$.
    \item $\mathcounterex{M}{w}{\mathdynbox{\pi}\psi}=\bigcup_{v\in N}\mleft( \mathsetextension{w\mathannotated{\pi}{f}v}  ++ \mathcounterex{M}{v}{\psi}\mright)$ 
	  for 
	  $N=\mathsetintension{v}{w\mathreduction{\pi}v \text{ and } {M, v}\not\models{\psi}}$
	  if ${M, w}\not\models{\mathdynbox{\pi}\varphi}$.
    \item $\mathcounterex{M}{w}{\mathrevbox{\pi}\psi}=\bigcup_{v\in B}\mleft( \mathsetextension{w\mathannotated{\pi}{b}v}  ++ \mathcounterex{M}{v}{\psi}\mright)$ 
	  for 
	  $B=\mathsetintension{v}{v\mathreduction{\pi}w \text{ and } {M, v}\not\models{\psi}}$
	  if ${M, w}\not\models{\mathrevbox{\pi}\varphi}$.
   \end{enumerate}

   We define $\mathiteration{M}{w}{\varphi}$ by
   \[
    \mathiteration{M}{w}{\varphi}=\mathof{\min}{\mathsetintension{\mathcountitex{w_{0}\dots \xi_{k-1}w_{k}}}{w_{0}\dots \xi_{k-1}w_{k}\in\mathcounterex{M}{w}{\varphi}}}.
   \]

   We define a set of executions $\mathmincounterex{M}{w}{\varphi}$ by
   \[
   \mathmincounterex{M}{w}{\varphi}=\mathsetintension{w_{0}\dots \xi_{k-1}w_{k}\in\mathcounterex{M}{w}{\varphi}}{\mathcountitex{w_{0}\dots \xi_{k-1}w_{k}}=\mathiteration{M}{w}{\varphi}}.
   \]

   We say that $\varphi$ \emph{is able to stay at $w$ in $M$}
   if there exists 
   $w_{0}\xi_{0}w_{1}\dots \xi_{k-1}w_{k}\in\mathmincounterex{M}{w}{\varphi}$
   such that $w_{0}=w_{1}$.
  \end{definition}

  \begin{lemma}
   \label[lemma]{lemma:it-imply}
   Let $M$ be a model.
   For a formula $\varphi\to\psi$,
   if ${M, w}\not\models{\varphi\to\psi}$,
   then 
   $\mathiteration{M}{w}{\varphi\to\psi}\geq\mathiteration{M}{w}{\psi}$.
  \end{lemma}

  \begin{proof}
   Let $M$ be a model. 
   Assume ${M, w}\not\models{\varphi\to\psi}$ for a formula $\varphi\to\psi$.

   Let $w_{0}\dots \xi_{k-1}w_{k}\in\mathmincounterex{M}{w}{\varphi\to\psi}$.
   Then, $w_{0}\dots \xi_{k-1}w_{k}\in\mathcounterex{M}{w}{\varphi\to\psi}$.
   By $\mathcounterex{M}{w}{\varphi\to\psi}=\mathcounterex{M}{w}{\psi}$,
   we have $w_{0}\dots \xi_{k-1}w_{k}\in\mathcounterex{M}{w}{\psi}$.
   Hence, we see
   \begin{align*}
    \mathiteration{M}{w}{\varphi\to\psi}
    &=\mathcountitex{w_{0}\dots \xi_{k-1}w_{k}} \\
    &\geq\mathiteration{M}{w}{\psi}.
   \end{align*}
  \end{proof}

  \begin{lemma}
   \label[lemma]{lemma:IT-monotone-dynbox}
   Let $M$ be a model. For a formula $\varphi$ and a program $\pi$,
   if ${M, w}\not\models{\mathdynbox{\pi}\varphi}$,
   then there exists $w'\in W$ such that ${M, w'}\not\models{\varphi}$,
   $\mathiteration{M}{w}{\mathdynbox{\pi}\varphi}\geq\mathiteration{M}{w'}{\varphi}$ and
   $w\mathreduction{\pi}w'$.
  \end{lemma}

  \begin{proof}
   Let $M$ be a model.
   Assume ${M, w}\not\models{\mathdynbox{\pi}\varphi}$
   for a formula $\varphi$ and a program $\pi$.

   Let $w_{0}\xi_{0}w_{1}\dots \xi_{k-1}w_{k}\in\mathmincounterex{M}{w}{\mathdynbox{\pi}\varphi}$.
   Then, $w_{0}\xi_{0}w_{1}\dots \xi_{k-1}w_{k}\in\mathcounterex{M}{w}{\mathdynbox{\pi}\varphi}$.
   Hence, we have $w\mathreduction{\pi}w_{1}$ and
   ${M, w_{1}}\not\models{\varphi}$.
   We also have 
   $w_{1}\dots \xi_{k-1}w_{k}\in\mathcounterex{M}{w_{1}}{\varphi}$.
   Let $w'=w_{1}$.
   Then,
   \begin{align*}
    \mathiteration{M}{w}{\mathdynbox{\pi}\varphi}
    &=\mathcountitex{w_{0}\xi_{0}w_{1}\dots \xi_{k-1}w_{k}} \\
    &=\sum_{i=0}^{k-1}\mathcountit{w_{i}}{\xi_{i}}{w_{i+1}} \\
    &=\mathcountit{w_{0}}{\mathannotated{\pi}{f}}{w_{1}}+\sum_{i=1}^{k-1}\mathcountit{w_{i}}{\xi_{i}}{w_{i+1}} \\
    &\geq \sum_{i=1}^{k-1}\mathcountit{w_{i}}{\xi_{i}}{w_{i+1}} \\
    &= \mathcountitex{w_{1}\dots \xi_{k-1}w_{k}} \\
    &\geq \mathiteration{M}{w'}{\varphi}.
   \end{align*}
   
  \end{proof}
  
  \begin{lemma}
   \label[lemma]{lemma:IT-monotone-revbox}
   Let $M$ be a model.
   For a formula $\varphi$ and a program $\pi$,
   if ${M, w}\not\models{\mathrevbox{\pi}\varphi}$,
   there exists $w'\in W$ such that ${M, w'}\not\models{\varphi}$,
   $\mathiteration{M}{w}{\mathrevbox{\pi}\varphi}\geq\mathiteration{M}{w'}{\varphi}$ and
   $w'\mathreduction{\pi}w$.
  \end{lemma}

  \begin{proof}
   In the similar way to \cref{lemma:IT-monotone-dynbox}.
  \end{proof}

  \begin{lemma}
   \label[lemma]{lemma:it-seq}
   Let $M$ be a model.
   For a formula $\varphi$ and programs $\pi_{0}$, $\pi_{1}$,
   if ${M, w}\not\models{\mathdynbox{\mathprogramseq{\pi_{0}}{\pi_{1}}}\varphi}$,
   then 
   $\mathiteration{M}{w}{\mathdynbox{\mathprogramseq{\pi_{0}}{\pi_{1}}}\varphi}\geq\mathiteration{M}{w}{\mathdynbox{\pi_{0}}\mathdynbox{\pi_{1}}\varphi}$.
  \end{lemma}

  \begin{proof}
   Let $M$ be a model. 
   Assume ${M, w}\not\models{\mathdynbox{\mathprogramseq{\pi_{0}}{\pi_{1}}}\varphi}$
   for a formula $\varphi$ and programs $\pi_{0}$, $\pi_{1}$.

   Let $w_{0}\xi_{0}w_{1}\dots \xi_{k-1}w_{k}\in\mathmincounterex{M}{w}{\mathdynbox{\mathprogramseq{\pi_{0}}{\pi_{1}}}\varphi}$.
   Then, $w_{0}\xi_{0}w_{1}\dots \xi_{k-1}w_{k}\in\mathcounterex{M}{w}{\mathdynbox{\mathprogramseq{\pi_{0}}{\pi_{1}}}\varphi}$.
   Hence, we have $w_{0}\mathreduction{\mathprogramseq{\pi_{0}}{\pi_{1}}}w_{1}$ and
   ${M, w_{1}}\not\models{\varphi}$.
   Then, 
 \begin{align*}
  \mathiteration{M}{w}{\mathdynbox{\mathprogramseq{\pi_{0}}{\pi_{1}}}\varphi}
  &=\mathcountitex{w_{0}\xi_{0}w_{1}\dots \xi_{k-1}w_{k}} \\
  &=\sum_{i=0}^{k-1}\mathcountit{w_{i}}{\xi_{i}}{w_{i+1}} \\
  &=\mathcountit{w_{0}}{\mathannotated{(\mathprogramseq{\pi_{0}}{\pi_{1}})}{f}}{w_{1}}+\sum_{i=1}^{k-1}\mathcountit{w_{i}}{\xi_{i}}{w_{i+1}} \\
  &=\mathof{\min}{\mathsetintension{\mathcountit{w_{0}}{\mathannotated{\pi_{0}}{f}}{w''}+\mathcountit{w''}{\mathannotated{\pi_{1}}{f}}{w_{1}}}{w_{0}\mathreduction{\pi_{0}}w'' \text{ and } w''\mathreduction{\pi_{1}}w_{1}}} \\
  &\phantom{=} \qquad +\sum_{i=1}^{k-1}\mathcountit{w_{i}}{\xi_{i}}{w_{i+1}}.
 \end{align*}

   Let $w'\in \maththeunderlyingset{M}$ be such that $w_{0}\mathreduction{\pi_{0}}w'$,
   $w'\mathreduction{\pi_{1}}w_{1}$, and
   \begin{gather*}
   \mathcountit{w_{0}}{\mathannotated{\pi_{0}}{f}}{w'}+\mathcountit{w'}{\mathannotated{\pi_{1}}{f}}{w_{1}} \\
    =\mathof{\min}{\mathsetintension{\mathcountit{w_{0}}{\mathannotated{\pi_{0}}{f}}{w''}+\mathcountit{w''}{\mathannotated{\pi_{1}}{f}}{w_{1}}}{w_{0}\mathreduction{\pi_{0}}w'' \text{ and } w''\mathreduction{\pi_{1}}w_{1}}}.     
   \end{gather*}
   Then,
   $w_{0}\mathannotated{\pi_{0}}{f}w'\mathannotated{\pi_{1}}{f}\xi_{1}w_{1}\dots \xi_{k-1}w_{k}\in\mathcounterex{M}{w}{\mathdynbox{\pi_{0}}\mathdynbox{\pi_{1}}\varphi}$.
   We have
   \begin{align*}
    \mathiteration{M}{w}{\mathdynbox{\mathprogramseq{\pi_{0}}{\pi_{1}}}\varphi}
    &=\mathcountit{w_{0}}{\mathannotated{\pi_{0}}{f}}{w'}+\mathcountit{w'}{\mathannotated{\pi_{1}}{f}}{w_{1}}+\sum_{i=1}^{k-1}\mathcountit{w_{i}}{\xi_{i}}{w_{i+1}} \\
    &=\mathcountitex{w_{0}\mathannotated{\pi_{0}}{f}w'\mathannotated{\pi_{1}}{f}\xi_{1}w_{1}\dots \xi_{k-1}w_{k}} \\
    &\geq \mathiteration{M}{w}{\mathdynbox{\pi_{0}}\mathdynbox{\pi_{1}}\varphi}.
   \end{align*}   
  \end{proof}

  \begin{lemma}
   \label[lemma]{lemma:it-ndc}
   Let $M$ be a model.
   For a formula $\varphi$ and programs $\pi_{0}$, $\pi_{1}$,
   if ${M, w}\not\models{\mathdynbox{\mathprogramndc{\pi_{0}}{\pi_{1}}}\varphi}$ and
   $\mathiteration{M}{w}{\mathdynbox{ \pi_{j} }\varphi}\leq\mathiteration{M}{w}{\mathdynbox{ \pi_{1-j} }\varphi}$ for $j\in\mathsetextension{0, 1}$,
   then 
   $\mathiteration{M}{w}{\mathdynbox{\mathprogramndc{\pi_{0}}{\pi_{1}}}\varphi}\geq\mathiteration{M}{w}{\mathdynbox{\pi_{j}}\varphi}$.
  \end{lemma}

  \begin{proof}
   Let $M$ be a model. 
   Assume ${M, w}\not\models{\mathdynbox{\mathprogramndc{\pi_{0}}{\pi_{1}}}\varphi}$
   for a formula $\varphi$ and programs $\pi_{0}$, $\pi_{1}$.
   We note that 
   $\mathiteration{M}{w}{\mathdynbox{\mathprogramndc{\pi_{0}}{\pi_{1}}}\varphi}$ is finite.

   Let $w_{0}\xi_{0}w_{1}\dots \xi_{k-1}w_{k}\in\mathmincounterex{M}{w}{\mathdynbox{\mathprogramndc{\pi_{0}}{\pi_{1}}}\varphi}$.
   Then, $w_{0}\xi_{0}w_{1}\dots \xi_{k-1}w_{k}\in\mathcounterex{M}{w}{\mathdynbox{\mathprogramndc{\pi_{0}}{\pi_{1}}}\varphi}$.
   Hence, we have $w\mathreduction{\mathprogramndc{\pi_{0}}{\pi_{1}}}w_{1}$ and
   ${M, w_{1}}\not\models{\varphi}$.
   Then, 
   \begin{align*}
    \mathiteration{M}{w}{\mathdynbox{\mathprogramndc{\pi_{0}}{\pi_{1}}}\varphi}
    &=\mathcountitex{w_{0}\xi_{0}w_{1}\dots \xi_{k-1}w_{k}} \\
    &=\sum_{i=0}^{k-1}\mathcountit{w_{i}}{\xi_{i}}{w_{i+1}} \\
    &=\mathcountit{w_{0}}{\mathannotated{(\mathprogramndc{\pi_{0}}{\pi_{1}})}{f}}{w_{1}}+\sum_{i=1}^{k-1}\mathcountit{w_{i}}{\xi_{i}}{w_{i+1}} \\
    &=\mathof{\min}{\mathcountit{w_{0}}{\mathannotated{\pi_{0}}{f}}{w_{1}}, \mathcountit{w_{0}}{\mathannotated{\pi_{1}}{f}}{w_{1}}} \\
    &\phantom{=} \qquad +\sum_{i=1}^{k-1}\mathcountit{w_{i}}{\xi_{i}}{w_{i+1}}.
   \end{align*}
   Assume
   $\mathcountit{w_{0}}{\mathannotated{\pi_{l}}{f}}{w_{1}}\leq\mathcountit{w_{0}}{\mathannotated{\pi_{1-l}}{f}}{w_{1}}$ 
   for $l\in\mathsetextension{0, 1}$.
   Since 
   $\mathiteration{M}{w}{\mathdynbox{\mathprogramndc{\pi_{0}}{\pi_{1}}}\varphi}$ is finite,
   $\mathof{\min}{\mathcountit{w_{0}}{\mathannotated{\pi_{0}}{f}}{w_{1}}, \mathcountit{w_{0}}{\mathannotated{\pi_{1}}{f}}{w_{1}}}$ is finite.
   Hence, $\mathcountit{w_{0}}{\mathannotated{\pi_{l}}{f}}{w_{1}}$ is finite.
   Then, we have $w_{0}\mathreduction{\pi_{l}}w_{1}$.
   Then,
   $w_{0}\mathannotated{\pi_{l}}{f}w_{1}\dots \xi_{k-1}w_{k}\in\mathcounterex{M}{w}{\mathdynbox{\pi_{l}}\varphi}$
   and 
   \[
   \mathcountitex{w_{0}\dots \xi_{k-1}w_{k}}=\mathcountitex{w_{0}\mathannotated{\pi_{l}}{f}w_{1}\dots \xi_{k-1}w_{k}}.
   \]
   Hence,
   \begin{align*}
    \mathiteration{M}{w}{\mathdynbox{\mathprogramndc{\pi_{0}}{\pi_{1}}}\varphi}
    &=\mathcountitex{w_{0}\mathannotated{\pi_{l}}{f}w_{1}\dots \xi_{k-1}w_{k}} \\
    &\geq\mathiteration{M}{w}{\mathdynbox{\pi_{l}}\varphi}.
   \end{align*}
   If $l=j$, then we have 
   $\mathiteration{M}{w}{\mathdynbox{\mathprogramndc{\pi_{0}}{\pi_{1}}}\varphi}\geq\mathiteration{M}{w}{\mathdynbox{\pi_{j}}\varphi}$.

   If $l\neq j$, then we have 
   $\mathiteration{M}{w}{\mathdynbox{\mathprogramndc{\pi_{0}}{\pi_{1}}}\varphi}\geq\mathiteration{M}{w}{\mathdynbox{\pi_{1-j}}\varphi}$.
   Because of $\mathiteration{M}{w}{\mathdynbox{ \pi_{1-j} }\varphi}\geq\mathiteration{M}{w}{\mathdynbox{ \pi_{j} }\varphi}$,
   we have 
   $\mathiteration{M}{w}{\mathdynbox{\mathprogramndc{\pi_{0}}{\pi_{1}}}\varphi}\geq\mathiteration{M}{w}{\mathdynbox{\pi_{j}}\varphi}$.
  \end{proof}

  \begin{lemma}
   \label[lemma]{lemma:it-test}
   Let $M$ be a model.
   For a formula $\mathdynbox{\mathprogramtest{\varphi}}\psi$,
   if ${M, w}\not\models{\mathdynbox{\mathprogramtest{\varphi}}\psi}$,
   then 
   $\mathiteration{M}{w}{\mathdynbox{\mathprogramtest{\varphi}}\psi}\geq\mathiteration{M}{w}{\psi}$.
  \end{lemma}

  \begin{proof}
   Let $M$ be a model. 
   Assume ${M, w}\not\models{\mathdynbox{\mathprogramtest{\varphi}}\psi}$ for a formula $\mathdynbox{\mathprogramtest{\varphi}}\psi$.

   Let $w_{0}\xi_{0}w_{1}\dots \xi_{k-1}w_{k}\in\mathmincounterex{M}{w}{\mathdynbox{\mathprogramtest{\varphi}}\psi}$.
   Then, $w_{0}\xi_{0}w_{1}\dots \xi_{k-1}w_{k}\in\mathcounterex{M}{w}{\mathdynbox{\mathprogramtest{\varphi}}\psi}$.
   Hence, we have $w_{0}\mathreduction{\mathprogramtest{\varphi}}w_{1}$ and
   $w=w_{0}=w_{1}$.
   We also have $w_{1}\dots \xi_{k-1}w_{k}\in\mathcounterex{M}{w}{\psi}$.
   Hence, we see
   \begin{align*}
    \mathiteration{M}{w}{\mathdynbox{\mathprogramtest{\varphi}}\psi}
    &=\mathcountitex{w_{0}\xi_{0}w_{1}\dots \xi_{k-1}w_{k}} \\
    &=\sum_{i=0}^{k-1}\mathcountit{w_{i}}{\xi_{i}}{w_{i+1}} \\
    &=\mathcountit{w_{0}}{\mathannotated{(\mathprogramtest{\varphi})}{f}}{w_{1}}+\sum_{i=1}^{k-1}\mathcountit{w_{i}}{\xi_{i}}{w_{i+1}} \\
    &\geq \sum_{i=1}^{k-1}\mathcountit{w_{i}}{\xi_{i}}{w_{i+1}} \\
    &=\mathcountitex{w_{1}\dots \xi_{k-1}w_{k}} \\
    &\geq\mathiteration{M}{w}{\psi}.
   \end{align*}
  \end{proof}

  \begin{lemma}
   \label[lemma]{lemma:it-base-case}
   Let $M$ be a model.
   For a formula $\varphi$ and a program $\pi$,
   if ${M, w}\not\models{\mathdynbox{\mathprogramit{\pi}}\varphi}$ holds and 
   $\mathdynbox{\mathprogramit{\pi}}\varphi$ is able to stay at $w$ in $M$,
   then $\mathiteration{M}{w}{\mathdynbox{\mathprogramit{\pi}}\varphi}\geq\mathiteration{M}{w}{\varphi}$.
  \end{lemma}

  \begin{proof}
   Let $M$ be a model.
   Assume that ${M, w}\not\models{\mathdynbox{\mathprogramit{\pi}}\varphi}$ holds and 
   $\mathdynbox{\mathprogramit{\pi}}\varphi$ is able to stay at $w$ in $M$.

   Because $\mathdynbox{\mathprogramit{\pi}}\varphi$ is able to stay at $w$ in $M$,
   there exists
   $w_{0}\xi_{0}w_{1}\dots \xi_{k-1}w_{k}\in\mathmincounterex{M}{w}{\mathdynbox{\mathprogramit{\pi}}\varphi}$ 
   such that $w_{0}=w_{1}$.
   Then, $w_{0}\xi_{0}w_{1}\dots \xi_{k-1}w_{k}\in\mathcounterex{M}{w}{\mathdynbox{\mathprogramit{\pi}}\varphi}$.
   Hence, we have $w\mathreduction{\mathprogramit{\pi}}w_{1}$ and
   ${M, w_{1}}\not\models{\varphi}$.
   We also have 
   $w_{1}\dots \xi_{k-1}w_{k}\in\mathcounterex{M}{w_{1}}{\varphi}$.
   Because of $w=w_{0}=w_{1}$, we have
   \begin{align*}
    \mathiteration{M}{w}{\mathdynbox{\mathprogramit{\pi}}\varphi}
    &=\mathcountitex{w_{0}\xi_{0}w_{1}\dots \xi_{k-1}w_{k}} \\
    &=\sum_{i=0}^{k-1}\mathcountit{w_{i}}{\xi_{i}}{w_{i+1}} \\
    &=\mathcountit{w_{0}}{\mathannotated{\mathprogramit{\pi}}{f}}{w_{1}}+\sum_{i=1}^{k-1}\mathcountit{w_{i}}{\xi_{i}}{w_{i+1}} \\
    &= \sum_{i=1}^{k-1}\mathcountit{w_{i}}{\xi_{i}}{w_{i+1}} \\
    &= \mathcountitex{w_{1}\dots \xi_{k-1}w_{k}} \\
    &\geq \mathiteration{M}{w}{\varphi}.
   \end{align*}
  \end{proof}

  \begin{lemma}
   \label[lemma]{lemma:it-step-case}
   Let $M$ be a model.
   For a formula $\varphi$ and a program $\pi$,
   if ${M, w}\not\models{\mathdynbox{\mathprogramit{\pi}}\varphi}$ holds and 
   $\mathdynbox{\mathprogramit{\pi}}\varphi$ is not able to stay at $w$ in $M$,
   then $\mathiteration{M}{w}{\mathdynbox{\mathprogramit{\pi}}\varphi}>\mathiteration{M}{w}{\mathdynbox{\pi}\mathdynbox{\mathprogramit{\pi}}\varphi}$.
  \end{lemma}

  \begin{proof}
   Let $M$ be a model.
   Assume that ${M, w}\not\models{\mathdynbox{\mathprogramit{\pi}}\varphi}$ holds and 
   $\mathdynbox{\mathprogramit{\pi}}\varphi$ is not able to stay at $w$ in $M$.

   Let $w_{0}\xi_{0}w_{1}\dots \xi_{k-1}w_{k}\in\mathmincounterex{M}{w}{\mathdynbox{\mathprogramit{\pi}}\varphi}$.
   Then,
   $\mathcountitex{w_{0}\dots \xi_{k-1}w_{k}}=\mathiteration{M}{w}{\mathdynbox{\mathprogramit{\pi}}\varphi}$.
   Hence, we have $w\mathreduction{\mathprogramit{\pi}}w_{1}$ and 
   ${M, w_{1}}\not\models{\varphi}$.
   Because $\mathdynbox{\mathprogramit{\pi}}\varphi$ is not able to stay at $w$ in $M$, we see $w\neq w_{1}$.
   Hence, $n=\mathof{\min}{\mathsetintension{m}{ w_{0}\mathreduction{\pi}^{m}w_{1} }}$ 
   for $n>0$.
   Let $v_{0}\mathannotated{\pi}{f}v_{1}\mathannotated{\pi}{f}v_{2}\dots\mathannotated{\pi}{f}v_{n}$ be an execution satisfying 
   $v_{0}=w_{0}$, $v_{n}=w_{1}$, and
   \[
    \mathcountitex{v_{0}\mathannotated{\pi}{f}v_{1}\mathannotated{\pi}{f}v_{2}\dots\mathannotated{\pi}{f}v_{n}}=\mathof{\min}{\mathsetintension{\sum_{i=0}^{n-1}\mathcountit{v'_{i}}{\mathannotated{\pi}{f}}{v'_{i+1}}}{
\begin{gathered}
 v'_{0}=w_{0}, v'_{n}=w_{1} \text{ and } \\ 
 v'_{i}\mathreduction{\pi}v'_{i+1} \text{ for all } i < n
\end{gathered} 
}}.
   \]
   Then, $w\mathreduction{\pi} v_{1} \mathreduction{\pi}^{n-1}w_{1}$.
   Because of ${M, w_{1}}\not\models{\varphi}$, 
   we have ${M, v_{1}}\not\models{\mathdynbox{\mathprogramit{\pi}}\varphi}$.
   Then, 
   $v_{1}\mathannotated{(\mathprogramit{\pi})}{f} w_{1}\dots \xi_{k-1}w_{k}\in\mathcounterex{M}{v_{1}}{\mathdynbox{\mathprogramit{\pi}}\varphi}$.
   We also have
   $w\mathannotated{\pi}{f}v_{1}\mathannotated{(\mathprogramit{\pi})}{f} w_{1}\dots \xi_{k-1}w_{k}\in\mathcounterex{M}{w}{\mathdynbox{\pi}\mathdynbox{\mathprogramit{\pi}}\varphi}$.
   Thus,
   \begin{align*}
    \mathiteration{M}{w}{\mathdynbox{\mathprogramit{\pi}}\varphi}
    &=\mathcountitex{w_{0}\xi_{1}w_{1}\dots \xi_{k-1}w_{k}} \\
    &=\sum_{i=0}^{k-1}\mathcountit{w_{i}}{\xi_{i}}{w_{i+1}} \\
    &=\mathcountit{w_{0}}{\mathannotated{(\mathprogramit{\pi})}{f}}{w_{1}}+\sum_{i=1}^{k-1}\mathcountit{w_{i}}{\xi_{i}}{w_{i+1}} \\
    &=n+\sum_{i=0}^{n-1}\mathcountit{v_{i}}{\mathannotated{\pi}{f}}{v_{i+1}}
    +\sum_{i=1}^{k-1}\mathcountit{w_{i}}{\xi_{i}}{w_{i+1}} \\
    &=n+\mathcountit{w_{0}}{\mathannotated{\pi}{f}}{v_{1}}
    +\sum_{i=1}^{n-1}\mathcountit{v_{i}}{\mathannotated{\pi}{f}}{v_{i+1}}
    +\sum_{i=1}^{k-1}\mathcountit{w_{i}}{\xi_{i}}{w_{i+1}} \\
    &> \mathcountit{w_{0}}{\mathannotated{\pi}{f}}{v_{1}} + n-1
    +\sum_{i=1}^{n-1}\mathcountit{v_{i}}{\mathannotated{\pi}{f}}{v_{i+1}}
    +\sum_{i=1}^{k-1}\mathcountit{w_{i}}{\xi_{i}}{w_{i+1}} \\
    &\geq \mathcountit{w_{0}}{\mathannotated{\pi}{f}}{v_{1}} 
    +\mathcountit{v_{1}}{\mathannotated{(\mathprogramit{\pi})}{f}}{w_{1}}
    +\sum_{i=1}^{k-1}\mathcountit{w_{i}}{\xi_{i}}{w_{i+1}} \\
    &=\mathcountitex{w\mathannotated{\pi}{f}v_{1}\mathannotated{(\mathprogramit{\pi})}{f} w_{1}\dots \xi_{k-1}w_{k}} \\
    &\geq \mathiteration{M}{w}{\mathdynbox{\pi}\mathdynbox{\mathprogramit{\pi}}\varphi}.
   \end{align*}
  \end{proof}

 \begin{lemma}
  \label[lemma]{lemma:strong-soundness-CGTPDL-proof}
  Let $\mleft(\mathdertree{D}, \mathcompanion{C}\mright)$ be a \CGTPDL\ proof.
  Any sequent occurring in $\mathdertree{D}$ is valid.
 \end{lemma}

  \begin{proof}
   Let $\mleft(\mathdertree{D}, \mathcompanion{C}\mright)$ be a \CGTPDL\ proof and
   $\Gamma\fCenter\Delta$ be a sequent occurring in $\mathdertree{D}$.
   To obtain a contradiction, suppose $M, w\not\models {\Gamma\fCenter\Delta}$ with a model
   $M=\mathtuple{W, \mathsequence{\mathreduction{\alpha}}{\alpha\in\maththesetofatomicprog}, V}$
   and $w \in W$.
   Then, $M, w\models {\Gamma}$ and $M, w\not\models {\varphi}$ for all $\varphi\in\Delta$.
     
   We inductively define an infinite path 
   $\mathsequence{\Gamma_{i} \fCenter \Delta_{i}}{0\leq i}$ 
   in $\mathproofgraph{\mleft(\mathdertree{D},  \mathcompanion{C}\mright)}$ and
   a sequence $\mathsequence{w_{i}}{0\leq i}$ of elements in $W$
   such that $M, w_{i}\not\models {\Gamma_{i}\fCenter\Delta_{i}}$ for each $i\geq 0$.
   
   Define ${\Gamma_{0} \fCenter \Delta_{0}}$ as ${\Gamma\fCenter\Delta}$. 
   We also define $w_{0}=w$.
   
   We consider the case $i>0$. $\Gamma_{i} \fCenter \Delta_{i}$ is defined
   according to the rule of which $\Gamma_{i-1} \fCenter \Delta_{i-1}$ is the conclusion.

   The rule is neither \rulename{Ax} nor \rulename{$\bot$}
   because of $M, w_{i-1}\not\models {\Gamma_{i-1}\fCenter\Delta_{i-1}}$.

   If the rule is \rulename{$\to$ R}, \rulename{Wk}, 
   \rulename{$\mathdynbox{\mathprogramseq{}{}}$ L},
   \rulename{$\mathdynbox{\mathprogramseq{}{}}$ R},
   \rulename{$\mathdynbox{ \mathprogramndc{}{} }$ L},
   \rulename{$\mathdynbox{\mathprogramitsy}$ L B},
   \rulename{$\mathdynbox{\mathprogramitsy}$ L S}, or
   \rulename{$\mathdynbox{ \mathprogramtestsy }$ R},
   then we define
   ${\Gamma_{i} \fCenter \Delta_{i}}$ as the only one assumption and
   $w_{i}=w_{i-1}$.

   Consider the case where the rule is \rulename{$\to$ L}. 
   Let $\Gamma_{i-1}\equiv{\Gamma, \varphi\to\psi}$ with a set of formulae $\Gamma$, and
   assume that $\varphi\to\psi$ is the principal formula.
   Define 
   ${\Gamma_{i}\fCenter\Delta_{i}}$ as 
   ${\Gamma\fCenter{\varphi, \Delta_{i-1}}}$ and $w_{i}=w_{i-1}$
   if $M, w_{i-1}\not\models \varphi$; otherwise
   we define 
   ${\Gamma_{i}\fCenter\Delta_{i}}$ as ${\Gamma, \psi\fCenter\Delta_{i-1}}$
   and $w_{i}=w_{i-1}$.

   Consider the case where the rule is \rulename{Cut}.
   Assume that $\varphi$ is the cut-formula.
   Define $\Gamma_{i}\fCenter\Delta_{i}$ as $\Gamma_{i-1}\fCenter{\varphi, \Delta_{i-1}}$
   and $w_{i}=w_{i-1}$ if $M, w_{i-1}\not\models \varphi$; otherwise
   we define $\Gamma_{i}\fCenter\Delta_{i}$ as 
   $\Gamma_{i-1}, \varphi\fCenter \Delta_{i-1}$ and $w_{i}=w_{i-1}$.

   Consider the case where the rule is \rulename{$\mathdynbox{ \, }$}.
   Let $\Gamma_{i-1}=\mathdynbox{\pi}\Gamma$ and 
   $\Delta_{i-1}={\mathdynbox{\pi}\varphi, \Delta}$
   with sets of formulae $\Gamma$ and $\Delta$, and
   assume that $\mathdynbox{\pi}\varphi$ is the principal formula.
   By \cref{lemma:IT-monotone-dynbox},
   there exists $w'\in W$ such that ${M, w'}\not\models{\varphi}$,
   $\mathiteration{M}{w_{i-1}}{\mathdynbox{\pi}\varphi}\geq\mathiteration{M}{w'}{\varphi}$ and
   $w_{i-1}\mathreduction{\pi}w'$.
   Define $\Gamma_{i}\fCenter\Delta_{i}$ as 
   $\Gamma \fCenter \varphi, \mathrevbox{\pi}\Delta$ and $w_{i}=w'$.

   Consider the case where the rule is \rulename{$\mathrevbox{ \, }$}.
   Let $\Gamma_{i-1}=\mathrevbox{\pi}\Gamma$ and 
   $\Delta_{i-1}={\mathrevbox{\pi}\varphi, \Delta}$
   with sets of formulae $\Gamma$ and $\Delta$, and
   assume that $\mathrevbox{\pi}\varphi$ is the principal formula.
   By \cref{lemma:IT-monotone-revbox},
   there exists $w'\in W$ such that ${M, w'}\not\models{\varphi}$,
   $\mathiteration{M}{w_{i-1}}{\mathdynbox{\pi}\varphi}\geq\mathiteration{M}{w'}{\varphi}$ and
   $w'\mathreduction{\pi}w_{i-1}$.
   Define $\Gamma_{i}\fCenter\Delta_{i}$ as 
   $\Gamma \fCenter \varphi, \mathdynbox{\pi}\Delta$ and $w_{i}=w'$.

   Consider the case where the rule is \rulename{$\mathdynbox{ \mathprogramndc{}{} }$ R}.
   Let $\Delta_{i-1}={\mathdynbox{ \mathprogramndc{\pi_{0}}{\pi_{1}} }\varphi, \Delta}$
   with a set of formulae $\Delta$, and
   assume that 
   $\mathdynbox{ \mathprogramndc{\pi_{0}}{\pi_{1}} }\varphi$ is the principal formula.
   Define $\Gamma_{i}\fCenter\Delta_{i}$ as $\Gamma \fCenter \Delta, \mathdynbox{ \pi_{j} }\varphi$
   and $w_{i}=w_{i-1}$
   if $\mathiteration{M}{w_{i-1}}{\mathdynbox{ \pi_{j} }\varphi}\leq\mathiteration{M}{w_{i-1}}{\mathdynbox{ \pi_{1-j} }\varphi}$ for $j\in\mathsetextension{0, 1}$. 

   Consider the case where the rule is \rulename{$\mathdynbox{ \mathprogramtestsy }$ L}.
   Let $\Gamma_{i-1}\equiv{\Gamma, \mathdynbox{ \mathprogramtest{\varphi} }\psi}$ 
   with a set of formulae $\Gamma$, and
   assume that $\mathdynbox{ \mathprogramtest{\varphi} }\psi$ is the principal formula.
   Define $\Gamma_{i}\fCenter\Delta_{i}$ as $\Gamma\fCenter{\varphi, \Delta_{i-1}}$
   and $w_{i}=w_{i-1}$
   if $M, w_{i-1}\not\models \varphi$; otherwise,
   $\Gamma_{i}\fCenter\Delta_{i}$ as $\Gamma, \psi\fCenter\Delta_{i-1}$
   and $w_{i}=w_{i-1}$.

   Consider the case where the rule is \rulename{C-s}.
   Let $\Delta_{i-1}\equiv{\mathdynbox{\mathprogramit{\pi}}\varphi, \Delta}$ 
   with a set of formulae $\Delta$, and
   assume that $\mathdynbox{\mathprogramit{\pi}}\varphi$ is the principal formula.
   Define $\Gamma_{i}\fCenter\Delta_{i}$ as $\Gamma_{i-1}\fCenter\varphi, \Delta_{i-1}$
   and $w_{i}=w_{i-1}$
   if $\mathdynbox{\mathprogramit{\pi}}\varphi$ is able to stay at $w_{i-1}$ in $M$; 
   otherwise, 
   $\Gamma_{i}\fCenter\Delta_{i}$ as 
   $\Gamma\fCenter{\mathdynbox{\pi}\mathdynbox{\mathprogramit{\pi}}\varphi, \Delta_{i-1}}$
   and $w_{i}=w_{i-1}$. 

   Since $\mathproofgraph{\mleft(\mathdertree{D},  \mathcompanion{C}\mright)}$
   satisfies the global trace condition,
   there exists an infinitely progressing trace following a tail of the path
   $\mathsequence{\Gamma_{i} \fCenter \Delta_{i}}{i\geq k}$ with some $k \geq 0$.
   Let $\mathsequence{\tau_{i}}{i\geq k}$ be 
   such an infinitely progressing trace.
   Consider the sequence 
   $\mathsequence{\mathiteration{M}{w_{i}}{\tau_{i}}}{i\geq k}$.

   We show that $\mathiteration{M}{w_{i}}{\tau_{i}}\geq\mathiteration{M}{w_{i+1}}{\tau_{i+1}}$
   holds for $i\geq k$, and
   in particular,
   $\mathiteration{M}{w_{i}}{\tau_{i}} > \mathiteration{M}{w_{i+1}}{\tau_{i+1}}$ 
   holds if $i$ is a progress point.
   Because of $M, w_{i}\not\models {\Gamma_{i}\fCenter\Delta_{i}}$,
   we have $M, w_{i}\not\models {\tau_{i}}$.

   We consider cases according to the rule of which 
   $\Gamma_{i} \fCenter \Delta_{i}$ is the conclusion. 

   The rule is neither \rulename{Ax} nor \rulename{$\bot$}
   because of $M, w_{i}\not\models {\Gamma_{i}\fCenter\Delta_{i}}$.

   If the rule is \rulename{$\to$ L}, \rulename{Wk}, \rulename{Cut},
   \rulename{$\mathdynbox{\mathprogramseq{}{}}$ L},
   \rulename{$\mathdynbox{ \mathprogramndc{}{} }$ L},
   \rulename{$\mathdynbox{\mathprogramitsy}$ L B},
   \rulename{$\mathdynbox{\mathprogramitsy}$ L S}, or
   \rulename{$\mathdynbox{ \mathprogramtestsy }$ L},
   then $\tau_{i+1}\equiv\tau_{i}$ and $w_{i+1}=w_{i}$.
   Hence, 
   $\mathiteration{M}{w_{i}}{\tau_{i}}=\mathiteration{M}{w_{i+1}}{\tau_{i+1}}$.
   Specifically,
   $\mathiteration{M}{w_{i}}{\tau_{i}}\geq\mathiteration{M}{w_{i+1}}{\tau_{i+1}}$.

   If the rule is \rulename{$\to$ R}, \rulename{$\mathdynbox{\mathprogramseq{}{}}$ R},
   \rulename{$\mathdynbox{ \mathprogramndc{}{} }$ R}, 
   \rulename{$\mathdynbox{ \mathprogramtestsy }$ R}, and \rulename{C-s}, and
   $i$ is not a rewriting point,
   then $\tau_{i+1}\equiv\tau_{i}$ and $w_{i+1}=w_{i}$.
   Hence, 
   $\mathiteration{M}{w_{i}}{\tau_{i}}=\mathiteration{M}{w_{i+1}}{\tau_{i+1}}$.
   Specifically,
   $\mathiteration{M}{w_{i}}{\tau_{i}}\geq\mathiteration{M}{w_{i+1}}{\tau_{i+1}}$.

   Assume that $i$ is a rewriting point.

   Consider the case where the rule is \rulename{$\to$ R}.
   In this case,
   $\tau_{i}$ is $\phi\to\psi$ and $\tau_{i+1}$ is $\psi$.
   By \cref{lemma:it-imply} and $w_{i}=w_{i+1}$, we have
   $\mathiteration{M}{w_{i}}{\varphi\to\psi}\geq\mathiteration{M}{w_{i+1}}{\psi}$.
   
   Consider the case where the rule is \rulename{$\mathdynbox{ \, }$}.
   In this case, 
   $\tau_{i}$ is $\mathdynbox{\pi}\varphi$ and $\tau_{i+1}$ is $\varphi$.
   By the definition of $\mathsequence{\Gamma_{i} \fCenter \Delta_{i}}{0\leq i}$ and 
   $\mathsequence{w_{i}}{0\leq i}$,
   we have
   $\mathiteration{M}{w_{i}}{\mathdynbox{\pi}\varphi}\geq\mathiteration{M}{w_{i+1}}{\varphi}$.

   Consider the case where the rule is \rulename{$\mathrevbox{ \, }$}.
   In this case,
   $\tau_{i}$ is $\mathrevbox{\pi}\varphi$ and $\tau_{i+1}$ is $\varphi$.
   By the definition of $\mathsequence{\Gamma_{i} \fCenter \Delta_{i}}{0\leq i}$ and 
   $\mathsequence{w_{i}}{0\leq i}$,
   we have
   $\mathiteration{M}{w_{i}}{\mathrevbox{\pi}\varphi}\geq\mathiteration{M}{w_{i+1}}{\varphi}$.

   Consider the case where the rule is \rulename{$\mathdynbox{\mathprogramseq{}{}}$ R}.
   In this case,
   $\tau_{i}$ is 
   $\mathdynbox{\mathprogramseq{\pi_{0}}{\pi_{1}}}\varphi$ and 
   $\tau_{i+1}$ is $\mathdynbox{\pi_{0}}\mathdynbox{\pi_{1}}\varphi$.
   By \cref{lemma:it-seq} and $w_{i}=w_{i+1}$, we have
   $\mathiteration{M}{w_{i}}{\mathdynbox{\mathprogramseq{\pi_{0}}{\pi_{1}}}\varphi}\geq\mathiteration{M}{w_{i+1}}{\mathdynbox{\pi_{0}}\mathdynbox{\pi_{1}}\varphi}$.

   Consider the case where the rule is \rulename{$\mathdynbox{ \mathprogramndc{}{} }$ R}.
   In this case, 
   $\tau_{i}$ is $\mathdynbox{ \mathprogramndc{\pi_{0}}{\pi_{1}} }\varphi$, and
   $\tau_{i+1}$ is $\mathdynbox{ \pi_{j} }\varphi$ for $j\in\mathsetextension{0, 1}$, where
   $\mathiteration{M}{w_{i}}{\mathdynbox{ \pi_{j} }\varphi}\leq\mathiteration{M}{w_{i}}{\mathdynbox{ \pi_{1-j} }\varphi}$.
   By \cref{lemma:it-ndc} and $w_{i}=w_{i+1}$, we have
   $\mathiteration{M}{w_{i}}{\mathdynbox{ \mathprogramndc{\pi_{0}}{\pi_{1}} }\varphi}\geq\mathiteration{M}{w_{i+1}}{\mathdynbox{\pi_{j} }\varphi}$.

   Consider the case where the rule is \rulename{$\mathdynbox{ \mathprogramtestsy }$ R}.
   In this case,
   $\tau_{i}$ is $\mathdynbox{ \mathprogramtest{\varphi} }\psi$ and $\tau_{i+1}$ is $\psi$.
   By \cref{lemma:it-test} and $w_{i}=w_{i+1}$, we have
   $\mathiteration{M}{w_{i}}{\mathdynbox{\mathprogramtest{\varphi} }\psi}\geq\mathiteration{M}{w_{i+1}}{\psi}$.

   Consider the case where the rule is \rulename{C-s}. 
   In this case, 
   $\tau_{i}$ is the principal formula $\mathdynbox{\mathprogramit{\pi}}\varphi$ 
   of the rule.
   There are two cases.
   
   Assume that $\mathdynbox{\mathprogramit{\pi}}\varphi$ is able to stay at $w_{i}$ in $M$.
   Then, $\tau_{i+1}$ is $\varphi$.
   By \cref{lemma:it-base-case} and $w_{i}=w_{i+1}$, we have
   $\mathiteration{M}{w_{i}}{\mathdynbox{\mathprogramit{\pi}}\varphi}\geq\mathiteration{M}{w_{i+1}}{\varphi}$.

   Assume that 
   $\mathdynbox{\mathprogramit{\pi}}\varphi$ is not able to stay at $w_{i}$ in $M$.
   Then, $\tau_{i+1}$ is $\mathdynbox{\pi}\mathdynbox{\mathprogramit{\pi}}\varphi$, and
   $i$ is a progress point.
   By \cref{lemma:it-step-case} and $w_{i}=w_{i+1}$, we have
   $\mathiteration{M}{w_{i}}{\mathdynbox{\mathprogramit{\pi}}\varphi}>\mathiteration{M}{w_{i+1}}{\mathdynbox{\pi}\mathdynbox{\mathprogramit{\pi}}\varphi}$.

   Because $\mathsequence{\tau_{i}}{i\geq k}$ is an infinitely progressing trace,
   $\mathsequence{\tau_{i}}{i\geq k}$ has infinitely many progress points.
   Hence,
   $\mathiteration{M}{w_{i}}{\tau_{i}}>\mathiteration{M}{w_{i+1}}{\tau_{i+1}}$
   holds for infinitely many $i$'s, which is a contradiction.
   Thus, $M\models {\Gamma\fCenter\Delta}$ holds for any model $M$.
  \end{proof}

  Now, we have the soundness of \CGTPDL\ immediately.

  \begin{proof}[Proof of \cref{theorem:soundness-CGTPDL}]
   By \cref{lemma:strong-soundness-CGTPDL-proof}.
  \end{proof}

\section{Failure of cut-elimination in \GTPDL\ and \CGTPDL}
\label{sec:failure-cut-elimination}
In this section, we show the failure of cut-elimination in \GTPDL\ and \CGTPDL.

We define some terms.
For a set of formulae $\Lambda$,
an instance of \rulename{Cut} is called $\Lambda$-cut
if the cut formula of the instance belongs to $\Lambda$.
Specifically, we call an $\mathFLcl{\Gamma\cup\Delta}$-cut, where
$\Gamma \fCenter \Delta$ is the conclusion, a \emph{Fischer-Ladner-cut}.
We say that a sequent $\Gamma\fCenter\Delta$ is \emph{Fischer-Ladner-cut provable}
in \GTPDL\ and \CGTPDL\
if there is an \GTPDL\ and \CGTPDL\ proof of $\Gamma\fCenter\Delta$ where
all instances of \rulename{Cut} are Fischer-Ladner-cuts, respectively.
We define a \emph{cut-free \GTPDL\ and \CGTPDL\ proof} as 
an \GTPDL\ and \CGTPDL\ proof where no instances of \rulename{Cut} occur, respectively. 
We say that a sequent $\Gamma\fCenter\Delta$ is \emph{cut-free provable} 
in \GTPDL\ and \CGTPDL\
if there is a cut-free \GTPDL\ and \CGTPDL\ proof of $\Gamma\fCenter\Delta$.
 
We give some sequents that are not cut-free provable in \GTPDL\ and \CGTPDL.
\begin{proposition}
  \label{prop:failure-of-cut-elimination-GTPDL}
  Let $p\in\maththesetofpropval$ and $\alpha\in\maththesetofatomicprog$.
  \begin{enumerate}
   \item $p, \mathdynbox{\alpha}\mathdynbox{\mathprogramit{\alpha}}p \fCenter\mathdynbox{\mathprogramit{\alpha}}p$ is \emph{not} cut-free provable in \GTPDL.
	\label{item:failure-GTPDL-prop-failure-of-cut-elimination}
	 \begin{enumerate}
	  \item Moreover, $p, \mathdynbox{\alpha}\mathdynbox{\mathprogramit{\alpha}}p \fCenter\mathdynbox{\mathprogramit{\alpha}}p$ is \emph{not} Fischer-Ladner-cut provable in \GTPDL.
	 \end{enumerate}
   \item $p, \mathdynbox{\alpha}\mathdynbox{\mathprogramit{\alpha}}p \fCenter\mathdynbox{\mathprogramit{\alpha}}p$ is cut-free provable in \CGTPDL
	 \label{item:cut-free-CGTPDL-prop-failure-of-cut-elimination}
   \item ${p}\fCenter{\mathdynbox{\alpha}\mathrevdia{\alpha}p}$
	 is provable in \GTPDL\ and \CGTPDL.
	 \label{item:neither-cut-free-provable-provable}
   \item ${p}\fCenter{\mathdynbox{\alpha}\mathrevdia{\alpha}p}$
	 is neither cut-free provable in \GTPDL\ nor \CGTPDL.
	 \label{item:neither-cut-free-provable}
  \end{enumerate}
 \end{proposition}

\begin{proof}
 Let $p\in\maththesetofpropval$ and $\alpha\in\maththesetofatomicprog$.
 Note that we can suppose that
 the sequent of the assumption is different from that of the conclusion
 for each instance of \rulename{Wk} without loss of generality.

 \noindent \cref{item:failure-GTPDL-prop-failure-of-cut-elimination}
 It suffices to show that $p, \mathdynbox{\alpha}\mathdynbox{\mathprogramit{\alpha}}p \fCenter\mathdynbox{\mathprogramit{\alpha}}p$ is not Fischer-Ladner-cut provable in \GTPDL.

 The rule in \GTPDL\ where 
 $p, \mathdynbox{\alpha}\mathdynbox{\mathprogramit{\alpha}}p \fCenter\mathdynbox{\mathprogramit{\alpha}}p$ 
 can occur as a conclusion is only \rulename{Wk} or \rulename{Cut}.
 Because no valid sequent can occur as an assumption 
 in an instance of \rulename{Wk} whose conclusion is
 $p, \mathdynbox{\alpha}\mathdynbox{\mathprogramit{\alpha}}p \fCenter\mathdynbox{\mathprogramit{\alpha}}p$, 
 the last rule of a \GTPDL\ proof of
 $p, \mathdynbox{\alpha}\mathdynbox{\mathprogramit{\alpha}}p \fCenter\mathdynbox{\mathprogramit{\alpha}}p$
 must be \rulename{Cut}.
 By
 $\mathFLcl{\mathsetextension{p, \mathdynbox{\alpha}\mathdynbox{\mathprogramit{\alpha}}p, \mathdynbox{\mathprogramit{\alpha}}p}}=\mathsetextension{p, \mathdynbox{\alpha}\mathdynbox{\mathprogramit{\alpha}}p, \mathdynbox{\mathprogramit{\alpha}}p}$,
 one of the assumptions of each Fischer-Ladner cut of
 $p, \mathdynbox{\alpha}\mathdynbox{\mathprogramit{\alpha}}p \fCenter\mathdynbox{\mathprogramit{\alpha}}p$
 is $p, \mathdynbox{\alpha}\mathdynbox{\mathprogramit{\alpha}}p \fCenter\mathdynbox{\mathprogramit{\alpha}}p$ itself.
 Hence, an instance of \rulename{Cut} which is not a Fischer-Ladner cut must occur 
 in a \GTPDL\ proof of
 $p, \mathdynbox{\alpha}\mathdynbox{\mathprogramit{\alpha}}p \fCenter\mathdynbox{\mathprogramit{\alpha}}p$.

 \noindent \cref{item:cut-free-CGTPDL-prop-failure-of-cut-elimination}
 A cut-free \CGTPDL\ proof of $p, \mathdynbox{\alpha}\mathdynbox{\mathprogramit{\alpha}}p \fCenter\mathdynbox{\mathprogramit{\alpha}}p$
 as follows:
 \begin{center}
  \begin{inlineprooftree}
   \AxiomC{}
   \RightLabel{\rulename{Ax}}
   \UnaryInfC{$p, \mathdynbox{\alpha}\mathdynbox{\mathprogramit{\alpha}}p \fCenter p$}

   \AxiomC{}
   \RightLabel{\rulename{Ax}}
   \UnaryInfC{$p, \mathdynbox{\alpha}\mathdynbox{\mathprogramit{\alpha}}p \fCenter \mathdynbox{\alpha}\mathdynbox{\mathprogramit{\alpha}}p$}

   \RightLabel{\rulename{C-s}}
   \BinaryInfC{$p, \mathdynbox{\alpha}\mathdynbox{\mathprogramit{\alpha}}p \fCenter\mathdynbox{\mathprogramit{\alpha}}p$}
  \end{inlineprooftree}.
 \end{center}

 \noindent \cref{item:neither-cut-free-provable-provable}
 A \GTPDL\ proof of ${p}\fCenter{\mathdynbox{\alpha}\mathrevdia{\alpha}p}$
 as follows:
 \begin{center}
  \begin{inlineprooftree}
   \AxiomC{}
   \RightLabel{\rulename{Ax}}
   \UnaryInfC{${p}\fCenter{p}$}
   \RightLabel{\rulename{$\lnot$ L}}
   \UnaryInfC{${p, \lnot p}\fCenter{}$}
   \RightLabel{\rulename{$\lnot$ R}}
   \UnaryInfC{${p}\fCenter{\lnot\lnot p}$}
   \RightLabel{\rulename{Wk}}
   \UnaryInfC{${p}\fCenter{\lnot\lnot p, \mathdynbox{\alpha}\mathrevdia{\alpha}p}$}

   \AxiomC{}
   \RightLabel{\rulename{Ax}}
   \UnaryInfC{${\mathrevbox{\alpha}\lnot p}\fCenter{\mathrevbox{\alpha}\lnot p}$}
   \RightLabel{\rulename{$\lnot$ R}}
   \UnaryInfC{${}\fCenter{\mathrevdia{\alpha}p, \mathrevbox{\alpha}\lnot p}$}
   \RightLabel{\rulename{$\mathdynbox{\;}$}}
   \UnaryInfC{${}\fCenter{\mathdynbox{\alpha}\mathrevdia{\alpha}p, \lnot p}$}
   \RightLabel{\rulename{$\lnot$ L}}
   \UnaryInfC{${\lnot\lnot p}\fCenter{\mathdynbox{\alpha}\mathrevdia{\alpha}p}$}
   \RightLabel{\rulename{Wk}}
   \UnaryInfC{${p, \lnot\lnot p}\fCenter{\mathdynbox{\alpha}\mathrevdia{\alpha}p}$}
   
   \RightLabel{\rulename{Cut}}
   \BinaryInfC{${p}\fCenter{\mathdynbox{\alpha}\mathrevdia{\alpha}p}$}
  \end{inlineprooftree}
 \end{center}

 This derivation tree is also a \CGTPDL\ proof.

 \noindent \cref{item:neither-cut-free-provable}
 The rule in \GTPDL\ where 
 ${p}\fCenter{\mathdynbox{\alpha}\mathrevdia{\alpha}p}$
 can occur as a conclusion is only \rulename{Wk} or \rulename{Cut}.
 Because no valid sequent 
 can occur as an assumption in an instance of \rulename{Wk} where
 ${p}\fCenter{\mathdynbox{\alpha}\mathrevdia{\alpha}p}$
 occurs as a conclusion,
 the last rule of a \GTPDL\ proof of
 ${p}\fCenter{\mathdynbox{\alpha}\mathrevdia{\alpha}p}$
 must be \rulename{Cut}.

 In the same way, we can show that  ${p}\fCenter{\mathdynbox{\alpha}\mathrevdia{\alpha}p}$
 is not cut-free provable in \CGTPDL.
\end{proof}

We note that
$p, \mathdynbox{\alpha}\mathdynbox{\mathprogramit{\alpha}}p \fCenter\mathdynbox{\mathprogramit{\alpha}}p$ 
is 
$\mathsetextension{\mathdynbox{\mathprogramit{\alpha}}\mathdynbox{\alpha}\mathdynbox{\mathprogramit{\alpha}}p}$-cut 
provable in \GTPDL\ (See \cref{fig:GTPDL-proof-not-FL-cut-provable}. \cite[p.387, Fig. 19]{Nishimura1979} is the same proof).

\begin{figure}[tb]
 \begin{prooftree}
  \AxiomC{}
  \RightLabel{\rulename{Ax}}
  \UnaryInfC{$\mathdynbox{\alpha}\mathdynbox{\mathprogramit{\alpha}}p \fCenter\mathdynbox{\alpha}\mathdynbox{\mathprogramit{\alpha}}p$}
  \RightLabel{\rulename{$\mathdynbox{\mathprogramitsy}$ L S}}
  \UnaryInfC{$\mathdynbox{\mathprogramit{\alpha}}p \fCenter\mathdynbox{\alpha}\mathdynbox{\mathprogramit{\alpha}}p$}
  \RightLabel{\rulename{$\mathdynbox{ \, }$}}
  \UnaryInfC{$\mathdynbox{\alpha}\mathdynbox{\mathprogramit{\alpha}}p \fCenter\mathdynbox{\alpha}\mathdynbox{\alpha}\mathdynbox{\mathprogramit{\alpha}}p$}
  \RightLabel{\rulename{$\mathdynbox{\mathprogramitsy}$ R}}
  \UnaryInfC{$\mathdynbox{\alpha}\mathdynbox{\mathprogramit{\alpha}}p \fCenter\mathdynbox{\mathprogramit{\alpha}}\mathdynbox{\alpha}\mathdynbox{\mathprogramit{\alpha}}p$}
  \RightLabel{\rulename{Wk}}
  \UnaryInfC{$p, \mathdynbox{\alpha}\mathdynbox{\mathprogramit{\alpha}}p \fCenter\mathdynbox{\mathprogramit{\alpha}}\mathdynbox{\alpha}\mathdynbox{\mathprogramit{\alpha}}p, \mathdynbox{\mathprogramit{\alpha}}p$}

  \AxiomC{}
  \RightLabel{\rulename{Ax}}
  \UnaryInfC{$p\fCenter p$}
  \RightLabel{\rulename{$\mathdynbox{\mathprogramitsy}$ L B}}
  \UnaryInfC{$\mathdynbox{\mathprogramit{\alpha}}p\fCenter p$}
  \RightLabel{\rulename{$\mathdynbox{ \, }$}}
  \UnaryInfC{$\mathdynbox{\alpha}\mathdynbox{\mathprogramit{\alpha}}p\fCenter\mathdynbox{\alpha}p$}
  \RightLabel{\rulename{Wk}}
  \UnaryInfC{$p, \mathdynbox{\alpha}\mathdynbox{\mathprogramit{\alpha}}p\fCenter\mathdynbox{\alpha}p$}
  \RightLabel{\rulename{$\mathdynbox{\mathprogramitsy}$ R}}
  \UnaryInfC{$p, \mathdynbox{\mathprogramit{\alpha}}\mathdynbox{\alpha}\mathdynbox{\mathprogramit{\alpha}}p\fCenter\mathdynbox{\mathprogramit{\alpha}}p$}
  \RightLabel{\rulename{Wk}}
  \UnaryInfC{$p, \mathdynbox{\alpha}\mathdynbox{\mathprogramit{\alpha}}p, \mathdynbox{\mathprogramit{\alpha}}\mathdynbox{\alpha}\mathdynbox{\mathprogramit{\alpha}}p\fCenter\mathdynbox{\mathprogramit{\alpha}}p$}

  \RightLabel{\rulename{Cut}}
  \BinaryInfC{$p, \mathdynbox{\alpha}\mathdynbox{\mathprogramit{\alpha}}p \fCenter\mathdynbox{\mathprogramit{\alpha}}p$}
 \end{prooftree}

 \caption{A \GTPDL\ proof of $p, \mathdynbox{\alpha}\mathdynbox{\mathprogramit{\alpha}}p \fCenter \mathdynbox{\mathprogramit{\alpha}}p$}
 \label{fig:GTPDL-proof-not-FL-cut-provable}
\end{figure}

 By \cref{prop:failure-of-cut-elimination-GTPDL}, we have the following theorem.
 \begin{theorem}[Failure of cut-elimination in \GTPDL\ and \CGTPDL]
  \label{thm:failure-of-cut-elimination}
  The following statements hold:
  \begin{enumerate}
   \item There exists a sequent not cut-free provable in \GTPDL\ but cut-free provable 
	 in \CGTPDL.
	 \label{item:GTPDL-not-cut-free-provable}
	 \begin{enumerate}
	  \item Moreover, there exists a sequent not Fischer-Ladner-cut provable in \GTPDL\
		but cut-free provable in \CGTPDL.
	 \end{enumerate}
   \item There exists a sequent such that it is provable in both \GTPDL\ and \CGTPDL,
	 but is cut-free provable in neither.
	 \label{item:CGTPDL-not-cut-free-provable}
  \end{enumerate}
 \end{theorem}

We do not find a sequent cut-free provable in \GTPDL\ but not in \CGTPDL.
If such a sequent exists, then some backwards modal operators must occur in 
the sequent because we have the cut-elimination theorem for the fragment of \CGTPDL\ 
obtained by removing backwards modal operators. We prove this theorem in the next section.

\section{Cut-elimination in a cyclic proof system for \PDL}
\label{sec:cut-elimination}
This section shows the cut-elimination property of a cyclic proof system for \PDL,
which is a fragment of \CGTPDL.

\subsection{\CGPDL : a cyclic proof system for \PDL}
This section defines a cyclic proof system for \PDL, \CGPDL.

\CGPDL\ is the cyclic proof system for \PDL\
obtained by removing backwards modal operators from \CGTPDL.
This is essentially a two-sided variant of cyclic-\LPD, described in \cite{Das2023}.
 The inference rules of \CGPDL\ are the same as that of \CGTPDL\ except for
 \rulename{$\mathdynbox{ \, }$} and \rulename{$\mathrevbox{ \, }$}.
 \rulename{$\mathrevbox{ \, }$} is removed and
 \rulename{$\mathdynbox{ \, }$} is replaced by \rulename{K}, 
 the following rule:
 \begin{center}
  \begin{inlineprooftree}
    \AxiomC{$\Gamma \fCenter \varphi$}
    \RightLabel{\rulename{K}}
    \UnaryInfC{$\Gamma', \mathdynbox{\pi}\Gamma \fCenter \mathdynbox{\pi}\varphi, \Delta$}
  \end{inlineprooftree}.
 \end{center}

 In the remainder of this section, no backwards modal operators $\mathrevbox{\pi}$
 occur in every formula; that is, we call a \PDL\ formula only ``a formula''.
 We define a \emph{\CGPDL\ (pre-)proof}, \emph{derivation graph}, \emph{trace}, and so on,
 in the same way as in \CGTPDL.

 Note that \rulename{K} is derivable in \CGTPDL\ as follows:
 \begin{center}
  \begin{inlineprooftree}
    \AxiomC{$\Gamma \fCenter \varphi$}
    \RightLabel{\rulename{$\mathdynbox{ \, }$}}
    \UnaryInfC{$\mathdynbox{\pi}\Gamma \fCenter \mathdynbox{\pi}\varphi$}
    \RightLabel{\rulename{Wk}}
    \UnaryInfC{$\Gamma', \mathdynbox{\pi}\Gamma \fCenter \mathdynbox{\pi}\varphi, \Delta$}
  \end{inlineprooftree}.
 \end{center}
 Hence, \CGPDL\ is a fragment of \CGTPDL.
 Therefore, we have the soundness theorem of \CGTPDL\ immediately.
\begin{theorem}[Soundness of \CGPDL]
 \label{theorem:soundness-CGPDL}
 If $\Gamma\fCenter\Delta$ is provable in \CGPDL,
 then ${\Gamma\fCenter\Delta}$ is valid.
\end{theorem}

To show the cut-elimination for \CGPDL,
we prove the following theorem. 
\begin{theorem}[Cut-free completeness of \CGPDL]
 \label{theorem:cut-free-completeness-CGPDL}
 If ${\Gamma\fCenter\Delta}$ is valid,
 then $\Gamma\fCenter\Delta$ is cut-free provable in \CGPDL.
\end{theorem}

 Under \cref{theorem:cut-free-completeness-CGPDL}, 
 we have the cut-elimination theorem of \CGPDL.

 \begin{corollary}[Cut-elimination of \CGPDL]
  If ${\Gamma\fCenter\Delta}$ is provable in \CGPDL,
  then ${\Gamma\fCenter\Delta}$ is cut-free provable in \CGPDL.
 \end{corollary}

 \begin{proof}
  Assume that ${\Gamma\fCenter\Delta}$ is provable in \CGPDL.
  By \cref{theorem:soundness-CGPDL}, ${\Gamma\fCenter\Delta}$ is valid.
  From \cref{theorem:cut-free-completeness-CGPDL},
  ${\Gamma\fCenter\Delta}$ is cut-free provable in \CGPDL.
 \end{proof}

 In the remainder of this section, we prove \cref{theorem:cut-free-completeness-CGPDL}.

\noindent \paragraph{The road map of the proof of \cref{theorem:cut-free-completeness-CGPDL}:}
We consider \emph{proof search games in \CGPDL} (\cref{def:proof-search-game}). 
Each game is defined for a sequent as a two-player \emph{finite} zero-sum game of 
perfect information between players $\Prover$ and $\Refuter$: 
If $\Prover$ has a winning strategy, 
then a cut-free \CGPDL\ proof of the sequent exists (\cref{lemma:Prover-win}). 
Conversely, if $\Refuter$ has a winning strategy, 
then a counter model of the sequent exists (\cref{lemma:Refuter-win}).
Because the game is a two-player \emph{finite} zero-sum game of perfect information,
we easily have its \emph{determinacy} (\cref{cor:determinacy}) 
by Zermelo's theorem (\cref{thm:Zermelo}).
\cref{lemma:Prover-win}, \cref{lemma:Refuter-win}, and \cref{cor:determinacy} imply
\cref{theorem:cut-free-completeness-CGPDL}.

\subsection{Proof search games in \CGPDL}
This section defines proof search games in \CGPDL.

For a sequent $\Gamma\fCenter\Delta$,
we define $\mathsubFLS{\Gamma\fCenter\Delta}=\mathsetintension{\Pi\fCenter\Sigma}{\Pi\cup\Sigma\subseteq\mathFLcl{\Gamma\cup\Delta}}$.
For a set of formulae $\Gamma$, we define
\[
\mathreducible{\Gamma} = \mathsetintension{\varphi\in\Gamma}{\varphi \text{ is of the form } \psi_{0}\to\psi_{1}, \mathdynbox{\mathprogramseq{\pi_{0}}{\pi_{1}}}\psi, \mathdynbox{\mathprogramndc{\pi_{0}}{\pi_{1}}}\psi, \mathdynbox{\mathprogramit{\pi}}\psi, \text{ or }  \mathdynbox{\mathprogramtest{\psi_{0}}}\psi_{1} }. 
\]
We call a formula in $\mathreducible{\Gamma}$ a \emph{reducible formula} in $\Gamma$.
A \emph{schedule} on a sequent $\Gamma\fCenter\Delta$ is defined as 
a linear order on $\mathFLcl{\Gamma\cup\Delta}$.

Fix a schedule $\leq_{\Gamma\fCenter\Delta}$ on each sequent $\Gamma\fCenter\Delta$ 
in the remainder of this section. 
For a set $S\subseteq\mathreducible{\Gamma\fCenter\Delta}$,
we write the maximum of $S$ in the schedule as $\mathmaxform{S}$.
The notation $\varphi<_{\Gamma\fCenter\Delta}\psi$ means that both
$\varphi\leq_{\Gamma\fCenter\Delta}\psi$ and $\varphi\not\equiv\psi$ hold.

Let $\maththesetofTr=\mathsetintension{T\subseteq\mathFLcl{\Gamma\cup\Delta}\times\mathsetextension{0, \dots, \mathcardinality{\mathFLcl{\Gamma\cup\Delta}}}}{\text{$\mathtuple{\varphi, n}, \mathtuple{\psi, n}\in T$ implies $\varphi\equiv\psi$}}$.
For $T\in\maththesetofTr$, 
we write $\mathfreshnum{T}$ for the minimum positive natural number of
$\mathsetintension{n\in\mathnatpos}{\mathtuple{\varphi, n}\notin T}$,
where $\mathnatpos$ is the set of positive natural numbers.
For $T\in\maththesetofTr$,
let $\mathof{T}{\varphi}$ denotes $\mathsetintension{n}{\mathtuple{\varphi, n}\in T}$.

\begin{definition}
 \label[definition]{def:proof-search-game}
 The \emph{proof search game} of $\Gamma\fCenter\Delta$ in \CGPDL,
 written as $\GameName{\Gamma\fCenter\Delta}$, is a two-player finite zero-sun game of 
 perfect information defined as follows:
 \begin{itemize}
  \item The \emph{players} are \Prover\ and \Refuter.
  \item The \emph{game board} is
	$\mathsubFLS{\Gamma\fCenter\Delta}\times\maththesetofTr\times\maththesetofUS$, 
	where $\maththesetofUS$ is the power set of
	$\mathreducible{\mathFLcl{\Gamma\cup\Delta}}$. 
	We call an element of the game board a \emph{position}.
	\begin{itemize}
	 \item We write $\mathtuple{\Pi\fCenter\Sigma; T; S}$
	       for a position with 
	       $\Pi\fCenter\Sigma\in\mathsubFLS{\Gamma\fCenter\Delta}$,
	       $T\in\maththesetofTr$ and $S\in\maththesetofUS$.
	 \item We call the first argument of each position 
	       the \emph{sequent of the position}.
	 \item We call the second argument of each position 
	       the \emph{track of the position}.
	 \item We call the third argument of each position 
	       the \emph{upcoming schedule of the position}.
	 \item Let 
	       $\GameSize=\mathcardinality{\mathsubFLS{\Gamma\fCenter\Delta}\times\maththesetofTr\times\maththesetofUS}$.
	\end{itemize}
  \item A \emph{turn function} $\lambda$, 
	which is a function from the game board to 
	$\mathsetextension{\Prover, \Refuter}$, and \emph{game rules}, 
	each of which is a pair of positions $\mathtuple{C, C'}$, 
	are defined as follows:
	\begin{itemize}
	 \item The case that
	       $C$ is of the form $\mathtuple{\Pi\fCenter\Sigma; T; S}$
	       and $(\Pi\cup\Sigma)\cap S\neq\emptyset$ with 
	       $\varphi\equiv\mathmaxform{(\Pi\cup\Sigma)\cap S}$:
	       \ifoup
	       \begin{itemize}
		\else
		\begin{description}
		 \fi
		\item[\rulename{G-$\to$ L}] 
			   If $\varphi\equiv\psi_{0}\to\psi_{1}$ and $\varphi\in\Pi$, 
			   then $\mathof{\lambda}{C}=\Refuter$ and
			   either 
			   $C'=\mathtuple{\Pi\fCenter\psi_{0}, \Sigma; T'; S\setminus \mathsetextension{\varphi}}$
			   or
			   $C'=\mathtuple{\Pi, \psi_{1}\fCenter\Sigma; T; S\setminus \mathsetextension{\varphi}}$,
			   where $T'=T\cup\mathsetextension{\mathtuple{\psi_{0}, \mathfreshnum{T}}}$
			   if $\psi_{0}\notin\Sigma$; $T'=T$ otherwise.
		\item[\rulename{G-$\to$ R}] 
			   If $\varphi\equiv\psi_{0}\to\psi_{1}$ and $\varphi\in\Sigma$, 
			   then
			   $C'=\mathtuple{\Pi, \psi_{0}\fCenter\psi_{1}, \Sigma; T'; S\setminus \mathsetextension{\varphi}}$, 
			   where $T'=\mleft(T\setminus \mathsetintension{\mathtuple{\varphi, n}}{n\in\mathof{T}{\varphi}}\mright)\cup\mathsetintension{\mathtuple{\psi_{1}, n}}{n\in\mathof{T}{\varphi}}$.
		\item[\rulename{G-$\mathdynbox{\mathprogramseq{}{}}$ L}] 
			   If $\varphi\equiv\mathdynbox{\mathprogramseq{\pi_{0}}{\pi_{1}}}\psi$ and 
			   $\varphi\in\Pi$, 
			   then
			   $C'=\mathtuple{\Pi, \mathdynbox{\pi_{0}}\mathdynbox{\pi_{1}}\psi\fCenter \Sigma; T; S\setminus \mathsetextension{\varphi}}$.
		\item[\rulename{G-$\mathdynbox{\mathprogramseq{}{}}$ R}] 
			   If $\varphi\equiv\mathdynbox{\mathprogramseq{\pi_{0}}{\pi_{1}}}\psi$ and 
			   $\varphi\in\Sigma$, 
			   then
			   $C'=\mathtuple{ \Pi\fCenter \mathdynbox{\pi_{0}}\mathdynbox{\pi_{1}}\psi, \Sigma; T'; S\setminus \mathsetextension{\varphi}}$,
			   where
			   $T'=\mleft(T\setminus \mathsetintension{\mathtuple{\varphi, n}}{n\in\mathof{T}{\varphi}}\mright)\cup\mathsetintension{\mathtuple{\mathdynbox{\pi_{0}}\mathdynbox{\pi_{1}}\psi, n}}{n\in\mathof{T}{\varphi}}$.
		\item[\rulename{G-$\mathdynbox{\mathprogramndc{}{}}$ L}]
			   If $\varphi\equiv\mathdynbox{\mathprogramndc{\pi_{0}}{\pi_{1}}}\psi$ 
			   and $\varphi\in\Pi$, 
			   then
			   $C'=\mathtuple{\Pi, \mathdynbox{\pi_{0}}\psi, \mathdynbox{\pi_{1}}\psi\fCenter \Sigma; T; S\setminus \mathsetextension{\varphi}}$.
		\item[\rulename{G-$\mathdynbox{\mathprogramndc{}{}}$ R}] 
			   If $\varphi\equiv\mathdynbox{\mathprogramndc{\pi_{0}}{\pi_{1}}}\psi$ and 
			   $\varphi\in\Sigma$, 
			   then
			   $\mathof{\lambda}{C}=\Refuter$ and either \unless\ifoup \\ \fi
			   $C'=\mathtuple{ \Pi\fCenter \mathdynbox{\pi_{0}}\psi, \Sigma; T'_{0}; S\setminus \mathsetextension{\varphi}}$
			   or
			   $C'=\mathtuple{ \Pi\fCenter \mathdynbox{\pi_{1}}\psi, \Sigma; T'_{1}; S\setminus \mathsetextension{\varphi}}$,
			   where 
			   $T'_{i}=\mleft(T\setminus \mathsetintension{\mathtuple{\varphi, n}}{n\in\mathof{T}{\varphi}}\mright)\cup\mathsetintension{\mathtuple{\mathdynbox{\pi_{i}}\psi, n}}{n\in\mathof{T}{\varphi}}$
			   for $i=0,1$.
		\item[\rulename{G-$\mathdynbox{\mathprogramit{}}$ L}] 
			   If $\varphi\equiv\mathdynbox{\mathprogramit{\pi}}\psi$ and $\varphi\in\Pi$, 
			   then
			   $C'=\mathtuple{ \Pi, \psi, \mathdynbox{\pi}\mathdynbox{\mathprogramit{\pi}}\psi\fCenter \Sigma; T ; S\setminus \mathsetextension{\varphi}}$.
		\item[\rulename{G-C-s}] 
			   If $\varphi\equiv\mathdynbox{\mathprogramit{\pi}}\psi$ and $\psi\in\Sigma$, 
			   then
			   $\mathof{\lambda}{C}=\Refuter$ and either \unless\ifoup \\ \fi
			   $C'=\mathtuple{ \Pi\fCenter \psi, \Sigma; T'_{0} ; S\setminus \mathsetextension{\varphi}}$
			   or
			   $C'=\mathtuple{ \Pi \fCenter \mathdynbox{\pi}\mathdynbox{\mathprogramit{\pi}}\psi, \Sigma; T'_{1} ; S\setminus \mathsetextension{\varphi}}$,
			   where 
			   $T'_{0}=\mleft(T\setminus \mathsetintension{\mathtuple{\varphi, n}}{n\in\mathof{T}{\varphi}}\mright)\cup\mathsetintension{\mathtuple{\psi, n}}{n\in\mathof{T}{\varphi}}$
			   and
			   $T'_{1}=\mleft(T\setminus \mathsetintension{\mathtuple{\varphi, n}}{n\in\mathof{T}{\varphi}}\mright)\cup\mathsetintension{\mathtuple{\mathdynbox{\pi}\mathdynbox{\mathprogramit{\pi}}\psi, n}}{n\in\mathof{T}{\varphi}}$.
		\item[\rulename{G-$\mathdynbox{\mathprogramtest{}}$ L}] 
			   If $\varphi\equiv\mathdynbox{\mathprogramtest{\psi_{0}}}\psi_{1}$ and 
			   $\varphi\in\Pi$, 
			   then
			   $\mathof{\lambda}{C}=\Refuter$ and either \unless\ifoup \\ \fi
			   $C'=\mathtuple{ \Pi\fCenter \psi_{0}, \Sigma; T' ; S\setminus \mathsetextension{\varphi}}$ or
			   $C'=\mathtuple{ \Pi, \psi_{1} \fCenter \Sigma; T ; S\setminus \mathsetextension{\varphi}}$,
			   where $T'=T\cup\mathsetextension{\mathtuple{\psi_{0}, \mathfreshnum{T}}}$
			   if $\psi_{0}\notin\Sigma$; $T'=T$ otherwise.
		\item[\rulename{G-$\mathdynbox{\mathprogramtest{}}$ R}] 
			   If $\varphi\equiv\mathdynbox{\mathprogramtest{\psi_{0}}}\psi_{1}$ 
			   and $\varphi\in\Sigma$, 
			   then
			   $C'=\mathtuple{ \Pi, \psi_{0} \fCenter \psi_{1}, \Sigma; T' ; S\setminus \mathsetextension{\varphi}}$,
			   where 
			   $T'=\mleft(T\setminus \mathsetintension{\mathtuple{\varphi, n}}{n\in\mathof{T}{\varphi}}\mright)\cup\mathsetintension{\mathtuple{\psi_{1}, n}}{n\in\mathof{T}{\varphi}}$.
		\ifoup
		\end{itemize}
		\else
		\end{description}
		\fi
	 \item The case that
	       $C$ is of the form $\mathtuple{\Pi\fCenter\Sigma, T, S}$ 
	       and $(\Pi\cup\Sigma)\cap S=\emptyset$:
	 \begin{itemize}
	  \item $\mathof{\lambda}{C}=\Prover$, and there are the following two rules:
		\ifoup
		\begin{enumerate}
		 \else
		\begin{description}
		 \fi
		\item[\rulename{G-K}] 
			   If $\mathtuple{\mathdynbox{\alpha}\varphi, n}\in T$
			   for a natural number $n$, then
			   $C'=\mathtuple{\Pi' \fCenter \varphi; \mathsetextension{\mathtuple{\varphi, 0}}; \mathreducible{\mathFLcl{\Gamma\cup\Delta}}}$
			   for $\alpha\in\maththesetofatomicprog$,
			   where  
			   $\Pi'=\mathsetintension{\psi}{\mathdynbox{\alpha}\psi\in\Pi}$.
			   \begin{itemize}
			    \item We say that $\Prover$ changes the track $0$ with 
				  the move $\mathtuple{C, C'}$
				  if $\mathtuple{\mathdynbox{\alpha}\varphi, 0}\notin T$ 
			   \end{itemize}
		 \item[\rulename{G-Retry}] 
			    $C'=\mathtuple{\Pi\fCenter\Sigma; T; \mathreducible{\mathFLcl{\Gamma\cup\Delta}}}$. 
		\ifoup
		\end{enumerate}
		\else
		\end{description}
		\fi
	 \end{itemize}
	       \end{itemize}
  \item A \emph{play} is a sequences of positions $C_{0}, \dots, C_{n}$ 
	with a natural number $n\leq\GameSize$ 
	satisfying the following conditions:
	\begin{enumerate}
	 \item The \emph{starting position} $C_{0}$ is 
	       $\mathtuple{\Gamma\fCenter\Delta; T; \mathreducible{\mathFLcl{\Gamma\cup\Delta}}}$,
	       where
	       $\Delta=\mathsetextension{\varphi_{1}, \dots, \varphi_{m}}$,
	       $T=\mathsetextension{\mathtuple{\varphi_{1}, 1}, \dots, \mathtuple{\varphi_{m}, m}}$,
	       and
	       $\varphi_{1} <_{\Gamma\fCenter\Delta} \varphi_{2} <_{\Gamma\fCenter\Delta} \dots <_{\Gamma\fCenter\Delta} \varphi_{m}$.
	 \item $\mathtuple{C_{i}, C_{i+1}}$ is an instance of game rules for $i<n$.
	 \item For $i<n$, if the sequent of $C_{i}$ is $\Pi\fCenter\Sigma$,
	       then $\Pi\cap\Sigma=\emptyset$.
	 \item For $i<n$, if the sequent of $C_{i}$ is $\Pi\fCenter\Sigma$,
	       then $\bot\notin\Pi$.
	 \item $i\neq j$ implies $C_{i}\neq C_{j}$ for all $i, j< n$.
	 \item $C_{n}$ satisfies some of the following conditions:
	       \begin{enumerate}
		\item The sequent of $C_{n}$ is $\Pi\fCenter\Sigma$ with 
	       $\Pi\cap\Sigma\neq\emptyset$.
		\item The sequent of $C_{n}$ is $\Pi\fCenter\Sigma$ with 
	       $\bot\in\Pi$.
		\item There exists a position $C_{m}$ for $m<n$ such that $C_{m}=C_{n}$.
		      \begin{itemize}
		       \item If $C_{m}=C_{n}$ for some $m<n$, then we call 
			     a play $C_{0}, \dots, C_{n}$ a \emph{cyclic play}.
		      \end{itemize}
	       \end{enumerate}
	\end{enumerate}
  \item We say that \emph{\Prover\ wins a play $C_{0}, \dots, C_{n}$} if some of the following conditions hold:
	\begin{enumerate}
	 \item The sequent of $C_{n}$ is $\Pi\fCenter\Sigma$ with 
	       $\Pi\cap\Sigma\neq\emptyset$.
	 \item The sequent of $C_{n}$ is $\Pi\fCenter\Sigma$ with 
	       $\bot\in\Pi$.
	 \item There exists a position $C_{m}$ for $m<n$ such that 
	       $C_{m}=C_{n}$ holds and
	       there exists a natural number $l$ such that
	       all the following conditions hold:
	       \ifoup
	       \begin{enumerate}
		\else
		\begin{description}
		\fi
		 \ifoup \item \else \item[(Progress condition)] \fi
			   There exists a natural number $i$ such that
			   $m\leq i\leq n$ holds and
			   $\psi_{i}\not\equiv\psi_{i+1}$
			   for $\mathtuple{\psi_{i}, l}\in T_{i}$ and
			   $\mathtuple{\psi_{i+1}, l}\in T_{i+1}$, where
			   $C_{i}=\mathtuple{ \Pi_{i} \fCenter \Sigma_{i}; T_{i}; S_{i}}$ 
			   and
			   $C_{i+1}=\mathtuple{ \Pi_{i+1} \fCenter \Sigma_{i+1}; T_{i+1}; S_{i+1}}$.
		\ifoup \item \else \item[(Tracking condition 1)] \fi
			   For all $m\leq i \leq n$ and 
			   $C_{i}=\mathtuple{ \Pi_{i} \fCenter \Sigma_{i}; T_{i}; S_{i}}$,
			   there exists a formula $\psi_{i}$ such that
			   $\mathtuple{\psi_{i}, l}\in T_{i}$.
		\ifoup \item \else \item[(Tracking condition 2)] \fi
			   $\Prover$ does not change the track $0$ with
			   $\mathtuple{C_{i}, C_{i+1}}$ for all $m\leq i \leq n$.
		\ifoup
		\end{enumerate}
		\else
		\end{description}
		\fi
	\end{enumerate}
  \item We say that \emph{\Refuter\ wins a play $C_{0}, \dots, C_{n}$} 
	if \Prover\ does not win the play.
 \end{itemize}
\end{definition}

 \Prover\ and \Refuter\ stand for \emph{Prover} and \emph{Refuter}.
 Note that \Prover\ can only choose the next move 
 when either \rulename{G-K} or \rulename{G-Retry} is applied, and
 \Refuter\ can only choose the next move
 when \rulename{G-$\to$ L}, \rulename{G-$\mathdynbox{\mathprogramndc{}{}}$ R},
 \rulename{G-C-s}, or \rulename{G-$\mathdynbox{\mathprogramtest{}}$ L} is applied.
 When any other rule is applied, the next move is determined uniquely.
 Hence, we do not have to define an active player in these cases. 

 $\mathof{\mathtree{T}}{\sigma}$ denotes a label of $\sigma$ 
 for a labelled tree $T$ and each node $\sigma$ of $\mathtree{T}$.

\ifoup
\begin{figure*}[tb]
\begin{minipage}[c]{0.49\linewidth}
\centering
  \begin{tikzpicture}[->]
   \node {$C_{0}$} [grow'=up]
     child {node (leaf1) {$C_{1}$} edge from parent node[left] {\rulename{G-C-s}} }
     child {node (comp3) {$C_{2}$} 
   child {node {$C_{3}$} 
   child {node {$C_{4}$} 
   child {node (leaf2) {$C_{5}$}  edge from parent node[left] {\rulename{G-C-s}} } 
   child {node (leaf3) {$C_{2}$}  edge from parent node[right] {\rulename{G-C-s}} } 
   edge from parent node[right] {\rulename{G-Retry}} 
   }
   edge from parent node[right] {\rulename{G-$\mathdynbox{\mathprogramtest{}}$ R}} 
   }
edge from parent node[right] {\rulename{G-C-s}} 
   };
   \node [above=1pt of leaf1] {$\Prover$ wins!};
   \node [above=1pt of leaf2] {$\Prover$ wins!};
   \node [above=1pt of leaf3] {$\Prover$ wins!};
   \draw [-, dashed] (leaf3) .. controls +(right:50pt) and +(right:115pt) .. (comp3);
  \end{tikzpicture}
  
 \caption{The game tree of $\GameName{p\fCenter \mathdynbox{\mathprogramit{\mathprogramtest{p}}}p}$.}
  \label{fig:game-tree-P}
\end{minipage}
\begin{minipage}[c]{0.49\linewidth}
\centering
  \begin{tikzpicture}[->]
   \node {$C'_{0}$} [grow'=up]
   child {node (comp4) {$C'_{1}$} 
   child {node {$C'_{2}$}
   child {node (leaf4) {$C'_{1}$}
   edge from parent node[left] {\scriptsize\rulename{G-C-s}}
   }
   child {node {$C'_{3}$}
   child {node (leaf5) {$C'_{4}$}
   edge from parent node[right, near end] {\tiny\rulename{G-$\mathdynbox{\mathprogramtest{}}$ R}}
   }
   edge from parent node[right] {\scriptsize\rulename{G-C-s}}
   } 
   edge from parent node[left] {\scriptsize\rulename{G-Retry}} 
   }
   edge from parent node[left] {\scriptsize\rulename{G-C-s}} }
   child {node {} edge from parent[draw=none]} 
   child {node {$C'_{5}$} 
   child {node (comp7) {$C_{3}$} 
   child {node {$C_{4}$} 
   child {node (leaf6) {$C_{5}$}  edge from parent node[left, near start] {\tiny\rulename{G-C-s}} 
   } 
   child {node {$C_{2}$}  
   child {node (leaf7) {$C_{3}$}
   edge from parent node[right]  {\scriptsize\rulename{G-$\mathdynbox{\mathprogramtest{}}$ R}}
   }
   edge from parent node[right] {\scriptsize\rulename{G-C-s}} } 
   edge from parent node[right] {\scriptsize\rulename{G-Retry}} 
   }
   edge from parent node[right] {\scriptsize\rulename{G-$\mathdynbox{\mathprogramtest{}}$ R}} 
   }
   edge from parent node[right] {\scriptsize\rulename{G-C-s}} 
   };
   \node [above=1pt of leaf4] {$\Refuter$ wins!};
   \node [above=1pt of leaf5] {$\Prover$ wins!};
   \node [above=1pt of leaf6] {$\Prover$ wins!};
   \node [above=1pt of leaf7] {$\Prover$ wins!};
   \draw [-, dashed] (leaf4) .. controls +(left:25pt) and +(left:80pt) .. (comp4);
   \draw [-, dashed] (leaf7) .. controls +(right:85pt) and +(right:80pt) .. (comp7);
  \end{tikzpicture}

 \caption{The game tree of $\GameName{\fCenter \mathdynbox{\mathprogramit{\mathprogramtest{p}}}p}$.}
 \label{fig:game-tree-R}
\end{minipage}
\end{figure*}
 \fi

 \begin{definition}[Game tree]
  The \emph{game tree} of $\GameName{\Gamma\fCenter\Delta}$
  is defined as a finite tree $\mathtree{T}$ of positions satisfying 
  the following conditions:
  \begin{enumerate}
   \item The root is labelled by the starting position
	 $\mathtuple{\Gamma\fCenter\Delta; T; \mathreducible{\mathFLcl{\Gamma\cup\Delta}}}$,
	 where
	 $\Delta=\mathsetextension{\varphi_{0}, \dots, \varphi_{m}}$,
	 $T=\mathsetextension{\mathtuple{\varphi_{0}, 0}, \dots, \mathtuple{\varphi_{m}, m}}$,
	 and
	 $\varphi_{0}<_{\Gamma\fCenter\Delta} \varphi_{1} <_{\Gamma\fCenter\Delta} \dots <_{\Gamma\fCenter\Delta} \varphi_{m}$.
   \item For any inner node $\sigma$, which is not a leaf but a node,
	 the set of its children is 
	 \[
	 \mathsetintension{\sigma'}{ \mathtuple{\mathof{\mathtree{T}}{\sigma}, \mathof{\mathtree{T}}{\sigma'}} \text{ is an instance of game rules}}.
	 \]
   \item For any inner node $\sigma$,
	 if the sequent of $\mathof{\mathtree{T}}{\sigma}$ is $\Pi\fCenter\Sigma$,
	 then $\Pi\cap\Sigma=\emptyset$ holds.
   \item For any inner node $\sigma$,
	 if the sequent of $\mathof{\mathtree{T}}{\sigma}$ is $\Pi\fCenter\Sigma$, 
	 then $\bot\notin\Pi$ holds.
   \item Let $\sigma_{0}, \dots, \sigma_{n}$ be a path,
	 where $\sigma_{0}$ is the root and $\sigma_{n}$ is a leaf.
	 $\mathof{\mathtree{T}}{\sigma_{i}}\neq \mathof{\mathtree{T}}{\sigma_{j}}$ for 
	 $i\neq j$ and $i, j< n$.
   \item Each leaf, which is labelled by $C$, satisfies some of the following conditions:
	 \begin{enumerate}
	  \item The sequent of $C$ is $\Pi\fCenter\Sigma$ with $\Pi\cap\Sigma\neq\emptyset$.
	  \item The sequent of $C$ is $\Pi\fCenter\Sigma$ with $\bot\in\Pi$.
	  \item There exists an ancestor such that 
		the position assigned to it is the same as $C$.
	 \end{enumerate}
  \end{enumerate}
 \end{definition}

 \unless\ifoup
 \begin{figure*}[tb]
  \centering
  \begin{subfigure}[t]{0.49\textwidth}
  \centering
  \hspace*{5.0em}
    \begin{tikzpicture}[->]
   \node {$C_{0}$} [grow'=up]
     child {node (leaf1) {$C_{1}$} edge from parent node[left] {\rulename{G-C-s}} }
     child {node (comp3) {$C_{2}$} 
   child {node {$C_{3}$} 
   child {node {$C_{4}$} 
   child {node (leaf2) {$C_{5}$}  edge from parent node[left] {\rulename{G-C-s}} } 
   child {node (leaf3) {$C_{2}$}  edge from parent node[right] {\rulename{G-C-s}} } 
   edge from parent node[right] {\rulename{G-Retry}} 
   }
   edge from parent node[right] {\rulename{G-$\mathdynbox{\mathprogramtest{}}$ R}} 
   }
edge from parent node[right] {\rulename{G-C-s}} 
   };
   \node [above=1pt of leaf1] {$\Prover$ wins!};
   \node [above=1pt of leaf2] {$\Prover$ wins!};
   \node [above=1pt of leaf3] {$\Prover$ wins!};
   \draw [-, dashed] (leaf3) .. controls +(right:50pt) and +(right:115pt) .. (comp3);
  \end{tikzpicture}
  
 \caption{The game tree of $\GameName{p\fCenter \mathdynbox{\mathprogramit{\mathprogramtest{p}}}p}$.}
  \label{fig:game-tree-P}
  \end{subfigure}
  \hfil
 \begin{subfigure}[t]{0.49\textwidth}
  \centering
    \begin{tikzpicture}[->]
   \node {$C'_{0}$} [grow'=up]
   child {node (comp4) {$C'_{1}$} 
   child {node {$C'_{2}$}
   child {node (leaf4) {$C'_{1}$}
   edge from parent node[left] {\scriptsize\rulename{G-C-s}}
   }
   child {node {$C'_{3}$}
   child {node (leaf5) {$C'_{4}$}
   edge from parent node[right, near end] {\tiny\rulename{G-$\mathdynbox{\mathprogramtest{}}$ R}}
   }
   edge from parent node[right] {\scriptsize\rulename{G-C-s}}
   } 
   edge from parent node[left] {\scriptsize\rulename{G-Retry}} 
   }
   edge from parent node[left] {\scriptsize\rulename{G-C-s}} }
   child {node {} edge from parent[draw=none]} 
   child {node {$C'_{5}$} 
   child {node (comp7) {$C_{3}$} 
   child {node {$C_{4}$} 
   child {node (leaf6) {$C_{5}$}  edge from parent node[left, near start] {\tiny\rulename{G-C-s}} 
   } 
   child {node {$C_{2}$}  
   child {node (leaf7) {$C_{3}$}
   edge from parent node[right]  {\scriptsize\rulename{G-$\mathdynbox{\mathprogramtest{}}$ R}}
   }
   edge from parent node[right] {\scriptsize\rulename{G-C-s}} } 
   edge from parent node[right] {\scriptsize\rulename{G-Retry}} 
   }
   edge from parent node[right] {\scriptsize\rulename{G-$\mathdynbox{\mathprogramtest{}}$ R}} 
   }
   edge from parent node[right] {\scriptsize\rulename{G-C-s}} 
   };
   \node [above=1pt of leaf4] {$\Refuter$ wins!};
   \node [above=1pt of leaf5] {$\Prover$ wins!};
   \node [above=1pt of leaf6] {$\Prover$ wins!};
   \node [above=1pt of leaf7] {$\Prover$ wins!};
   \draw [-, dashed] (leaf4) .. controls +(left:25pt) and +(left:80pt) .. (comp4);
   \draw [-, dashed] (leaf7) .. controls +(right:85pt) and +(right:80pt) .. (comp7);
  \end{tikzpicture}

 \caption{The game tree of $\GameName{\fCenter \mathdynbox{\mathprogramit{\mathprogramtest{p}}}p}$.}
 \label{fig:game-tree-R}
  \end{subfigure}

  \input{subfig-ex-game-tree-form}

  \caption{Examples of game trees.}
  \label{fig:game-tree}
 \end{figure*}
 \fi

 \ifoup
 There are examples of game trees in \cref{fig:game-tree-P} and \cref{fig:game-tree-R}, 
 where
 \begin{gather*}
 S_{0}=\mathsetextension{\mathdynbox{\mathprogramit{\mathprogramtest{p}}}p, \mathdynbox{\mathprogramtest{p}}\mathdynbox{\mathprogramit{\mathprogramtest{p}}}p}, \quad
 S_{1}=\mathsetextension{\mathdynbox{\mathprogramtest{p}}\mathdynbox{\mathprogramit{\mathprogramtest{p}}}p}, \\
 T_{0}=\mathsetextension{\mathtuple{\mathdynbox{\mathprogramit{\mathprogramtest{p}}}p, 1}}, \quad
 T_{1}=\mathsetextension{\mathtuple{\mathdynbox{\mathprogramtest{p}}\mathdynbox{\mathprogramit{\mathprogramtest{p}}}p, 1}}, \quad
 T_{2}=\mathsetextension{\mathtuple{p, 1}}, \\
 C_{0}=\mathtuple{p\fCenter \mathdynbox{\mathprogramit{\mathprogramtest{p}}}p;T_{0};S_{0}}, \ifoup \quad \else \\ \fi
 C_{1}=\mathtuple{p\fCenter p, \mathdynbox{\mathprogramit{\mathprogramtest{p}}}p;T_{2};S_{1}}, \ifoup \quad \else \\ \fi
 C_{2}=\mathtuple{p\fCenter \mathdynbox{\mathprogramtest{p}}\mathdynbox{\mathprogramit{\mathprogramtest{p}}}p, \mathdynbox{\mathprogramit{\mathprogramtest{p}}}p;T_{1};S_{1}}, \\
 C_{3}=\mathtuple{p\fCenter \mathdynbox{\mathprogramtest{p}}\mathdynbox{\mathprogramit{\mathprogramtest{p}}}p, \mathdynbox{\mathprogramit{\mathprogramtest{p}}}p;T_{0};\emptyset}, \ifoup \quad \else \\ \fi
 C_{4}=\mathtuple{p\fCenter \mathdynbox{\mathprogramtest{p}}\mathdynbox{\mathprogramit{\mathprogramtest{p}}}p, \mathdynbox{\mathprogramit{\mathprogramtest{p}}}p;T_{0};S_{0}}, \\
 C_{5}=\mathtuple{p\fCenter p, \mathdynbox{\mathprogramtest{p}}\mathdynbox{\mathprogramit{\mathprogramtest{p}}}p, \mathdynbox{\mathprogramit{\mathprogramtest{p}}}p;T_{2};S_{1}}, \\
 C'_{0}=\mathtuple{\fCenter \mathdynbox{\mathprogramit{\mathprogramtest{p}}}p;T_{0};S_{0}}, \ifoup \quad \else \\ \fi
 C'_{1}=\mathtuple{\fCenter p, \mathdynbox{\mathprogramit{\mathprogramtest{p}}}p;T_{2};S_{1}}, \ifoup \quad \else \\ \fi
 C'_{2}=\mathtuple{\fCenter p, \mathdynbox{\mathprogramit{\mathprogramtest{p}}}p;T_{2};S_{0}}, \\
 C'_{3}=\mathtuple{\fCenter \mathdynbox{\mathprogramtest{p}}\mathdynbox{\mathprogramit{\mathprogramtest{p}}}p, p, \mathdynbox{\mathprogramit{\mathprogramtest{p}}}p;T_{2};S_{1}}, \ifoup \quad \else \\ \fi
 C'_{4}=\mathtuple{p\fCenter \mathdynbox{\mathprogramtest{p}}\mathdynbox{\mathprogramit{\mathprogramtest{p}}}p, p, \mathdynbox{\mathprogramit{\mathprogramtest{p}}}p;T_{2};\emptyset}, \ifoup \quad \else \\ \fi
 C'_{5}=\mathtuple{\fCenter \mathdynbox{\mathprogramtest{p}}\mathdynbox{\mathprogramit{\mathprogramtest{p}}}p, \mathdynbox{\mathprogramit{\mathprogramtest{p}}}p;T_{1};S_{1}}.
  \end{gather*}
 \else There are examples of game trees in \cref{fig:game-tree}. \fi

 Obviously, each path from the root to a leaf on the game tree of
 $\GameName{\Gamma\fCenter\Delta}$ is one of plays.
 We write $\opponent{\player{P}}$ for the \emph{opponent} of $\player{P}$,
 defined as $\opponent{\Prover}=\Refuter$ and $\opponent{\Refuter}=\Prover$.

 We call a leaf $\sigma$ of the game tree of a proof search game a \emph{bud} if
 there exists a proper ancestor of $\sigma'$ such that
 $\sigma$ and $\sigma'$ are assigned to the same position.
 The \emph{corresponding companion} of a bud $\sigma$ is defined as its ancestor $\sigma'$
 such that $\sigma$ and $\sigma'$ are assigned to the same position.

 \begin{definition}[Winning strategy]
  \label[definition]{def:winning-strategy}
  Let $\player{P}$ be $\Prover$ or $\Refuter$.
  Let $\lambda$ be a turn function of $\GameName{\Gamma\fCenter\Delta}$.
  A \emph{winning strategy} of $\player{P}$ in $\GameName{\Gamma\fCenter\Delta}$ 
  is defined as a finite tree $\mathwinningstrategy{S}$ of positions 
  satisfying the following conditions:
  \begin{enumerate}
   \item $\mathwinningstrategy{S}$ is a subtree of the game tree such that 
	 the root of $\mathwinningstrategy{S}$ is
	 the root of the game tree of $\GameName{\Gamma\fCenter\Delta}$.
   \item For every inner node $\sigma$ of $\mathwinningstrategy{S}$, 
	 the following conditions hold:
	 \begin{enumerate}
	  \item If $\mathof{\lambda}{\mathof{\mathwinningstrategy{S}}{\sigma}}=\player{P}$,
		then the child of $\sigma$ in $\mathwinningstrategy{S}$ is only one.
	  \item If $\mathof{\lambda}{\mathof{\mathwinningstrategy{S}}{\sigma}}=\opponent{\player{P}}$, 
		then the children of $\sigma$ in $\mathwinningstrategy{S}$ are
		all the children of $\sigma$ in the game tree. 
	 \end{enumerate}
   \item Every path from the root to a leaf is a play $\player{P}$ wins.
  \end{enumerate}
 \end{definition}

 The winning strategy of $\Prover$ in 
 $\GameName{p\fCenter \mathdynbox{\mathprogramit{\mathprogramtest{p}}}p}$
 is the same as its game tree, \cref{fig:game-tree-P}.
 The winning strategy of $\Refuter$ in 
 $\GameName{\fCenter \mathdynbox{\mathprogramit{\mathprogramtest{p}}}p}$
 is in \cref{fig:winning-strategy-ex2}.

 \unless\ifoup
 \clearpage
 \fi

 \begin{figure*}[tb]
  \centering
  \begin{tikzpicture}[->]
   \node {$C'_{0}$} [grow'=up]
   child {node (comp8) {$C'_{1}$} 
   child {node {$C'_{2}$}
   child {node (leaf8) {$C'_{1}$}
   edge from parent node[left] {\rulename{G-C-s}}
   } 
   edge from parent node[left] {\rulename{G-Retry}} 
   }
   edge from parent node[left] {\rulename{G-C-s}} 
   };
   \node [above=1pt of leaf8] {$\Refuter$ wins!};
   \draw [-, dashed] (leaf8) .. controls +(right:20pt) and +(right:20pt) .. (comp8);
  \end{tikzpicture}
  
  \caption{The winning strategy of $\Refuter$ in  $\GameName{\fCenter \mathdynbox{\mathprogramit{\mathprogramtest{p}}}p}$}
  \label{fig:winning-strategy-ex2}
 \end{figure*}

 The following theorem is well-known.
 
 \begin{theorem}[Zermelo's theorem \cite{Zermelo1913}]
  \label{thm:Zermelo}
  One of the players has a winning strategy in a two-player finite zero-sum game of 
  perfect information. 
 \end{theorem}

 The proof of this theorem is in \cite{Van2014}, for example.
 By the consequence of this theorem, we have the following corollary.

 \begin{corollary}
  \label[corollary]{cor:determinacy}
  Either $\Prover$ or $\Refuter$ has a winning strategy in each proof search game.
 \end{corollary}

 To show \cref{theorem:cut-free-completeness-CGPDL},
 we show the following two propositions.

 \begin{proposition}
  \label[proposition]{lemma:Prover-win}
  $\Gamma\fCenter\Delta$ is cut-free provable in \CGPDL\
  if $\Prover$ has a winning strategy in the proof search game of $\Gamma\fCenter\Delta$.
 \end{proposition}

 \begin{proposition}
  \label[proposition]{lemma:Refuter-win}
  $\Gamma\fCenter\Delta$ is invalid
  if $\Refuter$ has a winning strategy in the proof search game of $\Gamma\fCenter\Delta$. 
 \end{proposition}

 We show \cref{lemma:Prover-win} and \cref{lemma:Refuter-win} in
 \cref{subsec:Prover-wins} and \cref{subsec:Refuter-wins}, respectively.

 \subsection{The case \Prover\ wins}
 \label{subsec:Prover-wins}
 This section proves \cref{lemma:Prover-win}.

 We write $\mathTr{C}$ for the track of a position $C$.
 We call an instance of game rules $\mathtuple{C, C'}$, where
 $\mathtuple{\varphi, n}\in\mathTr{C}$ and $\mathtuple{\psi, n}\in\mathTr{C'}$
 with $\varphi\not\equiv\psi$, a \emph{transforming pair with $n$}.
 We say that \emph{the track $m$ is continuous} in a path
 $\mathsequence{\sigma_{i}}{0\leq i< n}$ in a subgraph of the game tree $\mathtree{T}$ 
 if $\mathtuple{\varphi_{i}, m}\in\mathTr{\mathof{\mathtree{T}}{\sigma_{i}}}$ holds
 for all $0\leq i< n$.

 We note that each leaf $\sigma$ of a winning strategy $\mathwinningstrategy{S}_{\Prover}$
 of $\Prover$ in a proof search game satisfies exactly one of the following conditions:
 \begin{enumerate}
  \item The sequent of $\mathof{\mathwinningstrategy{S}_{\Prover}}{\sigma}$ is
	$\Pi\fCenter\Sigma$ with $\Pi\cap\Sigma\neq\emptyset$ or $\bot\in\Pi$.
  \item $\sigma$ is a bud, and there exists a natural number $n$ such that 
	the following three conditions hold:
  \begin{itemize}
   \item A transforming pair with $n$ occurs between $\sigma'$ and $\sigma$,
	 where $\sigma'$ is the corresponding companion of $\sigma$. 
   \item The track $n$ is continuous in a path from $\sigma'$ to $\sigma$,
	 where $\sigma'$ is the corresponding companion of $\sigma$. 
   \item $\Prover$ never changes the track $0$ in a path from $\sigma'$ to $\sigma$,
	 where $\sigma'$ is the corresponding companion of $\sigma$.
  \end{itemize}
 \end{enumerate}
 
 We write $\mathSeq{C}$ for the sequent of a position $C$.

 \begin{figure*}[tb]
  \centering
  \footnotesize
  \begin{inlineprooftree}
   \AxiomC{}
   \RightLabel{\rulename{Ax}}
   \UnaryInfC{$ p\fCenter p, \mathdynbox{\mathprogramit{\mathprogramtest{p}}}p $ \quad}

   \AxiomC{}
   \RightLabel{\rulename{Ax}}
   \UnaryInfC{$ p\fCenter p, \mathdynbox{\mathprogramtest{p}}\mathdynbox{\mathprogramit{\mathprogramtest{p}}}p, \mathdynbox{\mathprogramit{\mathprogramtest{p}}}p $ }

   \AxiomC{($\clubsuit$) $ p\fCenter \underline{\mathdynbox{\mathprogramtest{p}}\mathdynbox{\mathprogramit{\mathprogramtest{p}}}p}, \mathdynbox{\mathprogramit{\mathprogramtest{p}}}p $ \raisebox{0.5em}{\tikzmark{node-bud-CTPDL}}}

   \RightLabel{\rulename{C-s} \tikzmark{node-M-CTPDL}}
   \BinaryInfC{$p\fCenter \mathdynbox{\mathprogramtest{p}}\mathdynbox{\mathprogramit{\mathprogramtest{p}}}p, \underline{\mathdynbox{\mathprogramit{\mathprogramtest{p}}}p}$}
   \RightLabel{\rulename{Wk}}
   \UnaryInfC{$p\fCenter \mathdynbox{\mathprogramtest{p}}\mathdynbox{\mathprogramit{\mathprogramtest{p}}}p, \underline{\mathdynbox{\mathprogramit{\mathprogramtest{p}}}p}$}
   \RightLabel{\rulename{$\mathdynbox{\mathprogramtestsy}$ R}}
   \UnaryInfC{($\clubsuit$) $p\fCenter \underline{\mathdynbox{\mathprogramtest{p}}\mathdynbox{\mathprogramit{\mathprogramtest{p}}}p}, \mathdynbox{\mathprogramit{\mathprogramtest{p}}}p$ \raisebox{0.25em}{\tikzmark{node-companion-CTPDL}}}
   \RightLabel{\rulename{C-s}}
   \BinaryInfC{$ p\fCenter \mathdynbox{\mathprogramit{\mathprogramtest{p}}}p $}
  \end{inlineprooftree}
  \begin{tikzpicture}[remember picture, overlay, thick, relative, auto, rounded corners, line width=1.0pt]
   \coordinate (b) at ({pic cs:node-bud-CTPDL});
   \coordinate (c) at ({pic cs:node-companion-CTPDL});
   \coordinate (M) at ({pic cs:node-M-CTPDL});
   
   \draw[->] let \p1=($(M)-(b)$), \p2=($(c)-(b)$) in
   (b) -- ++(\x1 + 0.5em,0) -- ++(0,\y2) -- (c);
  \end{tikzpicture}
  \caption{The \CGPDL\ pre-proof constructed from the winning strategy of $\Prover$ in $\GameName{p\fCenter \mathdynbox{\mathprogramit{\mathprogramtest{p}}}p}$.}
  \label{fig:constructed-pre-proof-win}
 \end{figure*}

 \begin{definition}[The \CGPDL\ pre-proof constructed from a winning strategy]
  Let $\mathwinningstrategy{S}_{\Prover}$ be a winning strategy of $\Prover$ 
  in $\GameName{\Gamma\fCenter\Delta}$.

  We define a \CGPDL\ \emph{pre-proof constructed from $\mathwinningstrategy{S}_{\Prover}$}
  to be a pair $\mleft(\mathdertree{D}_{\mathwinningstrategy{S}_{\Prover}},  \mathcompanion{C}_{\mathwinningstrategy{S}_{\Prover}}\mright)$ as follows:
  \begin{enumerate}
   \item $\mathdertree{D}_{\mathwinningstrategy{S}_{\Prover}}$ is obtained 
	 from $\mathwinningstrategy{S}_{\Prover}$ by relabelling each node $\sigma$ of 
	 $\mathwinningstrategy{S}_{\Prover}$ 
	 with $\mathSeq{\mathof{\mathwinningstrategy{S}_{\Prover}}{\sigma}}$.
   \item For each bud $\sigma$, we define 
	 $\mathof{\mathcompanion{C}_{\mathwinningstrategy{S}_{\Prover}}}{\sigma}$ as
	 the corresponding companion of $\sigma$.

  \end{enumerate}
 \end{definition}

 By \cref{def:proof-search-game} and \cref{def:winning-strategy},
 the \CGPDL\ pre-proof constructed from a winning strategy is
 a cut-free \CGPDL\ pre-proof.
 The \CGPDL\ pre-proof constructed from a winning strategy of $\Prover$ in 
 $\GameName{p\fCenter \mathdynbox{\mathprogramit{\mathprogramtest{p}}}p}$ 
 is in \cref{fig:constructed-pre-proof-win}, where
 ($\clubsuit$) indicates the pairing of the companion with the bud.
 As we prove later, 
 the \CGPDL\ pre-proof constructed from a winning strategy is a \CGPDL\ proof 
 (\cref{lemma:proof-from-winning-strategy}). 
 The underlined formulae in \cref{fig:constructed-pre-proof-win} denote 
 the infinitely progressing trace for some tails of the infinite path.

 For a path $\mathsequence{\sigma_{i}}{0\leq i}$ in the derivation graph of 
 the \CGPDL\ pre-proof constructed from a winning strategy 
 $\mathwinningstrategy{S}_{\Prover}$, we say that
 \emph{the track $n$ is continuous} in the path if
 $\mathtuple{\varphi_{i}, n}\in\mathTr{\mathof{\mathwinningstrategy{S}_{\Prover}}{\sigma_{i}}}$
 holds for all $i\geq 0$.
 We also say that \emph{$\Prover$ never changes the track $0$} in the path if,
 for all $i\geq 0$,
 $\mathtuple{\mathof{\mathwinningstrategy{S}_{\Prover}}{\sigma_{i}}, \mathof{\mathwinningstrategy{S}_{\Prover}}{\sigma_{i+1}}}$
 is not an instance of \rulename{G-K} with which $\Prover$ change the track $0$.
 We note that, for a path $\mathsequence{\sigma_{i}}{0\leq i}$ in
 the derivation graph of the \CGPDL\ pre-proof constructed from 
 a winning strategy $\mathwinningstrategy{S}_{\Prover}$,
 if the track $n$ is continuous for some $n$ and $\Prover$ never changes the track $0$
 in the path $\mathsequence{\sigma_{i}}{0\leq i}$,
 then $\mathsequence{\varphi_{i}}{0\leq i}$ is a trace following the path, where
 $\mathtuple{\varphi_{i}, n}\in\mathTr{\mathof{\mathwinningstrategy{S}_{\Prover}}{\sigma_{i}}}$ for each $i\geq 0$.

 We write $(s_{0}, s_{1}, \dots, s_{n})^{\omega}$ for
 the infinite sequence $s_{0}, s_{1}, \dots, s_{n}, s_{0}, s_{1}, \dots, s_{n}, \dots \dots$.

 For a winning strategy $\mathwinningstrategy{S}$ in $\GameName{\Gamma\fCenter\Delta}$, 
 we define 
\[
 \mathCycle{\mathwinningstrategy{S}}=\mathsetintension{\mathsequence{\sigma_{i}}{0\leq i < n} }{ \sigma_{0}, \sigma_{1}, \dots, \sigma_{n} \text{ is a path in } \mathwinningstrategy{S} \text{ with } \mathof{\mathwinningstrategy{S}}{\sigma_{0}}=\mathof{\mathwinningstrategy{S}}{\sigma_{n}} }.
\]

 For a winning strategy $\mathwinningstrategy{S}_{\Prover}$ of $\Prover$ 
 in $\GameName{\Gamma\fCenter\Delta}$ and
 each $\mathsequence{\sigma_{i}}{0\leq i < n}\in\mathCycle{\mathwinningstrategy{S}_{\Prover}}$,
 $\Prover$ never changes the track $0$ in the path, and
 there exist natural numbers $m, j$ such that the track $m$ is continuous in the path and 
 $\mathtuple{\mathof{\mathwinningstrategy{S}_{\Prover}}{\sigma_{j}}, \mathof{\mathwinningstrategy{S}_{\Prover}}{\sigma_{j+1}}}$ 
 is a transforming pair with $m$. 

 We call an instance of \rulename{G-C-s} $\mathtuple{C, C'}$, where
 $\mathtuple{\mathdynbox{\mathprogramit{\pi}}\psi, n}\in\mathTr{C}$ and 
 $\mathtuple{\mathdynbox{\pi}\mathdynbox{\mathprogramit{\pi}}\psi, n}\in\mathTr{C'}$, 
 a \emph{progressing pair with $n$}.

 \begin{lemma}
  \label[lemma]{lem:infinite-path}
  Let $\mathwinningstrategy{S}_{\Prover}$ be a winning strategy of $\Prover$ 
  in $\GameName{\Gamma\fCenter\Delta}$ and 
  $\mleft(\mathdertree{D}_{\mathwinningstrategy{S}_{\Prover}}, \mathcompanion{C}_{\mathwinningstrategy{S}_{\Prover}}\mright)$ 
  be the \CGPDL\ pre-proof constructed from $\mathwinningstrategy{S}_{\Prover}$.
  Let $\mathproofgraph{\mleft(\mathdertree{D}_{\mathwinningstrategy{S}_{\Prover}}, \mathcompanion{C}_{\mathwinningstrategy{S}_{\Prover}}\mright)}$ be 
  the derivation graph of $\mleft(\mathdertree{D}_{\mathwinningstrategy{S}_{\Prover}},  \mathcompanion{C}_{\mathwinningstrategy{S}_{\Prover}}\mright)$.

  \begin{enumerate}
   \item For any infinite path $\mathsequence{\sigma_{i}}{0\leq i}$
	 in $\mathproofgraph{\mleft(\mathdertree{D}_{\mathwinningstrategy{S}_{\Prover}}, \mathcompanion{C}_{\mathwinningstrategy{S}_{\Prover}}\mright)}$,
	 there exists 
	 $\mathsequence{\sigma'_{i}}{0\leq i < n}\in\mathCycle{\mathwinningstrategy{S}_{\Prover}}$ 
	 such that
	 $\sigma'_{0}, \dots, \sigma'_{n-1}, \sigma'_{0}$ occurs in $\mathsequence{\sigma_{i}}{0\leq i}$
	 infinitely many times.
	 \label{item:existence-lem-infinite-path}
   \item For any infinite path $\mathsequence{\sigma_{i}}{0\leq i}$
	 in $\mathproofgraph{\mleft(\mathdertree{D}_{\mathwinningstrategy{S}_{\Prover}}, \mathcompanion{C}_{\mathwinningstrategy{S}_{\Prover}}\mright)}$,
	 there exist natural numbers $k$, $n$ such that
	 $\mathtuple{\varphi_{i}, n}\in\mathTr{\mathof{\mathwinningstrategy{S}_{\Prover}}{\sigma_{i}}}$
	 for all $i\geq k$ and
	 $\Prover$ never changes the track $0$ in $\mathsequence{\sigma_{i}}{k\leq i}$.
	 \label{item:tail-track-lem-infinite-path}
   \item For any infinite path $\mathsequence{\sigma_{i}}{0\leq i}$
	 in $\mathproofgraph{\mleft(\mathdertree{D}_{\mathwinningstrategy{S}_{\Prover}}, \mathcompanion{C}_{\mathwinningstrategy{S}_{\Prover}}\mright)}$,
	 there exist natural numbers $k$, $n$ such that
	 $\mathtuple{\varphi_{i}, n}\in\mathTr{\mathof{\mathwinningstrategy{S}_{\Prover}}{\sigma_{i}}}$
	 for all $i\geq k$,
	 $\Prover$ never changes the track $0$ in $\mathsequence{\sigma_{i}}{k\leq i}$,
	 and $\mathtuple{\mathof{\mathwinningstrategy{S}_{\Prover}}{\sigma_{j}}, \mathof{\mathwinningstrategy{S}_{\Prover}}{\sigma_{j+1}}}$ is
	 a transforming pair with $n$ for infinitely many $j\geq k$.
	 \label{item:tail-progress-lem-infinite-path}
   \item For any infinite path $\mathsequence{\sigma_{i}}{0\leq i}$
	 in $\mathproofgraph{\mleft(\mathdertree{D}_{\mathwinningstrategy{S}_{\Prover}}, \mathcompanion{C}_{\mathwinningstrategy{S}_{\Prover}}\mright)}$,
	 there exist natural numbers $k$, $n$ such that
	 $\mathtuple{\varphi_{i}, n}\in\mathTr{\mathof{\mathwinningstrategy{S}_{\Prover}}{\sigma_{i}}}$
	 for all $i\geq k$,
	 $\Prover$ never changes the track $0$ in $\mathsequence{\sigma_{i}}{k\leq i}$
	 and $\mathtuple{\mathof{\mathwinningstrategy{S}_{\Prover}}{\sigma_{j}}, \mathof{\mathwinningstrategy{S}_{\Prover}}{\sigma_{j+1}}}$ is
	 a progressing pair with $n$ for infinitely many $j\geq k$.
	 \label{item:tail-gt-lem-infinite-path}
  \end{enumerate}
 \end{lemma}

 \begin{proof}
  Let $\mathsequence{\sigma_{i}}{0\leq i}$ be an infinite path in 
  $\mathproofgraph{\mleft(\mathdertree{D}_{\mathwinningstrategy{S}_{\Prover}}, \mathcompanion{C}_{\mathwinningstrategy{S}_{\Prover}}\mright)}$.
  
  \noindent \cref{item:existence-lem-infinite-path} Obvious.

  \noindent \cref{item:tail-track-lem-infinite-path}
  We show that there exist natural numbers $k$, $n$ such that
  $\mathtuple{\varphi_{i}, n}\in\mathTr{\mathof{\mathwinningstrategy{S}_{\Prover}}{\sigma_{i}}}$
  for all $i\geq k$ and
  $\Prover$ never changes the track $0$ in $\mathsequence{\sigma_{i}}{k\leq i}$.

  Let $S=\mathsetintension{\mathsequence{\sigma'_{i}}{0\leq i < m} \in\mathCycle{\mathwinningstrategy{S}_{\Prover}}}{ \sigma'_{0}, \dots, \sigma'_{m} \text{ occurs in } \mathsequence{\sigma_{i}}{0\leq i} }$.
  By \cref{item:existence-lem-infinite-path}, $S$ is not empty.

  The proof is by induction on $\mathcardinality{S}$.

  Suppose $\mathcardinality{S}=1$.
  Let $S=\mathsetextension{\mathsequence{\sigma'_{i}}{0\leq i < m}}$.
  Then, $\sigma_{k}=\sigma'_{0}$ for some $k$.
  We show that $\mathsequence{\sigma_{i}}{k\leq i}$ is the infinite sequence
  $(\sigma'_{0}, \dots, \sigma'_{m-1})^{\omega}$. 
  Assume, for contradiction, that
  there exists $j> 0$ such that $\sigma_{k+j}$ does not occur in 
  $\mathsequence{\sigma'_{i}}{0\leq i < m}$ and
  $\sigma_{k+mj'+r}=\sigma'_{r}$ for all $mj'+r<j$, where $j'\geq 0$ and $0\leq r<m$.
  Since $\mathsequence{\sigma_{i}}{0\leq i}$ is an infinite path,
  $\mathsequence{\sigma_{i}}{k+j\leq i}$ is an infinite path.
  If no elements in $\mathCycle{\mathwinningstrategy{S}_{\Prover}}$ occur in $\mathsequence{\sigma_{i}}{k+j\leq i}$,
  it contradicts \cref{item:existence-lem-infinite-path}.
  Because $S=\mathsetextension{\mathsequence{\sigma'_{i}}{0\leq i < m}}$ holds,
  $\mathsequence{\sigma'_{i}}{0\leq i < m}$ must occur in 
  $\mathsequence{\sigma_{i}}{k+j\leq i}$.
  Hence, $\sigma_{k+j+k'}=\sigma'_{0}$ for some $k'$.
  However, since $\sigma_{k+j}$ is a descendant of $\sigma'_{0}$,
  another element of $\mathCycle{\mathwinningstrategy{S}_{\Prover}}$ occurs in $\mathsequence{\sigma_{i}}{k+j\leq i}$,
  which contradicts $S=\mathsetextension{\mathsequence{\sigma'_{i}}{0\leq i < m}}$.
  Hence, $\mathsequence{\sigma_{i}}{k\leq i}$ must be the infinite sequence
  $(\sigma'_{0}, \dots, \sigma'_{n-1})^{\omega}$. 
  Then, there exists a natural number $n$ such that
  the track $n$ is continuous in $\mathsequence{\sigma'_{i}}{0\leq i < m}$ and
  $\Prover$ never changes the track $0$ in $\mathsequence{\sigma_{i}}{k\leq i}$.
  Thus, there exist natural numbers $n$, $k$ such that
  $\mathtuple{\varphi_{i}, n}\in\mathTr{\mathof{\mathwinningstrategy{S}_{\Prover}}{\sigma_{i}}}$
  for all $i\geq k$ and
  $\Prover$ never changes the track $0$ in $\mathsequence{\sigma_{i}}{k\leq i}$.

  Suppose $\mathcardinality{S}>1$.
  If there exists $k$ such that
  $\mathtuple{\mathof{\mathwinningstrategy{S}_{\Prover}}{\sigma_{j}}, \mathof{\mathwinningstrategy{S}_{\Prover}}{\sigma_{j+1}}}$
  is not an instance of \rulename{G-K} for all $j\geq k$,
  then we have the statement immediately.

  Assume that 
  $\mathtuple{\mathof{\mathwinningstrategy{S}_{\Prover}}{\sigma_{j}}, \mathof{\mathwinningstrategy{S}_{\Prover}}{\sigma_{j+1}}}$
  is an instance of \rulename{G-K} for infinitely many $j$.
  Then, there exists $k'$ such that
  $\mathtuple{\varphi_{i}, 0}\in\mathTr{\mathof{\mathwinningstrategy{S}_{\Prover}}{\sigma_{i}}}$
  for all $i\geq k'$.
  Without loss of generality, we can assume 
  $\sigma_{k'}=\sigma'_{0}, \sigma_{k'+1}=\sigma'_{1}, \dots, \sigma_{k'+m}=\sigma'_{m}$ 
  for some $\mathsequence{\sigma'_{i}}{0\leq i < m}\in S$.
  If $\Prover$ never changes the track $0$ in $\mathsequence{\sigma_{i}}{k'\leq i}$,
  then we have the statement.
  Assume that $\Prover$ changes the track $0$ with
  $\mathtuple{\mathof{\mathwinningstrategy{S}_{\Prover}}{\sigma_{j}}, \mathof{\mathwinningstrategy{S}_{\Prover}}{\sigma_{j+1}}}$ for $j>k'$.
  Since $\mathwinningstrategy{S}_{\Prover}$ is a winning strategy of $\Prover$,
  $\Prover$ never changes the track $0$ in all $\mathsequence{\sigma'_{i}}{0\leq i < m}\in S$.
  Hence, we see that $\sigma_{j}, \sigma_{j+1}$ does not occur in all
  $\mathsequence{\sigma'_{i}}{0\leq i < m}\in S$.
  Therefore, 
  for each bud $\hat{\sigma}$, if $\hat{\sigma}$ is a descendant of $\sigma_{j+1}$,
  the corresponding companion of $\hat{\sigma}$ is a descendant of $\sigma_{j+1}$.
  Therefore, $\sigma_{i}$ is a descendant of $\sigma_{j+1}$ for all $i>j+1$.
  Let $k''\geq k'$ such that 
  $\sigma_{k''}, \dots, \sigma_{k''+m'}\in S$ and ${k''+m'+1} < j$ hold,
  and any element in $S$ does not occur between $\sigma_{k''+m'+1}$ and $\sigma_{j}$.
  Then, $\sigma_{j}$ is an ancestor of $\sigma_{k''+m'+1}(=\sigma_{k''})$.
  Because $\sigma_{i}$ is a descendant of $\sigma_{j+1}$ for all $i>j+1$, 
  we see that $\sigma_{k''}, \dots, \sigma_{k''+m'}$ does not 
  occur in $\mathsequence{\sigma_{i}}{j< i}$. From this, 
   \[
    \mathsetintension{\mathsequence{\sigma'_{i}}{0\leq i < m} \in\mathCycle{\mathwinningstrategy{S}_{\Prover}}}{ \sigma'_{0}, \dots, \sigma'_{n-1} \text{ occurs in } \mathsequence{\sigma_{i}}{j< i} }\subsetneq S.
   \]
  By the induction hypothesis, 
  there exist natural numbers $k\geq j$, $n$ such that
  $\mathtuple{\varphi_{i}, n}\in\mathTr{\mathof{\mathwinningstrategy{S}_{\Prover}}{\sigma_{i}}}$
  for all $i\geq k$ and
  $\Prover$ never changes the track $0$ in $\mathsequence{\sigma_{i}}{k\leq i}$.

  \noindent \cref{item:tail-progress-lem-infinite-path}
  By \cref{item:tail-track-lem-infinite-path},
  there exist natural numbers $k$, $n$ such that
  $\mathtuple{\varphi_{i}, n}\in\mathTr{\mathof{\mathwinningstrategy{S}_{\Prover}}{\sigma_{i}}}$
  for all $i\geq k$,
  $\Prover$ never changes the track $0$ in $\mathsequence{\sigma_{i}}{k\leq i}$.

  Assume that
  $\mathtuple{\mathof{\mathwinningstrategy{S}_{\Prover}}{\sigma_{j}}, \mathof{\mathwinningstrategy{S}_{\Prover}}{\sigma_{j+1}}}$
  is an instance of \rulename{G-K} for infinitely many $j\geq k$.
  Then, $n=0$. 
  We also see that each instance of \rulename{G-K}
  $\mathtuple{\mathof{\mathwinningstrategy{S}_{\Prover}}{\sigma_{j}}, \mathof{\mathwinningstrategy{S}_{\Prover}}{\sigma_{j+1}}}$ 
  is a transforming pair with $0$. Therefore, we have the statement.

  Assume that there exists $k'\geq k$ such that
  $\mathtuple{\mathof{\mathwinningstrategy{S}_{\Prover}}{\sigma_{j}}, \mathof{\mathwinningstrategy{S}_{\Prover}}{\sigma_{j+1}}}$
  is not an instance of \rulename{G-K} for all $j\geq k'$.
  Then, for all $k''\geq k'$ and each $j\geq k''$, if $\mathtuple{\varphi_{k''}, l}\in\mathTr{\mathof{\mathwinningstrategy{S}_{\Prover}}{\sigma_{k''}}}$,
  then there exists a formula $\varphi_{j}$ such that
  $\mathtuple{\varphi_{j}, l}\in\mathTr{\mathof{\mathwinningstrategy{S}_{\Prover}}{\sigma_{j}}}$.
  By \cref{item:existence-lem-infinite-path},
  there exists 
  $\mathsequence{\sigma'_{i}}{0\leq i < m}\in\mathCycle{\mathwinningstrategy{S}_{\Prover}}$ 
  such that
  $\sigma'_{0}, \dots, \sigma'_{m}, \sigma'_{0}$ occurs in $\mathsequence{\sigma_{i}}{k'\leq i}$
  infinitely many times.
  Then, a transforming pair with $n$ occurs between $\sigma'_{0}$ and $\sigma'_{m}$.
  Therefore, for infinitely many $j\geq k$,
  $\mathtuple{\mathof{\mathwinningstrategy{S}_{\Prover}}{\sigma_{j}}, \mathof{\mathwinningstrategy{S}_{\Prover}}{\sigma_{j+1}}}$ 
  is a transforming pair with $n$.
  
  \noindent \cref{item:tail-gt-lem-infinite-path}
  By \cref{item:tail-progress-lem-infinite-path},
  there exist natural numbers $k$, $n$ such that
  $\mathtuple{\varphi_{i}, n}\in\mathTr{\mathof{\mathwinningstrategy{S}_{\Prover}}{\sigma_{i}}}$
  for all $i\geq k$,
  $\Prover$ never changes the track $0$ in $\mathsequence{\sigma_{i}}{k\leq i}$, and 
  $\mathtuple{\mathof{\mathwinningstrategy{S}_{\Prover}}{\sigma_{j}}, \mathof{\mathwinningstrategy{S}_{\Prover}}{\sigma_{j+1}}}$ is
  a transforming pair with $n$ for infinitely many $j\geq k$.
  Assume, for contradiction, that there exists $k'$ such that
  no progressing pairs with $n$ occur in $\mathsequence{\sigma_{i}}{k'\leq i}$.
  Let $a_{i}=\mathlength{\varphi_{i}}$ for $i\geq k'$, 
  where $\mathtuple{\varphi_{i}, n}\in \mathTr{\mathof{\mathwinningstrategy{S}_{\Prover}}{\sigma_{i}}}$.
  Since $\mathtuple{\mathof{\mathwinningstrategy{S}_{\Prover}}{\sigma_{j}}, \mathof{\mathwinningstrategy{S}_{\Prover}}{\sigma_{j+1}}}$
  is a transforming pair with $n$ for infinitely many $j\geq k$ and
  no progressing pair with $n$ occur in $\mathsequence{\sigma_{i}}{k'\leq i}$,
  we have $a_{i}\geq a_{i+1}$ for all $i\geq k'$ and
  $a_{j}>a_{j+1}$ for infinitely many $j\geq k'$.
  Hence, $\mathsequence{a_{i}}{k'\leq i}$ is an infinite descending sequence
  of natural numbers.
  This is a contradiction.
 \end{proof}

 \begin{lemma}
  \label[lemma]{lemma:proof-from-winning-strategy}
  Let $\mathwinningstrategy{S}_{\Prover}$ be a winning strategy of $\Prover$ in 
  $\GameName{\Gamma\fCenter\Delta}$.
  The \CGPDL\ pre-proof constructed from $\mathwinningstrategy{S}_{\Prover}$
  is a \CGPDL\ proof.  
 \end{lemma}

 \begin{proof}
  Let $\mleft(\mathdertree{D}_{\mathwinningstrategy{S}_{\Prover}}, \mathcompanion{C}_{\mathwinningstrategy{S}_{\Prover}}\mright)$ 
  be the \CGPDL\ pre-proof constructed from $\mathwinningstrategy{S}_{\Prover}$ and
  $\mathproofgraph{\mleft(\mathdertree{D}_{\mathwinningstrategy{S}_{\Prover}}, \mathcompanion{C}_{\mathwinningstrategy{S}_{\Prover}}\mright)}$ be 
  the derivation graph of $\mleft(\mathdertree{D}_{\mathwinningstrategy{S}_{\Prover}},  \mathcompanion{C}_{\mathwinningstrategy{S}_{\Prover}}\mright)$.
  Since $\mleft(\mathdertree{D}_{\mathwinningstrategy{S}_{\Prover}}, \mathcompanion{C}_{\mathwinningstrategy{S}_{\Prover}}\mright)$ 
  is the \CGPDL\ pre-proof, it suffices to show that 
  $\mathproofgraph{\mleft(\mathdertree{D}_{\mathwinningstrategy{S}_{\Prover}}, \mathcompanion{C}_{\mathwinningstrategy{S}_{\Prover}}\mright)}$ satisfies
  the global trace condition.

  Fix $\mathsequence{\sigma_{i}}{0\leq i}$ be an infinite path in
  $\mathproofgraph{\mleft(\mathdertree{D}_{\mathwinningstrategy{S}_{\Prover}}, \mathcompanion{C}_{\mathwinningstrategy{S}_{\Prover}}\mright)}$.
  By \cref{lem:infinite-path} \cref{item:tail-gt-lem-infinite-path},
  there exist natural numbers $k$, $n$ and a sequence of formulae 
  $\mathsequence{\varphi_{i}}{i\geq k}$ such that
  $\mathtuple{\varphi_{i}, n}\in\mathTr{\mathof{\mathwinningstrategy{S}_{\Prover}}{\sigma_{i}}}$
  for $\mathsequence{\sigma_{i}}{i\geq k}$,
  $\Prover$ never changes the track $0$ in $\mathsequence{\sigma_{i}}{k\leq i}$,
  and $\mathtuple{\mathof{\mathwinningstrategy{S}_{\Prover}}{\sigma_{j}}, \mathof{\mathwinningstrategy{S}_{\Prover}}{\sigma_{j+1}}}$ is
  a progressing pair with $n$ for infinitely many $j\geq k$.
  Then, $\mathsequence{\varphi_{i}}{i\geq k}$ is a trace following
  $\mathsequence{\sigma_{i}}{i\geq k}$ and
  $\mathsequence{\varphi_{i}}{i\geq k}$ has infinitely many progress points.
  Hence,
  there exists an infinitely progressing trace following a tail of the infinite path.
  Therefore,
  $\mathproofgraph{\mleft(\mathdertree{D}_{\mathwinningstrategy{S}_{\Prover}}, \mathcompanion{C}_{\mathwinningstrategy{S}_{\Prover}}\mright)}$ satisfies
  the global trace condition.
 \end{proof}

 We are now ready to prove \cref{lemma:Prover-win}.

 \begin{proof}[Proof of \cref{lemma:Prover-win}.]
  Assume that 
  $\Prover$ has a winning strategy in the proof search game of $\Gamma\fCenter\Delta$.
  By \cref{lemma:proof-from-winning-strategy},
  there exists a cut-free \CGPDL\ proof of $\Gamma\fCenter\Delta$.
 \end{proof}

 \subsection{The case \Refuter\ wins}
 \label{subsec:Refuter-wins}
 This section proves \cref{lemma:Refuter-win}.

 Note that, 
 for each node $\sigma$ in a winning strategy $\mathwinningstrategy{S}_{\Refuter}$ 
 of $\Refuter$ in a proof search game, if $\Pi\fCenter\Sigma$ is the sequent of
 $\mathof{\mathwinningstrategy{S}_{\Refuter}}{\sigma}$,
 then $\Pi\cap\Sigma=\emptyset$ and $\bot\notin\Pi$.
 We also note that each leaf $\sigma$ of
 a winning strategy $\mathwinningstrategy{S}_{\Refuter}$ 
 of $\Refuter$ in a proof search game is a bud and satisfies 
 the following condition:
 \begin{quote}
  For any natural number $n$,
  no transforming pairs with $n$ occur in a path from the corresponding companion 
  of $\sigma$ to $\sigma$ 
  if $\Prover$ never changes the track $0$ and the track $n$ is continuous in the path.
 \end{quote}

 The \emph{graph} of a winning strategy $\mathwinningstrategy{S}_{\Refuter}$ of 
 $\Refuter$, 
 written as $\mathstgraph{\mathwinningstrategy{S}_{\Refuter}}$,
 is defined as the directed graph obtained from $\mathwinningstrategy{S}_{\Refuter}$ 
 by identifying each bud in $\mathwinningstrategy{S}_{\Refuter}$
 with its corresponding companion.
 We note that, for every node in $\mathstgraph{\mathwinningstrategy{S}_{\Refuter}}$,
 there exists its child.

For an infinite path $\mathsequence{\sigma_{i}}{0\leq i}$ in
$\mathstgraph{\mathwinningstrategy{S}_{\Refuter}}$,
we say that the track $n$ is continuous in the path if 
$\mathtuple{\varphi_{i}, n}\in\mathTr{\mathof{\mathwinningstrategy{S}_{\Refuter}}{\sigma_{i}}}$
for all $i\geq 0$.

 \begin{lemma}
  \label[lemma]{lemma:refuter-infinite}
  Assume that $\Refuter$ has a winning strategy $\mathwinningstrategy{S}_{\Refuter}$ 
  in $\GameName{\Gamma\fCenter\Delta}$. 

  For an infinite path $\mathsequence{\sigma_{i}}{0\leq i}$ in 
  $\mathstgraph{\mathwinningstrategy{S}_{\Refuter}}$,
  if the track $n$ is continuous in the path for some $n$ and
  $\Prover$ never changes the track $0$ in the path,
  then there exists $k\geq 0$ such that
  no transforming pairs with $m$ occur in $\mathsequence{\sigma_{i}}{k\leq i}$
  for any natural number $m$.
 \end{lemma}

 \begin{proof}
  Let $\mathsequence{\sigma_{i}}{0\leq i}$ be an infinite path in 
  $\mathstgraph{\mathwinningstrategy{S}_{\Refuter}}$.
  Assume that the track $n$ is continuous in the path for some $n$ and
  $\Prover$ never changes the track $0$.
  We show that there exists $k\geq 0$ such that
  no transforming pairs with $m$ occur in $\mathsequence{\sigma_{i}}{k\leq i}$
  for any natural number $m$.

  Let $S=\mathsetintension{\mathsequence{\sigma'_{i}}{0\leq i < l} \in\mathCycle{\mathwinningstrategy{S}_{\Refuter}}}{ \sigma'_{0}, \dots, \sigma'_{l} \text{ occurs in } \mathsequence{\sigma_{i}}{0\leq i} }$.
  Since the track $n$ is continuous for some $n$ and $\Prover$ never changes the track $0$ 
  in $\mathsequence{\sigma_{i}}{0\leq i}$, 
  $\Prover$ never changes the track $0$ and the track $n$ is continuous in any $\mathsequence{\sigma'_{i}}{0\leq i < l}\in S$.
  Hence, no transforming pairs with $m$ occur in any $\mathsequence{\sigma'_{i}}{0\leq i < l}\in S$
  for every natural number $m$.

  The proof is by induction on $\mathcardinality{S}$.

  Suppose $\mathcardinality{S}=1$.
  Let $S=\mathsetextension{\mathsequence{\sigma'_{i}}{0\leq i < l}}$.
  Then, $\sigma_{k}=\sigma'_{0}$ for some $k$.
  We show that $\mathsequence{\sigma_{i}}{k\leq i}$ is the infinite sequence
  $(\sigma'_{0}, \dots, \sigma'_{l-1})^{\omega}$. 
  Assume, for contradiction, that
  there exists $j> 0$ such that $\sigma_{k+j}$ does not occur in 
  $\mathsequence{\sigma'_{i}}{0\leq i < l}$ and
  $\sigma_{k+lj'+r}=\sigma'_{r}$ for all $lj'+r<j$, where $j'\geq 0$ and $0\leq r<l$.
  Since $\mathsequence{\sigma_{i}}{0\leq i}$ is an infinite path,
  $\mathsequence{\sigma_{i}}{k+j\leq i}$ is an infinite path.
  It is impossible that no elements in $\mathCycle{\mathwinningstrategy{S}_{\Refuter}}$ 
  occur in $\mathsequence{\sigma_{i}}{k+j\leq i}$.
  Then, $\mathsequence{\sigma'_{i}}{0\leq i < l}$ must occur in 
  $\mathsequence{\sigma_{i}}{k+j\leq i}$ because 
  $S=\mathsetextension{\mathsequence{\sigma'_{i}}{0\leq i < l}}$ holds.
  Hence, $\sigma_{k+j+k'}=\sigma'_{0}$ for some $k'$.
  However, since $\sigma_{k+j}$ is a descendant of $\sigma'_{0}$,
  another element of $\mathCycle{\mathwinningstrategy{S}_{\Refuter}}$ occurs in 
  $\mathsequence{\sigma_{i}}{k+j\leq i}$,
  which contradicts $S=\mathsetextension{\mathsequence{\sigma'_{i}}{0\leq i < l}}$.
  Therefore, $\sigma_{k+lj'+r}=\sigma'_{r}$ for all $j'\geq 0$ and $0\leq r<l$.
  Hence, $\mathsequence{\sigma_{i}}{k\leq i}$ must be the infinite sequence
  $(\sigma'_{0}, \dots, \sigma'_{l-1})^{\omega}$. 
  Since $\Prover$ never changes the track $0$ and the track $n$ is continuous in any $\mathsequence{\sigma'_{i}}{0\leq i < l}\in S$,
  no transforming pairs with $m$ occur in $\mathsequence{\sigma'_{i}}{0\leq i < l}$ for any $m$.
  Therefore, no transforming pairs with $m$ occur in $\mathsequence{\sigma_{i}}{k\leq i}$
  for any natural number $m$.

  Suppose $\mathcardinality{S}>1$.
  Assume that $\sigma_{k'}=\sigma'_{0}$, $\dots$, $\sigma_{k'+l-1}=\sigma'_{l-1}$, $\sigma_{k'+l}=\sigma'_{0}$ 
  for some $k'$ and $\mathsequence{\sigma'_{i}}{0\leq i < l}\in S$.
  If no transforming pairs with $m$ occur in $\mathsequence{\sigma_{i}}{k'\leq i}$
  for any natural number $m$, we have the statement.

  Assume that a transforming pair with $m$ occurs in $\mathsequence{\sigma_{i}}{k'\leq i}$.
  Let $\mathtuple{\mathof{\mathwinningstrategy{S}_{\Refuter}}{\sigma_{j}}, \mathof{\mathwinningstrategy{S}_{\Refuter}}{\sigma_{j+1}}}$ 
  be a transforming pair with $m$ for $j\geq k'$.
  Since no transforming pairs with $m$ occur in $\mathsequence{\sigma'_{i}}{0\leq i < l}$
  for any $m$, we have $j\geq k'+l$.
  Let $k''$ be a natural number such that $k'\leq k'' < j$ and
  $\sigma_{k''}=\sigma''_{0}$, $\sigma_{k''+1}=\sigma''_{1}$, $\dots$,  $\sigma_{k''+l'-1}=\sigma''_{l'-1}$, $\sigma_{k''+l'}=\sigma''_{0}$ hold
  for $\mathsequence{\sigma''_{i}}{0\leq i < l'}\in S$, and
  any element of $S$ does not occur between $\sigma_{k''+l'}$ and $\sigma_{j}$.
  Then, $\sigma_{j}$ is a descendant of $\sigma_{k''}$.
  Since no transforming pairs with $m$ occur in any $\mathsequence{\sigma'''_{i}}{0\leq i < l''}\in S$
  for every natural number $m$, we see that
  $\sigma_{j}, \sigma_{j+1}$ does not occur in any $\mathsequence{\sigma'''_{i}}{0\leq i < l''}\in S$.
  Hence, $\sigma_{i}$ is a descendant of $\sigma_{j+1}$ for any $i>j$.
  Therefore, $\mathsequence{\sigma''_{i}}{0\leq i < l'}$ does not occur in $\mathsequence{\sigma_{i}}{j< i}$.
  From this, 
   \[
    \mathsetintension{\mathsequence{\sigma'_{i}}{0\leq i < l} \in\mathCycle{\mathwinningstrategy{S}_{\Refuter}}}{ \sigma'_{0}, \dots, \sigma'_{l-1} \text{ occurs in } \mathsequence{\sigma_{i}}{j< i} }\subsetneq S.
   \]
  By the induction hypothesis, we have the statement.
 \end{proof}

 \begin{definition}[World in a winning strategy]
  \label[definition]{def:world-tree}
  Assume that $\Refuter$ has a winning strategy $\mathwinningstrategy{S}_{\Refuter}$ 
  in $\GameName{\Gamma\fCenter\Delta}$. 
 
  We define a \emph{world} in $\mathstgraph{\mathwinningstrategy{S}_{\Refuter}}$
  as an infinite path $\mathsequence{\sigma_{i}}{0\leq i}$ in 
  $\mathstgraph{\mathwinningstrategy{S}_{\Refuter}}$ satisfying
  all the following conditions:
  \begin{enumerate}
   \item $\sigma_{0}$ is the root of $\mathwinningstrategy{S}_{\Refuter}$ or 
	 $\mathtuple{\mathof{\mathwinningstrategy{S}_{\Refuter}}{\sigma'}, \mathof{\mathwinningstrategy{S}_{\Refuter}}{\sigma_{0}}}$ 
	 is an instance of \rulename{G-K}, where $\sigma'$ is the parent of $\sigma_{0}$.
   \item $\mathtuple{\mathof{\mathwinningstrategy{S}_{\Refuter}}{\sigma_{j}}, \mathof{\mathwinningstrategy{S}_{\Refuter}}{\sigma_{j+1}}}$ 
	 is not an instance of \rulename{G-K} for any $j\geq 0$.
  \end{enumerate}

  We write $\mathworldfirst{\mathworld{W}}$ for the first node in a world $\mathworld{W}$.
  The notation $\sigma\in\mathworld{W}$ means that the node $\sigma$ occurs in $\mathworld{W}$.
 \end{definition}

 For a position $C$,
 we write $\mathAnt{C}$ and $\mathCon{C}$ for the antecedent and the consequent
 of $\mathSeq{C}$, respectively.
 We also write $\mathUS{C}$ for the upcoming schedule of a position $C$.

 We call a node $\sigma$ in a winning strategy of $\Refuter$ a \emph{joint} if
 the position of $\sigma$ is of the form $\mathtuple{\Pi\fCenter\Sigma, T, S}$,
 where $(\Pi\cup\Sigma)\cap S=\emptyset$.
 Each branching point in a winning strategy of $\Refuter$ must be a joint.

 \begin{lemma}
  \label[lemma]{lemma:world-core}
  Assume that $\Refuter$ has a winning strategy $\mathwinningstrategy{S}_{\Refuter}$ 
  in $\GameName{\Gamma\fCenter\Delta}$. 
  Let $\mathworld{W}$ be a world in $\mathstgraph{\mathwinningstrategy{S}_{\Refuter}}$ and
  $\mathworld{W}=\mathsequence{\sigma_{i}}{0\leq i}$.

  \begin{enumerate}
   \item Joints occur in $\mathworld{W}$ infinitely many times.
	 \label{item:joint-lemma-world-core}
   \item If $i\leq j$,
	 then $\mathAnt{\mathof{\mathwinningstrategy{S}_{\Refuter}}{\sigma_{i}}}\subseteq\mathAnt{\mathof{\mathwinningstrategy{S}_{\Refuter}}{\sigma_{j}}}$
	 and
	 $\mathCon{\mathof{\mathwinningstrategy{S}_{\Refuter}}{\sigma_{i}}}\subseteq\mathCon{\mathof{\mathwinningstrategy{S}_{\Refuter}}{\sigma_{j}}}$.
	 \label{item:monotonicity-lemma-world-core}
   \item There exists $k\geq 0$ such that the following conditions:
	 \begin{itemize}
	  \item $\mathAnt{\mathof{\mathwinningstrategy{S}_{\Refuter}}{\sigma_{i}}}=\mathAnt{\mathof{\mathwinningstrategy{S}_{\Refuter}}{\sigma_{j}}}$
		and
	 $\mathCon{\mathof{\mathwinningstrategy{S}_{\Refuter}}{\sigma_{i}}}=\mathCon{\mathof{\mathwinningstrategy{S}_{\Refuter}}{\sigma_{j}}}$
		for all $i, j\geq k$.
	  \item No transforming pairs with $m$ occur in $\mathsequence{\sigma_{i}}{k\leq i}$
		for any natural number $m$.
	 \end{itemize}
	 \label{item:core-lemma-world-core}
  \end{enumerate}
 \end{lemma}

 \begin{proof}
  Assume that $\Refuter$ has a winning strategy $\mathwinningstrategy{S}_{\Refuter}$ 
  in $\GameName{\Gamma\fCenter\Delta}$.
  Let $\mathworld{W}$ be a world in $\mathstgraph{\mathwinningstrategy{S}_{\Refuter}}$ and
  $\mathworld{W}=\mathsequence{\sigma_{i}}{0\leq i}$.

  \noindent \cref{item:joint-lemma-world-core}
  Assume, for contradiction, no joints occur in $\mathsequence{\sigma_{i}}{k\leq i}$
  for some $k$.
  Then, neither instance of \rulename{G-Retry} nor \rulename{G-K} occurs in 
  $\mathsequence{\sigma_{i}}{k\leq i}$.
  Then, $\mathsequence{\mathcardinality{\mathUS{\mathof{\mathwinningstrategy{S}_{\Refuter}}{\sigma_{i}}}}}{k\leq i}$
  is an infinite descending sequence of natural numbers. This is a contradiction.

  \noindent \cref{item:monotonicity-lemma-world-core}
  Because no instances of \rulename{G-K} occur in $\mathsequence{\sigma_{i}}{0\leq i}$.

  \noindent \cref{item:core-lemma-world-core}
  Because $\mathAnt{\mathof{\mathwinningstrategy{S}_{\Refuter}}{\sigma_{i}}}\subseteq\mathFLcl{\Gamma\cup\Delta}$
  and $\mathCon{\mathof{\mathwinningstrategy{S}_{\Refuter}}{\sigma_{i}}}\subseteq\mathFLcl{\Gamma\cup\Delta}$
  for all $i\geq 0$,
  \cref{item:monotonicity-lemma-world-core} implies that there exists $k'\geq 0$ such that
  $\mathAnt{\mathof{\mathwinningstrategy{S}_{\Refuter}}{\sigma_{i}}}=\mathAnt{\mathof{\mathwinningstrategy{S}_{\Refuter}}{\sigma_{j}}}$
  and
  $\mathCon{\mathof{\mathwinningstrategy{S}_{\Refuter}}{\sigma_{i}}}=\mathCon{\mathof{\mathwinningstrategy{S}_{\Refuter}}{\sigma_{j}}}$
  for all $i, j\geq k'$.
  By \cref{lemma:refuter-infinite} there exists $k''\geq 0$ such that
  no transforming pairs with $m$ occur in $\mathsequence{\sigma_{i}}{k''\leq i}$
  for any natural number $m$.
  Let $k$ be the maximum of $\mathsetextension{k', k''}$.
  Then, $k$ is what we want.
 \end{proof}

 \begin{definition}[World core]
  Assume that $\Refuter$ has a winning strategy $\mathwinningstrategy{S}_{\Refuter}$ 
  in $\GameName{\Gamma\fCenter\Delta}$. 
  Let $\mathworld{W}$ be a world in $\mathstgraph{\mathwinningstrategy{S}_{\Refuter}}$ and
  $\mathworld{W}=\mathsequence{\sigma_{i}}{0\leq i}$.

  We define a \emph{core} of $\mathworld{W}$ as a joint $\sigma_{k}$ in $\mathworld{W}$
  such that the following conditions hold:
  \begin{itemize}
   \item $\mathAnt{\mathof{\mathwinningstrategy{S}_{\Refuter}}{\sigma_{i}}}=\mathAnt{\mathof{\mathwinningstrategy{S}_{\Refuter}}{\sigma_{j}}}$
	 and
	 $\mathCon{\mathof{\mathwinningstrategy{S}_{\Refuter}}{\sigma_{i}}}=\mathCon{\mathof{\mathwinningstrategy{S}_{\Refuter}}{\sigma_{j}}}$
	 for all $i, j\geq k$.
   \item No transforming pairs with $m$ occur in $\mathsequence{\sigma_{i}}{k\leq i}$
	 for any natural number $m$.
  \end{itemize}
 \end{definition}

 \cref{lemma:world-core} guarantees the existence of cores of each world.
 For a world $\mathworld{W}$, let $\mathworldcore{\mathworld{W}}$ denotes
 the first core of $\mathworld{W}$.
 We note that 
 $\mathAnt{\mathof{\mathwinningstrategy{S}_{\Refuter}}{\sigma}}\subseteq\mathAnt{\mathof{\mathwinningstrategy{S}_{\Refuter}}{\mathworldcore{\mathworld{W}}}}$
 and
 $\mathCon{\mathof{\mathwinningstrategy{S}_{\Refuter}}{\sigma}}\subseteq\mathCon{\mathof{\mathwinningstrategy{S}_{\Refuter}}{\mathworldcore{\mathworld{W}}}}$
 for all $\sigma\in\mathworld{W}$

 The notation $\sigma\mathancestorofcore\mathworld{W}$ means that 
 $\sigma$ occurs before $\mathworldcore{\mathworld{W}}$ in a world $\mathworld{W}$.

 \begin{lemma}
  \label[lemma]{lemma:core-property}
  Assume that $\Refuter$ has a winning strategy $\mathwinningstrategy{S}_{\Refuter}$ 
  in $\GameName{\Gamma\fCenter\Delta}$. 
  Let $\mathworld{W}$ be a world in $\mathstgraph{\mathwinningstrategy{S}_{\Refuter}}$ and
  $\mathworld{W}=\mathsequence{\sigma_{i}}{0\leq i}$. 

  \begin{enumerate}
   \item If $\mathtuple{\varphi, n}\in\mathTr{\mathof{\mathwinningstrategy{S}_{\Refuter}}{\sigma_{i}}}$
	 for some $n$ and $\varphi\in\mathreducible{\mathFLcl{\Gamma\cup\Delta}}$,
	 then $\sigma_{i}\mathancestorofcore\mathworld{W}$ 
	 \label{item:reducible-formula-lemma-core-property}
   \item If $\varphi\in\mathCon{\mathof{\mathwinningstrategy{S}_{\Refuter}}{\sigma_{i}}}$,
	 then there exists $j\leq i$ such that
	 $\mathtuple{\varphi, n}\in\mathTr{\mathof{\mathwinningstrategy{S}_{\Refuter}}{\sigma_{j}}}$
	 for some $n$.
	 \label{item:lemma-core-and-con-lemma-core-property}
   \item If $\varphi\in\mathCon{\mathof{\mathwinningstrategy{S}_{\Refuter}}{\mathworldcore{\mathworld{W}}}}$,
	 then there exists $\sigma_{j}\mathancestorofcore\mathworld{W}$ such that
	 $\mathtuple{\varphi, n}\in\mathTr{\mathof{\mathwinningstrategy{S}_{\Refuter}}{\sigma_{j}}}$
	 for some $n$.
	 \label{item:core-and-con-lemma-core-property}
  \end{enumerate}
 \end{lemma}

 \begin{proof} \phantom{We show each statements.}
  
  \noindent \cref{item:reducible-formula-lemma-core-property}
  Because no transforming pairs occur after $\mathworldcore{\mathworld{W}}$.

  \noindent \cref{item:lemma-core-and-con-lemma-core-property}
  By induction on $i$.

  \noindent \cref{item:core-and-con-lemma-core-property}
  By \cref{item:lemma-core-and-con-lemma-core-property}.
 \end{proof}

 \begin{definition}[The counter model constructed from a winning strategy]
  \label[definition]{def:counter-model-winning-strategy}
  Assume that $\Refuter$ has a winning strategy $\mathwinningstrategy{S}_{\Refuter}$ 
  in $\GameName{\Gamma\fCenter\Delta}$. 
 
  We define the \emph{counter model constructed from $\mathwinningstrategy{S}_{\Refuter}$},
  written as
  $M_{\mathwinningstrategy{S}_{\Refuter}}=\mathtuple{W_{\mathwinningstrategy{S}_{\Refuter}}, \mathsequence{{\mathreduction{\alpha}}_{\mathwinningstrategy{S}_{\Refuter}}}{\alpha\in\maththesetofatomicprog}, V_{\mathwinningstrategy{S}_{\Refuter}}}$, 
  as follows:
  \begin{enumerate}
   \item $W_{\mathwinningstrategy{S}_{\Refuter}}=\mathsetintension{\mathworld{W}}{\mathworld{W} \text{ is a world in } \mathstgraph{\mathwinningstrategy{S}_{\Refuter}} }$.
   \item $\mathworld{W}_{0}{\mathreduction{\alpha}}_{\mathwinningstrategy{S}_{\Refuter}}\mathworld{W}_{1}$ 
	 if and only if 
	 $\mathtuple{\mathof{\mathwinningstrategy{S}_{\Refuter}}{\mathworldcore{\mathworld{W}_{0}}}, \mathof{\mathwinningstrategy{S}_{\Refuter}}{\mathworldfirst{\mathworld{W}_{1}}}}$
	 is an instance of \rulename{G-K}, where
	 $\mathof{\mathwinningstrategy{S}_{\Refuter}}{\mathworldcore{\mathworld{W}_{0}}}=\mathtuple{\Pi\fCenter\Sigma; T; S}$,
	 $\mathof{\mathwinningstrategy{S}_{\Refuter}}{\mathworldfirst{\mathworld{W}_{1}}}=\mathtuple{\Pi' \fCenter \varphi; \mathsetextension{\mathtuple{\varphi, 0}}; \mathreducible{\mathFLcl{\Gamma\cup\Delta}}}$,
	 $\Pi'=\mathsetintension{\psi}{\mathdynbox{\alpha}\psi\in\Pi}$,
	 \label{item:reduction-def-counter-model-winning-strategy}
	 and $\mathtuple{\mathdynbox{\alpha}\varphi, n}\in T$.
   \item $p\in\mathof{V_{\mathwinningstrategy{S}_{\Refuter}}}{\mathworld{W}}$
	 if and only if 
	 $p\in \mathAnt{\mathworldcore{\mathworld{G}}}$.
	 \label{item:truth-fun-def-counter-model-winning-strategy}
  \end{enumerate}
 \end{definition}

 \begin{lemma}
  \label[lemma]{lemma:world-last}
  Assume that $\Refuter$ has a winning strategy $\mathwinningstrategy{S}_{\Refuter}$ 
  in $\GameName{\Gamma\fCenter\Delta}$. 
  Let $\mathworld{W}$ be a world in $\mathstgraph{\mathwinningstrategy{S}_{\Refuter}}$.

  \begin{enumerate}
   \item If $\psi_{0}\to\psi_{1}\in\mathAnt{\mathof{\mathwinningstrategy{S}_{\Refuter}}{\mathworldcore{\mathworld{W}}}}$, 
	 then either 
	 $\psi_{1}\in\mathAnt{\mathof{\mathwinningstrategy{S}_{\Refuter}}{\mathworldcore{\mathworld{W}}}}$ or 
	 $\psi_{0}\in\mathCon{\mathof{\mathwinningstrategy{S}_{\Refuter}}{\mathworldcore{\mathworld{W}}}}$.
	 \label{item:to-L-lemma-world-last}
   \item If $\mathdynbox{\mathprogramseq{\pi_{0}}{\pi_{1}}}\varphi\in\mathAnt{\mathof{\mathwinningstrategy{S}_{\Refuter}}{\mathworldcore{\mathworld{W}}}}$, 
	 then 
	 $\mathdynbox{\pi_{0}}\mathdynbox{\pi_{1}}\varphi\in\mathAnt{\mathof{\mathwinningstrategy{S}_{\Refuter}}{\mathworldcore{\mathworld{W}}}}$.
	 \label{item:seq-L-lemma-world-last}
   \item If $\mathdynbox{\mathprogramndc{\pi_{0}}{\pi_{1}}}\varphi\in\mathAnt{\mathof{\mathwinningstrategy{S}_{\Refuter}}{\mathworldcore{\mathworld{W}}}}$, 
	 then
	 $\mathdynbox{\pi_{0}}\varphi\in\mathAnt{\mathof{\mathwinningstrategy{S}_{\Refuter}}{\mathworldcore{\mathworld{W}}}}$ and
	 $\mathdynbox{\pi_{1}}\varphi\in\mathAnt{\mathof{\mathwinningstrategy{S}_{\Refuter}}{\mathworldcore{\mathworld{W}}}}$.
	 \label{item:ndc-L-lemma-world-last}
   \item If $\mathdynbox{\mathprogramit{\pi}}\varphi\in\mathAnt{\mathof{\mathwinningstrategy{S}_{\Refuter}}{\mathworldcore{\mathworld{W}}}}$,
	 then 
	 $\varphi\in\mathAnt{\mathof{\mathwinningstrategy{S}_{\Refuter}}{\mathworldcore{\mathworld{W}}}}$ and
	 $\mathdynbox{\pi}\mathdynbox{\mathprogramit{\pi}}\varphi\in\mathAnt{\mathof{\mathwinningstrategy{S}_{\Refuter}}{\mathworldcore{\mathworld{W}}}}$.
	 \label{item:it-L-lemma-world-last}
   \item If $\mathdynbox{\mathprogramtest{\psi_{0}}}\psi_{1}\in\mathAnt{\mathof{\mathwinningstrategy{S}_{\Refuter}}{\mathworldcore{\mathworld{W}}}}$, 
	 then either 
	 $\psi_{1}\in\mathAnt{\mathof{\mathwinningstrategy{S}_{\Refuter}}{\mathworldcore{\mathworld{W}}}}$ or
	 $\psi_{0}\in\mathCon{\mathof{\mathwinningstrategy{S}_{\Refuter}}{\mathworldcore{\mathworld{W}}}}$.
	 \label{item:test-L-lemma-world-last}
  \end{enumerate}
 \end{lemma}

 \begin{proof}
  Straightforward.
 \end{proof}

 \begin{lemma} 
  \label[lemma]{lemma:track-refuter}
  Assume that $\Refuter$ has a winning strategy $\mathwinningstrategy{S}_{\Refuter}$ 
  in the proof search game of $\Gamma\fCenter\Delta$. 
  Let $\mathworld{W}$ be a world in $\mathstgraph{\mathwinningstrategy{S}_{\Refuter}}$ and
  $\mathworld{W}=\mathsequence{\sigma_{i}}{0\leq i}$.

  \begin{enumerate}
   \item If $\mathtuple{\psi_{0}\to\psi_{1}, n}\in\mathTr{\mathof{\mathwinningstrategy{S}_{\Refuter}}{\sigma_{i}}}$, 
	 then there exists $j>i$ such that $\sigma_{j}\mathancestorofcore\mathworld{W}$,
     	 $\psi_{0}\in\mathAnt{\mathof{\mathwinningstrategy{S}_{\Refuter}}{\sigma_{j}}}$
	 and 
	 $\psi_{1}\in\mathCon{\mathof{\mathwinningstrategy{S}_{\Refuter}}{\sigma_{j}}}$.
     	 \label{item:to-R-lemma-track-refuter}
   \item For $\alpha\in\maththesetofatomicprog$,
	 if $\mathtuple{\mathdynbox{\alpha}\varphi, n}\in\mathTr{\mathof{\mathwinningstrategy{S}_{\Refuter}}{\sigma_{i}}}$,
	 then $\mathtuple{\mathdynbox{\alpha}\varphi, n}\in\mathTr{\mathof{\mathwinningstrategy{S}_{\Refuter}}{\sigma_{j}}}$ 
	 for all $j\geq i$.
	 \label{item:atomic-lemma-track-refuter}
   \item If $\mathtuple{\mathdynbox{\mathprogramseq{\pi_{0}}{\pi_{1}}}\varphi, n}\in\mathTr{\mathof{\mathwinningstrategy{S}_{\Refuter}}{\sigma_{i}}}$,
	 then there exists $j>i$ such that $\sigma_{j}\mathancestorofcore\mathworld{W}$
	 and
	 $\mathtuple{\mathdynbox{\pi_{0}}\mathdynbox{\pi_{1}}\varphi, n}\in\mathTr{\mathof{\mathwinningstrategy{S}_{\Refuter}}{\sigma_{j}}}$.
	 \label{item:seq-R-lemma-track-refuter}
   \item If $\mathtuple{\mathdynbox{\mathprogramndc{\pi_{0}}{\pi_{1}}}\varphi, n}\in\mathTr{\mathof{\mathwinningstrategy{S}_{\Refuter}}{\sigma_{i}}}$,
	 then there exists $j>i$ such that $\sigma_{j}\mathancestorofcore\mathworld{W}$ 
	 and
	 $\mathtuple{\mathdynbox{\pi_{l}}\varphi, n}\in\mathTr{\mathof{\mathwinningstrategy{S}_{\Refuter}}{\sigma_{j}}}$ 
	 for some $l=0, 1$.
	 \label{item:ndc-R-lemma-track-refuter}
   \item If $\mathtuple{\mathdynbox{\mathprogramit{\pi}}\varphi, n}\in\mathTr{\mathof{\mathwinningstrategy{S}_{\Refuter}}{\sigma_{i}}}$,
	 then there exists $j>i$ such that $\sigma_{j}\mathancestorofcore\mathworld{W}$
	 and either
	 $\mathtuple{\varphi, n}\in\mathTr{\mathof{\mathwinningstrategy{S}_{\Refuter}}{\sigma_{j}}}$ or 
	 $\mathtuple{\mathdynbox{\pi}\mathdynbox{\mathprogramit{\pi}}\varphi, n}\in\mathTr{\mathof{\mathwinningstrategy{S}_{\Refuter}}{\sigma_{j}}}$.
	 \label{item:it-R-lemma-track-refuter}
   \item If $\mathtuple{\mathdynbox{\mathprogramtest{\psi}}\varphi, n}\in\mathTr{\mathof{\mathwinningstrategy{S}_{\Refuter}}{\sigma_{i}}}$,
	 then there exists $j>i$ such that $\sigma_{j}\mathancestorofcore\mathworld{W}$,
     	 $\psi\in\mathAnt{\mathof{\mathwinningstrategy{S}_{\Refuter}}{\sigma_{j}}}$ and
	 $\mathtuple{\varphi, n}\in\mathTr{\mathof{\mathwinningstrategy{S}_{\Refuter}}{\sigma_{j}}}$.
	 \label{item:test-R-lemma-track-refuter}
  \end{enumerate}
 \end{lemma}

 \begin{proof}
  Straightforward
 \end{proof}

 \begin{lemma}
  \label[lemma]{lemma:counter-model-from-strategy}
  Assume that $\Refuter$ has a winning strategy $\mathwinningstrategy{S}_{\Refuter}$
  in $\GameName{\Gamma\fCenter\Delta}$.
  Let 
  $M=\mathtuple{W, \mathsequence{{\mathreduction{\alpha}}}{\alpha\in\maththesetofatomicprog}, V}$
  be the counter model constructed from $\mathwinningstrategy{S}_{\Refuter}$.
  Let $\lambda$ be a well-formed expression such that
  $\lambda\in{\mathFLcl{\Gamma\cup\Delta}}$ or
  $\mathdynbox{\lambda}\varphi\in{\mathFLcl{\Gamma\cup\Delta}}$ 
  with some formula $\varphi$. The following statements hold:
  \begin{enumerate}
   \item For ${\mathworld{W}}\in W$,
	 if $\lambda\in\mathAnt{\mathof{\mathwinningstrategy{S}_{\Refuter}}{\mathworldcore{\mathworld{W}}}}$, 
	 then $M, \mathworld{W}\models\lambda$.
	 \label{item:antecedent-lemma-counter-model-from-game}
   \item For ${\mathworld{W}}\in W$ and $\sigma\mathancestorofcore\mathworld{W}$,
	 if $\mathtuple{\lambda, n}\in\mathTr{\mathof{\mathwinningstrategy{S}_{\Refuter}}{\sigma}}$ with some $n$, 
	 then $M, \mathworld{W}\not\models\lambda$.
	 \label{item:consequent-lemma-counter-model-from-game}
   \item For ${\mathworld{W}}\in W$,
	 if $\mathdynbox{\lambda}\varphi\in\mathAnt{\mathof{\mathwinningstrategy{S}_{\Refuter}}{\mathworldcore{\mathworld{W}}}}$, 
	 then $\varphi\in\mathAnt{\mathof{\mathwinningstrategy{S}_{\Refuter}}{\mathworldcore{\mathworld{W'}}}}$
	 for any ${\mathworld{W'}}\in W$, where
	 ${\mathworld{W}}{\mathreduction{\lambda}}{\mathworld{W'}}$.
	 \label{item:dynbox-ant-lemma-counter-model-from-game}
   \item For ${\mathworld{W}}\in W$ and $\sigma\mathancestorofcore\mathworld{W}$,
	 if 
	 $\mathtuple{\mathdynbox{\lambda}\varphi, n}\in\mathTr{\mathof{\mathwinningstrategy{S}_{\Refuter}}{\sigma}}$ 
	 with some $n$, 
	 then there exists ${\mathworld{W'}}\in W$ 
	 such that ${\mathworld{W}}{\mathreduction{\lambda}}{\mathworld{W'}}$ and
	 there exists a path $\mathsequence{\sigma_{i}}{0\leq i\leq m}$ such that
	 the following conditions hold:
	 \label{item:dynbox-con-lemma-counter-model-from-game}
	 \begin{enumerate}
	  \item $\sigma_{0}=\sigma$ and $\sigma_{m}\mathancestorofcore\mathworld{W'}$.
		\label{item:path-item-dynbox-con-lemma-counter-model-from-game}
	  \item Either $\mathtuple{\varphi, 0}\in\mathTr{\mathof{\mathwinningstrategy{S}_{\Refuter}}{\sigma_{m}}}$
		or $\mathtuple{\varphi, n}\in\mathTr{\mathof{\mathwinningstrategy{S}_{\Refuter}}{\sigma_{m}}}$.
		\label{item:track-item-dynbox-con-lemma-counter-model-from-game}
	  \item If $n\neq 0$ holds and 
		$\mathtuple{\mathof{\mathwinningstrategy{S}_{\Refuter}}{\sigma_{j}}, \mathof{\mathwinningstrategy{S}_{\Refuter}}{\sigma_{j+1}}}$
		is an instance of \rulename{G-K} for $0\leq j\leq m$
		with $\sigma_{j}=\mathworldcore{\mathworld{W}}$, 
	 	then $\Prover$ never changes the track $0$ in $\mathsequence{\sigma_{i}}{j+1\leq i\leq m}$ and $\mathtuple{\varphi, 0}\in\mathTr{\mathof{\mathwinningstrategy{S}_{\Refuter}}{\sigma_{m}}}$;
		Otherwise,
		$\Prover$ never changes the track $0$ in $\mathsequence{\sigma_{i}}{0\leq i\leq m}$ and 
		$\mathtuple{\varphi, n}\in\mathTr{\mathof{\mathwinningstrategy{S}_{\Refuter}}{\sigma_{m}}}$.
		\label{item:path-and-track-item-dynbox-con-lemma-counter-model-from-game}
	 \end{enumerate}
  \end{enumerate}
 \end{lemma}

 \begin{proof}
  Assume that $\Refuter$ has a winning strategy $\mathwinningstrategy{S}_{\Refuter}$
  in $\GameName{\Gamma\fCenter\Delta}$.
  Let 
  $M=\mathtuple{W, \mathsequence{{\mathreduction{\alpha}}}{\alpha\in\maththesetofatomicprog}, V}$
  be the counter model constructed from $\mathwinningstrategy{S}_{\Refuter}$.
  Let $\lambda$ be a well-formed expression such that
  $\lambda\in{\mathFLcl{\Gamma\cup\Delta}}$ or
  $\mathdynbox{\lambda}\varphi\in{\mathFLcl{\Gamma\cup\Delta}}$ 
  with some formula $\varphi$. 
  We show the statements by induction on construction of $\lambda$.
  
  Consider the case $\lambda\equiv\bot$.
  In this case, there is nothing to prove for 
  \cref{item:dynbox-ant-lemma-counter-model-from-game} and
  \cref{item:dynbox-con-lemma-counter-model-from-game}.

  \noindent \cref{item:antecedent-lemma-counter-model-from-game}
  Let ${\mathworld{W}}\in W$.
  Since $\mathwinningstrategy{S}_{\Refuter}$ is a winning strategy of $\Refuter$,
  we have $\lambda\notin\mathAnt{\mathof{\mathwinningstrategy{S}_{\Refuter}}{\mathworldcore{\mathworld{W}}}}$.

  \noindent \cref{item:consequent-lemma-counter-model-from-game}
  Immediately.

  Consider the case $\lambda\equiv p$ for $p\in\maththesetofpropval$.
  In this case, there is nothing to prove for
  \cref{item:dynbox-ant-lemma-counter-model-from-game} and
  \cref{item:dynbox-con-lemma-counter-model-from-game}.

  \noindent \cref{item:antecedent-lemma-counter-model-from-game}
  Let ${\mathworld{W}}\in W$.
  Assume $\lambda\in\mathAnt{\mathof{\mathwinningstrategy{S}_{\Refuter}}{\mathworldcore{\mathworld{W}}}}$.
  By \cref{def:counter-model-winning-strategy} 
  \cref{item:truth-fun-def-counter-model-winning-strategy},
  $p\in\mathof{V}{\mathworld{W}}$.
  Hence, $M, {\mathworld{W}}\models\lambda$.

  \noindent \cref{item:consequent-lemma-counter-model-from-game}
  Let ${\mathworld{W}}\in W$ and $\sigma\mathancestorofcore\mathworld{W}$. Assume 
  $\mathtuple{\lambda, n}\in\mathTr{\mathof{\mathwinningstrategy{S}_{\Refuter}}{\sigma}}$.
  Then, $p\in\mathCon{\mathof{\mathwinningstrategy{S}_{\Refuter}}{\sigma}}$.
  Hence,
  $p\in\mathCon{\mathof{\mathwinningstrategy{S}_{\Refuter}}{\mathworldcore{\mathworld{W}}}}$.
  Since $\mathwinningstrategy{S}_{\Refuter}$ is a winning strategy of $\Refuter$,
  we have $p\notin\mathAnt{\mathof{\mathwinningstrategy{S}_{\Refuter}}{\mathworldcore{\mathworld{W}}}}$.
  By \cref{def:counter-model-winning-strategy} 
  \cref{item:truth-fun-def-counter-model-winning-strategy},
  $p\notin\mathof{V}{\mathworld{W}}$.
  Hence, $M, {\mathworld{W}}\not\models\lambda$.

  Consider the case $\lambda\equiv \psi_{0}\to\psi_{1}$. 
  In this case, there is nothing to prove for 
  \cref{item:dynbox-ant-lemma-counter-model-from-game} and
  \cref{item:dynbox-con-lemma-counter-model-from-game}.

  \noindent \cref{item:antecedent-lemma-counter-model-from-game}
  Let ${\mathworld{W}}\in W$.
  Assume $\lambda\in\mathAnt{\mathof{\mathwinningstrategy{S}_{\Refuter}}{\mathworldcore{\mathworld{W}}}}$.
  From \cref{lemma:world-last} \cref{item:to-L-lemma-world-last},
  $\psi_{1}\in\mathAnt{\mathof{\mathwinningstrategy{S}_{\Refuter}}{\mathworldcore{\mathworld{W}}}}$ or 
  $\psi_{0}\in\mathCon{\mathof{\mathwinningstrategy{S}_{\Refuter}}{\mathworldcore{\mathworld{W}}}}$.
  If $\psi_{0}\in\mathCon{\mathof{\mathwinningstrategy{S}_{\Refuter}}{\mathworldcore{\mathworld{W}}}}$,
  \cref{lemma:core-property} \cref{item:core-and-con-lemma-core-property} implies that
  there exists $\sigma\mathancestorofcore\mathworld{W}$ such that
  $\mathtuple{\psi_{0}, n}\in\mathTr{\mathof{\mathwinningstrategy{S}_{\Refuter}}{\sigma}}$ 
  with some $n$.
  By the induction hypothesis \cref{item:antecedent-lemma-counter-model-from-game} and
  \cref{item:consequent-lemma-counter-model-from-game}, either
  $M, {\mathworld{W}}\not\models\psi_{0}$ or $M, {\mathworld{W}}\models\psi_{1}$.
  Hence, $M, {\mathworld{W}}\models\lambda$.  

  \noindent \cref{item:consequent-lemma-counter-model-from-game}
  Let ${\mathworld{W}}\in W$ and $\sigma\mathancestorofcore\mathworld{W}$.
  Assume 
  $\mathtuple{\lambda, n}\in\mathTr{\mathof{\mathwinningstrategy{S}_{\Refuter}}{\sigma}}$.
  From \cref{lemma:track-refuter} \cref{item:to-R-lemma-track-refuter},
  $\psi_{0}\in\mathAnt{\mathof{\mathwinningstrategy{S}_{\Refuter}}{\sigma'}}$ and $\mathtuple{\psi_{1}, n}\in\mathTr{\mathof{\mathwinningstrategy{S}_{\Refuter}}{\sigma'}}$
  for some $\sigma'\in\mathworld{W}$.
  Because $\psi_{0}\in\mathAnt{\mathof{\mathwinningstrategy{S}_{\Refuter}}{\sigma'}}$
  for some $\sigma'\in\mathworld{W}$,
  we have $\psi_{0}\in\mathAnt{\mathworldcore{\mathworld{W}}}$.
  By the induction hypothesis \cref{item:consequent-lemma-counter-model-from-game,item:antecedent-lemma-counter-model-from-game}, 
  $M, {\mathworld{W}}\models\psi_{0}$ and $M, {\mathworld{W}}\not\models\psi_{1}$.
  Hence, $M, {\mathworld{W}}\not\models\lambda$. 

  Consider the case $\lambda\equiv \mathdynbox{\pi}\psi$.
  In this case, there is nothing to prove for 
  \cref{item:dynbox-ant-lemma-counter-model-from-game} and
  \cref{item:dynbox-con-lemma-counter-model-from-game}.

  \noindent \cref{item:antecedent-lemma-counter-model-from-game}
  Let ${\mathworld{W}}\in W$.
  Assume $\lambda\in\mathAnt{\mathof{\mathwinningstrategy{S}_{\Refuter}}{\mathworldcore{\mathworld{W}}}}$.
  By the induction hypothesis \cref{item:dynbox-ant-lemma-counter-model-from-game},
  $\psi\in\mathAnt{\mathof{\mathwinningstrategy{S}_{\Refuter}}{\mathworldcore{\mathworld{W'}}}}$
  for any $\mathworld{W'}\in W$, where 
  ${\mathworld{W}}{\mathreduction{\pi}}{\mathworld{W'}}$.
  From the induction hypothesis \cref{item:antecedent-lemma-counter-model-from-game},
  $M, {\mathworld{W'}}\models\psi$ for any $\mathworld{W'}\in W$, where
  ${\mathworld{W}}{\mathreduction{\pi}}{\mathworld{W'}}$.
  Therefore, $M, {\mathworld{W}}\models\lambda$.  

  \noindent \cref{item:consequent-lemma-counter-model-from-game}
  Let ${\mathworld{W}}\in W$ and $\sigma\mathancestorofcore\mathworld{W}$. Assume 
  $\mathtuple{\lambda, n}\in\mathTr{\mathof{\mathwinningstrategy{S}_{\Refuter}}{\sigma}}$.
  By the induction hypothesis \cref{item:dynbox-con-lemma-counter-model-from-game},
  there exist ${\mathworld{W'}}\in W$ and $\sigma'\mathancestorofcore\mathworld{W'}$
  such that ${\mathworld{W}}{\mathreduction{\lambda}}{\mathworld{W'}}$ and
  $\mathtuple{\psi, m}\in\mathTr{\mathof{\mathwinningstrategy{S}_{\Refuter}}{\sigma'}}$
  for some $m$.
  From the induction hypothesis \cref{item:consequent-lemma-counter-model-from-game},
  $M, {\mathworld{W'}}\not\models\psi$.
  Therefore, $M, {\mathworld{W}}\not\models\lambda$. 

  Consider the case $\lambda\equiv\alpha$ for $\alpha\in\maththesetofatomicprog$.
  In this case, there is nothing to prove for 
  \cref{item:antecedent-lemma-counter-model-from-game} and
  \cref{item:consequent-lemma-counter-model-from-game}.

  \noindent \cref{item:dynbox-ant-lemma-counter-model-from-game} 
  Assume $\mathdynbox{\lambda}\varphi\in\mathAnt{\mathof{\mathwinningstrategy{S}_{\Refuter}}{\mathworldcore{\mathworld{W}}}}$
  and ${\mathworld{W}}{\mathreduction{\lambda}}{\mathworld{W'}}$.
  Because of ${\mathworld{W}}{\mathreduction{\alpha}}{\mathworld{W'}}$, 
  \cref{def:counter-model-winning-strategy} \cref{item:reduction-def-counter-model-winning-strategy}
  implies that
  $\mathtuple{\mathof{\mathwinningstrategy{S}_{\Refuter}}{\mathworldcore{\mathworld{W}}}, \mathof{\mathwinningstrategy{S}_{\Refuter}}{\mathworldfirst{\mathworld{W'}}}}$
  is an instance of \rulename{G-K}, where
  $\mathof{\mathwinningstrategy{S}_{\Refuter}}{\mathworldcore{\mathworld{W}}}=\mathtuple{\Pi\fCenter\Sigma; T; S}$,
  $\mathof{\mathwinningstrategy{S}_{\Refuter}}{\mathworldfirst{\mathworld{W'}}}=\mathtuple{\Pi' \fCenter \varphi; \mathsetextension{\mathtuple{\varphi, 0}}; \mathreducible{\mathFLcl{\Gamma\cup\Delta}}}$,
  $\Pi'=\mathsetintension{\psi}{\mathdynbox{\alpha}\psi\in\Pi}$,
  and $\mathtuple{\mathdynbox{\alpha}\varphi, n}\in T$.
  Because of 
  $\mathdynbox{\lambda}\varphi\in\mathAnt{\mathof{\mathwinningstrategy{S}_{\Refuter}}{\mathworldcore{\mathworld{W}}}}$,
  we have $\varphi\in\Pi'$.
  Therefore,
  $\varphi\in\mathAnt{\mathof{\mathwinningstrategy{S}_{\Refuter}}{\mathworldcore{\mathworld{W'}}}}$.
  
  \noindent \cref{item:dynbox-con-lemma-counter-model-from-game} 
  Let ${\mathworld{W}}\in W$ and $\sigma\mathancestorofcore\mathworld{W}$.
  Assume 
  $\mathtuple{\mathdynbox{\lambda}\varphi, n}\in\mathTr{\mathof{\mathwinningstrategy{S}_{\Refuter}}{\sigma}}$ 
  with some $n$.
  By \cref{lemma:track-refuter} \cref{item:atomic-lemma-track-refuter},
  $\mathtuple{\mathdynbox{\alpha}\varphi, n}\in\mathTr{\mathof{\mathwinningstrategy{S}_{\Refuter}}{\mathworldcore{\mathworld{W}}}}$.
  Hence, there exists a child $\sigma'$ of $\mathworldcore{\mathworld{W}}$
  such that   $\mathtuple{\mathof{\mathwinningstrategy{S}_{\Refuter}}{\mathworldcore{\mathworld{W}}}, \mathof{\mathwinningstrategy{S}_{\Refuter}}{\sigma'}}$
  is an instance of \rulename{G-K}, and
  $\mathof{\mathwinningstrategy{S}_{\Refuter}}{\sigma'}=\mathtuple{\Pi' \fCenter \varphi; \mathsetextension{\mathtuple{\varphi, 0}}; \mathreducible{\mathFLcl{\Gamma\cup\Delta}}}$, 
  where $\Pi'=\mathsetintension{\psi}{\mathdynbox{\alpha}\psi\in\Pi}$.
  Let ${\mathworld{W'}}\in W$ such that $\sigma'=\mathworldfirst{\mathworld{W'}}$.
  Then, $\sigma'\mathancestorofcore\mathworld{W'}$.
  By \cref{def:counter-model-winning-strategy} \cref{item:reduction-def-counter-model-winning-strategy},
  we have ${\mathworld{W}}{\mathreduction{\alpha}}{\mathworld{W'}}$.
  Obviously, $\mathtuple{\varphi, 0}\in\mathTr{\mathof{\mathwinningstrategy{S}_{\Refuter}}{\sigma'}}$
  and a path with only one node $\sigma'$ satisfies that
  $\Prover$ never changes the track $0$ in the path.
  Hence, \cref{item:path-item-dynbox-con-lemma-counter-model-from-game,item:track-item-dynbox-con-lemma-counter-model-from-game,item:path-and-track-item-dynbox-con-lemma-counter-model-from-game} hold.

  Consider the case $\lambda\equiv\mathprogramseq{\pi_{0}}{\pi_{1}}$.
  In this case, there is nothing to prove for 
  \cref{item:antecedent-lemma-counter-model-from-game} and
  \cref{item:consequent-lemma-counter-model-from-game}.

  \noindent \cref{item:dynbox-ant-lemma-counter-model-from-game} 
  Assume $\mathdynbox{\lambda}\varphi\in\mathAnt{\mathof{\mathwinningstrategy{S}_{\Refuter}}{\mathworldcore{\mathworld{W}}}}$
  and ${\mathworld{W}}{\mathreduction{\lambda}}{\mathworld{W'}}$.
  Because of 
  ${\mathworld{W}}{\mathreduction{\mathprogramseq{\pi_{0}}{\pi_{1}}}}{\mathworld{W'}}$,
  there exists ${\mathworld{W''}}\in W$ such that
  ${\mathworld{W}}{\mathreduction{\pi_{0}}}{\mathworld{W''}}$ and
  ${\mathworld{W''}}{\mathreduction{\pi_{1}}}{\mathworld{W'}}$.
  By \cref{lemma:world-last} \cref{item:seq-L-lemma-world-last},
  $\mathdynbox{\pi_{0}}\mathdynbox{\pi_{1}}\varphi\in\mathAnt{\mathworldcore{\mathworld{W}}}$.
  From the induction hypothesis \cref{item:dynbox-ant-lemma-counter-model-from-game},
  $\mathdynbox{\pi_{1}}\varphi\in\mathAnt{\mathworldcore{\mathworld{W''}}}$.
  Then, the induction hypothesis \cref{item:dynbox-ant-lemma-counter-model-from-game} 
  implies $\varphi\in\mathAnt{\mathworldcore{\mathworld{W'}}}$.

  \noindent \cref{item:dynbox-con-lemma-counter-model-from-game} 
  Let ${\mathworld{W}}\in W$ and $\sigma\mathancestorofcore\mathworld{W}$.
  Assume that 
  $\mathtuple{\mathdynbox{\lambda}\varphi, n}\in\mathTr{\mathof{\mathwinningstrategy{S}_{\Refuter}}{\sigma}}$ 
  with some $n$.
  Let $\mathworld{W}=\mathsequence{\sigma^{\mathworld{W}}_{i}}{0\leq i}$ and $\sigma^{\mathworld{W}}_{i_{0}}=\sigma$.
  By \cref{lemma:track-refuter} \cref{item:seq-R-lemma-track-refuter},
  there exists $j_{0}>i_{0}$ such that $\sigma^{\mathworld{W}}_{j_{0}}\mathancestorofcore\mathworld{W}$ and
  $\mathtuple{\mathdynbox{\pi_{0}}\mathdynbox{\pi_{1}}\varphi, n}\in\mathTr{\mathof{\mathwinningstrategy{S}_{\Refuter}}{\sigma^{\mathworld{W}}_{j_{0}}}}$.
  From the induction hypothesis \cref{item:dynbox-con-lemma-counter-model-from-game},
  there exists ${\mathworld{W''}}\in W$ 
  such that ${\mathworld{W}}{\mathreduction{\pi_{0}}}{\mathworld{W''}}$ and
  there exists a path $\mathsequence{\sigma''_{i'}}{0\leq i'\leq m''}$ such that
  the following conditions hold:
  \begin{itemize}
   \item $\sigma''_{0}=\sigma^{\mathworld{W}}_{j_{0}}$ and $\sigma''_{m''}\mathancestorofcore\mathworld{W''}$.
   \item Either $\mathtuple{\mathdynbox{\pi_{1}}\varphi, 0}\in\mathTr{\mathof{\mathwinningstrategy{S}_{\Refuter}}{\sigma''_{m''}}}$
	 or $\mathtuple{\mathdynbox{\pi_{1}}\varphi, n}\in\mathTr{\mathof{\mathwinningstrategy{S}_{\Refuter}}{\sigma''_{m''}}}$.
   \item If $n\neq 0$ holds and 
	 $\mathtuple{\mathof{\mathwinningstrategy{S}_{\Refuter}}{\sigma''_{k}}, \mathof{\mathwinningstrategy{S}_{\Refuter}}{\sigma''_{k+1}}}$
	 is an instance of \rulename{G-K} for $0\leq k\leq m''$
	 with $\sigma_{k}=\mathworldcore{\mathworld{W}}$, 
	 then $\Prover$ never changes the track $0$ in $\mathsequence{\sigma''_{i'}}{k+1\leq i'\leq m''}$ and $\mathtuple{\mathdynbox{\pi_{1}}\varphi, 0}\in\mathTr{\mathof{\mathwinningstrategy{S}_{\Refuter}}{\sigma''_{m''}}}$;
	 Otherwise,
	 $\Prover$ never changes the track $0$ in $\mathsequence{\sigma''_{i'}}{0\leq i'\leq m''}$ and 
	 $\mathtuple{\mathdynbox{\pi_{1}}\varphi, n}\in\mathTr{\mathof{\mathwinningstrategy{S}_{\Refuter}}{\sigma''_{m''}}}$.
  \end{itemize}
  Because $\mathtuple{\mathdynbox{\pi_{1}}\varphi, n'}\in\mathTr{\mathof{\mathwinningstrategy{S}_{\Refuter}}{\sigma''_{m''}}}$ with some $n'$,
  the induction hypothesis \cref{item:dynbox-con-lemma-counter-model-from-game} implies
  that there exists ${\mathworld{W'}}\in W$ 
  such that ${\mathworld{W''}}{\mathreduction{\pi_{1}}}{\mathworld{W'}}$ and
  there exists a path $\mathsequence{\sigma'''_{i''}}{0\leq i''\leq m'''}$ such that
  the following conditions hold:
  \begin{itemize}
   \item $\sigma'''_{0}=\sigma''_{m''}$ and $\sigma'''_{m'''}\mathancestorofcore\mathworld{W'}$.
   \item Either $\mathtuple{\varphi, 0}\in\mathTr{\mathof{\mathwinningstrategy{S}_{\Refuter}}{\sigma'''_{m'''}}}$
	 or $\mathtuple{\varphi, n'}\in\mathTr{\mathof{\mathwinningstrategy{S}_{\Refuter}}{\sigma'''_{m'''}}}$.
   \item If $n'\neq 0$ holds and 
	 $\mathtuple{\mathof{\mathwinningstrategy{S}_{\Refuter}}{\sigma'''_{k'}}, \mathof{\mathwinningstrategy{S}_{\Refuter}}{\sigma'''_{k'+1}}}$
	 is an instance of \rulename{G-K} for $0\leq k'\leq m'''$
	 with $\sigma'''_{k'}=\mathworldcore{\mathworld{W''}}$, 
	 then $\Prover$ never changes the track $0$ in $\mathsequence{\sigma'''_{i''}}{k'+1\leq i''\leq m'''}$ and $\mathtuple{\varphi, 0}\in\mathTr{\mathof{\mathwinningstrategy{S}_{\Refuter}}{\sigma'''_{m'''}}}$;
	 Otherwise,
	 $\Prover$ never changes the track $0$ in $\mathsequence{\sigma'''_{i''}}{0\leq i''\leq m'''}$ and 
	 $\mathtuple{\varphi, n'}\in\mathTr{\mathof{\mathwinningstrategy{S}_{\Refuter}}{\sigma'''_{m'''}}}$.
  \end{itemize}
  We define $\sigma_{0}=\sigma$, $\dots$, $\sigma_{j_{0}-i_{0}}=\sigma^{\mathworld{W}}_{j_{0}}=\sigma''_{0}$,
  $\dots$, $\sigma_{j_{0}-i_{0}+m''}=\sigma''_{m''}=\sigma'''_{0}$, $\dots$, $\sigma_{j_{0}-i_{0}+m''+m'''}=\sigma'''_{m'''}$.
  Let $m={j_{0}-i_{0}+m''+m'''}$. Then, $\mathsequence{\sigma_{i}}{0\leq i\leq m}$ is 
  the sequence which we want.

  Consider the case $\lambda\equiv\mathprogramndc{\pi_{0}}{\pi_{1}}$.
  In this case, there is nothing to prove for 
  \cref{item:antecedent-lemma-counter-model-from-game} and
  \cref{item:consequent-lemma-counter-model-from-game}.

  \noindent \cref{item:dynbox-ant-lemma-counter-model-from-game} 
  Assume $\mathdynbox{\lambda}\varphi\in\mathAnt{\mathworldcore{\mathworld{W}}}$ and
  ${\mathworld{W}}{\mathreduction{\lambda}}{\mathworld{W'}}$.
  Because of 
  ${\mathworld{W}}{\mathreduction{\mathprogramndc{\pi_{0}}{\pi_{1}}}}{\mathworld{W'}}$,
  we have ${\mathworld{W}}{\mathreduction{\pi_{i}}}{\mathworld{W'}}$ 
  for some $i\in\mathsetextension{0, 1}$.
  By \cref{lemma:world-last} \cref{item:ndc-L-lemma-world-last},
  $\mathdynbox{\pi_{0}}\varphi\in\mathAnt{\mathworldcore{\mathworld{W}}}$
  and 
  $\mathdynbox{\pi_{1}}\varphi\in\mathAnt{\mathworldcore{\mathworld{W}}}$.
  From the induction hypothesis \cref{item:dynbox-ant-lemma-counter-model-from-game},
  $\varphi\in\mathAnt{\mathworldcore{\mathworld{W''}}}$
  for all ${\mathworld{W''}}\in W$
  with ${\mathworld{W}}{\mathreduction{\pi_{0}}}{\mathworld{W''}}$ or
  ${\mathworld{W}}{\mathreduction{\pi_{1}}}{\mathworld{W''}}$.
  Therefore, $\varphi\in\mathAnt{\mathworldcore{\mathworld{W'}}}$
  for ${\mathworld{W'}}\in W$.

  \noindent \cref{item:dynbox-con-lemma-counter-model-from-game} 
  Let ${\mathworld{W}}\in W$ and $\sigma\mathancestorofcore\mathworld{W}$.
  Assume that 
  $\mathtuple{\mathdynbox{\lambda}\varphi, n}\in\mathTr{\mathof{\mathwinningstrategy{S}_{\Refuter}}{\sigma}}$ 
  with some $n$.
  Let $\mathworld{W}=\mathsequence{\sigma^{\mathworld{W}}_{i}}{0\leq i}$ and $\sigma^{\mathworld{W}}_{i_{0}}=\sigma$.
  By \cref{lemma:track-refuter} \cref{item:ndc-R-lemma-track-refuter},
  there exists $j_{0}>i_{0}$ such that $\sigma^{\mathworld{W}}_{j_{0}}\mathancestorofcore\mathworld{W}$ and
  $\mathtuple{\mathdynbox{\pi_{l}}\varphi, n}\in\mathTr{\mathof{\mathwinningstrategy{S}_{\Refuter}}{\sigma_{j_{0}}}}$ 
  for some $l=0, 1$.
  From the induction hypothesis \cref{item:dynbox-con-lemma-counter-model-from-game},
  there exists ${\mathworld{W'}}\in W$ 
  such that ${\mathworld{W}}{\mathreduction{\pi_{l}}}{\mathworld{W'}}$ and
  there exists a path $\mathsequence{\sigma'_{i'}}{0\leq i'\leq m'}$ such that
  the following conditions hold:
  \begin{itemize}
   \item $\sigma'_{0}=\sigma^{\mathworld{W}}_{j}$ and $\sigma'_{m'}\mathancestorofcore\mathworld{W'}$.
   \item Either $\mathtuple{\varphi, 0}\in\mathTr{\mathof{\mathwinningstrategy{S}_{\Refuter}}{\sigma'_{m'}}}$
	 or $\mathtuple{\varphi, n}\in\mathTr{\mathof{\mathwinningstrategy{S}_{\Refuter}}{\sigma'_{m'}}}$.
   \item If $n\neq 0$ holds and 
	 $\mathtuple{\mathof{\mathwinningstrategy{S}_{\Refuter}}{\sigma'_{j'}}, \mathof{\mathwinningstrategy{S}_{\Refuter}}{\sigma_{j'+1}}}$
	 is an instance of \rulename{G-K} for $0\leq j'\leq m'$
	 with $\sigma_{j'}=\mathworldcore{\mathworld{W}}$, 
	 then $\Prover$ never changes the track $0$ in $\mathsequence{\sigma'_{i'}}{j'+1\leq i'\leq m'}$ and $\mathtuple{\varphi, 0}\in\mathTr{\mathof{\mathwinningstrategy{S}_{\Refuter}}{\sigma'_{m'}}}$;
	 Otherwise,
	 $\Prover$ never changes the track $0$ in $\mathsequence{\sigma'_{i'}}{0\leq i'\leq m'}$ and 
	 $\mathtuple{\varphi, n}\in\mathTr{\mathof{\mathwinningstrategy{S}_{\Refuter}}{\sigma'_{m'}}}$.
  \end{itemize}
  We define $\sigma_{0}=\sigma$, $\dots$, $\sigma_{j_{0}-i_{0}}=\sigma^{\mathworld{W}}_{j_{0}}=\sigma'_{0}$,
  $\dots$, $\sigma_{j_{0}-i_{0}+m'}=\sigma'_{m'}$.
  Let $m={j_{0}-i_{0}+m'}$. Then,  $\mathsequence{\sigma_{i}}{0\leq i\leq m}$ is 
  the sequence which we want.

  Consider the case $\lambda\equiv\mathprogramit{\pi}$.
  In this case, there is nothing to prove for 
  \cref{item:antecedent-lemma-counter-model-from-game} and
  \cref{item:consequent-lemma-counter-model-from-game}.

  \noindent \cref{item:dynbox-ant-lemma-counter-model-from-game} 
  Assume $\mathdynbox{\lambda}\varphi\in\mathAnt{\mathworldcore{\mathworld{W}}}$ and
  ${\mathworld{W}}{\mathreduction{\lambda}}{\mathworld{W'}}$.
  Because of ${\mathworld{W}}{\mathreduction{\mathprogramit{\pi}}}{\mathworld{W'}}$,
  there exists $n$ such that ${\mathworld{W}}{\mathreduction{\pi}^{n}}{\mathworld{W'}}$.
  We show $\varphi\in\mathAnt{\mathworldcore{\mathworld{W'}}}$ by induction on $n$.

  Consider the case $n=0$. In this case, $\mathworld{W'}=\mathworld{W}$.
  By \cref{lemma:world-last} \cref{item:it-L-lemma-world-last},
  $\varphi\in\mathAnt{\mathworldcore{\mathworld{W}}}$.
  Hence, $\varphi\in\mathAnt{\mathworldcore{\mathworld{W'}}}$.

  Consider the case $n>0$. In this case, there exists $\mathworld{W''}\in W$ such that
  ${\mathworld{W}}{\mathreduction{\pi}}{\mathworld{W''}}{\mathreduction{\pi}^{n-1}}{\mathworld{W'}}$.
  \cref{lemma:world-last} \cref{item:it-L-lemma-world-last} implies
  $\mathdynbox{\pi}\mathdynbox{\mathprogramit{\pi}}\varphi\in\mathAnt{\mathworldcore{\mathworld{W}}}$.
  From the induction hypothesis \cref{item:dynbox-ant-lemma-counter-model-from-game} on $\pi$,
  $\mathdynbox{\mathprogramit{\pi}}\varphi\in\mathAnt{\mathworldcore{\mathworld{W''}}}$.
  By the induction hypothesis on $n$,
  we have $\varphi\in\mathAnt{\mathworldcore{\mathworld{W'}}}$.

  \noindent \cref{item:dynbox-con-lemma-counter-model-from-game} 
  Let ${\mathworld{W}}\in W$ and $\sigma\mathancestorofcore\mathworld{W}$.
  Assume that 
  $\mathtuple{\mathdynbox{\lambda}\varphi, n}\in\mathTr{\mathof{\mathwinningstrategy{S}_{\Refuter}}{\sigma}}$ 
  with some $n$.
  Let $\mathworld{W}=\mathsequence{\sigma^{\mathworld{W}}_{i}}{0\leq i}$ and $\sigma^{\mathworld{W}}_{i_{0}}=\sigma$.
  Assume, for contradiction,
  the following statement holds for all ${\mathworld{W'}}\in W$, 
  where ${\mathworld{W}}{\mathreduction{\mathprogramit{\pi}}}{\mathworld{W'}}$:
  \ifoup \vspace{-1.0em} \fi
  \begin{quote}
   For all path $\mathsequence{\sigma_{i}}{0\leq i\leq m}$, neither
   $\mathtuple{\varphi, 0}\in\mathTr{\mathof{\mathwinningstrategy{S}_{\Refuter}}{\sigma_{m}}}$
   nor
   $\mathtuple{\varphi, n}\in\mathTr{\mathof{\mathwinningstrategy{S}_{\Refuter}}{\sigma_{m}}}$
   holds if the following condition holds:
   \begin{itemize}
    \item $\sigma_{0}=\sigma$ and $\sigma_{m}\mathancestorofcore\mathworld{W'}$ hold.
    \item If $n\neq 0$ and 
	  $\mathtuple{\mathof{\mathwinningstrategy{S}_{\Refuter}}{\sigma_{j}}, \mathof{\mathwinningstrategy{S}_{\Refuter}}{\sigma_{j+1}}}$
	  is an instance of \rulename{G-K} for $0\leq j\leq m$
	  with $\sigma_{j}=\mathworldcore{\mathworld{W}}$, 
	  then $\Prover$ never changes the track $0$ in $\mathsequence{\sigma_{i}}{j+1\leq i\leq m}$;
	  Otherwise,
	  $\Prover$ never changes the track $0$ in $\mathsequence{\sigma_{i}}{0\leq i\leq m}$.
   \end{itemize}
  \end{quote}

  \renewcommand*{\theenumi}{\textup{(}\roman{enumi}\textup{)}}
  \renewcommand*{\labelenumi}{\textup{(}\roman{enumi}\textup{)}}
  We show that there exists a sequence of worlds 
  $\mathsequence{\mathworld{W}_{k}}{0\leq k}$ and a sequence of finite paths 
  $\mathsequence{\mathsequence{\sigma^{(k)}_{i}}{0\leq i\leq m_{k}}}{0\leq k}$
  such that the following conditions hold:
  \begin{enumerate}
   \item ${\sigma^{(0)}_{0}}=\sigma$, ${\sigma^{(k+1)}_{0}}=\sigma^{(k)}_{m_{k}}$, and
	 ${\sigma^{(k)}_{0}}\mathancestorofcore\mathworld{W}_{k}$.
	 \label{item:node-proof-lemma-counter-model-from-game}
   \item ${\mathworld{W}_{k}}{\mathreduction{\pi}}{\mathworld{W}_{k+1}}$
	 for all $k\geq 0$.
	 \label{item:world-proof-lemma-counter-model-from-game}
   \item $\mathtuple{\mathdynbox{\mathprogramit{\pi}}\varphi, n_{k}}\in\mathTr{\mathof{\mathwinningstrategy{S}_{\Refuter}}{\sigma^{(k)}_{m_{k}}}}$
	 with either $n_{k}=0$ or $n_{k}=n$.
	 \label{item:track-proof-lemma-counter-model-from-game}
   \item If $n_{k}\neq 0$ and 
	 $\mathtuple{\mathof{\mathwinningstrategy{S}_{\Refuter}}{\sigma^{(k)}_{j_{k}}}, \mathof{\mathwinningstrategy{S}_{\Refuter}}{\sigma^{(k)}_{j_{k}+1}}}$
	 is an instance of \rulename{G-K} for $0\leq j_{k}\leq m_{k}$
	 with $\sigma_{j_{k}}=\mathworldcore{\mathworld{W}_{k}}$, 
	 then $\Prover$ never changes the track $0$ in $\mathsequence{\sigma^{(k)}_{i}}{j_{k}+1\leq i\leq m_{k}}$
	 and $\mathtuple{\mathdynbox{\mathprogramit{\pi}}\varphi, 0}\in\mathTr{\mathof{\mathwinningstrategy{S}_{\Refuter}}{\sigma^{(k)}_{m_{k}}}}$;
	  Otherwise,
	  $\Prover$ never changes the track $0$ in $\mathsequence{\sigma^{(k)}_{i}}{0\leq i\leq m_{k}}$ 
	 and $\mathtuple{\mathdynbox{\mathprogramit{\pi}}\varphi, n}\in\mathTr{\mathof{\mathwinningstrategy{S}_{\Refuter}}{\sigma_{m_{k}}}}$.
	 \label{item:never-change-proof-lemma-counter-model-from-game}
  \end{enumerate}

  \renewcommand*{\theenumi}{\textup{(}\arabic{enumi}\textup{)}}
  \renewcommand*{\labelenumi}{\textup{(}\arabic{enumi}\textup{)}}
  We inductively construct a sequence of worlds 
  $\mathsequence{\mathworld{W}_{k}}{0\leq k}$ and a sequence of finite paths 
  $\mathsequence{\mathsequence{\sigma^{(k)}_{i}}{0\leq i\leq m_{k}}}{0\leq k}$
  such that \crefrange{item:node-proof-lemma-counter-model-from-game}{item:never-change-proof-lemma-counter-model-from-game} hold.

  Consider the case $k=0$.
  Let $\mathworld{W}_{0}=\mathworld{W}$.
  By \cref{lemma:track-refuter} \cref{item:it-R-lemma-track-refuter},
  there exists $j_{0}>i_{0}$ such that $\sigma^{\mathworld{W}}_{j_{0}}\mathancestorofcore\mathworld{W}$
  and either
  $\mathtuple{\varphi, n}\in\mathTr{\mathof{\mathwinningstrategy{S}_{\Refuter}}{\sigma^{\mathworld{W}}_{j_{0}}}}$ or 
  $\mathtuple{\mathdynbox{\pi}\mathdynbox{\mathprogramit{\pi}}\varphi, n}\in\mathTr{\mathof{\mathwinningstrategy{S}_{\Refuter}}{\sigma^{\mathworld{W}}_{j_{0}}}}$.
  Because of ${\mathworld{W}}{\mathreduction{\mathprogramit{\pi}}}{\mathworld{W}}$ and
  the assumption for contradiction,
  $\mathtuple{\varphi, n}\notin\mathTr{\mathof{\mathwinningstrategy{S}_{\Refuter}}{\sigma^{\mathworld{W}}_{i}}}$ 
  for all $i\geq 0$.
  Hence, $\mathtuple{\mathdynbox{\pi}\mathdynbox{\mathprogramit{\pi}}\varphi, n}\in\mathTr{\mathof{\mathwinningstrategy{S}_{\Refuter}}{\sigma^{\mathworld{W}}_{j_{0}}}}$.
  From the induction hypothesis \cref{item:dynbox-con-lemma-counter-model-from-game},
  there exists ${\mathworld{W}_{1}}\in W$ 
  such that ${\mathworld{W}_{0}}{\mathreduction{\pi}}{\mathworld{W}_{1}}$ and
  there exists a path $\mathsequence{\sigma'_{i'}}{0\leq i'\leq m'}$ such that
  the following conditions hold:
  \begin{itemize}
   \item $\sigma'_{0}=\sigma^{\mathworld{W}}_{j_{0}}$ and $\sigma'_{m'}\mathancestorofcore\mathworld{W}_{1}$.
   \item $\mathtuple{\mathdynbox{\mathprogramit{\pi}}\varphi, n_{0}}\in\mathTr{\mathof{\mathwinningstrategy{S}_{\Refuter}}{\sigma'_{m'}}}$
	 with either $n_{0}=0$ or $n_{0}=n$.
   \item If $n\neq 0$ and 
	 $\mathtuple{\mathof{\mathwinningstrategy{S}_{\Refuter}}{\sigma'_{j'}}, \mathof{\mathwinningstrategy{S}_{\Refuter}}{\sigma'_{j'+1}}}$
	 is an instance of \rulename{G-K} for $0\leq j'\leq m'$
	 with $\sigma_{j'}=\mathworldcore{\mathworld{W}_{0}}$, 
	 then $\Prover$ never changes the track $0$ in $\mathsequence{\sigma'_{i'}}{j'+1\leq i'\leq m'}$
	 and $\mathtuple{\mathdynbox{\mathprogramit{\pi}}\varphi, 0}\in\mathTr{\mathof{\mathwinningstrategy{S}_{\Refuter}}{\sigma'_{m'}}}$;
	  Otherwise,
	  $\Prover$ never changes the track $0$ in $\mathsequence{\sigma'_{i}}{0\leq i'\leq m'}$ 
	 and $\mathtuple{\mathdynbox{\mathprogramit{\pi}}\varphi, n}\in\mathTr{\mathof{\mathwinningstrategy{S}_{\Refuter}}{\sigma'_{m'}}}$.
  \end{itemize}
  Define $\sigma^{(0)}_{0}=\sigma$, $\dots$, $\sigma^{(0)}_{j_{0}-i_{0}}=\sigma^{\mathworld{W}}_{j_{0}}=\sigma'_{0}$, $\dots$, $\sigma^{(0)}_{j_{0}-i_{0}+m'}=\sigma'_{m'}$.
  Let $m_{0}=j_{0}-i_{0}+m'$.

  Consider the case $k>0$. 
  By \cref{item:track-proof-lemma-counter-model-from-game},
  $\mathtuple{\mathdynbox{\mathprogramit{\pi}}\varphi, n_{k-1}}\in\mathTr{\mathof{\mathwinningstrategy{S}_{\Refuter}}{\sigma^{(k-1)}_{m_{k-1}}}}$
  with either $n_{k-1}=0$ or $n_{k-1}=n$.
  Let ${\sigma^{(k)}_{0}}={\sigma^{(k-1)}_{m_{k-1}}}$ and
  $\mathworld{W}_{k}$ be a world such that 
  ${\mathworld{W}_{k-1}}{\mathreduction{\pi}}{\mathworld{W}_{k}}$ and
  $\sigma^{(k)}_{0}\mathancestorofcore\mathworld{W}_{k}$.
  Let $\mathworld{W}_{k}=\mathsequence{\sigma^{\mathworld{W}_{k}}_{i}}{0\leq i}$ and
  ${\sigma^{\mathworld{W}_{k}}_{i_{k}}}={\sigma^{(k)}_{0}}$.
  By \cref{lemma:track-refuter} \cref{item:it-R-lemma-track-refuter},
  there exists $j_{k}>i_{k}$ such that $\sigma^{\mathworld{W}_{k}}_{j_{k}}\mathancestorofcore\mathworld{W}_{k}$
  and either
  $\mathtuple{\varphi, n_{k-1}}\in\mathTr{\mathof{\mathwinningstrategy{S}_{\Refuter}}{\sigma^{\mathworld{W}_{k}}_{j_{k}}}}$ or 
  $\mathtuple{\mathdynbox{\pi}\mathdynbox{\mathprogramit{\pi}}\varphi, n_{k-1}}\in\mathTr{\mathof{\mathwinningstrategy{S}_{\Refuter}}{\sigma^{\mathworld{W}_{k}}_{j_{k}}}}$.
  Because of ${\mathworld{W}}{\mathreduction{\mathprogramit{\pi}}}{\mathworld{W}_{k}}$ and
  the assumption for contradiction, we have
  $\mathtuple{\varphi, n_{k-1}}\notin\mathTr{\mathof{\mathwinningstrategy{S}_{\Refuter}}{\sigma^{\mathworld{W}_{k}}_{i}}}$ 
  for all $i\geq 0$.
  Hence, $\mathtuple{\mathdynbox{\pi}\mathdynbox{\mathprogramit{\pi}}\varphi, n_{k-1}}\in\mathTr{\mathof{\mathwinningstrategy{S}_{\Refuter}}{\sigma^{\mathworld{W}_{k}}_{j_{k}}}}$.
  From the induction hypothesis \cref{item:dynbox-con-lemma-counter-model-from-game},
  there exists ${\mathworld{W}_{k+1}}\in W$ 
  such that ${\mathworld{W}_{k}}{\mathreduction{\pi}}{\mathworld{W}_{k+1}}$ and
  there exists a path $\mathsequence{\sigma''_{i''}}{0\leq i''\leq m''}$ such that
  the following conditions hold:
  \begin{itemize}
   \item $\sigma''_{0}=\sigma^{\mathworld{W}_{k}}_{j_{k}}$ and $\sigma''_{m''}\mathancestorofcore\mathworld{W}_{k+1}$.
   \item $\mathtuple{\mathdynbox{\mathprogramit{\pi}}\varphi, n_{k}}\in\mathTr{\mathof{\mathwinningstrategy{S}_{\Refuter}}{\sigma''_{m''}}}$
	 with either $n_{k}=0$ or $n_{k}=n_{k-1}$.
   \item If $n_{k-1}\neq 0$ and 
	 $\mathtuple{\mathof{\mathwinningstrategy{S}_{\Refuter}}{\sigma''_{j''}}, \mathof{\mathwinningstrategy{S}_{\Refuter}}{\sigma''_{j''+1}}}$
	 is an instance of \rulename{G-K} for $0\leq j''\leq m''$
	 with $\sigma_{j''}=\mathworldcore{\mathworld{W}_{k}}$, 
	 then $\Prover$ never changes the track $0$ in $\mathsequence{\sigma''_{i''}}{j''+1\leq i''\leq m''}$
	 and $\mathtuple{\mathdynbox{\mathprogramit{\pi}}\varphi, 0}\in\mathTr{\mathof{\mathwinningstrategy{S}_{\Refuter}}{\sigma''_{m''}}}$;
	  Otherwise,
	  $\Prover$ never changes the track $0$ in $\mathsequence{\sigma''_{i}}{0\leq i''\leq m''}$ 
	 and $\mathtuple{\mathdynbox{\mathprogramit{\pi}}\varphi, n_{k-1}}\in\mathTr{\mathof{\mathwinningstrategy{S}_{\Refuter}}{\sigma''_{m''}}}$.
  \end{itemize}
  Define ${\sigma^{(k)}_{1}}={\sigma^{\mathworld{W}_{k}}_{i_{k}+1}}$, $\dots$, $\sigma^{(k)}_{j_{k}-i_{k}}=\sigma^{\mathworld{W}_{k}}_{j_{k}}=\sigma''_{0}$, $\dots$, $\sigma^{(k)}_{j_{k}-i_{k}+m''}=\sigma''_{m''}$.
  Let $m_{k}=j_{k}-i_{k}+m''$.

  By the construction,
  $\mathsequence{\mathworld{W}_{k}}{0\leq k}$ and a sequence of finite paths 
  $\mathsequence{\mathsequence{\sigma^{(k)}_{i}}{0\leq i\leq m_{k}}}{0\leq k}$ satisfies
  \crefrange{item:node-proof-lemma-counter-model-from-game}{item:never-change-proof-lemma-counter-model-from-game}.
  Let $\sigma_{i}=\sigma^{(k)}_{j}$ for $i=m_{0}+\dots +m_{k-1}+j$.
  Then, $\mathsequence{\sigma_{i}}{0\leq i}$ is an infinite path in which 
  transforming pairs occur infinitely many times.
  
  If $\mathtuple{\mathof{\mathwinningstrategy{S}_{\Refuter}}{\sigma_{j}}, \mathof{\mathwinningstrategy{S}_{\Refuter}}{\sigma_{j+1}}}$
  is an instance of \rulename{G-K} for $0\leq j$,
  then the track $0$ is continuous and
  $\Prover$ never changes the track $0$ in $\mathsequence{\sigma_{i}}{j\leq i}$;
  Otherwise, the track $n$ is continuous and
  $\Prover$ never changes the track $0$ in $\mathsequence{\sigma_{i}}{0\leq i}$.
  In both cases, we see that there exists an infinite path such that
  transforming pairs occur infinitely many times, but
  the track $n'$ is continuous for some $n'$ and
  $\Prover$ never changes the track $0$ in the path.
  This contradicts \cref{lemma:refuter-infinite}.
  Thus, we have \cref{item:dynbox-con-lemma-counter-model-from-game}.
  
  Consider the case $\lambda\equiv\mathprogramtest{\psi}$.
  In this case, there is nothing to prove for 
  \cref{item:antecedent-lemma-counter-model-from-game} and
  \cref{item:consequent-lemma-counter-model-from-game}.

  \noindent \cref{item:dynbox-ant-lemma-counter-model-from-game} 
  Assume $\mathdynbox{\lambda}\varphi\in\mathAnt{\mathworldcore{\mathworld{W}}}$ and
  ${\mathworld{W}}{\mathreduction{\lambda}}{\mathworld{W'}}$.
  Because of ${\mathworld{W}}{\mathreduction{\lambda}}{\mathworld{W'}}$,
  we have ${\mathworld{W'}}={\mathworld{W}}$ and
  $M, \mathworld{W} \models \psi$.
  By \cref{lemma:world-last} \cref{item:test-L-lemma-world-last},
  either $\varphi\in\mathAnt{\mathworldcore{\mathworld{W}}}$ or 
  $\psi\in\mathCon{\mathworldcore{\mathworld{W}}}$.
  If $\psi\in\mathCon{\mathof{\mathwinningstrategy{S}_{\Refuter}}{\mathworldcore{\mathworld{W}}}}$,
  \cref{lemma:core-property} \cref{item:core-and-con-lemma-core-property} implies that
  there exists $\sigma\mathancestorofcore\mathworld{W}$ such that
  $\mathtuple{\psi_{0}, n}\in\mathTr{\mathof{\mathwinningstrategy{S}_{\Refuter}}{\sigma}}$
  with some $n$.
  By the induction hypothesis \cref{item:consequent-lemma-counter-model-from-game},
  implies $M, \mathworld{W} \not\models \psi$.
  Because we have $M, \mathworld{W} \models \psi$,
  we see $\psi\notin\mathCon{\mathof{\mathwinningstrategy{S}_{\Refuter}}{\mathworldcore{\mathworld{W}}}}$.
  Hence, $\varphi\in\mathAnt{\mathworldcore{\mathworld{W}}}$.

  \noindent \cref{item:dynbox-con-lemma-counter-model-from-game} 
  Let ${\mathworld{W}}\in W$ and $\sigma\mathancestorofcore\mathworld{W}$.
  Assume that 
  $\mathtuple{\mathdynbox{\lambda}\varphi, n}\in\mathTr{\mathof{\mathwinningstrategy{S}_{\Refuter}}{\sigma}}$ 
  with some $n$.
  Let $\mathworld{W}=\mathsequence{\sigma^{\mathworld{W}}_{i'}}{0\leq i'}$ and $\sigma^{\mathworld{W}}_{i_{0}}=\sigma$.
  By \cref{lemma:track-refuter} \cref{item:test-R-lemma-track-refuter},
  there exists $j_{0}>i_{0}$ such that $\sigma_{j_{0}}\mathancestorofcore\mathworld{W}$ 
  holds, and both
  $\psi\in\mathAnt{\mathof{\mathwinningstrategy{S}_{\Refuter}}{\sigma_{j_{0}}}}$ and
  $\mathtuple{\varphi, n}\in\mathTr{\mathof{\mathwinningstrategy{S}_{\Refuter}}{\sigma_{j_{0}}}}$ hold.
  Because of $\psi\in\mathAnt{\mathof{\mathwinningstrategy{S}_{\Refuter}}{\sigma_{j_{0}}}}$,
  we have $\psi\in\mathAnt{\mathworldcore{\mathworld{W}}}$.
  By the induction hypothesis \cref{item:antecedent-lemma-counter-model-from-game},
  implies $M, \mathworld{W} \models \psi$.
  Hence, ${\mathworld{W}}{\mathreduction{\lambda}}{\mathworld{W}}$.
  We define $\sigma_{0}=\sigma$, $\dots$, $\sigma_{j_{0}-i_{0}}=\sigma^{\mathworld{W}}_{j_{0}}$.
  Let $m={j_{0}-i_{0}}$. Then, $\mathsequence{\sigma_{i}}{0\leq i\leq m}$ is 
  the sequence which we want.
 \end{proof}

  We have all the lemmata needed to show \cref{lemma:Refuter-win}.

 \begin{proof}[Proof of \cref{lemma:Refuter-win}.]
  Assume that
  $\Refuter$ has a winning strategy $\mathwinningstrategy{S}_{\Refuter}$
  in the proof search game of $\Gamma\fCenter\Delta$. 
  Let 
  $M=\mathtuple{W, \mathsequence{{\mathreduction{\alpha}}}{\alpha\in\maththesetofatomicprog}, V}$
  be the counter model constructed from $\mathwinningstrategy{S}_{\Refuter}$.
  Let $\mathworld{W}\in W$ such that $\mathworldfirst{\mathworld{W}}$ is
  the root of $\mathwinningstrategy{S}_{\Refuter}$.
  By \cref{lemma:counter-model-from-strategy},
  $M, \mathworld{W}\models\varphi$ for all $\varphi\in\Gamma$ and 
  $M, \mathworld{W}\not\models\psi$ for all $\psi\in\Delta$.
  Hence, $\Gamma\fCenter\Delta$ is invalid.
 \end{proof}

 \subsection{The proof of \cref{theorem:cut-free-completeness-CGPDL}}
 Now, we show \cref{theorem:cut-free-completeness-CGPDL}.

 \begin{proof}[Proof of \cref{theorem:cut-free-completeness-CGPDL}.]
  We show the contrapositive of \cref{theorem:cut-free-completeness-CGPDL}.

  Assume that 
  a cut-free \CGPDL\ proof of $\Gamma\fCenter\Delta$ does not exist.
  By \cref{lemma:Prover-win},
  $\Prover$ does not have a winning strategy in the proof search game of 
  $\Gamma\fCenter\Delta$.
  From \cref{cor:determinacy},
  $\Refuter$ has a winning strategy in the proof search game of $\Gamma\fCenter\Delta$.
  \cref{lemma:Refuter-win} implies that $\Gamma\fCenter\Delta$ is invalid.
 \end{proof}

\section{Conclusions}
\label{sec:conc}
We have introduced a sequent calculus, \GTPDL, and a cyclic proof system, \CGTPDL, 
for a variation of dynamic logic, \TPDL, and have proved the soundness and completeness of 
both systems.We have also shown that cut-elimination fails in \GTPDL\ and \CGTPDL.
We provide two sequents for counterexamples to cut-elimination in \GTPDL. 
One is not Fischer-Ladner-cut provable in \GTPDL\ but cut-free provable in \CGTPDL, 
while the other is not cut-free provable in either. 
We have defined a cyclic proof system \CGPDL\ for \PDL\ as a fragment of \CGTPDL, and
have proved its cut-free completeness and cut-elimination theorem
by considering a \emph{finite game}.
Since the cut-elimination property of \CGPDL\ holds,
the occurrence of a backwards modal operator is necessary in the counterexample to 
cut-elimination in \CGTPDL. 

Someone may think that we can derive the interpolation theorem of \PDL\ 
from the cut-elimination theorem in \CGPDL, as in \cite{Shamkanov2014}.
However, in our opinion, it is overly optimistic to expect, 
since the L\"{o}b rule is essential for the proof of \cite{Shamkanov2014} 
but does not hold in \PDL.
We need a new technique if we prove the interpolation theorem of \PDL\ from the cut-elimination theorem in \CGPDL.

\section*{Acknowledgements}
We would like to thank 
Koji Nakazawa, Daisuke Kimura, 
Takahiro Sawasaki,
Kenji Saotome, 
Taro Hiraoka, and
Takumi Sato 
for their valuable comments.

\bibliography{refs}

\end{document}